\documentclass[prx,twocolumn,showpacs,preprintnumbers,amsmath,amssymb,notitlepage]{revtex4-1}
\usepackage{graphicx}
\usepackage{ulem}
\UseRawInputEncoding

\begin{document}

\title{ Emergent space-time meets emergent quantum phenomena: observing a quantum phase transition in a moving sample  }
\author{ Fadi Sun$^{1}$ and Jinwu Ye$^{1,2}$   }
\affiliation{ $^{1}$ Institute for Theoretical Sciences, Westlake University, Hangzhou, 310024, Zhejiang, China  \\
$^{2}$  Department of Physics and Astronomy, Mississippi State University, MS, 39762, USA     }
\date{\today }


\begin{abstract}
 In material science, it was established that as the number of particles $ N $ in a material gets more and more,
 especially in the thermodynamic limit, various macroscopic quantum phenomena such as superconductivity, superfluidity, quantum magnetism, Fractional quantum Hall effects  and various quantum or topological phase transitions (QPT)
 emerge in such non-relativistic quantum many-body systems. There is always a reservoir which exchanges energy and particles with the material.
 This is the essence of P. W. Anderson's great insight `` More is different ''.
 However, there is still a fundamental component missing in this general picture:
 How the `` More is different '' becomes different in a moving inertial frame or a  moving sample ?
 Here we address this outstanding problem. 
 We propose there is an emergent space-time corresponding to any emergent quantum phenomenon, especially near a QPT.
 We demonstrate our claims by studying one of the simplest QPTs:
 Superfluid (SF)-Mott transitions of interacting bosons in a square lattice in a
 sample moving with a constant velocity $ \vec{v} $.
 We first elaborate the crucial difference between a moving sample and a moving inertial frame,  stressing the crucial roles played by a reservoir  in a grand canonical ensemble which is needed to study the SF-Mott transition in the first place.
 In this work, we mainly present the moving sample case and only discuss very briefly the moving inertial frame case.
 It is the moving which mixes the space and time. We also stress the important roles played by the underlying lattice. 
 Practically, we first construct two effective actions in a moving sample
 to study the SF-Mott QPT with the dynamic exponent $ z=1 $ and $ z=2 $ respectively, then explore them by
 applying various methods such as mean field analysis, field theory renormalization group, charge-vortex duality and scaling analysis.
 Then by putting the low velocity limit of the Lorentz transformation to teh order of $ (v/c_l)^2 $
 ( $ c_l $ is the speed of light ) in a lattice and projecting to the tight-binding limit,  we derive the two effective actions
 from a microscopic ionic lattice model, therefore establish the relations between the order parameter and the phenomenological parameters in the two effective actions and those in the boson Hubbard model and the velocity $ \vec{v} $.
 Armed by these relations, we map out the global phase diagram of the boson-Hubbard model in the moving sample.
 We find that the new emergent space-time structure leads to many new effects in the moving sample such as the change of the ground state (
 the Mott phase near the QPT turns into a SF phase, but not the other way around ),
 the change of the condensation momentum, the sign reverse of the Doppler shift in the excitation spectrum relative to the bare velocity $ v $,
 the emergence of new class of QPTs, the increase of the Kosterlize-Thouless (KT) transition temperature $ T_{KT} $ in the SF side, etc.
  These new effects are contrasted to the Doppler shift and temperature shift in a relativistic thermal quantum field theory, Unruh
  effects in an accelerating observer  and possible emergent curved space-time from the microscopic Sachdev-Ye-Kitaev models.
  The methods can be extended to study all the quantum and topological phase transitions  in a moving sample in any dimension.
  Doing various light or neutron scattering measurements in a moving sample may become
  an effective way not only  measure various intrinsic
  properties of the materials, tune various quantum and topological phases through phase transitions, but also probe the new emergent space-time structure near any QPT.  As a byproduct, we comment on the emergent space-time in the fractional Quantum Hall systems and the associated chiral edge
  states in a moving Hall bar to the order of $ v/c_l $. 
\end{abstract}

\maketitle

\section{ Introduction}
\begin{figure}[tbhp]
\centering
\includegraphics[width=.8 \linewidth]{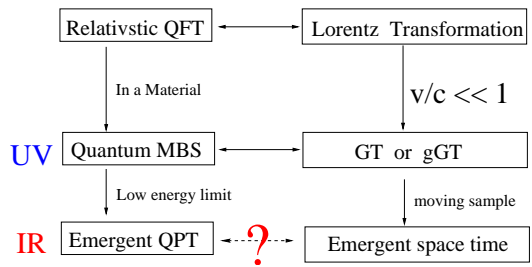}
\caption{ P. W. Anderson's " More is different "\cite{anderson} on emergent quantum phenomena need to be augmented
         by the corresponding emergent space-time denoted by the question mark to be  presented in this work.
          GT: Galileo transformation. gGT: generalized GT in the presence of an  external magnetic field ( see appendix A ).
          QFT: Quantum Field theory, QMS: Quantum many-body system ( microscopic level in the UV ).
          QPT: Quantum phase transitions ( effective level in the IR ).
          See also Fig.\ref{towerlevel}. One typical QPT is the SF-Mott QPT in a periodic lattice potential which is the main focus of
          the work. The GT in a moving sample in this case leads to an emergent space-time.
          Another is interacting electron system in an external magnetic field which may show FQH with topological order and associated edge states.
          The gGT in a moving Hall bar in this case leads to an emergent space-time to be studied in the appendix F-H.  }
\label{CMTmore}
\end{figure}

Poincare and Einstein's special theory of relativity tells us that there is no preferred inertial frame \cite{poincare}.
The Hamiltonian or Lagrangian takes identical form  and owns the same set of symmetries in all inertial frames.
The laws in two inertial frames are just related by the Lorentz transformation (LT) which lead to interesting phenomena such as
a moving stick becomes shorter, a running clock gets slower and relativistic Doppler effect, etc.
However, the story may change in materials
or atomic molecular and Optical (AMO) systems which explicitly break the Lorentz invariance.
There is indeed a preferred inertial frame where the substrate or a lattice holding the materials
or AMO system  is static. In this static frame, as advocated by P.W. Anderson " More is different "\cite{anderson},
one can observe various emergent quantum many body phenomena such as various symmetry breaking phases, topological phases and
Quantum or topological phase transitions (QPTs) between them.
The physical laws in two inertial frames are still related by the Lorentz transformation (LT) which
reduces to Galileo transformation (GT) at the lowest order in the expansion of the velocity over the speed of light. So the Hamiltonian or Lagrangian may take the largest symmetries in the static frame,
but take reduced symmetries in the moving sample.
How these emergent phenomena change when they are observed in the moving sample remains an outstanding open problem.
In this work, we will show that there is always an emergent space-time  corresponding to any
emergent quantum many body phenomena, especially near a QPT which is dramatically different than the bare space-time in a few particle cases ( Fig.\ref{CMTmore} ).
Our findings should
make up the missing component of P.W. Anderson's original great insight  " More is different " \cite{anderson}.

Quantum phase transitions (QPT) is one of the most fantastic phenomena in Nature \cite{sachdev,aue,wen}.
For example,  Superfluid to Mott transitions \cite{boson0,coldexp,bosonlattice,coldrev,z2,field2},
Anti-ferromagnetic state to Valence bond solid transition \cite{scaling,deconfined},
magnetic states to quantum spin liquid transition \cite{SLrev1,SLrev3,dimer1,dimer2,kit2,yao,yao2},
the magnetic transitions in itinerant systems \cite{Andy}
or quantum Hall (QH) to QH or insulator transitions \cite{CB,hlr,MSrev,field1,field3,blqh}, topological phase transitions of
non-interacting fermions or interacting systems \cite{kane,zhang,tenfold,wenrev}  are among the most popular QPTs.
In materials, one usually manipulates or controls a system by applying a magnetic field, electric field,
change doping, making a twist, adding a strain or pressure, to tune the system to go through various QPTs.
Ultra-cold atoms loaded on optical lattices can provide unprecedented experimental systems for the quantum simulations and manipulations
of some quantum phase and phase transitions. For example,  the Superfluid to Mott (SF-Mott) transition has  been successfully
realized by loading ultracold atoms in various optical lattice \cite{coldexp,bosonlattice,coldrev,rotation}.
However, due to the charge neutrality, it is difficult to manipulate the cold atom systems by applying a magnetic  or electric field
\cite{rotation}. Here, we study how the same QPT is observed in a moving sample and
show that it leads to new quantum phases through novel quantum phase transitions.
Most importantly, these new effects originate from the new space-time structure emerging from the QPT.
In this work, we focus on QPTs with symmetry breaking, topological phase transitions will be discussed in a separate
publication \cite{QAHinj}.

\begin{figure}[tbhp]
\centering
\includegraphics[width=.8 \linewidth]{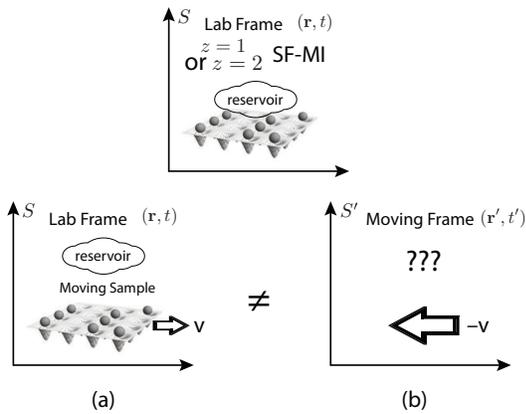}
\caption{ Top: The SF-Mott transition with the dynamic exponent $ z=1 $  or $ z=2 $ in Fig.\ref{phaseslattice0}
is realized  when the sample is static. There are two different experiments listed in the bottom.
(a) Moving the whole sample with a velocity $ \vec{v} $ including the optical lattice and the atoms loaded on it.
    This is the focus of this work presented in Fig.\ref{phaseslattice}.
    The effective phase diagrams for $ z=1 $ and $ z=2 $ in the moving sample is presented in Fig.\ref{phasesz1} and \ref{phasesz2} respectively.
    Some new effects are found to be: new quantum phases, the Doppler shifts in the excitation spectrum, novel QPTs and finite temperature phase transitions, etc.
   In a practical experimental scattering detection shown in Fig.\ref{detector},
   the sample is moving with a constant velocity, while the emitter and the receiver are static in the lab frame.
   This set-up may also be used to probe the new space-time structure emerging from the $ z=1 $  or $ z=2 $ QPT ( Fig.\ref{towerlevel} ).
(b) An observer moving with an opposite velocity $ - \vec{v} $.
    So the quantum phases and QPTs in the top is observed in a frame moving with a velocity $ -\vec{v} $
    with respect to the lab frame. The question marks in the moving sample show that it remains an open problem.
    It will be only briefly commented in the appendix L.
Note that if it were not due to the Reservoir shown explicitly in (a), (a) and (b) would be equivalent.
However, due to the existence of the reservoir in a grand canonical ensemble, there is a relative motion between the sample
and the reservoir in (a) but not in (b).
This subtle, but crucial difference makes our theory only apply to (a)
instead of (b). The case (b) will be briefly discussed in the last appendix.
In the following, we call the frame moving together with the sample as the co-moving frame. }
\label{frames}
\end{figure}

We take the boson-Hubbard (BH) model of interacting bosons at integer fillings in a square lattice as the simplest example
to show a QPT in a static sample Fig.\ref{frames}:
\begin{equation}
  H_{BH}  =  -t \sum_{ \langle ij \rangle } ( b^{\dagger}_{i} b_{j} + h.c. ) -
        \mu \sum_{i} n_{i} + \frac{U}{2} \sum_{i} n_{i} ( n_{i} -1 )
\label{boson}
\end{equation}
 which displays a SF-Mott transition around $ t/U \sim 1 $.
 Obviously, the very existence of the lattice breaks the Galileo and Lorentz invariance. It is also responsible
 for the existence of the Mott insulating state and the SF-Mott transition.
 Its known phase diagram \cite{boson0}  when the sample is static is shown in Fig.\ref{phaseslattice0}.
 Its new phase diagram in a moving sample with a given velocity $ v $  is found in this work to be in Fig.\ref{phaseslattice}.

Before starting, it is important to distinguish the two different cases (A) and (B) in Fig.\ref{frames}.
This is also a crucial new point when applying special relativity to a grand-Canonical ensemble.
Indeed. Einstein's original special relativity is about the relative motion between two point particles  (Fig. \ref{source} ),
NOT involving this kind of trinity of system+reservior combination, then the observer.
So despite the micro-canonical, canonical, grand-canonical ensemble make little differences in the thermodynamic limit
when the sample is static, they may make some observable differences in the moving sample in Fig.\ref{frames}a.
     The optical lattice is at rest in the lab frame, the ultra-cold atoms are loaded on top of it,
     but is observed in a sample ( Fig.\ref{frames}b,\ref{phasesz1}, \ref{phasesz2} and \ref{phaseslattice} ).
     ( the whole sample means  the lattice, the cold atoms and the trap )  moving along a track Fig.\ref{detector}.
    The status of the Mott and the  SF in the moving sample need to be determined by the effective action Eq.\ref{boostz1} and  Eq.\ref{boostz2}.
    We are particularly interested in how the Mott and SF near the QPT response to the boost, especially
    when it is beyond a critical velocity.
    We develop both microscopic calculations in the UV and the symmetry based effective actions in the IR to achieve such a goal which is summarized in Fig.\ref{phasesz1}, \ref{phasesz2} and Fig.\ref{phaseslattice}.

 To achieve Fig.\ref{phaseslattice}, we first take the effective action approach where we take the effective  boost
 velocity $ c $ as a phenomenological  input parameter.
 For $ z=1 $ case, when $ c $ is below a critical velocity, the Mott state and the SF state remain, but their
 excitation spectrum suffer a Doppler shift. However, in the SF side, when $ c$  increases above a critical velocity\cite{twovc} determined by $ v_y $,
 the SF becomes a boosted SF (BSF) carrying a finite momentum which spontaneously breaks the $ C $ symmetry ( also the $ CPT $ symmetry ), the transition from the SF to BSF  has an exotic, but exact dynamic exponent $ (z_x=3/2,3) $.
 There are both Doppler shifted Goldstone and Higgs mode in the SF phase, but only Doppler shifted Goldstone mode
 inside the BSF phase, no Higgs mode anymore due to the spontaneously broken $ C $ symmetry.
 In the Mott side, when $ c$  increases above a larger critical velocity than the one inside the SF \cite{twovc}, the Mott phase also turns into the BSF phase
 with the dynamic exponent $ z=2 $. We find a metric-crossing term which is
 a Type-II dangerously irrelevant operator (DIO) \cite{response} and leads to a Doppler shift term in both
 the Mott and the BSF near the $ z=2 $ transition line.
 We also evaluate  the conserved Noether currents in the three phases in both the lab frame and the co-moving frame
 from the  non-relativistic QFT with the higher-order derivatives.
 The BSF phase carries the conserved Noether currents due to the spontaneously broken $ C $ symmetry,
 but the Mott and SF phase do not. So the currents can be used as an order parameter to characterize the
 spontaneously breaking $ C $ symmetry  and also to distinguish BSF from the SF.
 By performing a field theory renormalization group (RG)
 developed for non-relativistic quantum field theory  up to two loops, we show that
 the QPT from the Mott to the SF in the moving sample is a
 new universality class. Despite it breaks the emergent Lorentz invariance explicitly,
 It still keeps  the charge conjugation ( C ) symmetry  which dictates $ z=1 $.
 We call such a new QPT a boosted 3D XY transition which differs from the conventional 3D XY due to its lack of Lorentz invariance.
 Charge-vortex duality transformation in the moving sample is also performed  to study the new QPT in the moving sample.
 Finite temperature properties of quantum,  classical ( $ \hbar \to 0 $ limit ) KT transition,  scaling functions of various physical quantities,
 especially their dependencies on the velocity $ c $ are explored.
 As a byproduct, by a quantum-classical correspondence, we show that the SF to the BSF transition with $ z=(3/2,3) $ may also describe the dynamic transition between the classical sound waves in a medium.
 For $ z=2 $ case when the sample is static, it has an emergent Galileo invariance. It still has the exact $ P $ and $ T $ symmetry, but
 the $ C $ symmetry was explicitly broken.  In the moving sample,
 the Mott to the SF with $ z=2 $ turns into the Mott to the BSF transition, still has the emergent GI with $ z=2 $, but with a shifted boundary favoring the BSF phase. Just from the symmetry principle, we identify one type-II dangerously irrelevant operator (DIO) which breaks the emergent GI, also the exact $ P $ and $ T $ symmetry, but keeps the $ PT $ symmetry, it leads to a Doppler shift term in the BSF phase.
 We also identify another type-II DIO which is nothing but the cross metric term.
 It also  breaks the emergent GI, the exact $ P $ and $ T $ symmetry, but keeps the $ PT $ symmetry and leads to a Doppler shift term
 in both the Mott phase and the BSF phase.

Then we put the GT in a lattice and derive the specific form of the boost in the tight-binding limit on a lattice.
By performing both non-perturbative and perturbative microscopic calculations on the boost terms,
we derive the effective actions with $ z= 1 $ and $ z=2 $ under the boost
and also establish the connections between the phenomenological parameters in the effective actions in the IR and the microscopic parameters in the boson Hubbard model on the lattice scale in the UV, especially the effective boost $ c $ in terms of the bare velocity $ v $.
These relations set-up the initial conditions for the Renormalization groups (RG) flow equations in the effective actions. We also identify
the relation between the original bosons on the lattice and the order parameters in the two effective actions.
By combining the results achieved from the effective actions in the IR and the microscopic calculations in UV in a lattice which are two complementary approaches,
we map out qualitatively the global phase diagram of the Boson Hubbard model in the moving sample.
Counter-intuitively, we find that due to the two steps shift of the Quantum critical point ( QCP ) to the Mott side,
a Mott insulating phase near the SF-Mott transition may become a BSF phase, but not the other way around.

By deriving the $ z=1 $ effective action in the moving sample, we determine the first step shift of the QCP to the Mott side.
we find a non-monotonic dependence of the effective boost $ c $ on the original bare velocity $ v $:
When $ v $ is below a critical velocity $ v_c $ which is solely determined  by the Wannier functions of the lattice system\cite{twovc}, the Doppler shift term is in the same direction as the original $ v $, vanishes at $ v=v_c $, but
changes sign and becomes opposite to the original $ v $ when $ v > v_c $.
Obviously, this surprising change of sign of the Doppler shift is due to the lattice effect.
We also find the microscopic values at the lattice scale of other phenomenological values such as the Goldstone velocity $ v_{y,z=1} $ and the tuning parameter $ r $. The former determines the KT transition temperature which is found to increase in the moving sample,
the latter determines the QPT boundary which is found to move to the Mott side.
Because $ |c | \ll v_{y,z=1} $ at the lattice scale and also any other scales by the RG analysis on the effective action\cite{twovc},
we conclude the BSF is not reachable in the moving sample, so does the
SF to the BSF transition with  the exact dynamic exponent $ (z_x=3/2,3) $.
However, in a separate work \cite{SSS}, one of the authors showed that it is reachable by directly driving the SF beyond a critical velocity.
For $ z=2 $, we find the microscopic values at the lattice scale of the phenomenological values such as
the effective mass $ v^2_{y,z=2}=1/2m $ and the effective chemical potential $ \mu $.
The former determines the KT transition temperature which is also found to increase in the moving sample.
The latter determines second shift of the QPT boundary to the Mott side favoring the BSF
which also has a non-monotonic dependence on the bare velocity $ v $.
Then we go further to extract the microscopic values at the lattice scale of the three leading type-II
dangerously irrelevant operators (DIO) identified from the general symmetry principle in the $ z=2 $ effective action.

 The combination of the effective actions and the microscopic calculations on the lattice scale lead to a rather complete understanding on the observation of the SF-Mott QPT when the sample is  moving with a given velocity $ v $ and
 culminated in the flagship Fig.\ref{phaseslattice}. Its main features are:
 After the two step's shifts of the QPT boundary, the Mott phase near the QPT in the Regime I and II turns into a SF phase,
 the change of the condensation momentum of $ k_{0L} $ in the first step  and $ k_{0L} + k_0 $ in the second step,
 the sign reverse of the Doppler shift in the excitation spectrum relative to the bare velocity $ v $ when $  v > v_c $,
 the new 3d boosted XY universality class at $ z=1 $, the increase of the KT transition temperature, etc.
 For a typical cold atom system such as $^{87}$Rb, we estimate the numerical value of $ v_c $ and find it is clearly reachable in the current cold atom systems. It tells that the shift of the QPT boundary, $ k_{0L} $,  the increase of the KT transition temperature $ T_c$
 are clearly within the current experimental reach. However, the Doppler shift  in the Mott and SF phase and $ k_0 $ are subleading effects
 which  are more challenging to measure
 ( However, they  are still much larger than the relativistic corrections which are at the order of $ (v/c)^2 $ ).
 But the fact that the Doppler shift reverse its sign at $ v=v_c $ still make the challenge not too hard to overcome
 in the current cold atom technology \cite{twovc}. By analyzing the applicabilities and limitations  of various current detection methods in a moving sample, we suggest that  doing measurements in a moving sample becomes an  effective way not only  measure various intrinsic and characteristic properties of the materials,
 but also to tune various quantum and topological phases and phase transitions,
 most importantly to probe the emergent space-time near a QPT.

 By analyzing the connections between the first quantization in the many-body wavefunctions
 and the effective actions in the second quantization on and off a lattice, also at different hierarchy energy levels, we demonstrate that
 new space-time structure emerge from a quantum or topological phase transition which leads to the above new effects in the moving sample.
 Of course, due to the WW's no-go theorem \cite{WW}, it can only be a emergent flat space-time within the same space-time dimension.
 Any possible emergent curved space-time can only be achieved  from an extra dimension such as $ AdS_{d+1}/CFT_d $ correspondence.
 Then we contrast these novel effects in a moving sample with some known effects in the relativistic QFT.
 Contrasts to the Doppler shifts in a relativistic quantum field theory, the temperature effect in a finite temperature relativistic QFT,  Unruh
 effects in an accelerating  observer and emergent AdS geometry from the boundary  at $ d=1,2 $ are made.
 The hierarchy of the emergency from the ionic lattice model in the bottom-up approach here is contrasted to that from
 the string theory to the standard model/general relativity in the up-bottom approach.
 Our systematic and complete methods can be extended to study all the quantum or topological phase transitions in any dimension.
 As a byproduct, from the GT and GI point of view, we comment on the two competing theories describing the gapless compressive state
 at $ \nu=1/2 $: the HLR theory versus Son's Dirac fermion theory. Then we explore the emergent space-time in
 the chiral edge state of some simple  Abelian Fractional Quantum Hall state.
 We conclude emergent particles in a lattice system transform very much differently than the elementary particles in particle physics.


The rest of the paper is organised as follows:
In the  sections II, we will discuss the three QPTs along the three path I,II,III in Fig.\ref{phasesz1} respectively.
Then by employing the field theory renormalization group developed to study non-relativistic QFT, we will
study the new universality class of the boosted Mott to SF transition along path I in Sec.III.
We will evaluate the conserved currents in all the three phases in Fig.\ref{phasesz1}, then
perform charge-vortex duality transformation in the moving sample along the path I in Sec.IV.
We derive the scaling functions of various physical quantities at a finite temperature  along the three paths in Sec.V
and also its classical limit $ \hbar \to 0 $ leading the KT transition in the moving sample.
In Sec.VI, we present the $ z=2 $ case in a moving sample and also contrast with the $ z=1 $ case discussed in the previous sections.
In Sec.VII, we put GT in a lattice, then derive the boost form in the tight binding limit.
We explore the bare space-time encoded in the many-body wavefunctions in the first quantization
and the emergent space-time encoded in the tight-binding limit in the second quantization.
In Sec.VIII, we derive the effective actions with $ z= 1 $ and $ z=2 $ under the boost.
The derivation not only provide the physical meanings of the order parameters and
the phenomenological parameters in the effective actions in terms of the original ones
in the microscopic Hamiltonian, but also bring new insights to the emergent space-time structure near the two QPTs.
In Sec.IX, we contrast our findings  to the relativistic Doppler effects, Unruh effects by an accelerating observer,
comment on the lack of Lorentz invariance and the  emergency of bulk AdS geometry from the boundary
in $ AdS_{d+1}/CFT_d $ at $ d=1,2 $ with $ d=1 $ corresponds to the SYK model on the boundary.
Especially we derive the temperature LT law which should close its long dispute.
In Sec.X, we analyze applicable experimental detections  and observable effects in a moving sample.
Conclusions and perspectives are presented in Sec.XI. In several appendices, we perform systematic investigations on
GT in various quantum  and classical systems and also clarify some confusing treatments on GT in some previous literatures.
In Appendix A, we clarify the relations among the Galileo transformation (GT), generalized GT, low velocity expansion of
Lorentz transformation (LT) in terms of $ v/c $ upto the order of $ (v/c_l)^2 $, Lorentz transformation and Pioncare group. 
We also contrast the expansion with the well known $ 1/N, 1/S $ or $ 1/\hbar $ expansion.
In Appendix B, we develop the Hamiltonian formalism which is complementary to
the Lagrangian formalism thoroughly used in the main text.
In Appendix C, we analyze the Galileo invariance in a single particle Schrodinger equation in the first quantization, 
the enlargement of a Fermi surface and
the breaking of Galileo invariance by a spin-obit coupling in the second quantization.
In Appendix D, we study the excitation spectrum in a SF away from the integer fillings and stress the crucial differences than
that at the integer fillings discussed in the main text.
In appendix F and G, we apply our formalism to study GT on many body systems in an external magnetic field, then fractional quantum Hall effects (FQH ).
In appendix H, we apply the GT to study a moving SF  to
examine  the propagating chiral edge mode of a bulk FQH in different inertial frames.
In the next 3 appendices, we address the evolution from high energy physics to materials, 
the difference between an open system and a closed system, how to determine the  temperature consistently in both QFT and curved space-time.
In the final section, we discuss the case (B) briefly.

\section{ The effective phase diagram of $ z=1 $ in a moving sample }

  Here we first focus on the 2d superfluid (SF) to Mott transitions \cite{boson0,z2,field2,bosonlattice}
  in Fig.\ref{phaseslattice} with  the  dynamic exponent $ z=1 $,
  then study  the one with  $ z=2 $ in Fig.\ref{phaseslattice} in Sec.VI.

  The effective action consistent with all the symmetries is:
\begin{align}
	\mathcal{S}_{L,z=1} & =\int d\tau d^2x
	[|\partial_\tau\psi|^2+v_x^2|\partial_x\psi|^2+v_y^2|\partial_y\psi|^2
	+ r |\psi|^2   \nonumber   \\
    & + u |\psi|^4+ \cdots ]
\label{z1}
\end{align}
 Where $ \psi $ is the complex order parameter which has
 the effective mass $ r $ tunes the SF-Mott transition with $ z=1 $.
 $ r > 0,  \langle \psi \rangle =0 $ is in the Mott state which respects the $ U(1) $ symmetry,
 $ r < 0, \langle \psi \rangle \neq 0 $ is in the SF state which breaks the $ U(1) $ symmetry.
 It is related to the microscopic parameters in Eq.\ref{boson} by $ r \sim (t/U)_c - t/U $ at a fixed chemical potential $ \mu=U/2 $ ( see also
 Fig.\ref{phaseslattice} ).

 After the scaling away the $ v_x, v_y $, it has an emergent Lorentz invariance
 ( appendix A ) whose characteristic velocity is
 the intrinsic velocity $ v_x, v_y $ instead of the speed of light $ c_l $ in the relativistic quantum field theory \cite{light}.
 The Lorentz transformation reduces to the Galileo transformation when $ v/c_l \ll 1 $  ( Appendix A ).
 It has a exact Time-reversal symmetry $ T $: $ \psi(\vec{x},t) \rightarrow \psi(\vec{x},-t) $ and $ i \to -i $.
 It  also has a parity $ P $ under which $ \psi(\vec{x},t) \rightarrow \psi(-\vec{x}, t) $ and dictates $ \omega_{\pm} ( \vec{k} )
 = \omega_{\pm} ( -\vec{k} ) $.
 It also has a particle-hole (PH) ( because it is called charge conjugation ( C ) symmetry
 in the relativistic quantum field theory. So we adopt this notation in the following )
 under $ \psi(\vec{x},t) \rightarrow \psi^{*}(\vec{x},t) $ which dictates the
 particle spectrum is related to that of a hole by $ \omega_{+} ( \vec{k} )=  \omega_{-} ( -\vec{k} ) $,

 Note that the $ P $ and $ T $ are discrete space-time symmetry, while the $ C $ is the discrete internal symmetry.
 The GT is just a space-time translation which can be considered as the mixing of  the $ P $ and $ T $,
 so it breaks them separately, but keeps their combination $ P T $. It is this mixing of the space and time which leads to 
 novel effects in a moving sample.
 It does not touch any internal symmetry, so it still keeps the $ C $ symmetry. However, the space-time in the effective
 action Eq.\ref{z1} is a emergent space-time which is a coarse-grained version of the bare space-time of the original boson Hubbard model
 Eq.\ref{boson}. Simultaneously, the bosonic field $ \psi( \vec{x},t) $ living in the emergent space-time could also be very much different than
 the original bosons in Eq.\ref{boson}. their connections can only be established by microscopic calculations on the lattice which will be achieved in Sec.VIII.

The lattice breaks the Galileo invariance and is static in the lab frame.
Then one try to detect these SF-Mott transitions when the whole sample is moving with the velocity $ \vec{c}=c \hat{y} $ along the $ y- $ axis with respect to the lab frame. This moving sample could be a fast-moving train or space-craft/satellite in the space.
To address possible new  physics in the moving sample,  one just perform a Galileo transformation
$ y^{\prime}= y + ct, t^{\prime}=t $ to the moving sample where the prime means the co-moving frame and no-prime means the lab frame.
In the real time, it implies $ \partial_y \to \partial^{\prime}_y,
\partial_t \to \partial^{\prime}_t + c \partial^{\prime}_y $.
In the imaginary time $ \tau= it $, it implies $ \partial_y \to \partial^{\prime}_y,
\partial_\tau \to \partial^{\prime}_\tau -i c \partial^{\prime}_y $.
Because the effective action when the sample is static has the emergent "Lorentz" invariant ( Appendix A ) instead of  Galileo invariant, so the Galileo
transformation may lead to some dramatic effects.

 Boosting the action Eq.\ref{z1} with $ \partial_y \to \partial^{\prime}_y,
\partial_\tau \to \partial^{\prime}_\tau -i c \partial^{\prime}_y $ and
 $ \psi( \vec{x}, t) \to \psi( \vec{x}^{\prime} - \vec{c} t^{\prime}, t^{\prime})
 = \psi^{\prime}( \vec{x}^{\prime}, t^{\prime}) $, using the invariant  space-time measure
 $ \int d\tau d^2x = \int d\tau^{\prime} d^2 x^{\prime} $ and the functional measure
 $ \int D\psi D\psi^{*} = \int D\psi^{\prime} D\psi^{\prime *} $ lead to the following effective action in
 the lab frame when the sample is moving ( for notational simplicity, we drop the $ \prime $ ) :
\begin{align}
    \mathcal{S}_{M,z=1}
	 =\int d\tau d^2x
	[(\partial_\tau\psi^*-ic\partial_y\psi^*)
	(\partial_\tau\psi-ic\partial_y\psi)
                   \nonumber   \\
	 +v_x^2|\partial_x\psi|^2
	+v_y^2|\partial_y\psi|^2
	+ r |\psi|^2+ u |\psi|^4+ \cdots ]
\label{boostz1}
\end{align}
 where the space-time is related by the intrinsic velocity $ v_x, v_y $ instead of the speed of light $ c_l $ in the relativistic QFT.
 One need to stress \cite{wrong,addmore}
 $ (\partial_\tau\psi^*-ic\partial_y\psi^*)(\partial_\tau\psi-ic\partial_y\psi) \neq |(\partial_\tau\psi-ic\partial_y\psi)|^2
 = (\partial_\tau\psi^*-ic\partial_y\psi^*)(\partial_\tau\psi + ic\partial_y\psi) $. See Sec.VIII for the derivation
 from the lattice model Eq.\ref{boson}.
 Because the particle-number remains conserved when the sample is moving, it still keeps the exact $ U(1) $ symmetry and the $ C $ symmetry.
 It breaks the emergent Lorentz invariance, the $ T $ and the $ P $,
 but keeps its combination the latter two which can be called $ PT $.
 The $ C $  dictates the particle spectrum is related to that of a hole by $ \omega_{+} ( \vec{k} )=  \omega_{-} ( -\vec{k} ) $.
 It is this C which plays crucial roles in all the calculations, especially the boson-vortex duality transformation,
 and the RG analysis along the $ z=1 $ line in Fig.\ref{phasesz1}b.  So it still keeps CPT symmetry.
 Indeed, the action Eq.\ref{boostz1} has the largest symmetries when the sample is static $ c=0 $, but takes  the reduced symmetries
    when the sample is moving $ c \neq 0 $. So the static frame is indeed the preferred frame.
    In this work, we show that the quantum phase and phase transitions
    observed when the sample is moving  display quite different phenomena than those observed when the sample is static.

   However, we like to stress that Eq.\ref{boostz1} is achieved just from symmetry principle, so it is not known how this effective boost
   $ c $ in this effective action related to the bare boost $ v $ relative to the underlying lattice. It is also not known if the other phenomenological parameters such as $ v_x, v_y $ with the velocity dimension also depends on $ v $. If yes, what is the dependence.
   This important question can only be addressed by  performing a microscopic calculations on the original boson Hubbard model Eq.\ref{boson} and will be achieved in Sec.VII and VIII.
   In Sec.I-VI, we will simply take $ c $ as a given parameter and also assume all the other parameters are independent of $ c $,
   then work out its phase diagram when the sample is moving in Fig.\ref{phasesz1} for $ z=1 $ and Fig.\ref{phasesz2} for $ z=2 $.
   This effective action approach is interesting on its own, because it may be applied to many other systems such as the
   SOC system in a Zeeman field \cite{response}.
   Then in Sec.VII and VIII, by performing the GT on the lattice model, we will establish the relations between these phenomenological parameters and the microscopic
   parameters $ v $ and those in the boson Hubbard model Eq.\ref{boson}.
   The phenomenological effective action, RG analysis, charge-vortex duality and scaling functions approach in Sec.I-VI and the microscopic calculations on a lattice in Sec.VII-VIII
   are complementary to each other. Their combination will lead to the global phase diagram of
   Eq.\ref{boson} when the sample is moving with the velocity $ v $ relative to the lattice shown in Fig.\ref{phaseslattice}a.

\subsection{ Mean field phase diagram }

\begin{figure}[tbhp]
\centering
\includegraphics[width=.8\linewidth]{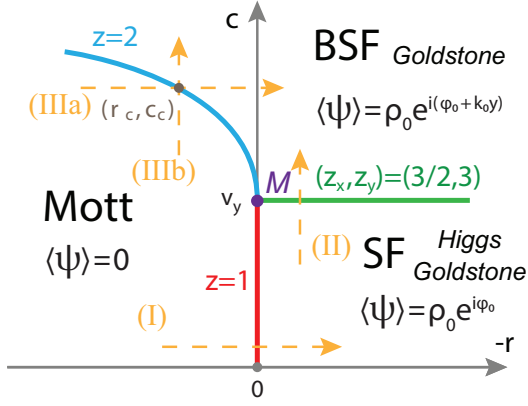}
\caption{  The phase diagram of the effective action Eq.\ref{boostz1} with $ z=1 $  as a function of $ r $ and $ c $
when the sample is moving. For the boson-Hubbard model Eq.\ref{boson}, $ r \sim (t/U)_c - t/U $ at a fixed chemical potential $ \mu=U/2 $.
The two tuning parameters are the effective mass $ r $ and the boost $ c $. $ c=0 $ axis recovers that when the sample is static.
The BSF is the new phase which spontaneously breaks the $ C $ symmetry and also the only phase carrying a flowing current.
The Mott, SF and  Boosted SF (BSF) phase meet at the multi-critical point $ M $.
The Goldstone modes exist in both the SF and BSF, but the Higgs mode exists only in the SF protected by the $ C $ symmetry,
disappears in the BSF phase due to its spontaneously breaking of the $ C $ symmetry.
As shown in Sec.VII-1, the order parameter to distinguish between SF and BSF  can be taken as the bilinear currents Eq.\ref{threecurrents}.
The three QPTs along the three paths I with $ z=1 $, II with $ z=(3/2,3) $ and IIIa or IIIb with $ z=2 $
are presented in the following sections. All the four paths can be scanned by scattering measurements shown in Fig.\ref{detector}.
 The boost $ 0 < c < v $ is exactly marginal which stands for the Doppler shift inside the SF phase, so the $ z=1 $ line is a line of fixed points in the boosted 3D XY universality class.
The CFT of this novel M point remains to be explored.
Here we assume the boost $ c $  and the intrinsic velocity $ v_y $ as independent parameters.
The microscopic evaluations in Sec.VII-VIII show that
the $ (3/2, 3 ) $ transition line is not reachable for  the microscopic boson Hubbard model in Fig.\ref{phaseslattice0}.
But as shown in \cite{SSS}, it is reachable by directly driving a SF in the class-2 mentioned in the introduction.   }
\label{phasesz1}
\end{figure}

 Expanding Eq.\ref{boostz1} leads to
\begin{align}
	\mathcal{L}_{M,z=1}&=
	|\partial_\tau\psi|^2-i2c\partial_\tau\psi^*\partial_y\psi
	+v_x^2|\partial_x\psi|^2+(v_y^2-c^2)|\partial_y\psi|^2     \nonumber   \\
	&+a|\partial_y^2\psi|^2 + b |\partial_y \psi|^4
	+ r |\psi|^2+ u |\psi|^4 + \cdots
\label{expand}
\end{align}
where the higher derivative or higher order $a,b >0$ terms are not important when the sample is static \cite{addmore0},
but may become important when the sample is moving,
especially near the new quantum phase transitions
in Fig.\ref{phasesz1}. In principle, one need also add $ d |\psi|^2 |\partial_y \psi|^2 $ to Eq.\ref{expand}, but it will not change
the physical results, so for the notational simplicity, we omit it in the following.

When writing in terms of the metric $ g_{\mu,\nu} \partial_{\mu} \psi^{*} \partial_{\nu} \psi $
in $ (\tau, y ) $ space-time in Eq.\ref{expand},
it is $ g_{\tau, \tau}=1,  g_{\tau, y}=g_{y, \tau}=-i c, g_{y, y}= v^2_y-c^2 $.
The crossing ( or off-diagonal ) metric component $ g_{\tau, y} $ is the only new term when the sample is moving generated by the Galileo transformation which does not appear  in Eq.\ref{z1} when the sample is staic. It is this new term which represents the new space-time structure near the $ z=1 $ QPT
and plays important roles when the sample is moving.
Because the boost $ c $ adds a new tuning parameter,
both $\gamma =v_y^2-c^2 $ and $ \mu$ can change sign and tune various QPTs in Fig.\ref{phasesz1}.

 Taking the mean field ansatz $ \langle \psi \rangle =\sqrt{\rho}e^{i(\phi+k_0y)}$ leads to the energy density
\begin{align}
	E[\rho,k_0]=[(v_y^2-c^2) k_0^2+a k_0^4 + r ]\rho+ u \rho^2
\end{align}
Minimizing $E[\rho,\phi,k_0]$ with respect to $\rho$ and $k_0$ results in
\begin{align}
	k_0^2
	=\begin{cases}
		0, & c^2 < v_y^2 \\
		\frac{c^2-v_y^2}{2a},  & c^2 > v_y^2 \\
	\end{cases},~~~~~~~~~~~~~
\end{align}
and
\begin{align}
	\rho_0
	=\begin{cases}
		0, & r > 0 \text{ and } c^2 < v_y^2 \\
		-\frac{r }{2U}, & r < 0 \text{ and } c^2 < v_y^2 \\
		0, & r > \frac{(c^2-v_y^2)^2}{4a} \text{ and } c^2 > v_y^2\\
		-\frac{r}{2U}+\frac{(c^2-v_y^2)^2}{8aU},
		& r < \frac{(c^2-v_y^2)^2}{4a} \text{ and } c^2 > v_y^2 \\
	\end{cases}
\label{k0}
\end{align}

The phase diagram is summarized in Fig.\ref{phasesz1}. Due to the $ C $ symmetry inside the Mott phase, $ \omega_{+} ( \vec{k} )= \omega_{-} ( -\vec{k} ) $, one can look at the instability from either particle or hole band $\omega_{\pm}=\sqrt{v_x^2k_x^2+v_y^2k_y^2}\pm ck_y$.

\subsection{ The Mott to SF transition along the path I with $ z= 1 $. }

   Along the path I in Fig.\ref{phasesz1}b,  at the mean-field level, we can substitute $\psi\to\sqrt{\rho_0}e^{i\phi_0}$ into the boosted effective action Eq.\ref{boostz2}
\begin{align}
    \mathcal{S}= r \rho_0+ u \rho_0^2
\end{align}
When $  r > 0$, it is in the Mott phase with $ \langle \psi \rangle =0$.
When $ r <  0$, it is in the SF phase with  $ \langle \psi \rangle=\sqrt{\rho_0}e^{i\phi_0}$
where $\rho_0=-r/2u$ and $\phi_0$ is a arbitrary angle due to the  $ U(1)$ symmetry.

{\sl (a) The Mott state: }

   In the Mott phase, $ r > 0$, one can write $\psi=\psi_R+i \psi_I$ as its real part and imaginary part
and expand the action upto second order
\begin{align}
    \mathcal{L}_{MI}
	& =
	\sum_{\alpha=R,I}
	[(\partial_\tau \psi_\alpha
	-ic\partial_y \psi_\alpha)^2
	+v_x^2(\partial_x  \psi_\alpha)^2
	+v_y^2(\partial_y   \psi_\alpha)^2    \nonumber   \\
	& + r ( \psi_\alpha)^2 + \cdots]
\end{align}
which lead to 2 degenerate gapped modes with the effective mass $ r > 0 $:
\begin{align}
	\omega_{R,I}=\sqrt{ r +v_x^2k_x^2+v_y^2k_y^2}-ck_y
\label{Mottz1}
\end{align}
which indicates the dynamic exponent $ z=1 $.

{\sl (b) The SF phase: }

In the SF phase, $ r < 0 $, we can write the fluctuations in the polar coordinates
$\psi=\sqrt{\rho_0+\delta\rho}e^{i(\phi_0+\phi)}$
and expand the action up to the second order in the fluctuations:
\begin{eqnarray}
    \mathcal{L}
	& = &
	[(\partial_\tau-ic\partial_y)\delta\rho]^2
	+[v_x^2(\partial_x\delta\rho_+)+v_y^2(\partial_y\delta\rho)]
	+4\rho_0U(\delta\rho)^2  \nonumber   \\
	& +  & \rho_0^2[(\partial_\tau-ic\partial_y)\phi]^2
	+\rho_0^2[v_x^2(\partial_x\phi)^2
		 +v_y^2(\partial_y\phi)^2] + \cdots
\label{GoldHiggsact}
\end{eqnarray}
  where $ \cdots $ means the coupling between the two modes.

  Due to the $ C $ symmetry dictating $ z=1 $ ( See Sec.VI ), one finds one gapless Goldstone $ \phi $
  mode and one gapped Higgs $ \delta\rho $ mode:
\begin{eqnarray}
   \omega_{\text{G}}
	& = & \sqrt{v_x^2k_x^2+v_y^2k_y^2}-c k_y    \nonumber   \\
    \omega_{\text{H}}
	& = & \sqrt{4\rho_0 u +v_x^2k_x^2+v_y^2k_y^2}-c k_y
\label{GoldHiggs}
\end{eqnarray}

Note that it is the $ C $ symmetry which ensures the separation of the real part from the imaginary part when $ r > 0 $ in the Mott phase in Eq.\ref{Mottz1} and the separation of the Higgs mode from the Goldstone mode when $ r < 0 $ in the SF phase in  Eq.\ref{GoldHiggs}.
Intuitively, one can say the two degenerate gapped modes in Eq.\ref{Mottz1} turn into the Goldstone mode and the Higgs mode in Eq.\ref{GoldHiggs}
through the $ z=1 $ QPT from the Mott phase to the SF phase.

{\sl (c) The QCP: a SF-Mott transition still with $ z=1 $ when the sample is moving: RG analysis }

If putting $ c=0 $ in the effective action Eq.\ref{boostz1}, it is nothing but a 3D XY universality class
with the critical exponents $ \nu=0.67, \eta=0.04 $ which is emergent
Lorentz invariant. The $ z=1 $ is protected by the Lorentz invariance at $ c=0 $.
The interaction $ u $ term is relevant at $ c=0 $ and controlled by the 3D XY fixed point.
Any $ c > 0 $ breaks the Lorentz invariance explicitly. So the action is neither Lorentz invariant nor Galileo invariant.
The $ c $ term is exact marginal suggesting a line of fixed points.
The RG flow of $ u $ along the fixed line is determined by RG calculations in Sec.VI.
It was shown that despite  the lack of Lorentz invariance at $ c \neq 0 $, the $ C $ symmetry still detects
the dynamic exponent $ z=1 $. The Mott to the SF transition at $ 0 < c < v_y $ remains in the 3D XY universality class.

One can also look at the $ z=1 $ QCP from 3d CFT point of view:
at $ c=0 $, it has the emergent pseudo-Lorentz invariance, respects the exact C, P and T
separately ( then CPT ) and scale invariance  with $ z=1 $. It is a fixed point in the 3D XY universality class.
Any $ 0 < c < v $ just adds a marginal direction to  the 3d CFT: it  breaks P and T
separately,  but still keeps C, PT ( then CPT ) and scale invariance with $ z=1 $.
Note that Lorentz invariance always implies $ z=1 $, but not otherwise.
Of course, in the CFT in materials, the LI is always a pseudo one ( Appendix A ). In the CFT
in the string theory such as SUSY Yang-Mills ( see also the conclusion section ), the LI is the real one.
Despite the $ z=1 $ line has a $ c=0 $ limit, the other two QPTs have no $ c=0 $ counter-parts, so can only be observed
when the sample is moving. As stressed in the introduction, just from the symmetry point of view,
there is indeed a preference frame where the lattice is at rest, and there is an enhanced symmetry.
In a sharp contrast, in the relativistic QFT, all the inertial frames are related by LT,
all the physical laws take the same form, therefore have the same symmetry.

\subsection{ The SF to BSF  quantum Lifshitz transition along path II with $ (z_x, z_y)=(3/2,3) $. }

 Inside the SF phase at a fixed $ r < 0  $, as $ c $ increases along the path II in Fig.\ref{phasesz1}b,
 there is a quantum Lifshitz  transition from the SF phase to the BSF driven by the boost
 of the Goldstone mode in Eq.\ref{GoldHiggsact}. Because the gapped Higgs mode remains un-critical across
 the transition, one can simply drop it. In fact, as shown below, the Higgs mode disappears in the BSF side due to the explicit
 $ C $ symmetry breaking inside the BSF.
 Although the Goldstone mode to the quadratic order in Eq.\ref{GoldHiggsact} is enough
 inside the SF phase, when studying  the transition to the BSF, one must incorporate higher derivative terms and
 also higher order terms in Eq.\ref{expand} to the Goldstone mode in Eq.\ref{GoldHiggsact}.
 A simple symmetry analysis leads to the following   bosonic quantum Lifshitz transition from the SF to BSF in terms of the phase degree of freedom ( which can also be derived by substituting $\psi=\sqrt{\rho_0+\delta\rho}e^{i(\phi_0+\phi)}$ into Eq.\ref{expand}, then integrating out the Higgs mode $ \delta \rho $ in Eq.\ref{GoldHiggsact} ):
\begin{align}
    \mathcal{L}_{SF-BSF}
	&=(\partial_\tau\phi-ic\partial_y\phi)^2
	+v_x^2(\partial_x\phi)^2
	+v_y^2(\partial_y\phi)^2    \nonumber   \\
	&+a(\partial_y^2\phi)^2
	+b(\partial_y\phi)^4+ \cdots
\label{boosttoright}
\end{align}
where $a,b>0$ terms come from those in Eq.\ref{expand}.
It has the Translational symmetry $ \vec{x}  \to \vec{x} + \vec{a} $ and
the $ U(1)_c $ symmetry $ \phi \to \phi + \phi_0 $. Note that $ \gamma= v^2_y- c^2 $ is the tuning parameter
which drives the quantum Lifshitz transition
from the SF phase to the BSF phase \cite{morederivative}.
A simple scaling shows that when $ z=1 $ inside the SF phase $ [a]=-2, [b]=-3 $, so they are irrelevant
inside the SF phase, but become important near the SF-BSF transition as to be shown in the following.

The mean-field state can be written as $\phi=\phi_0+k_0 y$.
Substituting it to the effective action Eq.\ref{boosttoright} leads to:
\begin{align}
    \mathcal{L}_{0}
	\propto(v_y^2-c^2)k_0^2+bk_0^4
\label{pmk0}
\end{align}

At a low boost $c^2 < v_y^2 $, $k_0=0$ is in the SF phase.
At a high boost $ r<0,  c^2 > v_y^2 $
\begin{equation}
   k_0^2=(c^2-v_y^2)/2b
\label{rhighboost}
\end{equation}
 which shows the BSF phase has the modulation $ \pm k_0 $ along the $ y-$ axis.

Note that the numerical value of $ k_0 $ in Eq.\ref{rhighboost} is different than that listed in Eq.\ref{k0}.
That should not disturbing at all, because the former applies near the SF-BSF transition, the latter  applies near the $ z=2 $ Mott-SF transition to be discussed in the next section.


{\sl (a)  The excitation spectrum in the SF phase  }

At a low boost $c^2<v_y^2$ inside the SF phase, the quantum phase fluctuation can be written as $\phi=\phi_0+\phi$.
It breaks the $ U(1)_c $ symmetry, but still keeps the translational symmetry.
Expanding the action Eq.\ref{boosttoright} upto the second order  leads to:
\begin{align}
    \mathcal{L}_{SF}
	=(\partial_\tau\phi-ic\partial_y\phi)^2
	+v_x^2(\partial_x\phi)^2
	+v_y^2(\partial_y\phi)^2
\label{s2cant}
\end{align}
   which reproduces the gapless Goldstone mode in Eq.\ref{GoldHiggs} inside the SF phase:
\begin{align}
    \omega_\mathbf{k}
	=\sqrt{v_x^2 k_x^2+v_y^2k_y^2}-ck_y
\label{sfdis}
\end{align}
 which is consistent with the Goldstone mode in Eq.\ref{GoldHiggs}. The Higgs mode was dropped at very beginning, so can not be seen in
 Eq.\ref{s2cant}.

{\sl (b)  The spontaneous $ C $ symmetry breaking in the BSF phase  }

Due to the exact $ C $ symmetry, $ \pm k_0 $ in Eq.\ref{pmk0} are related by the $ C $ symmetry, so the ground state
could take $ \pm $ sign or its any linear combination. To determine the ground state, we take the most general  mean field ansatz:
\begin{equation}
 \langle\psi\rangle=\sqrt{\rho}(c_1e^{ik_0y}+c_2e^{-ik_0y})
\label{c1c2}
\end{equation}
with $|c_1|^2+|c_2|^2=1$.

  Substituting Eq.\ref{c1c2} into Eq.\ref{expand}, integration over the space kills the oscillating parts and
  leads to its energy density:
\begin{align}
	E[\rho,k_0,c_1,c_2]=[(v_y^2-c^2)k_0^2+ak_0^4 + r ]\rho      \nonumber   \\
    +(1+2|c_1|^2|c_2|^2) (u+ bk^4_0) \rho^2
\end{align}
$ u >0$ and $ b > 0 $ dictates the minimization condition $c_1c_2=0$. So the ground-state is either
$\langle\psi\rangle=\sqrt{\rho}e^{ik_0y}$ or
or $\langle\psi\rangle=\sqrt{\rho}e^{-ik_0y}$ which implies the spontaneously breaking of the $ C $ symmetry.
We call such a $ C $ symmetry broken SF state the Boosted superfluid (BSF) \cite{putative}.

{\sl (c)  The excitation spectrum in the BSF phase  }

Inside the BSF phase, the quantum phase fluctuations in one of the two solutions  can be written as $ \phi=\phi_0+k_0y+\phi $.
It still keeps the diagonal symmetry $ y \to y + a $ and also $  \phi \to \phi -k_0 a $.
So the symmetry breaking is
\begin{equation}
     U(1)_T \times U(1)_c  \to [ U(1)_T \times U(1)_c ]_D
\label{u1u1}
\end{equation}
which still leads to just one Goldstone mode. This symmetry breaking pattern is different than that in the SF phase $ U(1)_c \to 1 $.

 Expanding the action upto the second order in the phase fluctuations leads to
\begin{align}
    \mathcal{L}_{BSF}
	=(\partial_\tau\phi-ic\partial_y\phi)^2
	+v_x^2(\partial_x\phi)^2
	+(v_y^2+6bk_0^2)(\partial_y\phi)^2
\label{s2ic}
\end{align}
   which leads to the gapless Goldstone mode inside the BSF phase:
\begin{equation}
    \omega_\mathbf{k}=\sqrt{v_x^2 k_x^2+(3c^2-2v_y^2)k_y^2+ak_y^4}-ck_y
\label{bsfdis}
\end{equation}
where one can see  $3c^2-2v_y^2>= 2(c^2-v_y^2)+ c^2 > c^2 $ when $c^2>v_y^2$,
thus the $\omega_\mathbf{k}$ is stable in BSF phase.
Due to the spontaneous $ C $ symmetry breaking, the Higgs mode may not even exist anymore in the BSF phase.
This result will also be confirmed further in the next section from the Mott to the BSF transition.

{\sl (d) The exotic QCP scaling with the dynamic exponents $ (z_x=3/2, z_y=3 )  $ }

    It is instructive to expand the first kinetic term in Eq.\ref{boosttoright} as:
\begin{align}
    \mathcal{L}
	& =Z(\partial_\tau\phi)^2
	-2i c \partial_\tau\phi\partial_y\phi
	+v_x^2(\partial_x\phi)^2 + \gamma (\partial_y \phi)^2      \nonumber   \\
	& +a(\partial_y^2\phi)^2
	+b(\partial_y\phi)^4]
\label{ab}
\end{align}
  where $ Z $ is introduced to keep track of the
  renormalization of  $ (\partial_\tau\phi)^2 $, $ \gamma= v^2_y- c^2 $ is the tuning parameter.

  The scaling $ \omega \sim k^3_y, k_x \sim k^2_y $ leads to the exotic dynamic exponents $ (z_x=3/2, z_y=3 )  $.
  Then one can get the scaling dimension of $ [\gamma]=2 $ which is relevant, as expected, to tune the transition,
  but $ [Z]=[b]=-2 < 0 $, so both are leading irrelevant operators\cite{morederivative} which
  determine the finite $ T $ behaviours and corrections to the leading scalings.
  Setting $ Z=b=0 $ in Eq.\ref{ab} leads to the Gaussian fixed action  at the QCP where $ \gamma=0 $.
  Exotically and interestingly, it is the crossing matric  $ g_{\tau, y}=g_{y, \tau}=-i c $ in Eq.\ref{ab} which dictates the
  quantum  dynamic scaling near the QCP. It is a direct reflection of the new emergent space-time near the $ z=1 $ QPT.
  As a byproduct, the results achieved here can also be applied to study the classical dynamic phase transitions
  in sound waves in a medium to be discussed in Appendix B.

\subsection{ The Mott to BSF Transition along the path III  with $ z=2 $.  }

When $k_0\neq0$ along the path III in Fig.\ref{phasesz1}b, it is convenient to introduce
the new order parameter $\psi=\tilde{\psi}e^{ik_0y}$,
then the original action Eq.\ref{expand} can be expressed in terms of $\tilde{\psi}$
\begin{eqnarray}
	\mathcal{L}
	& = & |\partial_\tau\tilde{\psi}|^2
	-2ck_0\tilde{\psi}^*\partial_\tau\tilde{\psi}
	-i2c\partial_\tau\tilde{\psi}^*\partial_y\tilde{\psi}
	+v_x^2|\partial_x\tilde{\psi}|^2    \nonumber  \\
	& + & (v_y^2-c^2+6ak_0^2)|\partial_y\tilde{\psi}|^2  - i2k_0(v_y^2-c^2+2ak_0^2)\tilde{\psi}^*\partial_y\tilde{\psi}    \nonumber  \\
	& + &  [r + (v_y^2-c^2) k_0^2 + a k_0^4]|\tilde{\psi}|^2+U|\tilde{\psi}|^4+ \cdots
\end{eqnarray}
where $\cdots$ denotes all the possible high-order derivative term.

Setting $i2k_0(v_y^2-c^2+2ak_0^2)\tilde{\psi}^*\partial_y\tilde{\psi}=0$ leads to
$k_0^2=\frac{c^2 -v_y^2}{2a}$ which is the same as Eq.\ref{k0} at $ c^2 > v_y^2 $.
As shown in the last section, due to the spontaneous $ C $ symmetry breaking in the BSF, one can only take
one of the $ \pm k_0 $ value.

By using $ 2ak^2_0=c^2-v_y^2$, one can simplify the above action to
\begin{align}
	\mathcal{L} & =
	|\partial_\tau\tilde{\psi}|^2	
	-i2c\partial_\tau\tilde{\psi}^*\partial_y\tilde{\psi}
	+2ck_0\tilde{\psi}\partial_\tau\tilde{\psi}^*
	+v_x^2|\partial_x\tilde{\psi}|^2     \nonumber  \\
	 & +4ak_0^2|\partial_y\tilde{\psi}|^2
	+(r-a k_0^4)|\tilde{\psi}|^2+ u |\tilde{\psi}|^4
\label{z1z2square}
\end{align}
 where one can observe $ k_0 \neq 0 $ leads to a linear derivative term $ \tilde{\psi}\partial_\tau\tilde{\psi}^* $. It dictates the dynamic exponent $ z=2 $.

{\sl (a) Scaling analysis near the $ z=2 $ QCP  }

Simple scaling analysis shows that the first term $|\partial_\tau\tilde{\psi}|^2$ is irrelevant with scaling dimension $ -2 $,
the second term is the metric crossing term $\partial_\tau\tilde{\psi}^*\partial_y\tilde{\psi}$ which
is irrelevant with the scaling dimension $ -1 $,
the third  ( linear derivative ) term $\tilde{\psi}^*\partial_\tau\tilde{\psi}$ leads to $ z=2 $.

After only keeping the leading irrelevant term which is the metric crossing term, we arrive at the effective action:
\begin{align}
	\mathcal{L} & =
	Z_1\tilde{\psi}^*\partial_\tau\tilde{\psi}	
	-iZ_2\partial_\tau\tilde{\psi}^*\partial_y\tilde{\psi}
	+\tilde{v}_x^2|\partial_x\tilde{\psi}|^2
	+\tilde{v}_y^2|\partial_y\tilde{\psi}|^2     \nonumber  \\
	& -\tilde{\mu}|\tilde{\psi}|^2 + u|\tilde{\psi}|^4
\label{z1z2}
\end{align}
 where $Z_1=-2ck_0 $, $Z_2=2c$, $\tilde{v}_x^2=v_x^2$, $\tilde{v}_y^2=4ak_0^2=2(c^2-v_y^2)$.
 It is the effective chemical potential:
\begin{align}
 \tilde{\mu} & =-r+ak_0^4 =-(r-r_c),   \nonumber  \\
  r_c & =a k^4_0  = \frac{ ( c^2- v^2_y )^2 }{4a}  > 0
\label{tildemuz2}
\end{align}
 which tunes the Mott to BSF transition.
 As shown in Path-IIIa and IIIb, there are two independent ways to tune $\tilde{\mu} $:
 Vertical path-IIIa, at a fixed $ r > 0 $, one increases $ c $, therefore $ k_0 $
 or Horizontal path-IIIb, at a fixed $ \gamma < 0 $, one increases $ -r $.

 Now we focus on near the $ z=2 $ quantum critical line.
 Due to $ [Z_2]=-1 $, $ Z_2 $ metric crossing term gets to zero quickly under the RG flows, so can be treated very small $ Z_2 \ll Z_1 $.
 In the following, we will use this fact to simplify the excitation spectrum in the Mott and BSF phase and
 also stress the roles of the leading irrelevant operator $ Z_2 $.

{\sl (b) Excitations in the Mott phase:  }

In the Mott phase, $\tilde{\mu} < 0$ and $\rho_0=0$, the excitation spectrum is
\begin{align}
	\omega
     &=\frac{ -\tilde{\mu}+\tilde{v}_x^2k_x^2+\tilde{v}_y^2k_y^2}{Z_1-Z_2k_y}  \nonumber  \\
	 & =\frac{\tilde{v}_y^2k_*^2 - \tilde{\mu}}{Z_1-Z_2k_*}
	+\frac{\tilde{v}_x^2k_x^2}{Z_1-Z_2k_*}   \nonumber  \\
	 &+\frac{(Z_1^2\tilde{v}_y^2 - Z_2^2\tilde{\mu})(k_y-k_*)^2}{(Z_1-Z_2k_*)^3}
	+\cdots
\end{align}
where $k_*=\frac{Z_1}{Z_2}
[1-\sqrt{1-(Z_2^2\tilde{\mu})/(v_y^2Z_1^2)}] $.
Consider the $Z_2\ll Z_1 $ limit, the result is simplified as
\begin{align}
	\omega=Z^{-1}_1[ -\tilde{\mu}
	+\tilde{v}_x^2k_x^2+\tilde{v}_y^2(k_y-k_{*}  )^2 ]
\label{Mottz2}
\end{align}
   where $ k_{*}= \frac{Z_2\tilde{\mu}}{2 Z_1 v_y^2} $ which vanishes at the phase boundary  $ \tilde{\mu}=0 $.
   The excitation spectrum has one minimum at $ (0,0) $ right at the QCP, but at $ (0, k_{*} ) $ away from it inside the Mott phase.
   It indicates the condensation at $ k_0 $ with the dynamic exponent $ z=2 $.

   It can be contrasted to Eq.\ref{Mottz1} where the Doppler shift term
   appears outside of the square root indicating $ z=1 $. It contains also both particle and hole excitations.
   Here it appears just as a shift in $ k_y $ implying $ z=2 $,
   it only contains the particle excitation spectrum, while the hole excitation is at much higher energy, so
   can be dropped. It is the leading irrelevant metric-crossing term $ Z_2 $ which leads to
   the shift of the minimum away from the origin.

{\sl (c) Excitations in the BSF phase:  }

In the BSF phase, $\tilde{\mu} > 0$ and $\rho_0 >0$,
the excitation spectrum are:
\begin{align}
	\omega_\pm
	=Z^{-1}_{1} [ \pm\sqrt{4 u \rho_0[\tilde{v}_x^2 k_x^2+( \tilde{v}_y^2+ u \rho_0Z_2^2)k_y^2]}
	+2 u \rho_0Z_2 k_y ]
\label{BSFz2}
\end{align}
  which is always stable inside the BSF phase.
  It shows the Doppler shift term
  $ \pm $ corresponds to the P and H excitations respectively.
  Due to $ z=2 $ QCP, the magnitude and phase are conjugate to each other, so
  the Higgs mode inside the SF phase in Fig.\ref{phasesz1} does not exist anymore inside the BSF phase.
  As argued below Eq.\ref{bsfdis},  this fact is also due to the spontaneous $ C $ symmetry breaking inside the BSF phase.

  The Doppler shift term $ \sim  u \rho_0 c $
  inside the BSF phase near the $ z=2 $ line can be contrasted to that $ =c $ near the $ z=(3/2,3) $ line in Eq.\ref{bsfdis},
  also that $ =c $ inside the SF phase in Eq.\ref{sfdis}.
  The Doppler shift term $ = c $ in Eq.\ref{Mottz1} inside the Mott phase near the $ z=1 $ line can also be contrasted to
  that inside the Mott phase $ k_{*} $ in Eq.\ref{Mottz2}  near the $ z=2 $ line.
  These facts show that at the effective action level, as soon as the Doppler shift in Eq.\ref{Mottz1}
  inside the Mott phase near the $ z=1 $ line is given, then the Doppler shifts in all the other phases and regimes are fixed
  and connected by the Multi-critical (M) point. They could be renormalized to different values.
  As shown in the microscopic calculations in Sec.VII and VIII,
  this value $ c $ in the Mott phase is given in Eq.\ref{vxvytc} in terms of the bare boost velocity respect to the lattice $ v $,
  and also the microscopic parameters $ t_0, t_{b1}, t_{b2} $ determined by the Wannier functions.
  So it could even change sign when $ v $ moves past $ v_c=t_0/t_{b1} $.

  Eq.\ref{Mottz2} and Eq.\ref{BSFz2} show that it is the type-II dangerously irrelevant metric crossing term $ Z_2 $ which leads to
  the Doppler shift term in the Mott and BSF phase respectively. They can be contrasted to Eq.\ref{Mottz1} along the path-I
  and Eq.\ref{bsfdis} along the path-II respectively.
  So we reach consistent results from the $ z=2 $ line and the $ z=(3/2, 3 ) $ line in Fig.\ref{phasesz1}.

\section{  RG analysis on the boosted Mott to SF transition  along path I. }

  So far, we analyzed the effective action  Eq.\ref{expand} by mean field theory + Gaussian fluctuations.
  The results may be valid well inside the phases, but will surely break down near the QPTs.
  It becomes important to study the nature of the QPTs by performing renormalization group (RG) analysis.
  Unfortunately, the conventional Wilsonian momentum shell method  seems in-applicable to study the RG when the sample is moving.
  Very fortunately,  the field theory methods developed for non-relativistic quantum field theory in \cite{field1,field2,field3}
  by one of the authors can be effectively applied when the sample is moving.
 Following this method, we will perform the RG to investigate the nature of the QCP from Mott to SF in Fig.\ref{phasesz1} along the path I
 in Fig.\ref{phasesz1}. We stress the important roles played by the $ C $ symmetry.

  If setting $ c=0 $ in Eq.\ref{expand}, it is nothing but a 3D XY model. It has an emergent  Lorentz invariance, also
  C-conjugation ( PH symmetry ), Time reversal and Parity symmetry, therefore satisfies CPT theorem.
  However, any $ c \neq 0 $  breaks the emergent  Lorentz invariance, T- and P- symmetry, but keeps the C- symmetry, also CPT symmetry.
  In the following, we will show that it has the same critical exponent as the 3D XY model.
  However, because  $ c $ is exactly marginal, it is still a new universality class we name boosted 3D XY model.
  It is constructive to compare with inverted XY model which also has the same critical exponent as the 3D XY model.
  But it has a local $ U(1) $ gauge  invariance in contrast to the latter  which only have a global $ U(1) $ invariance.
  In fact, they are dual to each other as to be studied in the following section.

\subsection{ RG of the self-energy at two -loops  }

\begin{figure}[tbhp]
\centering
\includegraphics[width=.8 \linewidth]{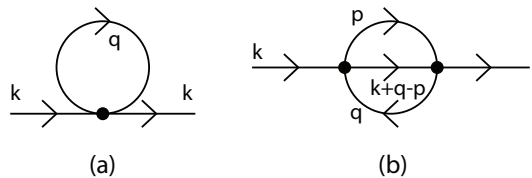}
\caption{ The self-energy of the complex bosons to one loop (a) and  two loops (b) of Eq.\ref{boostz1}. The line indicates the propagator Eq.\ref{prop0}. The arrow gives the creation of the complex bosons.
The dot stands for the interaction $ u $. $ k= ( k_0, \vec{k} ) $ means the 3 momentum.   }
\label{selfenergy}
\end{figure}

   In the following, by applying the method \cite{field2,field1,field3} developed for the field renormalization RG for non-relativistic QFT,
   the idea is to split the integrals to frequency and momentum, then always perform the integral over the frequency first,
   then perform dimensional regularization in the momentum space only.
   However, in relativistic QFT, due to the Euclidian invariance, such a splitting is not necessary,
   the frequency and momentum can be combined into a 4 momentum.  The advantage of this method over the traditional
   Wilsonian  momentum shell method is that it is systematic expansion going to any loops.

   Because the field theory method only focus on canceling  the UV divergence at $ T=0 $, so one can just look at the massless case at the QCP
   $ r=0 $. From Eq.\ref{boostz1}, one can identify the bare single particle ( boson ) Green function in $ ( \vec{k}, \omega_n ) $ space:
\begin{align}
    G_0  ( \vec{k}, \omega_n ) & = \langle \psi(\vec{k}, \omega_n) \psi^{*}(\vec{k}, \omega_n) \rangle
                      \nonumber  \\
     & = \frac{1}{ -( i \omega_n -c k_y )^2 + k^2_x  + k^2_y }    \nonumber  \\
     & =\frac{i}{2k}[ \frac{1}{ \omega_n +i \epsilon_{+}( \vec{k} ) } - \frac{1}{ \omega_n - i \epsilon_{-}( \vec{k} ) } ]
\label{prop0}
\end{align}
   where $ \epsilon_{\pm}( \vec{k} )= k \pm c k_y \geq 0 $ is the particle-hole excitation energy respectively.
   The sum of them is $  \epsilon_{+}( \vec{k} ) + \epsilon_{-}( - \vec{k} )= 2 \epsilon( \vec{k} )= 2 k $.
   The $ C $ symmetry dictates $ \epsilon_{+}( \vec{k} )= \epsilon_{-}( - \vec{k} ) > 0 $ which plays crucial roles in the RG.
   At $ c=0 $, it reduces to the usual PH symmetry which dictates $ \epsilon_{+}( \vec{k} )= \epsilon_{-}( \vec{k} )
   = \epsilon_{-}( -\vec{k} ) $.

   We first look at the boson self-energy at one -loop Fig.\ref{selfenergy}a.
\begin{align}
   (4a)  & = \int \frac{d^d q}{ (2 \pi)^d } \int \frac{d \nu}{ 2 \pi} G( \vec{q}, \nu )   \nonumber  \\
   & = \int \frac{d^d q}{ (2 \pi)^d } \int \frac{d \nu}{ 2 \pi}
   \frac{i}{2q}[ \frac{1}{ \nu +i \epsilon_{+}( \vec{q} ) } - \frac{1}{ \nu - i \epsilon_{-}( \vec{q} ) } ]   \nonumber  \\
   & = \int \frac{d^d q}{ (2 \pi)^d }\frac{1}{2q}=0
\label{3a}
\end{align}
  where we first perform the integral over the frequency, then doing dimensional regularization in the momentum space.
  One can see that at one-loop order, $ c $ does not even appear, so it is identical to the relativistic case.

  One loop is trivial. To find a non-vanishing anomalous dimension for the boson field, one must get to
  two loops Fig.\ref{selfenergy}b :
\begin{align}
   &(4b) =  \int \int
   G( \vec{q}, q_0 ) G( \vec{p}, p_0 ) G( \vec{k}+ \vec{p} - \vec{q}, k_0+p_0-q_0 )
                        \nonumber  \\
   & =    \int
  \frac{1}{2q}  \frac{1}{2p} \frac{1}{2 | \vec{k} +\vec{p} -\vec{q} | }
  \frac{ q + p+ | \vec{k} +\vec{p} -\vec{q} | }{ ( k_0 + i c k_y )^2 +   ( q + p+ | \vec{k} +\vec{p} -\vec{q} | )^2 }
                     \nonumber  \\
   &=  \frac{u^2}{ 4 \pi^2 \epsilon } [ ( k_0 + i c k_y )^2 + k^2 ] + \cdots
\label{UVfield}
\end{align}
   where $ \int \int =\int \frac{d^d q}{ (2 \pi)^d } \int \frac{d q_0}{ 2 \pi} \int \frac{d^d p}{ (2 \pi)^d } \int \frac{d p_0}{ 2 \pi} $
   and $ \int  = \int \frac{d^d q}{ (2 \pi)^d } \int \frac{d^d p}{ (2 \pi)^d } $.
   We only list the field renormalization UV divergence and   $ \cdots $ means the UV finite parts.
   when one perform the frequency integral, one must pick two poles at the two opposite side of the frequency integral
   to get a non-vanishing answer. Putting $ c=0 $ gives back to the  relativistic case.
   Because $ c $ always appears in the combination of $ k_0 + i c k_y $, so the UV divergency is identical to that of the
   $ c=0 $ case. It gives the identical anomalous dimension to the $ c=0 $ case.
   So the dynamic exponent $ z=1 $ at least to two loops. In fact, we expect that
   the $ C $ symmetry dictates $ z=1 $ is exact to all loops in Fig.\ref{phasesz1}.

\subsection{ RG of the interaction at one -loop  }

\begin{figure}[tbhp]
\centering
\includegraphics[width=.8\linewidth]{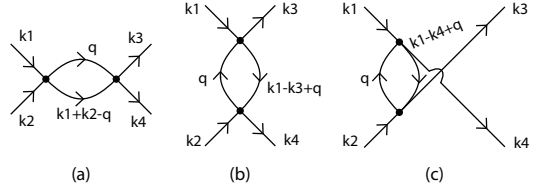}
\caption{  The renormalization of the interaction vertex of Eq.\ref{boostz1} upto one loop.  }
\label{interactions}
\end{figure}

  Now we move to the interaction vertex Fig.\ref{interactions}.
  Setting the external frequency and momentum  $ \omega= \omega_1 + \omega_2, \vec{k}= \vec{k}_1 + \vec{k}_2 $, Fig.\ref{interactions}a
  can be written as:
\begin{eqnarray}
   (5a) & = & \int \frac{d^d q}{ (2 \pi)^d } \int \frac{d \nu}{ 2 \pi} G( \vec{q}, \nu ) G( \vec{k} - \vec{q}, \omega- \nu )
                       \nonumber  \\
  & = & \int \frac{d^d q}{ (2 \pi)^d } \frac{1}{2q | \vec{k} -\vec{q} |  }
   \frac{ q + | \vec{k} -\vec{q} | }{ ( \omega + i c k_y )^2 +   ( q + | \vec{k} -\vec{q} | )^2 }
                        \nonumber  \\
  &= & \frac{u^2}{ 8 \pi^2 \epsilon } + \cdots
\end{eqnarray}
   where    $ \cdots $ means the UV finite parts.
   when one perform the frequency integral, one must pick two poles at the two opposite side of the frequency integral
   to get a non-vanishing answer. Putting $ c=0 $ gives back to the  relativistic case.
   Because $ c $ always appears in the combination of $ \omega + i c k_y $, so the UV divergency is identical to that of the
   $ c=0 $ case. One can get a similar expression in Fig.\ref{interactions}(b), (c) cases.
   So the $ \beta $ function $ \beta( u ) $ is identical to the $ c=0 $ case. This shows that the boost $ c $ is exactly marginal
   with always the combination $ \omega + i c k_y $ appearing in all the physical quantities ( see Fig.\ref{phasesz1} and Eq.\ref{scalingz1} ).

   Despite the critical exponent is the same as the $ c=0 $ case, various physical quantities at a finite $ T $  still depend on $ c $ as
   calculated in the following.

\subsection{ Finite temperature RG }

  Following the method developed in \cite{field2,field1,field3}, we can also study the RG at a finite temperature.
  The strategy is that even for a relativistic QFT at $ T=0 $, any finite temperature breaks the Lorentz invariance,
  so the imaginary time direction has to be treated separately from the space, the summation over the imaginary frequency
  has to be performed first before doing the dimensional regularization in the momentum space only.

\begin{figure}[!htb]
    \includegraphics[width=0.8\linewidth]{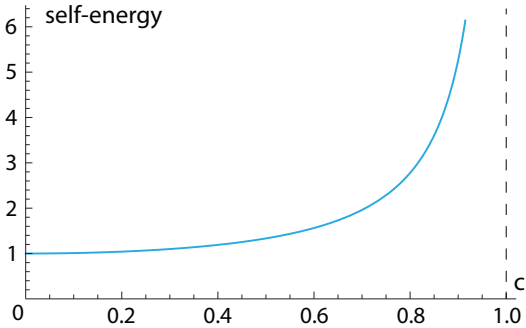}
    \caption{The boson self-energy at one-loop and at $ d_u=3 $ and a fixed temperature as a function of $ 0 < c < 1 $.
    It is a monotonically increasing function in $c$, and diverges as $1/(1-c)$ when $c\to 1$.  }
\label{CN}
\end{figure}

  Now we look at the boson self-energy at one-loop ( or the mass renormalization ) and a finite temperature Fig.\ref{selfenergy}a, Eq.\ref{3a} becomes:
\begin{align}
  & (4a) ( T )  =  u \int \frac{d^d q}{ (2 \pi)^d } \frac{1}{\beta} \sum_{i \nu_n }
   \frac{1}{2q}[ \frac{1}{ i \nu - \epsilon_{+}( \vec{q} ) } - \frac{1}{ i \nu + \epsilon_{-}( \vec{q} ) } ]   \nonumber  \\
   & =  u [ \int \frac{d^d q}{ (2 \pi)^d }\frac{1}{2q} + 2 \int \frac{d^d q}{ (2 \pi)^d }
    \frac{1}{2 q} \frac{1}{ e^{ \beta \epsilon_{+}( \vec{q} ) }-1 }  ]       \nonumber  \\
   & =   2 u (k_B T)^2 \int \frac{d^3 q}{ (2 \pi)^3 } \frac{1}{2 q}\frac{1}{ e^{q+ cq_y}-1 },~~~~~~ d=3
\label{3aT}
\end{align}
  where we evaluate the last line at the upper critical dimension $ d_u=3 $,
  drop the first term which is the $ T=0 $ result in Eq.\ref{3a}. So
  $ c $ does appear at any $ T > 0 $. Setting $ c=0 $ recovers the result when the sample is static
  $  u (k_B T)^2/2 \pi^2 \int^{\infty}_0 \frac{ x dx }{ e^x-1} = u (k_B T)^2/12 $.
  For a finite $c$, the numerical evaluation of the Eq.\ref{3aT} is shown in Fig.\ref{CN}.
  It is plotted in the unit of $u(k_BT)^2/12$.

  The interaction $ u $ is marginally irrelevant at $ d_u=3 $ and will lead to logarithmic corrections to
  Eq.\ref{3aT}.  For $ d=2 < d_u=3 $, the integral in Eq.\ref{3aT} is IR divergent, this is expected because the Gaussian fixed point flows to
  the Wilson-Fisher fixed point, then it goes to the $ d=2 $ scaling  analysis at Sec.VII.

  Fig.\ref{selfenergy}b  ( or the wavefunction renormalization ) can also be similarly evaluated at a finite $ T $,
  It is evaluated in Eq.\ref{UVfield} at $ T=0 $ where the pole structure in the six terms
  considerably simplify the final UV divergent answer. But at a  finite $ T $, all the six terms contribute due to the boson distribution factor.
  The  cancellation of the $ T=0 $ UV divergence in Eq.\ref{UVfield} by the counter-terms leads to a finite answer at $ T > 0 $.

\section{ The  Noether current and the charge -vortex duality along the path I}

  So far, we analyzed the effective action  Eq.\ref{expand} by mean field theory + Gaussian fluctuations.
  Here we will study it by the non-perturbative duality transformation.
  It was well known that there is a charge-vortex duality at $ z=1 $ case when the sample is static \cite{pq1,dual1,dual2}.
  The Mott insulating phase is due to the condensation of vortices from the SF side.
  The charge -vortex duality when the sample is moving provides a non-perturbative proof of $ z =1 $  from the Mott to SF transition
  along the path I when $ c < v_y $ tuned by $ r $. It also provides an exact proof that the boost $ c $ is exactly marginal
  with always the combination $ \omega + i c k_y $ appearing in all the physical quantities ( see Fig.\ref{phasesz1}b ).
  But one need to study the conserved Noether current first before
  investigating the charge vortex duality. In Sec.A, we will also pay special attentions to the higher derivative terms such as the $ a, b $ terms in Eq.\ref{expand}.

\subsection{  The conserved Noether current in the lab and co-moving }

   The global $ U(1) $ symmetry $ \psi \rightarrow \psi e^{ i \chi} $ leads to the conserved Noether current
   $ \tilde{J}_\mu=( J_\tau, J_x, \tilde{J}_y ) $ in Eq.\ref{boostz1}:
\begin{eqnarray}
   J_\tau & = & i( \psi^{*} \tilde{\partial}_\tau \psi- \psi \tilde{\partial}_\tau \psi^{*} )
                         \nonumber   \\
   & = &  i[ ( \psi^{*} \partial_\tau \psi- \psi \partial_\tau \psi^{*} ) -ic ( \psi^{*} \partial_y \psi- \psi \partial_y \psi^{*} ) ]
                          \nonumber   \\
   J_x & = &  i v^2_x ( \psi^{*} \partial_x \psi- \psi \partial_x \psi^{*} )
                          \nonumber   \\
   \tilde{J}_y & = &  i  v^2_y ( \psi^{*} \partial_y \psi- \psi \partial_y \psi^{*} )  -ic J_{\tau} = J_y-ic J_{\tau}
\label{threecurrents}
\end{eqnarray}
  which is odd under the C, but even under PT, so odd under the CPT \cite{currenttrick}.
  Both $ J_\tau $ and $ \tilde{J}_y $ contain the effects from the boost $ c $.

  It also show the current along the $ y $ direction  $ \tilde{J}_y $ is the sum of the intrinsic one $ J_y $ and the one due to the boost
  $ -ic J_{\tau} $. They satisfy
\begin{equation}
  \partial_\tau J_\tau + \partial_x J_x+  \partial_y (J_y - ic J_\tau) =0
\label{way2}
\end{equation}
  which is equivalent to Eq.\ref{way1}.  This equivalence gives the physical meaning of the particle-hole
  3-currents $ {J}_\mu=( J_\tau, J_x, J_y ) $ introduced in the charge-vortex duality.

  In short, there are two sets of conserved currents $ ( J_\tau, J_x, \tilde{J}_y ) $ in the lab frame and
  $ ( J_\tau, J_x, J_y ) $ in the co-moving frame with the sample. They are related by the GT listed in Eq.\ref{threecurrents} for $ z=1 $.
  In fact, as shown below, Eq.\ref{threecurrentsz2} for $ z=2 $  and  Eq.\ref{threecurrentsz1z2}
  for including both $ z=1 $ and $ z=2 $ are identical to Eq.\ref{threecurrents},
  this is because the GT is independent of $ z $ which is the nature consequence of the space-time GT. It turns out the latter set serves better in
  characterizing different phases.

{\sl 1. The Noether current in the Mott, SF and BSF phases }

  Now we evaluate the mean field 3-currents in all the three phases in Fig.\ref{phasesz1}.
  Plugging in $ \langle \psi \rangle=\sqrt{\rho_0} e^{i k_0 y } $ into Eq.\ref{threecurrents} leads to \cite{groundexcitation}:
\begin{eqnarray}
    J_\tau & = & i 2 k_0 c \rho_0,   \nonumber  \\
    J_x & = & 0    \nonumber  \\
    \tilde{J}_y & = & 2 k_0 \rho_0 ( c^2-v^2_y ) ,~~~J_y= -2 k_0 \rho_0 v^2_y
\label{z1current0}
\end{eqnarray}
   where the factor of $ i $ is due to the imaginary time $ \tau=i t $\cite{wrong}.
   Due to the vanishing of either $ \rho_0 $ or $ k_0 $ in the Mott and SF phase, the currents vanish in both phases.

   Near the $ z=2 $ line in Fig.\ref{phasesz1},   $ k_0= \pm \sqrt{ \frac{c^2-v^2_y}{2a} } $ listed in Eq.\ref{k0} for the BSF phase.
   As shown in Sec.V, due to the $ C $ symmetry breaking inside the BSF phase, one can only pick one of the $ \pm $.
   The Mott state with $ \rho_0=0 $ and the SF with $ k_0=0 $ carry nothing dictated by the $ C $ symmetry.
   The BSF near the $ z=2 $ line  starts to carry both the density
   $ J_{\tau} $ and the current  $ \tilde{J}_y $  which comes from both the particle and the hole.
   Note that the $ a $ term in Eq.\ref{boostz1} also contribute to the conserved current.
   So to evaluate the current in the BSF phase, one may need to consider the contributions from the higher order derivative terms.

 {\sl 2. The contribution from the higher derivative terms in the BSF phase }

   In general, any quantum field theory contains a high order derivative  such as
   $ L( \psi, \partial_\mu \psi, \partial_\mu  \partial_\nu \psi ) $, the high derivative term \cite{lovelock} also contributes to
   the conserved Noether current:
\begin{align}
    J^{a}_\mu & = \frac{ \partial {\cal L} }{ \partial ( \partial_\mu \psi_i ) } \frac{ \delta \psi_i }{ \delta \omega_a}
        +  \frac{ \partial {\cal L} }{ \partial (\partial_\mu \partial_\nu \psi_i) } \partial_\nu ( \frac{ \delta \psi_i }{ \delta \omega_a} )
         \nonumber  \\
       & -[\partial_\nu \frac{ \partial {\cal L} }{ \partial (\partial_\mu \partial_\nu \psi_i) }]  \frac{ \delta \psi_i }{ \delta \omega_a}
\label{currenthigh}
\end{align}
    where the sum over $ i $ is assumed.

    Near the $ z=2 $ line in Fig.\ref{phasesz1}, when adding back $ a $ term's contribution:
\begin{equation}
    J_y[a]= -i a  ( \psi^{*} \partial^3_y \psi- \psi \partial^3_y \psi^{*} )= - 2a k^3_0 \rho_0
\label{currentja}
\end{equation}
   one finds $ \tilde{J}_y =2a k^3_0 \rho_0,~~~J_y= - k_0 \rho_0 ( c^2 + v^2_y ) $.

   Near the $ z=(3/2,3) $ line in Fig.\ref{phasesz1}, $ k_0= \pm \sqrt{ \frac{c^2-v^2_y}{2b} } $ listed in Eq.\ref{rhighboost}.
   Again due to the $ C $ symmetry breaking inside the BSF phase, one can only pick one of the $ \pm $.
   The SF state  with $ k_0=0 $ carries nothing dictated by the $ C $ symmetry.
   However, the $ b $ term in Eq.\ref{boostz1} also contribute to the conserved current.
   When adding back $ b $ term's contribution:
\begin{equation}
    J_y[b]= i 2b | \partial_y \psi |^2  ( \psi^{*} \partial_y \psi- \psi \partial_y \psi^{*} )
\label{currentjb}
\end{equation}
   one finds $ \tilde{J}_y =0,~~~J_y= -2 k_0 \rho_0 c^2 $.


{\sl 3.  The Noether current  serve as the order parameter to  distinguish BSF from the SF }

   In short, the bi-linear currents Eq.\ref{threecurrents} which is odd under C, but even under PT ( so odd under CPT )
   can be taken as the order parameter to distinguish the SF from the BSF: in the former, the $ C $ is respected, the currents vanish.
   in the latter, the $ C $ is spontaneously broken, the currents shown in Eq.\ref{z1current0} does not vanish.
   More precisely, it is  $ J_y $ which acts as the order parameter to distinguish the BSF from the SF phase  in Fig.\ref{phasesz1}.
   In fact, it contains both the magnitude $ \rho_0 $ and the phase $ k_0 $.
   Of course, $ \psi $ remains the order parameter to distinguish the Mott from the SF or BSF.
   Also note that despite there is only one Goldstone mode in each phase, the symmetry breaking patterns of
   the two phases are different as shown in Eq.\ref{u1u1} respectively.

\subsection{ The charge-vortex duality  along the Path-I. }

 We will investigate the duality first in the boson picture, then from the dual vortex picture.
 Both lead to consistent results. Because we focus along the Path-I in Fig.\ref{phasesz1},
 we can ignore the higher order terms such as the $ a $ and $ b $ term in Eq.\ref{expand}.

{\sl 1. Duality transformation in the boson picture  }

  We start from the hard-spin representation Eq.\ref{boosttoright}:
\begin{equation}
    \mathcal{L}_b
	=(\partial_\tau\theta-ic\partial_y\theta)^2	+ v_x^2(\partial_x\theta)^2	+ v_y^2(\partial_y\theta)^2
\label{bosonhard}
\end{equation}
 where the angle $ \theta $ includes both the spin-wave and vortex excitation.
 To simplify the transformation, one can scale away $ v_x k_x \rightarrow k_x, v_y k_y \rightarrow k_y $, so $ c \rightarrow c/v_y < 1 $ near the Mott to the SF transition.

 To perform the duality transformation, one can decompose $ \theta \rightarrow \theta + \phi $ which stands for the
 the spin-wave and vortex respectively.
 Introducing the 3-currents $ J_\mu=( J_\tau, J_x,J_y ) $ to decouple the three quadratic terms leads to:
\begin{align}
    \mathcal{L}_b
	& = \frac{1}{2} J^2_\mu + i J_\tau (\partial_\tau \theta-ic\partial_y\theta  + \partial_\tau\phi -ic\partial_y \phi )	
                         \nonumber   \\
    & + i J_x (\partial_x\theta + \partial_x \phi )	+ i J_y( \partial_y\theta + \partial_y\phi )
\label{currents3}
\end{align}

 Then integrating out $ \theta $ leads to the conservation of the three current:
\begin{equation}
  (\partial_\tau- i c \partial_y ) J_\tau + \partial_x J_x+  \partial_y J_y=0
\label{way1}
\end{equation}
   which is equivalent to Eq.\ref{way2}. This equivalence shows that the 3-currents $ J_\mu=( J_\tau, J_x,J_y ) $
   is nothing but the ones listed in Eq.\ref{threecurrents} directly derived from the Noether theorem.

 According to Eq.\ref{way1} and \ref{way2}, there are two equivalent ways to proceed the duality transformation
 one is to introduce the three derivatives $ \tilde{\partial}_\mu=( \partial_\tau- i c \partial_y, \partial_x, \partial_y ) $
 in the co-moving frame with the sample,
 so Eq.\ref{way1} can be written as $ \tilde{\partial}_\mu J_\mu =0 $
 or  introduce the three current $ \tilde{J}_\mu=( J_\tau, J_x, J_y-i c J_\tau ) $ in the lab frame,
 so Eq.\ref{way2} can be written as $ \partial_\mu \tilde{J}_\mu =0 $.
 It turns out the first way in the co-moving frame with the sample is  more convenient, so we take it in the following.
 Note that the derivatives along the three directions still commute with each other in this co-moving frame,
 so Eq.\ref{way1} implies
\begin{equation}
     J_{\mu}= \epsilon_{\mu \nu \lambda} \tilde{\partial}_{\nu} a_{\lambda}
\end{equation}
  where $  a_{\lambda} $ is a non-compact $ U(1) $ gauge field.

  Then Eq.\ref{currents3} reduces to
\begin{equation}
    \mathcal{L}_v
	= \frac{1}{4} \tilde{f}^2_{\mu \nu} + i 2 \pi a_\mu \tilde{j}^{v}_{\mu}
\label{vortexcurrent}
\end{equation}
   where $ \tilde{f}_{\mu \nu}= \tilde{\partial}_\mu a_\nu- \tilde{\partial}_\nu a_\mu  $ is the gauge invariant field strength
   ( see below Eq.\ref{boostvortex} ) and
   $ \tilde{j}^{v}_{\mu}=  \frac{1}{2 \pi}  \epsilon_{\mu \nu \lambda} \tilde{\partial}_{\nu}  \tilde{\partial}_{\lambda} \phi  $ is the vortex current.

   Now we introducing the dual complex order parameter $ \psi_v $
   and considering $ \tilde{\partial}_\mu=( \partial_\tau- i c \partial_y, \partial_x, \partial_y ) $,
   Eq.\ref{vortexcurrent} can be written in terms of $ \psi_v $. It leads to Eq.\ref{boostvortex} where we explicitly wrote  $ \tilde{\partial}_\mu=( \partial_\tau- i c \partial_y, \partial_x, \partial_y ) $  out in the kinetic term,
   but only keep it implicitly in $ \tilde{f}_{\mu \nu} $.

{\sl 2. Duality transformation in the vortex picture }

   When the sample is static, one can perform the well known charge -vortex duality on Eq.\ref{z1}:
\begin{align}
    \mathcal{L}_{v}
	& =|(\partial_\tau -i a_{\tau} )\psi_v |^2 + v^2_{vx} |(\partial_x-i a_x)\psi_v|^2 + v^2_{vy} |(\partial_y-i a_y)\psi_v|^2  \nonumber  \\
	& + r_v|\psi_v|^2+ u_v |\psi_v |^4+ \frac{1}{4} f^2_{\mu \nu} + \cdots
\label{vortex}
\end{align}
    When $ r_v < 0 $, it is in the Mott phase $ \langle \psi_v \rangle \neq 0 $,
    $ r_v > 0 $, it is in the SF phase $ \langle \psi_v \rangle = 0 $.

    In addition to the emergent Lorentz invariance, the T, PH and P symmetry, the Global $ U(1) $ symmetry of the boson
    is promoted to the local ( gauge ) symmetry $ \psi_v \rightarrow \psi_v e^{ i \chi },  a_{\mu} \rightarrow  a_{\mu} + \partial_\mu \chi $.
    The gauge invariance is completely independent of the  emergent Lorentz invariance, it is also much more robust than the emergent Lorentz invariance in materials or AMO systems.

    Then going to the moving sample by  substituting $ \partial_\tau \rightarrow  \partial_\tau -ic\partial_y $
    into Eq.\ref{vortex} leads to:
\begin{align}
    \mathcal{L}_{v}
	& =(\partial_\tau -ic\partial_y -i a_{\tau} )\psi^*_v (\partial_\tau -ic\partial_y + i a_{\tau} )\psi_v
	+ |(\partial_x-i a_x)\psi_v|^2    \nonumber  \\
    & + |(\partial_y-i a_y)\psi_v|^2   + r_v|\psi_v|^2+ u_v |\psi_v |^4+ \frac{1}{4} \tilde{f}^2_{\mu \nu} + \cdots
\label{boostvortex}
\end{align}
 where $ \tilde{f}_{\mu \nu}= \tilde{\partial}_\mu a_\nu- \tilde{\partial}_\nu a_\mu  $.
  As stressed in \cite{wrong}, one can see the sign difference between the boost $ ic \partial_y $ and the time component of the gauge field
  $ a_\tau $ in the first term. This different  structure in the boost and gauge field could be important in the lattice version
  of Eq.\ref{boostvortex} to be discussed in the conclusion section.
  The boost also generalize the original $ U(1) $ gauge invariance
  $ \psi_v \rightarrow \psi_v e^{ i \chi },  a_{\mu} \rightarrow  a_{\mu} + \partial_\mu \chi $ in the lab frame to
  the $ \tilde{U}(1) $ gauge invariance  in the co-moving frame:
\begin{align}
   \psi_v \rightarrow \psi_v e^{ i \chi },~~~
  a_{\mu} \rightarrow  a_{\mu} + \tilde{\partial}_\mu \chi
\label{tildeu1}
\end{align}

    To perform  the duality transformation, it is convenient to get to
    the hard spin representation of Eq.\ref{boostvortex} ( For the notational convenience in the following,
    we replace $ a_{\mu} $ in Eq.\ref{vortex} and \ref{boostvortex} by $ A_{\mu} $ in Eq.\ref{vortexhard} ):
\begin{equation}
    \mathcal{L}_v
	=(\partial_\tau\theta-ic\partial_y\theta-A_\tau )^2	+ (\partial_x\theta- A_x )^2	+ (\partial_y\theta-A_y )^2
    + \frac{1}{4} \tilde{F}^2_{\mu \nu}
\label{vortexhard}
\end{equation}
 where the angle $ \theta $ includes both the spin-wave and vortex excitation.

 Following the similar procedures as done in the boson representation
 (1) decomposing $ \theta \rightarrow \theta + \phi $ which stands for
 the "spin-wave" and "vortex" respectively.
 (2) Introducing the 3-currents $ J_\mu=( J_\tau, J_x,J_y ) $ to decouple the three "quadratic"  terms in Eq.\ref{vortexhard}.
 (3) Integrating out $ \theta $ leads to the conservation of the three " boson" current: $ \tilde{\partial}_\mu J_\mu =0 $
  which implies $  J_{\mu}= \epsilon_{\mu \nu \lambda} \tilde{\partial}_{\nu} a_{\lambda} $
  where $  a_{\lambda} $ is a non-compact $ U(1) $ gauge field.

  Then we reach:
\begin{equation}
    \mathcal{L}_b
	= \frac{1}{4} \tilde{F}^2_{\mu \nu}-i A_{\mu} \epsilon_{\mu \nu \lambda} \tilde{\partial}_{\nu} a_{\lambda} + i 2 \pi a_\mu \tilde{j}^{v}_{\mu} +  \frac{1}{4} \tilde{f}^2_{\mu \nu}
\label{bosoncurrent}
\end{equation}
   where $ \tilde{f}_{\mu \nu}= \tilde{\partial}_\mu a_\nu- \tilde{\partial}_\nu a_\mu  $ is the $ \tilde{U}(1) $ gauge invariant field strength and
   $ \tilde{j}^{v}_{\mu}=  \frac{1}{2 \pi}  \epsilon_{\mu \nu \lambda} \tilde{\partial}_{\nu}  \tilde{\partial}_{\lambda} \phi  $ is the "vortex" current.

   Now integrating out $ A_{\mu} $ leads to a mass term for $ a_\nu $.
\begin{equation}
    \mathcal{L}_b
	=  i 2 \pi a_\mu \tilde{j}^{v}_{\mu} +  \frac{1}{4} \tilde{f}^2_{\mu \nu} + \frac{1}{2} ( a_\mu )^2
\label{gaugemass}
\end{equation}
    where the mass term makes the Maxwell term in-effective in the low energy limit.
    Note that this "vortex " current in the vortex representation is nothing but the original boson current in the boson representation.
    Now using $ \tilde{\partial}_\mu=( \partial_\tau- i c \partial_y, \partial_x, \partial_y ) $, it leads back to Eq.\ref{bosonhard}.
    After introducing the dual complex order parameter $ \psi $ which is nothing but the original boson
    leads back to Eq.\ref{boostz1}.

  It is easy to see Eq.\ref{vortex} has two sectors, the vortex degree of freedoms $ \psi_v $ and the gauge field $ a_{\mu} $.
  Then as shown in Eq.\ref{boostvortex}, going to a co-moving frame adds a boost to both sectors.
  So far, our boson-vortex duality is limited to the path I with $ c < v_y $ in Fig.\ref{phasesz1}, it would be interesting to push it to path II and path III.
  So the boost could trigger instabilities in the two sectors respectively.
  Note that the artificial gauge field here is different from the Electro-Magnetism (EM), the former's intrinsic velocity is $ v \ll c_l $, the latter is just the speed of light $ c_l $. So one can safely use the GT in the former,
  but must use the LT first in the latter, then keep upto the linear term in $ v/c_l $ when
  taking the small $ v/c_l $ limit as shown in Appendix F and G.
  It is worth to note that the boson-vortex duality transformation
  focus on only low energy sector, so the Higgs mode may not be seen in such a duality transformation. However,it was shown in Sec.IV,
  the Higgs mode is irrelevant anyway from the SF to BSF transition.
  It remains interesting to achieve Fig.\ref{phasesz1} from  the vortex representation.

\subsection{ Galileo transformation of the dual gauge field in the vortex picture }

  The $ \tilde{U}(1) $ gauge invariance  Eq.\ref{tildeu1} indicates that under the GT:
\begin{align}
  y^{\prime} & = y+ c t,~~~ t^{\prime}=t     \nonumber   \\
  \partial_\tau & = \partial_{\tau^{\prime}} -i c \partial_{y^{\prime}},~~~\partial_y=\partial_{y^{\prime}}
\label{GTy}
\end{align}
  where we define the imaginary time $ \tau=it $. If one define the dual gauge field as:
\begin{equation}
   \tilde{a}_0 = a_0 + i c a_y,~~~~  \tilde{a}_x= a_x,~~~~\tilde{a}_y= a_y
\label{GTdual}
\end{equation}
  Then the $ \tilde{U}(1) $ gauge invariance  Eq.\ref{tildeu1} can be implemented as
\begin{align}
   \psi_v \rightarrow \psi_v e^{ i \chi },~~~
  \tilde{a}_{\mu} \rightarrow  \tilde{a}_{\mu} + \partial_\mu \chi
\label{tildegauge}
\end{align}
  which means  $  \tilde{a}_{\mu}  $ just transforms as the $ U(1) $ dual gauge field.

  In fact, as to be shown in Appendix F, Eq.\ref{tildegauge} is nothing but the GT of the gauge field
  which, just like the currents,  is the nature consequence of the space-time GT.
  This can also be seen by looking how the $ \tilde{U}(1) $ field strength
  $ \tilde{f}_{\mu \nu}= \tilde{\partial}_\mu a_\nu- \tilde{\partial}_\nu a_\mu  $
  transform under the GT. By using the definition Eq.\ref{GTdual}, one can find
\begin{equation}
   \tilde{f}_{0 \alpha}=f_{0 \alpha} + ic f_{\alpha y},~~~\tilde{f}_{\alpha \beta }=f_{\alpha \beta }
\label{GTdualfield}
\end{equation}
  where $ f_{\mu \nu}= \partial_\mu \tilde{a}_\nu- \partial_\nu \tilde{a}_\mu  $ are expressed
  in terms of  $ \tilde{a}_\mu  $. The extra term  $ ic f_{\alpha y} $ is due to the fact the Maxwell term is not GI.
  It is well known that the Maxwell term is dual to the SF Goldstone mode \cite{cbtwo}, so the SF Goldstone mode is not GI either,
  consistent with the results achieved in Sec.II-B.

  At $ 2+1 $ d, $ f_{\mu \nu} $ can be expressed as $ E_{\alpha}=f_{0 \alpha}, B_z =  f_{xy} $. Then Eq.\ref{GTdualfield}
  can be expressed as:
\begin{equation}
   \tilde{E}_{x}=E_x + ic B_z,~~~ \tilde{E}_{y}=E_{y },~~~\tilde{B}_{z}=B_z
\label{GTdualEB}
\end{equation}
  which is nothing but how the artificial dual " EM "  field transforms under the GT boost in the Euclidean space-time.
  The extra term  $ ic B_z $ is due to the fact the Maxwell term is not GI. However, as shown in
  the appendix F, G, H, the Chern-Simon term is GI.

  The artificial dual gauge field here is described by Maxwell term. But its GT can also be applied to Chern-Simon
  gauge field in the bulk FQH and its associated edge properties in real time formalism in the appendix F,G,H.

\section{ Finite temperature properties and quantum critical scalings  }

\begin{figure}[tbhp]
\centering
\includegraphics[width=0.9 \linewidth]{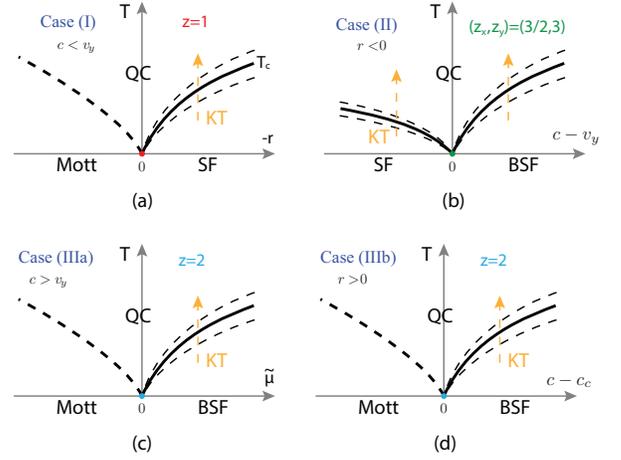}
\caption{  The finite temperature phase diagram along the path (I), (II) and (III)a and (III)b at $ d=2 $ in the Fig.\ref{phasesz1}.
 The SF density $ \rho_s \sim (-r)^{\nu}, \nu=0.67 $ across the Mott-SF transition with $ z=1 $ at $ T=0 $ in (a),
 $ \rho_s \sim |c-v_y | $ across the SF-BSF transition with $ z=(3/2,3) $  at $ T=0 $ in (b),
 then $ \rho_s \sim \tilde{\mu} \sim -( r -r_c )  $  or $ \rho_s \sim ( c -c_c )  $  upto logarithmic correction with $ z=2 $ at $ T=0 $ in (c) and (d) respectively. The solid line in the SF or BSF side represents the finite temperature KT transition.
 The two dashed lines sandwiching the KT line stands for the narrow window of classical fluctuation regimes which are
 squeezed to zero at the QCP at $ T=0 $. The dashed line
 in the Mott regime indicates the crossover from the Mott phase to the Quantum critical regime.  (a) and (c) are tuned by the chemical potential $ \mu $ (b) and (d)  are tuned by the boost. The interaction $ u $ is relevant near $ z=1 $ leading to the 3D XY class, marginally irrelevant
 near $ z=2 $ leading to logarithmic corrections. Despite $ z=(3/2,3) $ is a Gaussian fixed point with two leading irrelevant operators, it is the interaction which leads to the existence of the SF, BSF and the QPT between them.
 The Lyapunov exponent \cite{SY,kittalk,syk2} in the QC regime when the sample is moving is another important physical quantity to be explored in a future publication. As shown in the microscopic calculations Eq.\ref{Tcz1}, only (a) is reachable,
 the KT transition temperature $ T_c $ on the SF side, in fact, increases when the sample is moving as shown in Fig.\ref{phaseslattice}c.   }
\label{finiteT}
\end{figure}

 In the previous two sections, we perform the RG analysis and non-perturbation duality  transformation on the effective action Eq.\ref{boostz1}.
 Here, we derive the finite temperature effects and quantum critical scaling functions of various physical quantizes along the three paths
 in Fig.1 and Fig.2 when the sample is moving. Even we may not be able to get all the analytic expressions in some cases,
 we still stress the important effects of the boost $ c $ and compare to those when the sample is static.

\subsection{ Near $ z=1 $ QCP along the path I }

As shown in Sec.VI and VII, the Mott to the SF transition at $ 0 < c < v $ is in the same universality class as that when the sample is static,
namely in the 3D XY class with the critical exponent $ z=1, \nu=0.67, \eta=0.04   $.
The boost $ 0 < c < v  $ is exactly marginal
with always the combination $ \omega + i c k_y $ appearing in all the physical quantities ( see Fig.\ref{phasesz1}b ).
Armed with these facts,  one can write down the scaling functions for the Retarded single particle Green function,
the Retarded density-density correlation function, compressibility and the specific heat in the Mott side near the QCP in Fig.\ref{finiteT}a:
\begin{eqnarray}
   &  & G^R ( \vec{k}, \omega )  =
   {\cal A} ( \frac{\hbar \sqrt{v_x v_y} }{k_B T} )^2  ( \frac{ k_B T }{\Delta } )^{\eta} \Psi ( \frac{ \hbar \tilde{\omega} }{ k_B T}, \frac{ \hbar \tilde{k} }{  k_B T }, \frac{ \Delta }{ k_B T } )
     \nonumber  \\
   &  & \chi^R ( \vec{k}, \omega )  =
   \frac{ k_B T }{ \hbar v_x v_y } \Phi ( \frac{ \hbar \tilde{\omega} }{ k_B T}, \frac{ \hbar \tilde{k}  }{  k_B T  }, \frac{ \Delta }{ k_B T } )
    \nonumber  \\
   &   & \kappa  =  \frac{ k_B T }{ \hbar^2 v_x v_y } A ( \frac{ \Delta }{ k_B T } ),
    ~~~~~~~ C_v =  \frac{ T^2}{ \hbar^2 v_x v_y  }  B (  \frac{ \Delta }{ k_B T } )
\label{scalingz1}
\end{eqnarray}
 where $ {\cal A } \sim  r^{ \eta \nu} $ is the single particle residue,  $ \Delta \sim r^{ \nu} $ is the
 Mott gap inside the Mott phase, $ \tilde{\omega}=\omega - c k_y $ is the Doppler shifted frequency when the sample is moving,
 $ \tilde{k}= \sqrt{ v^2_x k^2_x +  v^2_y k^2_y } $ is the scaled momentum and
 the compressibility is defined by $ \kappa= \lim_{\vec{k} \rightarrow 0 }  \chi^R ( \vec{k}, \omega=0 ) $.
 From the two conserved quantities, one can also form the Wilson ratio $ W=\kappa T/C_v $.

 In the SF side near the QCP,  $ \frac{ \Delta }{ k_B T } $ need to be replaced by $ \frac{ \rho_s }{ k_B T } $
 where $ \rho_s \sim  | r |^{ ( d+z-2 ) \nu} \sim |r |^{\nu} $
 is the SF density. As shown in Fig.\ref{finiteT}a and Sec.V-B-3, all the scaling functions in the SF side
 will run into singularity at a finite $ T=T_{KT} $ signaling the finite KT transition.
 In the narrow window near the KT transition, the scaling function reduce to those of classical KT transition
 shown in Eq.\ref{KT}.

From the effective action Eq.\ref{boostz1}, one can evaluate the three scaling functions explicitly.
From the three currents Eq.\ref{threecurrents}, following the method developed in \cite{lightatom2},
one can also evaluate the retarded density-density correlation function $ \kappa^R ( \vec{k}, \omega_n ) $.
Note that the scaling functions $ \Psi $ and $ \Phi $ are identical to those when the sample is static with  $ c=0 $
studied in  Ref.\cite{lightatom1,lightatom2}.
The $ c $ dependence is absorbed into the scaling variable $ \tilde{\omega}=\omega - c k_y $.
However, the scaling functions $ A $ and $ B $ do depend on $ c $ explicitly due to the frequency summation
and the momentum  integral ( see Eq.\ref{3aT} ).

From Eq.\ref{boostz1}, one can identify the single particle ( boson ) Green function in $ ( \vec{k}, \omega_n ) $ space in
the Mott side $ r > 0 $:
\begin{eqnarray}
   G_0 ( \vec{k}, \omega_n ) & = & \frac{1}{ -( i \omega_n -c k_y )^2 + \tilde{k}^2  + r }
                         \nonumber  \\
    & = &  \frac{-1}{ ( i \omega_n - \epsilon_{+}( \vec{k} ) )( i \omega_n +  \epsilon_{-}( \vec{k} ) ) }
\label{propz1}
\end{eqnarray}
   where $ \epsilon_{\pm}( \vec{k} ) = \sqrt{ \tilde{k}^2 + r } \pm c k_y  > 0 $ well inside the Mott phase $ r > 0 $.

From Eq.\ref{boostz1}, one can also get the free energy density well inside the Mott:
\begin{equation}
   f= 2 k_B T \int \frac{ d^d \vec{k} }{ (2\pi)^d } \log( 1-e^{- \beta \epsilon_{+}( \vec{k} ) } )
   + \int \frac{ d^d \vec{k} }{ (2\pi)^d }\epsilon_{+}( \vec{k} )
\label{freez1}
\end{equation}
  where we have used the $ C $ symmetry  $ \epsilon_{+}( \vec{k} )= \epsilon_{-}( -\vec{k} ) $ to
  get rid of the hole excitation spectrum in favor or that of the particle. The last term is the ground state energy at $ T=0 $.

  From the free energy, one can immediately evaluate the specific heat:
  From Eq.\ref{boostz1}, one can also get the free energy well inside the Mott:
\begin{equation}
   C_v= - T \frac{ \partial^2 f }{ \partial T^2} =
   \int \frac{ d^d \vec{k} }{ (2\pi)^d } \frac{ e^{ \beta \epsilon_{+}( \vec{k} ) }} { ( e^{ \beta \epsilon_{+}( \vec{k} )}-1)^2 }
   ( \frac{ \epsilon_{+}( \vec{k} ) }{k_B T } )^2
\label{specz1}
\end{equation}
  Plugging in the $ \epsilon_{+}( \vec{k} ) $ in the Mott phase leads to  $ C_v (T) $ in Eq.\ref{scalingz1} well inside
  the Mott side where one can just use the mean field theory $ \Delta= r $.
  Eq.\ref{specz1} breaks down near the QCP.
  However, one can apply the simple scaling analysis $ C_v \sim T^{d/z} \sim T^2 $ for the $ z=1 $ QCP.
  As explained below Eq.\ref{scalingz1}, the coefficient does depend on $ c $.

   The dynamic density-density response function is
\begin{equation}
    \chi(\vec{k},i\omega_n)
    =-T\sum_{\vec{q}, i \nu_m } G_0( \vec{k}+ \vec{q}, i\omega_n+i\nu_m ) G_0( \vec{q}, i \nu_m )
\end{equation}
  which can be similarly evaluated. So the Wilson ratio can also be obtained.

  One can compute the momentum carried by the quasi-particle:
\begin{equation}
   \frac{\vec{P}_{p} }{V}= \int \frac{ d^d \vec{k} }{ (2\pi)^d } \hbar \vec{k} f( \epsilon_{+}( \vec{k} ) )
\label{driftp}
\end{equation}
  As demonstrated in Eq.\ref{vxvytc} in Sec.VIII-B, the drift velocity $ \vec{c} $ is quite small, so Eq.\ref{driftp} can be simplified to:
\begin{equation}
   \frac{\vec{P}_{p} }{V}= \vec{c} \frac{ \hbar^2 }{ 6 \pi^2 } \int^{\infty}_{0} k^4
   [ - \frac{ \partial f( \epsilon ( \vec{k} ) ) }{ \partial \epsilon ( \vec{k} ) }] \sim  \vec{c} T^4
\label{driftSF}
\end{equation}
  inside a SF phase. It will be exponentially suppressed inside the Mott phase due to its gap.

  One can compute the momentum carried by the quasi-hole:
\begin{align}
   \frac{\vec{P}_{h} }{V}  = - \int \frac{ d^d \vec{k} }{ (2\pi)^d } \hbar \vec{k} f( -\epsilon_{-}( \vec{k} ) )
               =  \int \frac{ d^d \vec{k} }{ (2\pi)^d } \hbar \vec{k} f( -\epsilon_{+}( \vec{k} ) )
\label{drifth}
\end{align}

   Adding the contributions from both the particle and hole lead to
\begin{align}
   \frac{ \vec{P}_{h} + \vec{P}_{h} }{V} & = - \int \frac{ d^d \vec{k} }{ (2\pi)^d } \hbar \vec{k}
       [  f( \epsilon_{+}( \vec{k} ) ) +  f( -\epsilon_{+}( \vec{k} ) ) ]
                                             \nonumber   \\
              & = - \int \frac{ d^d \vec{k} }{ (2\pi)^d } \hbar \vec{k} =0
\label{driftph}
\end{align}
  which tells the P and H carry just opposite momentum,
  therefore the  total momentum  $ \vec{P} =\vec{P}_{h} + \vec{P}_{h}=0 $ just vanishes at any temperature.
  However, due to their opposite charges,  they contribute equally to the current in Eq.\ref{threecurrents}.

\subsection{  Near $ ( z_x=3/2, z_y=3 ) $ QCP along the path II  }

Here the fundamental degree of freedoms is the phase $ \phi $ in Eq.\ref{boosttoright} of the boson $ \psi= \sqrt{\rho_0} e^{i \phi } $.
If dropping the leading irrelevant operators $ [Z]=[b]=-2 $ and neglecting the vortex excitations,  it becomes a Gaussian theory.
So we first calculate the phase-phase correlation function, then the boson-boson correlation functions.
In contrast to the $ z=1 $ case discussed in the last subsection, this QPT only happens when the sample is moving, no $ c=0 $ analog.

{\sl 1. Correlation functions in the Quantum regimes}

 On both sides, the phase-phase correlation function:
\begin{equation}
   \langle \phi( -\vec{k}, -\omega_n) \phi (\vec{k}, \omega_n) \rangle
    = \frac{-1}{ ( i \omega_n - \epsilon_{+}( \vec{k} ) )( i \omega_n +  \epsilon_{-}( \vec{k} ) ) }
\label{propz1phase}
\end{equation}
   where $ \epsilon_{\pm}( \vec{k} ) = \sqrt{v^2_x k^2_x  + v^2_y k^2_y + a k^4_y } \pm c k_y  > 0 $,
   in the SF $ v_y > c $, at the QCP $ v_y = c $,  $ v^2_y \to 3 c^2- 2  v^2_y $ in the BSF phase $ v_y < c $.
   $ \epsilon_{+}( \vec{k} ) + \epsilon_{-}( \vec{k} )= 2 \epsilon( \vec{k} )  = 2 \sqrt{v^2_x k^2_x  +
   v^2_y k^2_y + a k^4_y }  $ is an even function of $ \vec{k} $ where $ c $ drops out.

   One can find the single particle ( boson ) Green function:
\begin{eqnarray}
   G ( \vec{x}, \tau ) & = & \langle \psi(  \vec{x}, \tau  ) \psi^{*} (0,0) \rangle  =  \rho_0 e^{- g( \vec{x}, \tau  ) }
       \nonumber  \\
   g( \vec{x}, \tau  ) & = & \frac{1}{\beta} \sum_{ i \omega_n }  \int \frac{ d^d \vec{k} }{ (2\pi)^d }
     \frac{ 1- e^{i ( \vec{k} \cdot \vec{x} - \omega_n \tau ) } }{ 2 \epsilon( \vec{k} ) }
              \nonumber  \\
    & \times & [\frac{1}{ i \omega_n - \epsilon_{+}( \vec{k} ) } - \frac{1}{ i \omega_n +  \epsilon_{-}( \vec{k} ) } ]
\label{gGSF}
\end{eqnarray}
   The frequency  integral can be done first by paying the special attention to $ \tau=0=\beta $ at any finite $ T $:
\begin{eqnarray}
   g( \vec{x}, \tau  ) & = &  \int \frac{ d^d \vec{k} }{ (2\pi)^d } \frac{1} { 2 \epsilon( \vec{k} ) }
    [ \frac{  e^{i  \vec{k} \cdot \vec{x} - \epsilon_{+}( \vec{k} ) \tau  } - e^{ - \beta \epsilon_{+}( \vec{k} ) }  }
      { 1- e^{ - \beta \epsilon_{+}( \vec{k} ) } }
                              \nonumber  \\
      &- & \frac{   e^{ \beta \epsilon_{-}( \vec{k} ) }
      -  e^{i  \vec{k} \cdot \vec{x} + \epsilon_{-}( \vec{k} ) \tau  } }   { e^{ \beta \epsilon_{-}( \vec{k} ) }-1 } ]
\label{gany}
\end{eqnarray}
 where one can use the $ C $ symmetry   $ \epsilon_{+}( \vec{k} )=\epsilon_{-}( -\vec{k} ) $ to get an expression only
 in terms of  $ \epsilon_{+}( \vec{k} ) $.

 Now we look at several special cases:

 (1)   Putting $ T=0 $ leads to
\begin{eqnarray}
   g( \vec{x}, \tau  )  =  \int \frac{ d^d \vec{k} }{ (2\pi)^d }
   \frac{ e^{i \vec{k} \cdot \vec{x} - \epsilon_{+}( \vec{k} ) \tau  } -1 } { 2 \epsilon( \vec{k} ) }
\end{eqnarray}

   Its equal time at $ d=2 $ is $ g(\vec{x}, 0) =\int \frac{ d^2 \vec{k} }{ (2\pi)^2 }
   \frac{ e^{i \vec{k} \cdot \vec{x} } -1 } { 2 \epsilon( \vec{k} ) }\sim 1/ |x| $  is independent of $ c $,
   $ G( \vec{x},0 ) \sim \rho_0 e^{ -1/|\vec{x}| } $.
   but its auto- correlation ( equal-space ) $ g( 0, \tau  )  =  \int \frac{ d^2 \vec{k} }{ (2\pi)^2 }
   \frac{ e^{ - \epsilon_{+}( \vec{k} ) \tau  } -1 } { 2 \epsilon( \vec{k} ) } $ does depend on $ c $.

 (2 )   Putting  equal-time $ \tau=\beta $ in Eq.\ref{gany} leads to the equal time :
\begin{eqnarray}
   g( \vec{x}, 0  )  =  \int \frac{ d^d \vec{k} }{ (2\pi)^d } \frac{  e^{i  \vec{k} \cdot \vec{x} } - 1} { 2 \epsilon( \vec{k} ) }
    [ \frac{1} { e^{ \beta \epsilon_{+}( \vec{k} ) } -1 }- \frac{1} { e^{ -\beta \epsilon_{-}( \vec{k} ) } -1 } ]
\label{ganyx}
\end{eqnarray}
  Putting  $  \epsilon_{+}( \vec{k} )= \epsilon_{-}( \vec{k} )= \epsilon( \vec{k} ) $, one recovers the
  well-known $ c=0 $ result $ g( \vec{x}, 0  )  =  \int \frac{ d^d \vec{k} }{ (2\pi)^d } \frac{  e^{i  \vec{k} \cdot \vec{x} } - 1} { 2 \epsilon( \vec{k} ) } \coth \frac{ \beta \epsilon( \vec{k} ) }{2}  $.

  In the BSF phase, Eq.\ref{gGSF} should be replaced by
\begin{equation}
  G ( \vec{x}, \tau )  =  \langle \psi(  \vec{x}, \tau  ) \psi^{*} (0,0) \rangle
  =  \rho_0 e^{ i k_0 y } e^{- g( \vec{x}, \tau  ) }
\label{modquan}
\end{equation}
  where in the $ g( \vec{x}, \tau ) $, one just replaces $ v^2_y \rightarrow 3c^2- 2 v^2_y $. It is the modulation
  $ e^{ i k_0 y } $ in the correlation function which distinguishes the BSF from the SF.

{\sl 2. The thermodynamic quantities in the quantum regimes }

  Applying Eq.\ref{specz1} to the SF or BSF side, one can find the specific heat:
\begin{equation}
   C_v= \frac{T^2}{v_x v_y } \int  \frac{ d^2 \vec{k} }{ (2\pi)^2 }  \frac{ e^{ k+ \alpha k_y}  ( k+ \alpha k_y)^2 }
   {  ( e^{ k+ \alpha k_y} - 1 )^2 }= f(\alpha) T^2
 \label{specz23}
\end{equation}
  where $ \alpha= c/v_y < 1 $  in the SF side and $ \alpha= c/\sqrt{3c^2-2 v^2_y} < 1 $ in the BSF side.
  This is consistent with the scaling 
 $   C_v \sim T^{d/z} \sim f(\alpha) T^2 $ with $ z= 1 $.

  At the QCP $ f( \alpha=1 ) $  diverges where we find
\begin{align}
   & C_v  = \frac{T}{v_x v_y } \int^{\infty}_{- \infty}  \frac{ d k_x }{ 2\pi  } \int^{\infty}_{0}  \frac{ d k_y }{ 2\pi  }
    \frac{ e^{ g(k_x, k_y, \alpha ) } g^2(k_x, k_y, \alpha )  }
   {  ( e^{ g(k_x, k_y, \alpha ) } - 1 )^2 } + O ( T^2 )
           \nonumber \\
   & g(k_x, k_y, \alpha )= \frac{ k^2_x + \alpha k^4_y }{ 2 k_y }
 \label{specz23QCP}
 \end{align}
  where the integral over $ k_y $ is only half of the line, the other half contributes to the subleading $ T^2 $ term.
  It is consistent with the scaling 
\begin{equation}  
   C_v \sim T ^{1/z_x + 1/z_y } \sim T 
\label{specz323}
\end{equation}    
    with $ ( z_x=3/2, z_y=3 ) $ at the QCP.

  The superfluid density near the SF to BSF transition scales as:
\begin{equation}
    \rho_s \sim v_x \gamma \sim | c-v_y | \sim | \alpha-1 |
\label{rhos}
\end{equation}
  So all the scaling functions can be written in terms of $ \rho_s/k_BT $ on both sides.

{\sl 3. In the classical regime near the KT transition: the $ \hbar \to 0 $ limit }

So far, we only look at the quantum effects of the boost $ \partial_{\tau} \rightarrow \partial_{\tau} -ic \partial_{y} $.
Now we look at its classical effects originating from such a substitution, namely, the $ \hbar \to 0 $ limit.
At a finite temperature, setting the quantum fluctuations ( the $ \partial_{\tau} $ term ) vanishing, in Eq.\ref{s2cant} or Eq.\ref{s2ic},
then both equations reduce to
 \begin{align}
    \mathcal{S}_{KT}
	=\frac{1}{k_B T} \int d^2r
	 [ v_x^2(\partial_x\phi)^2 + \gamma^2 (\partial_y\phi)^2 + a (\partial^2_y\phi)^2 ]
\label{KT}
\end{align}
 where $ \gamma^2= v^2_y- c^2 $ inside the SF phase and  $ \gamma^2=2(c^2-v_y^2) $ inside the BSF phase.
 By setting the $ \partial_{\tau} $ term  vanishing, the crucial crossing metric  terms also vanish. This facts suggest that
 the effects of boosts is mainly quantum effects, but still have important classical remanent effects encoded in the
 coefficient $ \gamma $ in Eq.\ref{KT}.
 It indicates the finite temperature phase transition is still in Kosterlize-Thouless (KT) universality class
 with a reduced $ T_{KT} \sim \rho_s \sim | c-v_y | $ shown in Eq.\ref{rhos}.


  The classical boson correlation functions in the narrow window around the classical KT transition line show algebraic decay order,
  just like those  in the classical KT transition.
\begin{eqnarray}
    G ( \vec{x} ) & = & \langle \psi(  \vec{x}  ) \psi^{*} (0 ) \rangle  =  \rho_0 e^{- g( \vec{x} ) } = ( |x|/a )^{- \frac{ T}{ \rho_s } }
       \nonumber  \\
    g ( \vec{x} ) &= &  k_B T \int \frac{ d^d \vec{k} }{ (2\pi)^d }
    \frac{ e^{i \vec{k} \cdot \vec{x} } -1 } { v^2_x k^2_x + \gamma^2 k^2_y } \sim  \frac{ T}{ \rho_s } \ln |x|/a
\end{eqnarray}
   where $ a $ is the lattice constant ( not confused with the higher order coefficient $ a $ in Eq.\ref{KT} ).
   It can be contrasted to the quantum regimes listed in Eq.\ref{gGSF} and \ref{gany}.
   As shown in Eq.\ref{modquan}, inside the KT transition above the BSF phase has the modulation factor $ e^{i k_0 y } $.

   Of course, the classical regime around the classical KT transition line squeezes to zero at the QCP $ \gamma=0 $ as shown in Fig.\ref{finiteT}.

\subsection{ Near the $ z=2 $ QCP along the path III }

 As shown in the previous sections and Fig.\ref{phasesz1}b,
 the $ k_0 $ of BSF is also exactly marginal which stands for the ordering wavevector in the BSF,
 so the $ z=2 $ line is a line of fixed points all in 2d zero density SF-Mott transition class with the exact critical exponent $ z=1, \nu=1/2, \eta=0 $ subject to logarithmic corrections at the upper critical dimension $ d=2 $.
 The RG flow is along the constant contour of $ k_0 $. Armed with these facts, one can write down
 the scaling functions for the Retarded single particle Green function,
 the Retarded density-density correlation function, compressibility and the specific heat near the QCP in Fig.\ref{finiteT}c, d:
\begin{eqnarray}
   & & G^R ( \vec{k}, \omega_n )  =
   e^{ i k_0 y } \frac{\hbar}{ k_B T} \Psi ( \frac{ \hbar  Z_1 \omega  }{ k_B T}, \frac{ \hbar \tilde{k}  }{ \sqrt{ k_B T} },
    \frac{ \tilde{\mu} }{ k_B T } )
     \nonumber  \\
   &  & \chi^R ( \vec{k}, \omega_n )  =
   \frac{ 1 }{ \hbar v_x \tilde{v}_y }
   \Phi ( \frac{ \hbar  Z_1 \omega  }{ k_B T}, \frac{ \hbar \tilde{k}  }{ \sqrt{ k_B T} },  \frac{ \tilde{\mu} }{ k_B T } )
    \nonumber  \\
   & & \kappa  =  \frac{1}{ \hbar v_x \tilde{v}_y } A ( \frac{ \tilde{\mu} }{ k_B T } ),
   ~~~~~~ C_v  =  \frac{ T }{ v_x \tilde{v}_y}  B (  \frac{ \tilde{\mu} }{ k_B T } )
\label{scalingz2}
\end{eqnarray}
  where $ Z_1 \omega $ is the scaled frequency
  with $ Z_1=-2ck_0 $, $ \tilde{k}= \sqrt{ v^2_x k^2_x + \tilde{v}^2_y  k^2_y } $ is the scaled momentum
  with $ \tilde{v}^2_y= 2 (c^2-v^2_y) $,
  the effective chemical potential $ \tilde{\mu}=-( r-r_c), r_c=a k^4_0 >0  $ is listed in Eq.\ref{tildemuz2}.
  The fact that $ c $ is exactly marginal, so does $ k_0 $ is reflected in the arguments of the scaling functions.
  The characteristic frequency scales as $ \Delta \sim \tilde{\mu}^{z\nu} \sim  \tilde{\mu} $ upto some logarithmic correction.
  Note the dramatic changes from the $ z=1 $ scaling sets in Eq.\ref{scalingz1}  to the $ z=2 $ scaling sets in Eq.\ref{scalingz2}.
 Again, the scaling functions $ \Psi $ and $ \Phi $ are identical to that listed in Ref.\cite{z2,lightatom1,lightatom2}.
 The $ c $ dependence is absorbed into the 3 scaling variables $ Z_1\omega, \tilde{k}  $ and $ \tilde{\mu} $.
 However, the scaling functions $ A $ and $ B $ do depend on $ c $ explicitly due to the frequency summation
 and the momentum  integral.

 In the BSF side near the QCP,  $ \frac{ \tilde{\mu} }{ k_B T } $ need to be replaced by $ \frac{ \tilde{\rho}_s }{ k_B T } $ where $ \tilde{\rho}_s \sim \tilde{\mu}^{ (d+z-2) \nu } \sim  \tilde{\mu}=-(r -r_c) $ upto some logarithmic correction
 is the SF density in the BSF. As shown in Fig.\ref{finiteT}c,d, all the scaling functions in the BSF side
 runs into a singularity at a finite $ T=T_{KT} $ signaling the finite KT transition.
 In the narrow window near the KT transition, the scaling function reduce to those of classical KT transition \cite{CFT12} shown in Eq.\ref{KT}.

  Note that it is the effective chemical potential $ \tilde{\mu}=-( r-r_c), r_c=a k^4_0 >0  $ listed in Eq.\ref{tildemuz2} which
  tunes the Mott to the BSF transition. So one can either tune the bare mass $ r $ or the boost velocity $ c $
  to tune the transition ( Fig.\ref{finiteT}c or Fig.\ref{finiteT}d ) respectively. Especially, at a fixed
  $ r > 0 $ inside the Mott state, one can tune it into the BSF just by increasing the boost velocity.
  The energy come from boosting the sample. The effects of the dangerously irrelevant $ Z_2 $ ( the metric crossing term ) has not been considered in the scaling function near the QCP,
  but become important inside the two phases as shown in Sec.IV.

  In summary, one can see the importance of the metric crossing term which represents the
  new space-time structure emerging from the QPT: it is marginal, dominant and irrelevant
  near the $ z=1 $, $ z=(3/2, 3 ) $ and $ z=2 $ QCP respectively. The first has the $ c=0 $ limit when the sample is static, the latter two do not have, so
  only happen when the sample is moving.
  The interaction $ u $ is relevant and marginally irrelevant
  near the $ z=1 $ and $ z=2 $ respectively. Of course, despite the SF to BSF transition with $ z=(3/2, 3 ) $ is
  a Gaussian one, it is the interaction which leads to the very existence of the SF, BSF
  and the QPT between the two.

\section{ The effective phase diagram of $ z=2 $ when the sample is moving}


  In this section, we study the $ z=2 $ SF-Mott transitions in a 
  moving sample and contrast to the $ z=1 $ ones addressed in the previous sections.
  We also analyze the intrinsic relations between the Galileo transformation
  and the global $ U(1) $ symmetry breaking in the SF phase where the number is not conserved anymore.


  At integer fillings and in the absence of the $ C $ ( or PH ) symmetry, the SF-Mott transition in Eq.\ref{boson} when the sample is static
  ( Fig.\ref{frames}a and Fig.\ref{phaseslattice} )
  can be described by the $ z=2 $ effective action
\begin{align}
	\mathcal{S}_{L,z=2}& =\int d\tau d^2r
	[Z_1 \psi^* \partial_\tau \psi +v_x^2|\partial_x\psi|^2+v_y^2|\partial_y\psi|^2	
                   \nonumber  \\
    & -\mu_p |\psi|^2+ u |\psi|^4 + \cdots ]
\label{z2}
\end{align}
 where  the space-time is related by  $ z=2 $, the chemical potential $ \mu_p $ tunes the SF-Mott transition,
 $ \mu_p < 0,  \langle \psi \rangle =0 $ is in the Mott state which respects the $ U(1) $ symmetry,
 $ \mu_p > 0, \langle \psi \rangle \neq 0 $ is in the SF state which breaks the $ U(1) $ symmetry.
 It is related to the microscopic parameters in Eq.\ref{boson} by $ \mu_p \sim  t/U- (t/U)_c  $ at a fixed chemical potential $ \mu_p $
 or $ \mu_p \sim \mu- \mu_c $ at a fixed  $ t/U $ ( Fig.\ref{phaseslattice} ).
 In the following, for the notational simplicity, we set $ Z_1=1 $ and also drops the subscript $ p $
 ( but still keeps in mind its difference than the bare chemical potential in the boson Hubbard model Eq.1 ) .
 We will put back $ Z_1 $ in the Sec.VI-B where we also consider the irrelevant $ |\partial_\tau \psi|^2 $ term.

\subsection{ Emergent Galileo invariance }

 Eq.\ref{z2} explicitly breaks the $ C $ symmetry, but has the $ P $ and $ T $ symmetry ( therefore no such thing like CPT
 as in the $ z=1 $ emergent pseudo-Lorentz theory ).
 It has an emergent Galileo invariance. So we expect performing a Galileo transformation to a moving sample should not change its form.
 This is indeed the case as demonstrated in the following.
 Performing the Galileo transformation described in Sec.I leads to the following effective action when the sample is moving ( Fig.\ref{frames}b )
 ( again, for the notational simplicity, we drop the $ \prime $ when the sample is moving  ):
\begin{align}
	\mathcal{S}_{M,z=2} & =\int d\tau d^2r
	[\psi^*(\partial_\tau-ic\partial_y)\psi
	+v_x^2|\partial_x\psi|^2+v_y^2|\partial_y\psi|^2
                     \nonumber  \\
	& -\mu|\psi|^2+u|\psi|^4 + \cdots ]
\label{boostz2}
\end{align}
 which breaks the $ P $ and $ T $ symmetry separately, but keeps its combination.

 The mean field ansatz $\psi=\sqrt{\rho}e^{i(\phi+k_0y)}$ leads to the energy density
\begin{align}
	E[\rho,k_0]
	=(c k_0+v_y^2k_0^2-\mu)\rho+ u \rho^2
\end{align}

  Minimizing $E[\rho,k_0]$ with respect to $\rho$ and $k_0$ results in
\begin{align}
	k_0=-\frac{c}{2v_y^2},~~~~~~
	\rho=
	\begin{cases}
	    0,&\mu<-\frac{c^2}{4v_y^2}\\
	    \frac{\mu}{2u}+\frac{c^2}{8 u v_y^2},&\mu>-\frac{c^2}{4v_y^2}\\
	\end{cases}
\label{z2k0}
\end{align}
   It is easy to see that due to the explicit C- symmetry breaking of $ z=2 $ action Eq.\ref{boostz2}, the sign of $ k_0 $
   is automatically given. This is in sharp contrast to  in Eq.\ref{k0} and Eq.\ref{rhighboost} in the $ z=1 $ case
   where the sign of $ k_0 $  is determined by the spontaneous $ C $-symmetry breaking.

 It is convenient to introduce the new order parameter $ \psi=\tilde{\psi}e^{ik_0y} $,
 then $\partial_y\psi=e^{ik_0y}(\partial_y+ik_0)\tilde{\psi}$ and
\begin{align}
	\mathcal{L} & =\tilde{\psi}^*\partial_\tau\tilde{\psi}
	-i(c+2k_0v_y^2)\tilde{\psi}^*\partial_y\tilde{\psi}
	+v_x^2|\partial_x\tilde{\psi}|^2
	+v_y^2|\partial_y\tilde{\psi}|^2
                      \nonumber  \\
	& -(\mu-ck_0-v_y^2k_0^2)|\tilde{\psi}|^2+u|\tilde{\psi}|^4 + \cdots
\end{align}

Setting $i(c+2k_0v_y^2)\tilde{\psi}^*\partial_y\tilde{\psi}=0$ leads to $k_0=-\frac{c}{2v_y^2}$ and
\begin{align}
	\mathcal{L}& =\tilde{\psi}^*\partial_\tau\tilde{\psi}
	+v_x^2|\partial_x\tilde{\psi}|^2
	+v_y^2|\partial_y\tilde{\psi}|^2   -(\mu+v_y^2k_0^2)|\tilde{\psi}|^2
     \nonumber  \\
    &+ u |\tilde{\psi}|^4+ \cdots
\label{z2prime}
\end{align}
which as expected, remains the same as original $z=2$ theory when the sample is static
after identifying the chemical potential when the sample is moving
$ \tilde{\mu}=\mu+v_y^2k_0^2$. After identifying $ v^2_y=1/2m $ ( which is not a dimension of velocity anymore as in the $ z=1 $ case ), then $ \tilde{\mu}=\mu+v_y^2k_0^2 = \mu + \frac{m c^2}{2} $.

\begin{figure}[tbhp]
\centering
\includegraphics[width=0.8 \linewidth]{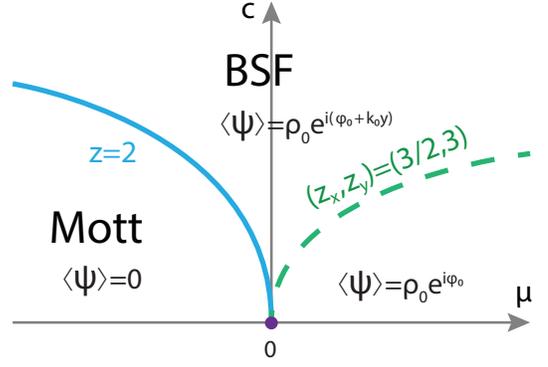}
\caption{  The phase diagram of action Eq.\ref{boostz2} with $ z=2 $ observed when the sample is moving Fig.\ref{frames}b
 as a function of $\mu$ and $c$. For Eq.\ref{boson}, $ \mu_p \sim  t/U- (t/U)_c  $ at a fixed chemical potential $ \mu $
 or $ \mu_p \sim \mu- \mu_c $ at a fixed  $ t/U $ in Fig.\ref{phaseslattice0}.
 $ c=0 $ recovers to that when the sample is static. Due to the explicit C- symmetry breaking,  no Higgs mode inside the BSF phase.
 The blue line with $ z=2 $ is in the same universality class as the corresponding $ z=2 $ line
 in Fig.\ref{phasesz1}.
 The dashed line means that it does not exist in the microscopic boson Hubbard model Eq.\ref{boson},
 just crashes onto the $ \mu $ axis. But it exists when directly boosting the SF  shown in \cite{SSS}.}
\label{phasesz2}
\end{figure}

Setting $ \tilde{\mu}=0 $ leads to the $ z=2 $ phase boundary when the sample is moving:
\begin{equation}
   \mu= - \frac{c^2}{4 v^2_y} < 0
\label{z2left}
\end{equation}
  which gives the $ z=2 $ line in Fig.\ref{phasesz2}.

     In summary, Eq.\ref{z2} is invariant  under the Galileo transformation:
\begin{equation}
   \tilde{\psi} = \psi e^{-ik_0y},~~~~   \tilde{\mu}= \mu + \frac{m c^2}{2} = \mu+ \frac{c^2}{4 v^2_y}
\label{z2inv}
\end{equation}
  which can be contrasted to the Lorentz transformation in Eq.\ref{spin012} for relativistic QFT.
  Obviously, the functional measure
  $ \int D\psi D\psi^{*} = \int D \tilde{\psi} D \tilde{\psi}^{*} $ in the path-integral in Eq.\ref{z2prime}.
  The reason for this absorption should be traced back to the fact that the boost term
  $ \psi^*(-ic\partial_y)\psi = - \frac{ ic }{2} ( \psi^* \partial_y \psi- \partial_y \psi^*  \psi ) $ in Eq.\ref{boostz2}  is nothing but
  a conserved current along the $ \hat{y} $ direction due to the $ U(1) $ symmetry, so can be absorbed by the transformation Eq.\ref{z2inv}.

 In the Mott phase $ \mu < 0 $  with a Mott gap $ - \mu $, increasing the boost, the Mott gap decreases
 until to zero signifying the QPT to the BSF phase.
 The scaling analysis in Sec.II-C and Sec.V-C with $ z=2 $ also hold here.
 Compared to Fig.\ref{phasesz1}, the $ z=1 $ line is absent in Fig.\ref{phasesz2}.

{\sl 1. The remanent of dangerously irrelevant $ T $ and $ P $ breaking terms when the sample is moving }

It is important to observe that if Eq.\ref{z2} had exact Galileo-invariance, this would be the end of story.
The action keeps its exact Galileo-invariance in any inertial frame. However, Eq.\ref{z2} has only emergent Galileo-invariance,
so this is not the end of story. Just from general symmetry principle, there should be many other terms which break the emergent Galileo-invariance.
Performing GT on these terms will still lead to some remanent terms which break the $ T $ and $ P $ symmetry, but keeps their combination $ PT $.
These terms also break the emergent Galileo-invariance when the sample is moving. Despite they are irrelevant
near the $ z=2 $ SF-Mott transition when the sample is moving, they are still important in the SF and Mott phases by leading to the
corresponding Doppler shift term. In terms of the terminology proposed in \cite{response}, they are type-II dangerously irrelevant:
namely they are irrelevant near the QPT, does not change the ground states on the two sides of the QPT either, but change the form of
the excitations on the two ground states. In the following two subsection, we explicitly discuss
two types of type-II dangerously irrelevant operators allowed by the symmetry:
the first one is either higher order in the order parameter $ \psi $ or higher order in the derivative  which
lead to the Doppler shift term inside the Boost SF only.
the second one is a crossing metric term which is still quadratic  in the order parameter $ \psi $ and
leads to the Doppler shift term inside both the Mott phase and in the Boost SF phase.
All the 3 terms break the $ T $ and $ P $ symmetry, but keeps their combination $ PT $, proportional to the boost $ c $,
have the same scaling dimension $ -1 $, so are three equally sub-leading irrelevant operators.

\subsection{ The Doppler shift in the BSF phase due to a dangerously irrelevant $ P $ or $ T $ breaking term  }


As said in the introduction, the microscopic system Eq.\ref{boson}
is not Galileo invariant. To see the effects of the boost which break the Galileo invariance, one must add
some boosting terms to Eq.\ref{z2} which are irrelevant near the $ z=2 $ QCP, but
break the Galileo invariance explicitly. This will be achieved in this and the next section.

 Just from the general symmetry principle, one need to incorporate two leading irrelevant terms into Eq.\ref{z2prime}:
\begin{align}
	\mathcal{L}& =\tilde{\psi}^*\partial_\tau\tilde{\psi}
	+v_x^2|\partial_x\tilde{\psi}|^2
	+v_y^2|\partial_y\tilde{\psi}|^2
	-(\mu+v_y^2k_0^2)|\tilde{\psi}|^2
                               \nonumber  \\
 & + u |\tilde{\psi}|^4  + i V | \tilde{\psi} |^2 \tilde{\psi}^*\partial_y \tilde{\psi}
 + iw \tilde{\psi}^*\partial^3_y \tilde{\psi} + \cdots
\label{z2primeleading}
\end{align}
    both terms break the $ P $ and $ T $ symmetry, but keeps their combination $ P T $. Both
    have the scaling dimension $ [V]=[w]=-1 $, so are dangerously irrelevant.
    The $ \cdots $ means  the always allowed terms $ a | \partial^2_y \tilde{\psi}|^2 + b | \partial_y \tilde{\psi}|^4 +
    d |\tilde{\psi}|^2 | \partial_y \tilde{\psi}|^2 $. They are less relevant than
    the two terms kept. Of course, the two C- symmetry breaking terms are excluded in the $ z=1 $ case Eq.\ref{boostz1}, but
    are allowed here at the $ z=2 $ case.
    Our microscopic calculations in Sec.VIII indeed find the two terms in terms of the microscopic parameters as
    $ V= 8 c \frac{ u_h }{ \Delta_h } ( \frac{ \lambda }{ \Delta_h } )^4 \propto c $
    and $ w= c= \alpha t_{b1} v $  listed in Eq.\ref{paction}, Eq.\ref{VwZ2} and Eq.\ref{vxvytc}.
    In the following, for the notational simplicity, we drop the $ \tilde{} $.

Plugging the mean field ansatz $\psi=\sqrt{\rho}e^{i\phi}$ into the action Eq.\ref{z2primeleading} leads to the energy density
\begin{align}
	E[\rho,\phi]
	=-\mu\rho+ u \rho^2
\end{align}
Minimizing $E[\rho,\phi]$ with respect to $\rho$
results in
\begin{align}
	\rho_0=
	\begin{cases}
	    0,&\mu<0\\
	    \frac{\mu}{2u},&\mu>0\\
	\end{cases}
\end{align}
In the superfluid phase, $\mu>0$ and $\rho_0>0$,
then $\psi=\sqrt{\rho_0+\delta\rho}e^{i\phi}$
and
\begin{align}
	\mathcal{L}_{M:SF}[ \delta\rho, \phi ] & =
	i\delta\rho\partial_\tau\phi
	+\frac{1}{4\rho_0}[v_x^2(\partial_x\delta\rho)^2+v_y^2(\partial_y\delta\rho)^2]
                          \nonumber  \\
	& + \rho_0[v_x^2(\partial_x\phi)^2+v_y^2(\partial_y\phi)^2]
	+ u (\delta\rho)^2     \nonumber   \\
    & + 2 V \rho_0 \delta\rho\partial_y \phi + w \rho_0 ( \partial_y \phi )^3 + \cdots
\label{droplinear}
\end{align}
 where, in addition to $ i\delta\rho\partial_\tau\phi $, the newly added last two terms also break the C- symmetry
 $ \phi \to -\phi $ in the $ ( \delta \rho, \phi) $ representation.
 It is tempting to include $ a \phi \partial^3_y \phi $ term, but this term is a total derivative term, so can be dropped.
 It shows $ ( \delta \rho, \phi) $ becomes a conjugate variable, so no Higgs mode, in sharp contrast to the
 SF phase in the $ z=1 $ case presented in Eq.\ref{GoldHiggsact}. Of course, as stressed in Sec.IV, despite the
 existence of the Higgs mode inside the SF near the $ z=1 $ line, it plays no roles in the SF to BSF transition
 in Fig.\ref{phasesz1}. Due to the spontaneous $ C $ symmetry breaking, it disappears in the BSF anyway.

 Integrating out $\delta\rho$ lead to the action when the sample is moving:
\begin{align}
	\mathcal{L}_{M:SF}[ \phi]& =
	\frac{1}{2 u }[(\partial_\tau-i 2 \rho_0 V \partial_y)\phi]^2
	+\rho_0[v_x^2(\partial_x\phi)^2
                \nonumber  \\
    &+v_y^2(\partial_y\phi)^2] + w \rho_0 ( \partial_y \phi )^3
\label{z2cubic}
\end{align}
which takes the same form as Eq.\ref{s2cant} and
leads to the exotic superfluid Goldstone mode when the sample is moving:
\begin{align}
	\omega=\sqrt{2 u \rho_0(v_x^2k_x^2+v_y^2k_y^2)}-2 \rho_0 V k_y
\label{z2SFV}
\end{align}

 As shown in Sec.III, the SF to BSF transition in Fig.\ref{phasesz1} is solely driven by the Goldstone mode,
 the Higgs mode is irrelevant. Here due to the explicit C- symmetry breaking, the Higgs mode does not exist in the first place.
 However, due to the smallness of $ \rho_0 $ near the QPT, also the smallness of $ V $
 listed below in Eq.\ref{paction},  Eq.\ref{z2SFV} is always stable, no QPT is possible.

 From Eq.\ref{droplinear}, one can also integrate out $ \phi $ to get an effective action in terms of $ \delta \rho $
\begin{align}
	\mathcal{L}_{M:SF}[ \delta \rho]  =
	\frac{1}{2} \delta \rho ( - k, - \omega )
[  - \frac{ (i \omega-2 \rho_0 V k_y)^2 }{ \rho_0 ( v^2_x k^2_x +  v^2_y k^2_y )  }
         + U ] \delta \rho( k, \omega )
\label{ddsfbsf}
\end{align}
which leads to the density-density correlation function ( or sound mode in the SF ):
\begin{align}
	& \chi( k, \omega )= \langle  \delta \rho ( - k, - \omega )  \delta \rho( k, \omega ) \rangle
                      \nonumber  \\
    & = \frac{ \rho_0 ( v^2_x k^2_x +  v^2_y k^2_y ) }{
       -(i \omega- 2 \rho_0 V k_y)^2 + \rho_0 U ( v^2_x k^2_x +  v^2_y k^2_y ) }
\end{align}
  whose pole, under the analytic continuation $ i \omega \to \omega + i \delta $ also leads to Eq.\ref{z2SFV}.
  Again, this is because the $ \delta \rho $ is always conjugate to the phase $ \phi $ through the QPT from the SF to the BSF.
  This is in sharp contrast the $ z=1 $ case where the $ \delta \rho $ is the Higgs mode which simply decouples from the
  Goldstone mode in the long-wave-length limit.


\subsection{ The  dangerously irrelevant $ P $ or $ T $ breaking metric-crossing term    }


In fact, in Eq.\ref{droplinear}, we drop the two linear derivative terms
$ \frac{1}{2} \partial_\tau \delta \rho + \rho_0 i \partial_\tau \phi $.
This is justified because the topological vortex excitations are irrelevant inside the SF phase, so the phase windings
in the phase $ \phi $ can be safely dropped. However, one must check if it remains so under the GT.
Under the boost, they become  $ \frac{1}{2} \partial_\tau \delta \rho + \rho_0 i \partial_\tau \phi
- \frac{i}{2} c \partial_y \delta \rho + \rho_0 c \partial_y \phi $.
Again, the two extra terms generated by the boost are still linear in $ \partial_y $, so still vanish after the integration by parts.
So the result achieved in the last section remain valid.

   Let us consider the typical irrelevant second order derivative term which breaks the Galileo invariance explicitly:
\begin{align}
	|\partial_\tau\psi|^2
	 & \to [(\partial_\tau-ic\partial_y)\psi^*][(\partial_\tau-ic\partial_y)\psi]
                        \nonumber  \\
	& =|\partial_\tau\psi|^2-i2c\partial_\tau\psi^*\partial_y\psi
	-c^2|\partial_y\psi|^2
\end{align}
 where  the metric-crossing term breaks the $ P $ and $ T $, but still keeps their combination $ PT $.
 Its coefficient is also the boost $ c $.

 Just from the general symmetry principle,  a more complete effective action than Eq.\ref{boostz2} is
\begin{align}
	\mathcal{L}_{M} & =
	Z_1\psi^*(\partial_\tau-ic\partial_y)\psi
	+Z_2(\partial_\tau-ic\partial_y)\psi^*(\partial_\tau-ic\partial_y)\psi
     \nonumber  \\
     &+v_x^2|\partial_x\psi|^2+v_y^2|\partial_y\psi|^2-\mu|\psi|^2+u|\psi|^4+ \cdots
\label{Z1Z2}
\end{align}
 Our microscopic calculations in Sec.VIII indeed find such a  term in terms of the microscopic parameters as
 $ Z_2= \frac{\lambda^2}{\Delta^3_h} $ listed below Eq.\ref{pactionlong} and in Eq.\ref{VwZ2}.

 Plugging in the mean field ansatz $\psi=\sqrt{\rho_0}e^{i(\phi+k_0y)}$ leads to the energy density
\begin{align}
	E[\rho,k_0]
	=[Z_1ck_0+(v_y^2-Z_2 c^2)k_0^2 -\mu]\rho +  u \rho^2
\end{align}
   In the following, we will drop the $ b $ term which can be shown to be irrelevant in the following discussions.

 Similar to Sec.IV, near the $ z=2 $ QCP, due to $ [Z_2]=-2 $, one can study the $ Z_2 \ll Z_1 $ limit which holds away from
 the $ z=1 $ tip in Fig.\ref{phaseslattice}. The minimization with respect to $k_0$ and $\rho$ leads to the saddle point solution
\begin{align}
	k_0 & =-\frac{Z_1 c}{2(v_y^2-Z_2 c^2)},
     \nonumber  \\
   \rho_0 & =\max(0,\frac{\mu-Z_1 ck_0-(v_y^2-Z_2 c^2)k_0^2}{2U})
\label{z2k0rho}
\end{align}
  which leads to a small correction to Eq.\ref{z2k0}.

  The critical line between the Mott $(\rho_0=0)$ and the BSF $(\rho_0>0)$ are given by
\begin{align}
	\mu_c=-\frac{Z_1^2c^2}{4(v_y^2-Z_2c^2)}
\label{z2leftcorr}
\end{align}
  which leads to a small correction to Eq.\ref{z2left}.
  The mean-field phase diagram is qualitatively the same as $Z_2=0$ case with
  the slight change of $ k_0 $ and the phase boundary $ \mu_c $ listed in Eq.\ref{z2k0rho} and Eq.\ref{z2leftcorr}.

  When $k_0\neq 0$, it is convenient to introduce a new order parameter  $\psi=\tilde{\psi}e^{ik_0y}$, Eq.\ref{Z1Z2} becomes:
\begin{align}
	\mathcal{L}_{M} & =
	(Z_1-2Z_2 c k_0)\tilde{\psi}^*\partial_\tau\tilde{\psi}
	+Z_2|\partial_\tau\tilde{\psi}|^2
	-i2Z_2 c\partial_\tau\tilde{\psi}^*\partial_y\tilde{\psi}
          \nonumber  \\	
	 & + v_x^2|\partial_x\tilde{\psi}|^2
	+(v_y^2-Z_2 c^2)|\partial_y\tilde{\psi}|^2
                       \nonumber  \\
     & -(\mu-Z_1ck_0-v_y^2k_0^2
    +Z_2 c^2k_0^2)|\tilde{\psi}|^2
               \nonumber  \\
	& -i(Z_1c+2k_0v_y^2-2Z_2 k_0c^2)\tilde{\psi}^*\partial_y\tilde{\psi} + u |\tilde{\psi}|
\end{align}
  Setting the linear in $ k_y $ term $Z_1c+2k_0v_y^2-2Z_2 k_0c^2=0$ recovers
  the $k_0$ in Eq.\ref{z2k0rho}. One reaches the final action in terms of $\tilde{\psi}$:
\begin{align}
	\mathcal{L} & =
	(Z_1\!-\!2Z_2 c k_0)\tilde{\psi}^*\partial_\tau\tilde{\psi}
	+Z_2|\partial_\tau\tilde{\psi}|^2
	-i2Z_2 c\partial_\tau\tilde{\psi}^*\partial_y\tilde{\psi}	
                 \nonumber  \\
	& + v_x^2|\partial_x\tilde{\psi}|^2
	+(v_y^2\!-\!Z_2 c^2)|\partial_y\tilde{\psi}|^2
	-\tilde{\mu}|\tilde{\psi}|^2+U|\tilde{\psi}|^4
\end{align}
where $\tilde{\mu} =\mu+\frac{Z_1^2c^2}{4(v_y^2-Z_2c^2)}$
and $Z_1-2Z_2 c k_0=\frac{Z_1v_y^2}{v_y^2-Z_2 c^2}  >0$. Setting $\tilde{\mu}=0 $ recovers Eq.\ref{z2leftcorr}.

   Now we arrive at the same form as the effective action Eq.\ref{z1z2square}. Similar scaling
   analysis below Eq.\ref{z1z2square} apply here also.
   After dropping the more irrelevant term $ |\partial_\tau\tilde{\psi}|^2 $ and keep only the leading irrelevant metric-crossing term,
   we arrive at the same action as Eq.\ref{z1z2}:
\begin{align}
	\mathcal{L}&=
	\tilde{Z}_1\tilde{\psi}^*\partial_\tau\tilde{\psi}	
	-i\tilde{Z}_2\partial_\tau\tilde{\psi}^*\partial_y\tilde{\psi}
	+\tilde{v}_x^2|\partial_x\tilde{\psi}|^2
	+\tilde{v}_y^2|\partial_y\tilde{\psi}|^2
                         \nonumber  \\
	&-\tilde{\mu}|\tilde{\psi}|^2+U|\tilde{\psi}|^4+ \cdots
	\big)
\label{tildeZ2}
\end{align}
  where $ \tilde{Z}_1 =Z_1-2Z_2ck_0$, $ \tilde{Z}_2 =-2Z_2c$, $\tilde{v}_x^2=v_x^2$, $\tilde{v}_y^2=v_y^2-Z_2c^2$.
  Then the discussions following Eq.\ref{z1z2} apply here also. Of course, setting $ \tilde{Z}_1=1, \tilde{Z}_2=0 $ recovers Eq.\ref{z2prime}.

  So we conclude that it is the  metric crossing term $  \tilde{Z}_2 $  which leads to
  the shift of the value $ k_0 $ and the critical chemical potential $ \mu_c $.
  As shown in Sec.II-D, it is this metric crossing term which leads to the Doppler shift term in both the Mott and BSF phases.
  Its contribution to the conserved Noether currents will be addressed in the following sub-section.
  The  metric crossing $  \tilde{Z}_2 $ term   has the same scaling dimension as the $ V $ and $ w $ term
  in Eq.\ref{z2primeleading} with $ [\tilde{Z}_2]=[V]=[b]=-1 $.
  All the three terms break $ T $ and $ P $, but keep their combinations $ PT $, their
  coefficients are all proportional to $ c $.
  So considering the $  \tilde{Z}_2 $ term will not change the results achieved in Sec.A qualitatively except that:
  this is the only metric term which couples the space and time, also the only term which leads to the Doppler shift in the Mott phase
  listed in Eq.\ref{Mottz2}.

\subsection{ The conserved Noether current with $ z=2 $ in the lab and co-moving }

{\sl 1. The emergent Galileo invariant case }

   The global $ U(1) $ symmetry $ \psi \rightarrow \psi e^{ i \chi} $ leads to the conserved Noether current
   $ \tilde{J}_\mu=( J_\tau, J_x, \tilde{J}_y ) $ in Eq.\ref{boostz2}:
\begin{eqnarray}
   J_\tau & = & -i \psi^{*} \psi
                          \nonumber   \\
   J_x & = &  i v^2_x ( \psi^{*} \partial_x \psi- \psi \partial_x \psi^{*} )
                          \nonumber   \\
   \tilde{J}_y & = &  i v^2_y ( \psi^{*} \partial_y \psi- \psi \partial_y \psi^{*} )  -c \psi^{*} \psi = J_y-ic J_{\tau}
\label{threecurrentsz2}
\end{eqnarray}
  which also show the current along the $ y $ direction  $ \tilde{J}_y $ is the sum of the intrinsic one $ J_y $ and the one due to the boost
  $ -ic J_{\tau} $, identical to the last equation in Eq.\ref{threecurrents}. They are PT even \cite{currenttrick} and satisfy
\begin{equation}
  \partial_\tau J_\tau + \partial_x J_x+  \partial_y (J_y - ic J_\tau) =0
\label{way2z2}
\end{equation}
  which is identical to Eq.\ref{way2}.

  Now we evaluate the 3-currents in all the three phase in Fig.\ref{phasesz2}.
  Plugging in $ \langle \psi \rangle=\sqrt{\rho_0} e^{i k_0 y } $ into Eq.\ref{threecurrentsz2} leads to:
\begin{eqnarray}
    J_\tau & = & -i \rho_0,   \nonumber  \\
    J_x & = & 0    \nonumber  \\
    \tilde{J}_y & = & -2 v^2_y k_0 \rho_0 - c \rho_0
\label{z2current}
\end{eqnarray}
  where again the factor of $ i $ in $ J_{\tau} $ is due to the imaginary time.

   The Mott state  with $ \rho_0=0 $ carries no current.
   This is expected, because the Mott has neither P- or H- carriers, so no current.
   Near the $ z=2 $ line in Fig.\ref{phasesz2}, substituting $ k_0= -\frac{ c}{ 2 v^2_y} < 0  $ into
   Eq.\ref{z2current}, so the BSF near the $ z=2 $ line
   carries  $  J_\tau  = -i \rho_0, J_x=0 $ and $ \tilde{J}_y = 0, J_y= -2 v^2_y k_0 \rho_0  $
   which means the intrinsic current just cancels that due to the boost.
   We believe this result is exact  due to the emergent Galileo invariance.
   Similar to $ z=1 $ case, $ J_y= -2 v^2_y k_0 \rho_0  $ in the co-moving frame can be used to distinguish the Mott and the BSF phase.


{\sl 2. The metric crossing $ Z_2 $ term  }

   It is important to stress that the current Eq.\ref{threecurrents} for the $ z=1 $ case studied in Sec.V is particle-hole current,
   but here for $ z=2 $, Eq.\ref{z2current} is either particle or hole current ( but not both ).
   By adding the leading irrelevant $ Z_2 $ term in Eq.\ref{Z1Z2}, one can also evaluate the Noether 3-currents in Eq.\ref{Z1Z2}:
\begin{eqnarray}
   J_\tau & = & -i Z_1 \psi^{*} \psi + i Z_2
   ( \psi^{*} \tilde{\partial}_\tau \psi- \psi \tilde{\partial}_\tau \psi^{*} )
                                \nonumber  \\
    & = & -i Z_1 \psi^{*} \psi
    + i Z_2 [ ( \psi^{*} \partial_\tau \psi- \psi \partial_\tau \psi^{*} ) -ic ( \psi^{*} \partial_y \psi- \psi \partial_y \psi^{*} ) ]
                          \nonumber   \\
   J_x & = &  i v^2_x ( \psi^{*} \partial_x \psi- \psi \partial_x \psi^{*} )
                          \nonumber   \\
   \tilde{J}_y & = &  i  v^2_y ( \psi^{*} \partial_y \psi- \psi \partial_y \psi^{*} )  -ic J_{\tau} = J_y-ic J_{\tau}
\label{threecurrentsz1z2}
\end{eqnarray}
  Setting $ Z_1=0, Z_2=1 $ or  $ Z_1=1, Z_2=0 $ recovers Eq.\ref{threecurrents}  or
   Eq.\ref{threecurrentsz2} respectively \cite{currenttrick}.

  The first equation shows that $ J_\tau $ consists of the particle ( or hole ) current Eq.\ref{z2current} with the weight $ Z_1 $ and
  the particle-hole current Eq.\ref{threecurrents}  with a small weight $ Z_2 \ll Z_1 $.
  The second shows that $ J_x $ remains the same.
  The third  shows the current along the $ y $ direction
  $ \tilde{J}_y $ is still the sum of the intrinsic one $ J_y $ and the one due to the boost $ -ic J_{\tau} $.
  So this feature remains the same as in Eq.\ref{threecurrents} ( or  Eq.\ref{threecurrentsz2} ).
  Eq.\ref{way2} ( or Eq.\ref{way2z2} ) follows automatically.

 Plugging in $ \langle \psi \rangle=\sqrt{\rho_0} e^{i k_0 y } $ into Eq.\ref{threecurrentsz1z2} leads to:
\begin{eqnarray}
    J_\tau & = & -i Z_1 \rho_0 + i Z_2 2 k_0 c \rho_0= -i \rho_0 [Z_1-Z_2 2 k_0 c],   \nonumber  \\
    J_x & = & 0    \nonumber  \\
    \tilde{J}_y & = & -2 v^2_y k_0 \rho_0 - c \rho_0[Z_1-Z_2 2 k_0 c]
\label{z2currentz1z2}
\end{eqnarray}
   which need to be evaluated at the corrected  $ k_0 $ in Eq.\ref{z2k0rho}.

   The Mott phase with $ \rho_0=0 $ carries no current.
   In the BSF phase near the $ z=2 $ line, plugging in the  $ k_0 $ in Eq.\ref{z2k0rho},
   one finds $ J_\tau=- i \rho_0 ( \frac{ Z_1 v^2_y }{v^2_y-Z_2 c^2 } ), J_x=0 $ and $  \tilde{J}_y=0, J_y= -2 v^2_y k_0 \rho_0  $.
   Again $ J_y= -2 v^2_y k_0 \rho_0  $ in the co-moving frame can still be used to distinguish the Mott and the BSF phase.
   When considering all the possible terms which break the $ T $, $ P $ ( still keeps $ PT $ ) and
   the  emergent Galileo invariance ( all of them are irrelevant near the $ z=2 $ line, but become important away from it ),
   we conclude that the physical picture listed below Eq.\ref{z2current} reached simply by setting $ Z_2=0 $ remains valid.


   Similarly, as outlined in Sec.V, one can also achieve the phase diagram Fig.\ref{phasesz2} from
   the charge-vortex duality performed when the sample is moving.
   Namely, the $ z=2 $ line is driven by the instability in the vortex degree of freedoms, while the $ z=(3/2,3) $ line is
   the instability in the dual gauge degree of freedoms.


\section{ Putting the Galileo transformation in a lattice }

 We study the Galileo transformation in a non-relativistic many body system in a periodic potential.
 We will make the first attempt to do it directly on the BH model Eq.\ref{boson} and point out its limitations..
 Then we get to the many body ionic model which leads to the BH model under lattice projection.
 We study the ionic model in the first quantization, then we transfer it to the second quantization language and project it to the Hubbard model
 in a tight binding limit in a square lattice. Finally we describe the emergent space-time in the low energy limit in the second quantization.


\subsection{  Galileo transformation in the boson Hubbard model: its form and un-certainties  }

  To perform a GT on the lattice, the first thing to do is to do it directly on the BH model Eq.\ref{boson}.
  The action corresponding to the boson Hubbard model Eq.\ref{boson} is:
\begin{align}
	\mathcal{S}_{BH}=\int d\tau \sum_i b_i^\dagger \partial_\tau b_i
	+H_{BH}[b_i^\dagger, b_i]
\end{align}
In the continuum theory, a Galilean boost with a constant velocity $\vec{v}_b$ will lead to
$\partial_\tau\to\partial_\tau+i\vec{v}_b\cdot\vec{\nabla}$.
In the lattice theory, one need replace the spatial derivative $\partial_{\hat{\mu}}$
by its discrete version (lattice derivative) $\Delta_{\hat{\mu}}$
via $\partial_{\hat{\mu}}\phi_i\to
\Delta_{\hat{\mu}}\phi_i=a^{-1}(\phi_{i+\hat{\mu}}-\phi_i)+ \cdots $
and $\Delta_{\hat{\mu}}^*\phi_i=a^{-1}(\phi_i-\phi_{i-\hat{\mu}}) + \cdots $,
where $a$ is the lattice constant where $ \cdots $ means that one should add infinite number of higher order terms
which still lead to the same contume limit

Thus, under a Galilean boost, the lattice action can be written as
\begin{align}
	\mathcal{L}_{BH,b}&=\sum_i b_i^\dagger \partial_\tau b_i
	+i \sum_i \sum^{\infty}_{n=1 } [ t_{bn,x} b_i^\dagger b_{i+nx}+t_{bn,y} b_i^\dagger b_{i+ny} ]
                                     \nonumber  \\
	&+  h.c. + H_{BH}[b_i^\dagger,b_i]
\label{BoostedBHform}
\end{align}
  which sets up the form of the Boosted BH model.

  In principle, one need to include the infinite sum of terms. The simplest thing to do is to include only the $ n=1 $ NN term
  $ i (t_{b1,x} b_i^\dagger b_{i+x}+t_{b1,y} b_i^\dagger b_{i+y}-h.c.) $
  where $t_{b,x}=\hbar v_{b,x}/(2a)$ and $t_{b,y}=\hbar v_{b,y}/(2a)$  has the energy dimension.
  However, this leading term is proportional to the conserved current term, so can be absorbed by the unitary transformation
  Eq.\ref{k0phase} into the hopping term in
  $ H_{BH}[b_i^\dagger,b_i] $ as shown in Sec.VIII-A. So one need also consider also the subleading term $ n=2 $ NNN current term
  in the series. Unfortunately,  one can not determine the ratio of $ t_{b2}/t_{b1} $ just from the substitution.
  One need to repeat the derivation from the ionic model to the BH model to determine the whole series in Eq.\ref{BoostedBHform}.
  This will be achieved in the following sections.

\subsection{  Galileo transformation in the many-body wavefunction in a periodic potential: bare space-time in the first quantization }

  Eq.\ref{Seq0} can be easily generalized to $ N \gg 1 $ fermionic or bosonic systems in a one-body potential
  $ V_1(x) $ and also a short-range or long-range two-body ( Coulomb ) interaction $ V_2( x_i- x_j ) $.
  When the sample is static, the many-body wavefunction satisfies ( ignoring the lattice vibrations, SOC, etc ):
\begin{align}
  & i \hbar \frac{ \partial \Psi(x_1,x_2,\cdots x_N) }{ \partial t }  =
   [  -\frac{\hbar^2}{ 2 m} \sum_{i}  \frac{\partial^2 }{ \partial x^2_i }+  \sum_i  V_1(x_i)
                      \nonumber  \\
   & + \sum_{i<j} V_2(x_i-x_j ) ] \Psi(x_1,x_2,\cdots x_N)
\label{Seq0Nlattice}
\end{align}
 where   $  V_1( \vec{x} )=V_1 ( \vec{x } + \vec{a} ) $ where the  $ \vec{a} $ is a lattice vector stands for the lattice potential.
 $ V_1( \vec{x}   )= \sum_{\vec{R}} v( \vec{x} -\vec{R} ) $ where $ \vec{R} $ stand for the positions of all the ions
 in a periodic lattice \cite{retardation}.
 For the Coulomb interaction $  v( \vec{x} -\vec{R} )= - \frac{Ze^2}{| \vec{x}_i-\vec{R} | } $ and
 $ V_2(x_i-x_j )=  - \frac{e^2}{| \vec{x}_i-\vec{x}_j | } $.
 Setting $ V_1(x_i)=0 $ and  $ V_2(x_i-x_j ) $ as Wan der-Waals interaction
 recovers the charge neutral Helium 4 case which is clearly Galileo invariant.

  The many-body wavefunction when the sample is static are related to that when the sample is moving $ \vec{v} $ by:
\begin{align}
   \Psi( x_i,t)  =  e^{ i ( k_0 \sum_{i} x^{\prime}_i- N E_0 t^{\prime}/\hbar )  }
   \Psi^{\prime} ( x^{\prime}_i, t^{\prime} )
\label{psi0Nion}
\end{align}
 where $  \Psi( x_i,t)= \Psi( x^{\prime}_i- vt^{\prime}, t^{\prime}) $ and $  k_0= -\frac{ m v}{\hbar}, E_0= \frac{ \hbar^2 k^2_0 }{ 2 m }= \frac{1}{2} m v^2 $.
 Note that, when the sample is moving, the ions are also moving, so $ \vec{R}=\vec{R}^{\prime}- \vec{v}t^{\prime} $,
 then $ V_1(x_i)=\sum_{\vec{R}} v( \vec{x} -\vec{R} )= \sum_{\vec{R}^{\prime}} v( \vec{x}^{\prime} -\vec{R}^{\prime} )
 =V_1(x^{\prime}_i) $ breaks the translational invariance up to a lattice constant $ \vec{a} $, but is still
 Galileo invariant \cite{impuritypotential}. Furthermore,
 $   V_2( x_1- x_2 )= e^2/|x_1-x_2|= V_2( x^{\prime}_1- x^{\prime}_2 ) $.
 Note that both $ \vec{x} -\vec{R}=\vec{x}^{\prime} -\vec{R}^{\prime} $ and $ x_1- x_2=x^{\prime}_1- x^{\prime}_2 $ are
 GI. It is easy to check that Eq.\ref{psi0Nion} including the sign of $ k_0 $ agrees with Eq.\ref{z2inv}.

 When the sample is moving, the many-body wavefunction $ \Psi^{\prime} ( x^{\prime}_i, t^{\prime} ) $ satisfies :
\begin{align}
  &  i \hbar \frac{ \partial \Psi^{\prime} ( x^{\prime}_i, t^{\prime} ) }{ \partial t^{\prime} } =
   [  -\frac{\hbar^2}{ 2 m} \sum_{i}  \frac{\partial^2 }{ \partial x^{\prime 2}_i }
                                     \nonumber  \\
  &  +  \sum_i  V_1(x^{\prime}_i)+ \sum_{i<j} V_2(x^{\prime}_i- x^{\prime}_j ) ] \Psi^{\prime} ( x^{\prime}_i, t^{\prime} )
\label{Seq0Nlatticeprime}
\end{align}
  which takes the identical form as Eq.\ref{Seq0Nlattice}.

 Despite microscopic model Eq.\ref{Seq0Nlattice} is {\em formally } GI, the wavefunction exponential factor
 $ e^{ i ( k_0 \sum_{i} x^{\prime}_i- N E_0 t^{\prime}/\hbar )  } $ in Eq.\ref{psi0Nion} in the first quantization  may still have some observable experimental consequences.
 As advocated by P. W. Anderson, "More is different ".
 In the thermodynamic limit $ N \rightarrow \infty $, there could be emergent phenomena such as quantum and topological phases
 and phase transitions. Especially, in the superfluids or superconductors, the global $ U(1) $ symmetry is broken,
 the number $ N $ is not even conserved. There could be some interference effects between different number  $ N $ sectors.
 Unfortunately, it is hard to retrieve such effects  in the mathematical formal form of Eq.\ref{Seq0Nlattice}  and Eq.\ref{Seq0Nlatticeprime} in the first quantization. In fact, it is even difficult to extract the well known Doppler shift term in the
 quasi-particle excitation which may correspond to $ N \to N+1 $ in the ground state and the first excited state in Eq.\ref{psi0Nion}.
 To proceed further, one may resort to second quantization and quantum field theory approach to be developed in the following.

\subsection{  Galileo transformation in the effective action in the second quantization in a periodic potential }

 Eq.\ref{Seq0Nlattice} can be written in the second quantization language:
\begin{align}
 & {\cal H}  = \int d^2 x \psi^{\dagger}( \vec{x} )
   [  -\frac{\hbar^2}{ 2 m} \nabla^2 + V_1( \vec{x} ) - \mu ] \psi( \vec{x} )
                \nonumber  \\
   &  +  \int d^2 x_1 d^2 x_2  \psi^{\dagger}( \vec{x}_1 )\psi( \vec{x}_1 )
   V_2(x_1-x_2 ) \psi^{\dagger}( \vec{x}_2 )\psi( \vec{x}_2 )
\label{Seq0Nlattice2}
\end{align}
  where the single-body lattice potential $ V_1( \vec{x} ) $ and the two-body interaction $ V_2(x_1-x_2 ) $ are automatically
  incorporated into the kinetic term and the interaction term respectively.
  By adding the chemical potential $ \mu $, we also change the canonical ensemble with a fixed number of particles
  $ N $ in the first quantization to
  the grand canonical ensemble in the second quantization.

  Following the step leading from Eq.\ref{z1} to Eq.\ref{boostz1}, one can obtain the effective action when the sample is moving
  ( for notational simplicity, we drop the $ \prime $ when the sample is moving ):
\begin{align}
 & {\cal H}  = \int d^2 x \psi^{\dagger}( \vec{x} )
   [  -\frac{\hbar^2}{ 2 m} \nabla^2 + V_1( \vec{x} )- \mu -iv ( \partial_x + \partial_y ) ] \psi( \vec{x} )
               \nonumber  \\
 &   +  \int d^2 x_1 d^2 x_2  \psi^{\dagger}( \vec{x}_1 )\psi( \vec{x}_1 )
   V_2(x_1-x_2 ) \psi^{\dagger}( \vec{x}_2 )\psi( \vec{x}_2 )
\label{Seq0Nlattice2prime}
\end{align}
   which hold not only in the periodic lattice potential case, also random impurity case.
   Formally at this level, the boost term may also be written in the driving current form
   $ -\frac{iv}{2} (\psi^{\dagger}( \vec{x} )\partial_x  \psi( \vec{x} )- \psi( \vec{x} ) \partial_x \psi^{\dagger}( \vec{x} ) ) $.
   However, as shown in the following, under the lattice projection, the boost and the driving current takes the similar form
   in the NN term, but differs in the NNN term \cite{currentx}.

 It is easy to show that Eq.\ref{Seq0Nlattice2prime} takes the same form as Eq.\ref{Seq0Nlattice2}
 after formally making the GT listed in Eq.\ref{z2inv} \cite{evenso}. The wavefunction exponential factor
 in Eq.\ref{psi0Nion} in the first quantization is equivalent to the shift of the momentum and
 the chemical potential as listed in Eq.\ref{z2inv} in the second quantization. Indeed,
 $ - \mu \int d^2 x \psi^{\dagger}( \vec{x} ) \psi( \vec{x} ) = - \mu N $ where $ N $ is the number of particles
 in Eq.\ref{psi0Nion} in the canonical ensemble.   For small number of $ N $, this is the end of story.
 Unfortunately, it is still difficult to see what are the effects of such a GT in the present second quantization scheme
 in the thermodynamic limit $ N \to \infty $. This should not be too surprising, because as advocated by P. W. Anderson,
 `` More is different '', there are many emergent quantum or topological phenomena such as quantum magnetism,
 superconductivity, superfluidity, etc which can nowhere be seen in such a formal model Eq.\ref{Seq0Nlattice}.
 This kind of  model looks complete, exact to some formal level, but not effective to see any emergent phenomena (
 See the comparison between Fig.\ref{towerlevel} and Fig.\ref{towerlevelm} ).
 So only when proceeding further to get an effective model in terms of relevant low degree of freedoms to describe the emergent quantum or topological phenomenon,
 one may start to see the real effects of such a GT on the emergent phenomena.

\subsection{ The emergent space-time in the low energy limit in the second quantization }
  In a continuum system without underlying lattice such as Helium 4, by setting $ V_1(x_i)=0 $,  one can
  directly proceed further from Eq.\ref{Seq0Nlattice2prime}, see \cite{SSS}.
   In a periodic lattice,  the field operator $ \psi( \vec{x} ) = \sum_n \psi_n( \vec{x} ) $ can be  expanded either in the Wannier basis
   $ \psi_n( \vec{x} ) = \sum_i b_i \phi_n ( \vec{x}-\vec{R}_i ) $ or in the Bloch basis $ \psi_n( \vec{x} )
   = \sum_{\vec{k}} b_{\vec{k}} \phi_{n \vec{k}} ( \vec{x} ) $ where $ n $ is the band index, $ \vec{k} $ is
   the crystal momentum in the BZ. The Wannier functions satisfy the Wannier equation:
\begin{align}
   \hat{h}  \phi_n( \vec{x}-\vec{R}_i )
  & = \sum_{\vec{R}} E_n( \vec{R}_i- \vec{R} )  \phi_n ( \vec{x}-\vec{R} ),
   \nonumber  \\
  \hat{h} & = -\frac{\hbar^2}{ 2 m} \nabla^2 + V_1( \vec{x} )
\label{hphiE}
\end{align}
 where  $ E_n(\vec{k})= \sum_{\vec{R}} e^{i \vec{k} \cdot \vec{R} } E_n(  \vec{R} ) $  is the tight-binding energy dispersion
 satisfying $ \hat{h} \phi_{n \vec{k}} ( \vec{x} )=E_n(\vec{k}) \phi_{n \vec{k}}( \vec{x} ) $. One can transfer between the two basis by
 $ \phi_{n \vec{k}}( \vec{x} )= \sum_{\vec{R}} e^{i \vec{k} \cdot \vec{R} }  \phi_n ( \vec{x}-\vec{R} ) $.

   Here, in the tight-binding limit, it is convenient to use
   the Wannier basis confined to the $ n=s $ band and drop all the higher bands $ l=2,3,\cdots $.
   Substituting  $ \psi( \vec{x} ) = \sum_i b_i \phi ( \vec{x}-\vec{R}_i ) $ into Eq.\ref{Seq0Nlattice2prime}
   leads to:
\begin{align}
  {\cal H} & =  \sum_{ij} b^{\dagger}_i b_j t_{ij} - \mu \sum_{i} b^{\dagger}_i b_i
     + \sum_{ij,kl} b^{\dagger}_i b_j b^{\dagger}_k b_l  V_{ij,kl}
       \nonumber    \\
   & + \sum_{ij} b^{\dagger}_i b_j t_{b,ij}
\label{pathprime}
\end{align}
   where
\begin{align}
  t_{ij} & = \int d^2 x \phi^{*}( \vec{x}-\vec{R}_i )
   [  -\frac{\hbar^2}{ 2 m} \nabla^2 + V_1( \vec{x} ) ] \phi( \vec{x}-\vec{R}_j )
        \nonumber   \\
  t_{b,ij}& =(-iv) \int d^2 x \phi^{*}( \vec{x}-\vec{R}_i )
    [ \frac{ \partial}{\partial x } +  \frac{ \partial}{\partial y } ]\phi( \vec{x}-\vec{R}_j )
\label{tijtbij}
\end{align}
  and $ V_{ij,kl}= \int d^2 x_1 d^2 x_2  \phi^{*}( \vec{x}_1-\vec{R}_i )\phi( \vec{x}_1-\vec{R}_j )
   V_2(x_1-x_2 )  \phi^{*}( \vec{x}_2-\vec{R}_k )\phi( \vec{x}_2-\vec{R}_l ) $.
   It indeed takes the form Eq.\ref{BoostedBHform} expected from the general ground.

  By using Eq.\ref{hphiE}, one can see  $ t_{ij}= E_s( \vec{R}_j- \vec{R}_i) $,  the first term can be written as
  $ \sum_{ij} b^{\dagger}_i b_j  E_s( \vec{R}_j- \vec{R}_i )
  = \sum_i b^{\dagger}_i b_i E_s( \vec{R}=0 ) + \sum_{<ij>} b^{\dagger}_i b_j  E_s( \vec{R} =\pm a \hat{x}, \pm a \hat{y} ) + \cdots $
  where the first term renormalizes the chemical potential $ - \mu \to -\mu + E_s( \vec{R}=0 ) $, the second, third...are the NN, NNN,.....hopping respectively.
  In the tight-binding limit, it maybe justified to consider only the on-site term and the NN hopping term \cite{vbsdetect}.
  For the interaction term, one may just keep the on-site term. Then for fermions or bosons,
  Eq.\ref{pathprime} reduces to the fermionic \cite{spd} or Boson Hubbard model Eq.\ref{boson} respectively.
  Setting $ v=0 $ recovers back to the static sample case. Obviously, one can start to see the Mott phase, SF phase and
  a quantum phase transition from Mott to the SF, so  this is a more effective model than the formal model Eq.\ref{Seq0Nlattice2prime}.
  Similarly, one may start to see the effects of GT on this more effective model.

  In the following, due to the exchange symmetry between $ x $ and $ y $, we only examine the current along the x-bond.
  The identical argument applies to the y- bond.
  Now we investigate the second ( boost ) term which can be written as the current form
  $ J_{ij,x}= -\frac{i v}{2} \int d^2 x [ \phi^{*}( \vec{x}-\vec{R}_i )
   \frac{ \partial}{\partial x } \phi( \vec{x}-\vec{R}_j )-
  \phi( \vec{x}-\vec{R}_j ) \frac{ \partial}{\partial x } \phi^{*}( \vec{x}-\vec{R}_i ) ] $ as alerted below Eq.\ref{Seq0Nlattice2prime}.
  For the S-wave band, one can take $ \phi( \vec{x}-\vec{R}_j ) $ to be real and only depends on its magnitude,
  so that $ \phi( \vec{x}-\vec{R}_j ) = \phi( | \vec{x}-\vec{R}_j | ) $. Then the boost term in Eq.\ref{pathprime} can be written as:
\begin{equation}
 H^{S}_{bx}=-iv\sum_{ij} b^{\dagger}_i b_j \int d^2 x \phi( |\vec{x}| )
 [ \alpha \frac{ \partial}{\partial x } + \beta \frac{ \partial}{\partial y } ] \phi( |\vec{x}-\vec{R}_{ij} | ),
\label{Hblattice}
\end{equation}
  where $ \vec{R}_{ij}= \vec{R}_i-\vec{R}_j $ and
  $ (\alpha,\beta) $ indicate any generic boost direction. For simplicity, we take the boost along the $ x- $ direction in the following.
  A simple reflection symmetry analysis under $ \vec{x} \to -\vec{x} $  shows that only when $ \vec{R}_{ij}= n a \hat{x}, n=1,2, \cdots $ is along the x-bond direction, $ H_b $ is non-vanishing.
  It is easy to see that there is no on-site term  $ n=0 $ with $ i=j $, the NN term with $ n=1 $  is nothing but
  a conserved Noether current along the $ \hat{x} $ direction due to the $ U(1) $ symmetry,
  so can be absorbed into the NN hopping term by a unitary transformation ( See Sec.VIII ).
  So one may need go to at least the $ n=2 $ NNN current term  to see all the effects of the boost:
\begin{equation}
 H^{S}_{bx}=-iv[ t_{b1} \sum_{i} b^{\dagger}_i b_{i + \hat{x} } + t_{b2} \sum_{i} b^{\dagger}_i b_{i + 2 \hat{x} }] + h. c
\label{Hblattice12}
\end{equation}
   where $ t_{b1}, t_{b2} $ are completely determined by Wannier functions, so independent of the boost velocity $ v $.
   So it just keeps the first two terms in  Eq.\ref{BoostedBHform} expected from the general ground.

   Of course, one may also keep the NN $ t_1 $ and NNN $ t_2 $ hopping term in  Eq.\ref{pathprime},
   due to the different integrands involving in the matrix elements,
   $ t_{b2}/t_{b1} \neq t_2/t_1 $. So adding $ t_2 $ back will not change the physics qualitatively.
   However, if it is a conserved current, then $ t_{b2}/t_{b1} = t_2/t_1 $, so the current is completely determined by the
   kinetic term, so can be absorbed to it order by order in the NN, NNN,....\cite{currentx}.
   hoppings just a single unitary transformation like Eq.\ref{k0phase}, but not the boost which can only
   be absorbed in the NN term by chance, but not
   the higher orders.  This is the main difference between the boost and a driving current as alerted below Eq.\ref{Seq0Nlattice2prime}.

   In the following section, we will show that the $ n=1 $ term behaves like a current term which introduces
   a new basis and also changes the parameters of the Hamiltonian in the new basis.
   One need also transform the $ n=2 $ term into the new basis and find it plays the role of the boost.


\begin{figure}[tbhp]
\centering
\includegraphics[width=.6\linewidth]{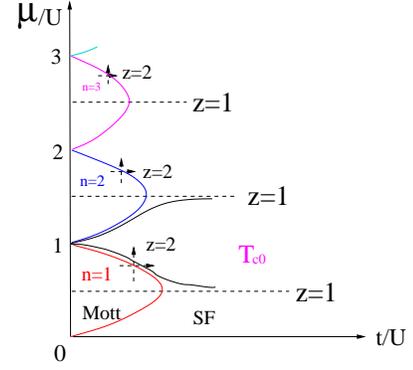}
\caption{ The well known phase diagram \cite{boson0} of the Boson Hubbard model Eq.\ref{boson} when the sample is static.
 The Mott insulating phase resides inside the $ n=1,2,3,\cdots $ lobes. The superfluid (SF) phase takes the other space.
 There is a particle-hole (PH) ( or charge conjugation $ C $ ) symmetry along the horizontal dashed lines going through the tip of the lobes,
 the Mott-SF transition has the dynamic exponent $ z=1 $.
 There is no PH symmetry away from the horizontal dashed lines, the Mott-SF transition has the dynamic exponent $ z=2 $.
 The solid lines  emerging from the joint point between
 $ n $-th and $ n+1 $-th Mott lobe delineates  the contours of constant density with $ n + \epsilon $
 and $  n +1- \epsilon $ respectively.
 The $ z=1 $ and $ z=2 $ SF-Mott transitions have emergent "Lorentz " invariance ( Appendix A ) and emergent Galileo invariance respectively.
 As shown in Fig.\ref{phasesz1}, \ref{phasesz2}, they response very differently to the boost of the moving sample.}
\label{phaseslattice0}
\end{figure}

\begin{figure}[tbhp]
\centering
\includegraphics[width=.8\linewidth]{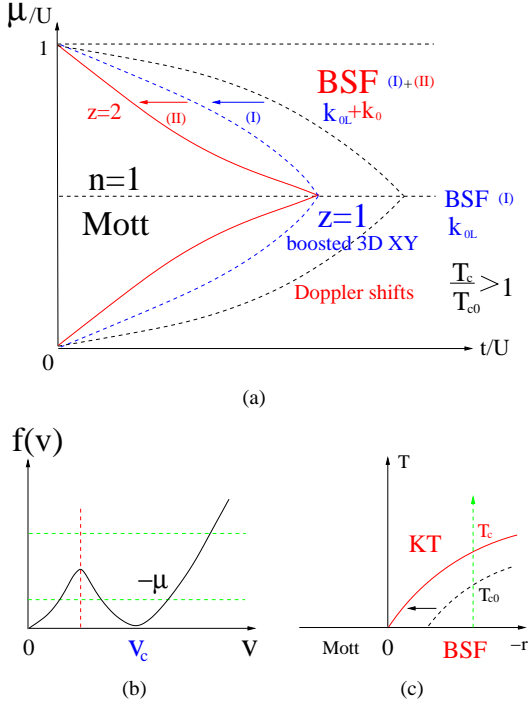}
\caption{  (a) The phase diagram when the sample is moving with a fixed velocity $ \vec{v} $.
 For simplicity, we only draw the Mott lobe at $ n=1 $. $ n > 1 $ can be similarly constructed.
 It is achieved by combining Fig.\ref{phasesz1} and Fig.\ref{phasesz2} with the relations between the phenomenological parameters
 and the microscopic parameters in Eq.\ref{relationz1} and  Eq.\ref{relationz2}.
 The black dashed line just copy the $ n=1 $ lobe when the sample is static on the left.
 It evolves into the red solid line when the sample is moving by two steps.
 The step (I) to  the blue dashed line is due to $ H_{b1} $ with $ k_{0L} $ of the ground state in Eq.\ref{k0phase}.
 The $ z=1 $ QPT is in the new boosted 3D XY class with no emergent Lorentz invariance.
 The step (II) to the red solid line from the blue dashed line is due to $ H_{b2} $ with $ k_0 $ of the
 ground state in Eq.\ref{z2k0} and Eq.\ref{z2k0lattice}.
 The shape of the boundary is given by  Eq.\ref{z2boundary} with the non-monotonic  $ f(v) $ shown in Fig.\ref{phaseslattice}b.
 So the phase boundary shift in step (II) is also non-monotonic reflecting the underlying lattice structure.
 The Doppler shifts in the excitation spectrum in the BSF phase is given in Eq.\ref{GoldHiggs} and Eq.\ref{vxvytc}
 for $ z=1 $ and  in Eq.\ref{z2SFV} and Eq.\ref{VwZ2} for $ z=2 $ respectively. Their numerical values are given in Sec.XI-3.
 The $ z=(3/2, 3) $ line from the SF to the BSF transition in Fig.\ref{phasesz1} is not reachable
 in the bosonic Hubbard model, but reachable by  directly driving the SF  as analyzed in \cite{SSS}.
 The relations between the original bosons and the order parameters in the $ z=1 $ and $ z=2 $ effective theories
 are given in Eq.\ref{z1order} and Eq.\ref{z2order} respectively.
 (c) The KT transition temperature $ T_c $ in the SF side  increases when the sample is moving as shown in Eq.\ref{Tcz1} and Eq.\ref{Tcz2}.
 See also Fig.\ref{finiteT}a. }
\label{phaseslattice}
\end{figure}

\section{  The Global phase diagram of the Mott-SF transition in a lattice observed when the sample is moving   }

 In this section, we extend the derivation \cite{derivez1} of Eq.\ref{z1} from the boson Hubbard model Eq.\ref{boson} when the sample is static
to the boost case and derive Eq.\ref{boostz1} from Eq.\ref{boson} plus Eq.\ref{Hblattice12} when the sample is moving.
The derivation process is highly instructive and bring new insights to the emergent space-time from both the $ z=2 $ and the $ z=1 $ QCP.
Furthermore, it also establishes the relation between phenomenological parameters in Eq.\ref{boostz1} and \ref{boostz2}
and the microscopic parameters in Eq.\ref{boson}. When combining this relation with
the phase diagram Fig.\ref{phasesz1} and Fig.\ref{phasesz2} achieved by the effective actions in terms of the phenomenological parameters,
one reach the global phase diagram shown in Fig.\ref{phaseslattice}a,b.

\subsection{ The exact non-perturbative and perturbative treatments on the NN an  NNN boost term respectively}

Here we will still take the "divide and conquer " strategy \cite{CF} to examine the NN and NNN boost terms in Eq.\ref{Hblattice12}.
For the notional simplicity, we set $ v=1 $ first ( will put it back above Eq.\ref{alphasign} )
and study the boost along the x-bond. The $ t_{b1} $ boost term  in Eq.\ref{Hblattice12} can be combined with
the NN hopping term, since $t_0+it_{b1}=\sqrt{t_0^2+t_{b1}^2}e^{-ik_{0L} }$ with $ \tan k_{0L}=- t_{b1}/t_0 $,
one can introduce a lattice version of GT,
\begin{align}
    \tilde{b}_i=b_ie^{i k_{0L} i_y}
\label{k0phase}
\end{align}
  It maybe important to stress that the boost term $ t_{b1} $ term looks like a gauge field putting on the link.
  However, it was known any gauge field which generates a flux $ f $ through each plaquette breaks the lattice translational symmetry
  and generates non-commutative magnetic space group \cite{pq1,dual1,dual2}.
  However, all the boost terms keep the lattice translational symmetry, so
  if the current can be viewed as a gauge field on the link, it can only do so by a "pure gauge" which generates no flux.
  Indeed, for the particular case like $ H_{b1} $, this " pure gauge" can be transformed away by the "gauge" transformation
  Eq.\ref{k0phase}.

 Then Hamiltonian $H_1=H_0+H_{b1}$ can be rewritten in the same form as the original Bose Hubbard model Eq.\ref{boson}:
\begin{align}
    H_1&=-\sum_{i} (t_{0}\tilde{b}_{i}^\dagger \tilde{b}_{i+x}
                +\sqrt{t_0^2+t_{b1}^2}\tilde{b}_{i}^\dagger \tilde{b}_{i+y}+h.c.)-\mu\sum_i \tilde{b}_i^\dagger \tilde{b}_i
                \nonumber  \\
               & +U\sum_i \tilde{b}_i^\dagger \tilde{b}_i^\dagger \tilde{b}_i \tilde{b}_i
\label{absorb}
\end{align}
  with just the hopping anisotropy along the $ y $ bond.
  While the interaction $ U $ and the chemical potential $ \mu $ stay the same.
  So it does not change any symmetry of the Hamiltonian.
  Effectively it favors the SF over the Mott as one increases the boost.

  In fact, one can see the physics in Eq.\ref{absorb} easily when expanding:
\begin{align}
\sqrt{t_0^2 +t_{b1}^2} & e^{i k_{0L}} [ b_{i}^\dagger b_{i+y} +h.c]=
\sqrt{t_0^2+t_{b1}^2}[ \cos k_{0L} ( b_{i}^\dagger b_{i+y} +h.c)
              \nonumber  \\
 & + i \sin k_{0L}  ( b_{i}^\dagger b_{i+y} - h.c) ]
\end{align}
  where the first term is just the kinetic energy, the second is the conserved current.
  This expansion naturally explains why the first boost $ t_{b1} $ term can be absorbed by the unitary transformation Eq.\ref{k0phase}.
  While the other form of the boost such as the $ t_{b2} $ term in Eq.\ref{Hblattice12} may not.

  After the non-perturbation treatment of the $ t_{b1} $, we discuss the possible perturbation handling of  $ t_{b2} $.
  Substituting Eq.\ref{k0phase} to the $ t_{b2} $ term in Eq.\ref{Hblattice12} leads to:
\begin{align}
   & H_{b2}  =-i t_{b2} e^{i 2 k_{0L} } \sum_{i} \tilde{b}_{i}^\dagger \tilde{b}_{i+2x} + h.c.   \nonumber  \\
           & =-i  t_{b2} \cos 2 k_{0L} \tilde{b}_{i}^\dagger \tilde{b}_{i+2x}
           +  t_{b2} \sin 2 k_{0L} \tilde{b}_{i}^\dagger \tilde{b}_{i+2x}   + h.c.
\label{absorbtb2}
\end{align}
  where $  \cos 2 k_{0L} = \frac{t^2_0 - t^2_{b1} }{ t^2_0 + t^2_{b1} } $.
  The first term is odd in $ \vec{k} $, so even it could be small, it plays very important roles.
  The second term is even in $ \vec{k} $, so can be dropped relative to the dominant hopping term in Eq.\ref{absorb}.
  After inserting back the boost velocity $ v $ as $ t_{b1} \to t_{b1} v, t_{b2} \to t_{b2} v $,
  we will measure its strength relative to $ t_{b1} $ as:
\begin{equation}
   t_b= t_{b2} \cos 2 k_{0L}= \alpha_b v t_{b1},~~~~  \alpha_b=  ( \frac{v^2_c -v^2  }{ v^2_c + v^2  } ) \frac{t_{b2}}{t_{b1}} \ll 1,
\label{alphasign}
\end{equation}
  where $ v_c= t_0/t_{b1} $ displayed in Eq.\ref{vc}.
  Because $ \frac{t_{b2}}{t_{b1}} $ is independent of the boost $ v $,
  so $ \alpha_b $ changes sign when $ v  $ is tuned to pass  $ v_c $.

  Obviously, the discrete GT transformation can be easily extended to the boosts along  both $x $ and $ y $ bonds with
\begin{align}
     H_{b12}&=-iv t_{b1x}\sum_{i} (b_{i}^\dagger b_{i+y}-h.c.)
     \nonumber  \\
     &-iv t_{b1y}\sum_{i} (b_{i}^\dagger b_{i+y}-h.c.)
\end{align}
   Then the single particle spectrum takes the form
\begin{align}
    \epsilon_k
    &=-2t_0(\cos k_x+\cos k_y)
    \nonumber  \\
    &+2v t_{b1x} \sin k_x +2 v t_{b1y} \sin k_y
\label{bxby}
\end{align}
  which develops a single minimum at $ K=(k_{0Lx}, k_{0Ly} ) $ where
  $ k_{0Lx}=-\arctan( v t_{b1x}/t_0),
   k_{0Ly}=-\arctan( v t_{b1y}/t_0)  $.

  Then similar to Eq.\ref{k0phase}, the boosts can be transformed away by
\begin{align}
    \tilde{b}_i=b_i e^{i ( k_{0Lx} i_x + k_{0Ly} i_y ) }
\end{align}
  Then one may treat $ H_{b2} $  along both $ x- $ and $ y- $ bond  perturbatively by constructing an effective action.
  One may also just treat $ H_{b3} $ similarly in Eq.\ref{threeb} in the appendix D.

  In the following sections, we will address $ H_1 + H_{b2} $.
  Because $ H_{b1} $ has been absorbed into $ H_0 $, $ H_{b2} $ will be explored by effective actions in the continuum limit.
  This is the "divide and conquer" strategy to deal with the emergent phenomena.
  One can treat $ H_1 + H_{b12} $ similarly \cite{sloppy}. Note that Eq.\ref{k0phase} with $ \tan k_{0L}=-v/v_c $ can only be achieved
  by the microscopic calculations, not to be reflected in the effective action approaches in the Sec.II-VI.

\subsection{ The derivation of Eq.\ref{boostz1} and Eq.\ref{boostz2}
 from the lattice model: the emergent complex bosons above the Mott phase  }

We will do the calculations in the $ \tilde{} $  basis in Eq.\ref{absorb} and \ref{absorbtb2}. For notational simplicity, we will drop the tilde.
In the strong coupling limit $ U/t \gg 1 $,  the ground state is just the Mott state
$ | Mott \rangle = \prod_i ( b^{\dagger}_i )^N | 0 \rangle $. Its lowest excited state is either
one particle or one hole at the site $ i $:
\begin{equation}
 |p \rangle_i= b^{\dagger}_i | Mott \rangle,~~~~~ |h \rangle_i= b_i | Mott \rangle
\label{Mottb}
\end{equation}
  where one can see the $ p $ and $ h $ is defined with respect to the Mott state.
  Both of which annihilate the Mott state $ p  | Mott \rangle = h  | Mott \rangle =0 $.
  Because the Mott state is the " vacuum " state of both $ p $ and $ h $, so they are very much different  than
  the original boson creation $ b^{\dagger}_i $ and annihilation operator $ b_i $.

At $ t/U =0 $, the excitation  energy of the particle and hole are
$ E^0_p= U(1/2- \alpha) $ and $ E^0_h= U(1/2 + \alpha) $ respectively where $ \alpha=\mu-1/2 $ is
the re-defined chemical potential. Now we turn on  a weak hopping  and also a weak boost Eq.\ref{Hblattice}.
We first construct the translational invariant one particle and one hole eigen-state:
\begin{equation}
  | p \rangle = \frac{1}{\sqrt{N}} \sum_i e^{i \vec{k} \cdot \vec{R}_i } |p \rangle_i,~~~~
  | h \rangle = \frac{1}{\sqrt{N}} \sum_i e^{i \vec{k} \cdot \vec{R}_i } |h \rangle_i
\label{phstates}
\end{equation}

 Then by a first-order perturbation in both weak hopping  and weak boost,
 one can find the particle and hole eigen-spectrum in the continuum limit:
\begin{align}
  E_{p/h} &=E^0_{p/h}- 2t( \cos k_x+ \cos k_y) -2 t_b \sin 2 k_x
           \nonumber  \\
  & \sim \Delta_{p/h} + \frac{k^2}{2m} -c k_x + c k^3_x + \cdots
\label{phenergy}
\end{align}
  where $ t_b $ is given in Eq.\ref{alphasign}, the particle/hole excitation gap
  $ \Delta_{p/h}= U(1/2 \mp \alpha)-4t $. It is important to stress the crucial difference between
  the hopping and the boost in Eq.\ref{phenergy}:
  the former is even in the momentum, while the latter is odd. So the specific form of the hopping term and the current term
  may not be important, only their symmetry matter and lead to the same low energy form in Eq.\ref{phenergy}.

  Using Eq.\ref{alphasign} and taking the low energy limit, we find
  the effective mass and the boost velocity in the effective action Eq.\ref{phaction} can be expressed in terms of the 4 microscopic quantities
  on the lattice: the hopping $ t_0 $, the NN and NNN overlapping $ t_{b1}, t_{b2} $ and the boost $ v $:
\begin{align}
   v^2_x & = v^2_y= \frac{1}{2m}=ta^2, t = t_{b1} \sqrt{  v^2_c + v^2  },
                    \nonumber  \\
   c & = v \alpha_b t_{b1}a
\label{vxvytc}
\end{align}
   where we have put back the lattice constant $ a $ to make the dimension consistent, so they can be numerically evaluated
   in the experimental section X. It will be shown in the following that  $ v^2_y= \frac{1}{2m}=t a^2 $
   ( not the dimension of velocity yet, see Eq.\ref{relationz1} for $ z=1 $ which carries the velocity dimension and
   Eq.\ref{relationz2} for $ z=2 $ which does not respectively ) sets up
   the scale for the superfluid density $ \rho_s $ in the SF side for both $ z=1 $ and $ z=2 $,
   therefore the KT transition temperature in the SF side.
   While $ c $ is the Doppler shift term.

   So Eq.\ref{alphasign} and Eq.\ref{vxvytc} show that the characteristic velocity scale
   near the SF to Mott QPT is:
\begin{equation}
    v_c=t_0/t_{b1}
\label{vc}
\end{equation}
   which is completely determined by the Wannier functions of the lattice system. So it is
   completely different than the $ v^{O}_c $ and $ v^{SF}_c $ in the continuum in the case 1 and case 2 mentioned above \cite{twovc}.
   As listed in Eq.\ref{tijtbij} and Eq.\ref{tttt},
   $ t_0 $ stands for the kinetic energy of the hopping when the sample is static. 
   $ t_{b1} $ means the current generated by the boost of the moving sample,
   their ratio gives the only characteristic velocity scale of the system.
   Indeed, the Doppler shift $c $ also changes sign when $ v > v_c $:
   when $ v < v_c $, the Doppler shift term is in the same direction as the original $ v $.
   At $ v=v_c $, simply no shift, $ v > v_c $, it changes sign and becomes opposite to the original $ v $.
   Obviously, this surprising change of sign of the Doppler shift is due to the Mott state in Eq.\ref{Mottb}
   and experimental detectable in Sec.XI-2.

  Extrapolating Eq.\ref{phenergy} to larger hopping leads to the following simple physical picture:
  for $ \alpha > 0 $, the $ p $ condenses first at $ t/U \sim 1/2- \alpha $,
  $ \alpha < 0 $, the $ h $ condenses first at $ t/U \sim 1/2 + \alpha $, for
  $ \alpha=0 $, both condense at the same time at $ t/U \sim 1/2 $.
  These critical values will be slightly modified by various effects for the $ z=1 $ and $ z=2 $ respectively to be discussed in the following.
  The first-order perturbation can be pushed to the third order \cite{thirdorder} without changing the estimates qualitatively.

  In the path-integral language, one can simply write down
  the action to describe such a particle/hole condensation process as;
\begin{align}
  & {\cal L}[p, h] = p^{\dagger} ( \partial_\tau-ic \partial_x + \Delta_p-\frac{1}{2m} \nabla^2 ) p
                             \nonumber  \\
                &  + h^{\dagger} ( \partial_\tau-ic \partial_x + \Delta_h-\frac{1}{2m} \nabla^2 ) h
                              \nonumber  \\
                 & - \lambda ( p^{\dagger} h^{\dagger} + p h )
                 +  u_p ( p^{\dagger} p )^2 +  u_h ( h^{\dagger} h )^2 +   \cdots
\label{phaction}
\end{align}
  where  $ \lambda $ creates the particle-hole pair on two neighbouring sites,
  $ u_p, u_h $ stand for the on-site interaction of the $ p $ and $ h $ respectively and
  $ \cdots $ stand for all the possible higher order interactions  among the particles and holes.
  Due to the diluteness of the p- or h- near the QPT, we expect $ u_p, u_h \ll U $.

  Despite it is not known what is the precise mathematical  relation between
  $ (p, h), (p^{\dagger}, h^{\dagger} ) $ and the original $ (  b^{\dagger}_i, b_i) $,
  the original $ U(1) $ symmetry of the bosons $ b_i \to b_i e^{i \theta}  $ in Eq.\ref{boson} implies the $ U(1) $ symmetry
  of Eq.\ref{phaction} under which $ p \to p e^{i \theta}, h \to h e^{-i \theta} $. It dictates the conservation
  of $ p^{\dagger} p- h^{\dagger} h $. Despite the C-transformation \cite{Cdefine} in Eq.\ref{boson} can not be explicitly expressed in terms of
  $ (  b^{\dagger}_i, b_i) $, it can be easily expressed in the $ ( p, h) $ space as
\begin{equation}
   p \leftrightarrow h
\label{Csymmetry}
\end{equation}

  In view of the transformation in Eq.\ref{z2inv}, it is tempesting to do the
  simultaneous transformation of $ p $ and  $ h $ as $ p= \tilde{p} e^{i k_0 x }, h= \tilde{h} e^{i k_0 x } $.
  However, the p-h pair term becomes
  $ - \lambda e^{-i 2 k_0 x } \tilde{p}^{\dagger} \tilde{h}^{\dagger} +  h.c $ which becomes $ x- $ dependent
  in the $ \tilde{p}, \tilde{h} $ basis, so not easy to deal with.
  If had the boost term in the hole part in Eq.\ref{phaction} reversed its sign, the
  simultaneous transformation of $ p $ and  $ h $ would be  $ p= \tilde{p} e^{i k_0 x }, h= \tilde{h} e^{-i k_0 x } $,
  the p-h pair term would stay the same  in the $ \tilde{p}, \tilde{h} $ basis, then Eq.\ref{phaction} would own the Galileo Invariance.
  If had the p-h pair term changed to $ \lambda ( p^{\dagger} h + p h^{\dagger} ) $, it would have the same effects.
  So it is the Mott state and the p-h pair creation process in Eq.\ref{Mottb} which break the GI.
  This important fact is crucial to explore the emergent space-time from both the $ z=2 $ and the $ z=1 $ QPT.

  Without losing any generality, one can always make $ \lambda $ to be positive, so $ \lambda \sim t $.
  Note that the $ t_{b2} $ term, due to its odd in $ \vec{k} $, making negligible contributions to $ \lambda $ in the long wavelength limit.
  In the following, we set $ \lambda \to \lambda t $, so $ \lambda \sim 1 $ is dimensionless.

\subsubsection{  The emergent space-time from the $ z=1 $ theory }

  When $ \alpha = 0 $, $ \Delta_p=\Delta_h=\Delta_0=U/2-4t $ dictated by the particle-hole( C-) symmetry,
  so the particle and hole condense at the same time, so one must treat them at equal footing.
  After performing the unitary transformation:
\begin{equation}
    \Psi= \frac{1}{\sqrt{2} } ( p+ h^{\dagger} ),~~~~\Pi=\frac{1}{\sqrt{2} } ( p- h^{\dagger} )
\label{unitaryt}
\end{equation}
   Eq.\ref{phaction} can be written as:
\begin{align}
   {\cal L}[ \Psi, \Pi] & = \Psi^{\dagger} ( \partial_\tau-ic \partial_x) \Pi + \Pi^{\dagger} ( \partial_\tau-ic \partial_x) \Psi
                                            \nonumber  \\
    & +   \Psi^{\dagger}(-\frac{1}{2m} \nabla^2 +  \Delta_0 - \lambda t ) \Psi
                                              \nonumber  \\
   & +  \Pi^{\dagger}(-\frac{1}{2m} \nabla^2  + \Delta_0 + \lambda t ) \Pi + \cdots
\label{PP}
\end{align}
  which indicate $  ( \Psi^{\dagger}, \Pi) $ or $  ( \Pi^{\dagger}, \Psi) $ become conjugate variables.
  Under the $ U(1) $ symmetry, $ \Psi \to \Psi e^{ i \theta }, \Pi \to \Pi e^{ i \theta } $.
  Under the C-transformation $ \Psi \leftrightarrow \Psi^{\dagger}, \Pi \leftrightarrow -\Pi^{\dagger} $.
  Obviously, Eq.\ref{PP} enjoys both the  $ U(1) $ symmetry and the C-symmetry.

   Obviously when $ \Psi $ becomes critical, $ \Pi $ remains massive, so integrating out $ \Pi $ leads to
\begin{align}
  & {\cal L}[ \Psi]= Z_2 ( \partial_\tau-ic \partial_x) \Psi^{\dagger} ( \partial_\tau-ic \partial_x) \Psi
   \nonumber  \\
    & +  \Psi^{\dagger}(-\frac{1}{2m} \nabla^2 +  \Delta_0 - \lambda t ) \Psi +  u | \Psi|^4 + \cdots
\label{Psi}
\end{align}
  where $ Z_2=1/(\Delta_0+ \lambda t ) $.  It also enjoys both the  $ U(1) $ symmetry and the C-symmetry.
  After scaling out the factor $ Z_2 $ which is independent of $ c $  in the long wavelength limit $ {\cal L}[ \Psi]/Z_2 $, one can see
  it is nothing but the effective action Eq.\ref{boostz1} describing the  $ z=1 $ SF-Mott transition.

  Scaling out the factor $ Z_2 $ leads to the relation between the phenomenological parameters in the effective action
  and the microscopic quantities on the lattice scale:
\begin{align}
   v^2_{y,z=1} & \sim t a^2 [ U/2 + (\lambda -4 ) t ],
       \nonumber  \\
   r & =\Delta^2_0 - \lambda^2 t^2
\label{relationz1}
\end{align}
  where $ v_{y,z=1} $ depends on the interaction $ U $ explicitly and carries the dimension of velocity and
  can be contrasted to Eq.\ref{relationz2} with $ z=2 $ case which does not.

  Now we need to apply this connection to Fig.\ref{phasesz1}.
  As said below Eq.\ref{vxvytc},  $ r =0 $ controls the QPT boundary shift for $ z=1 $.
  While $ v^2_{y,z=1} $ controls the KT transition temperature $ T_c $ in the SF side.
  Setting $ r=0 $ leads to the critical value of $ (t/U)_c $ as:
\begin{equation}
    (t/U)_c = \frac{1}{2(4+ \lambda)}
\label{curvez1}
\end{equation}
  When taking into account the effective $ t $ in Eq.\ref{vxvytc}, we find the Step-I shift of the QPT boundary
  to the Mott side  \cite{sloppy}  at the tip with $ z=1 $ in Fig.\ref{phaseslattice}a.
  This Step-I shift is determined by the scale $ v_c $.

   At the QCP $ r=0 $, setting $ U/2= (\lambda +4)t $
   leads to $ v^2_{y,z=1}=2 \lambda ( t^2_0 + v^2 t^2_{b1} ) a^2 $.
   Because $ c= \alpha_b v t_{b1} a $ with $ \alpha_b \ll \lambda \sim 1 $ in Eq.\ref{alphasign}, so
\begin{equation}
    | c | \ll v_{y,z=1}
\label{clessless}
\end{equation}
   at any $ v $ on the lattice scale.
   It sets up the initial condition on the lattice scale for the RG flow.
   As shown in Sec.III, both $ c $ and  $ v_{y,z=1} $ are exactly marginal, so Eq.\ref{clessless}
   hold at any scale along the RG flow. So we conclude that the exotic $ z=(3/2,3) $ QPT line is not reachable, it can only move along
   the Path I in Fig.\ref{phasesz1}.
   In the strong coupling limit at a fixed $ U $, we start from in the Mott phase,  gradually increases $ t $ to
   approach the QCP, then pass into the SF in Fig.\ref{phaseslattice}. In this approach, Eq.\ref{relationz1} indicates that at a fixed $ U $,
   the intrinsic velocity $  v_{y,z=1} $ increases as $ t/U $ goes through the SF phase.

  As stressed in Sec.VIII-A, the relation Eq.\ref{k0phase} between the original boson on the lattice and the order parameter in
  the effective actions can not be  achieved just from the effective actions, but can be established by the microscopic calculations.
  For the $ z=1 $ effective theory, Eq.\ref{unitaryt} shows that:
\begin{equation}
   \Psi =e^{ -i k_{0L} x }\tilde{\Psi}
\label{z1order}
\end{equation}
  which can be detected in the Time of Flight (TOF) and the  scattering experiments outlined in Sec.X.

  Setting $ \partial_\tau $ vanishing in Eq.\ref{Psi}, one obtains:
\begin{equation}
   k_B T_c = Z_2 ( v^2_{y,z=1} - c^2 )
\label{Tcz10}
\end{equation}

  Then taking the ratio to get rid of the lattice constant $ a $ leads to:
\begin{equation}
   \frac{T_c}{T_{c0}}|_{z=1} = \sqrt{ 1 + (\frac{v}{v_c})^2 } -  \frac{ v^2 \alpha^2_b t^2_{b1} }{ [ U/2 + (\lambda -4 ) t ] t_0 }
\label{Tcz1}
\end{equation}
  where $ T_{c0} $ is the KT transition temperature when the sample is static.
  At the QCP, it simplifies to
  $ \frac{T_c}{T_{c0}} = \sqrt{ 1 + (\frac{v}{v_c})^2 } -  \frac{ v^2 \alpha^2_b t^2_{b1} }{ 2 \lambda t t_0 } $.
  It can be contrasted to Eq.\ref{KT} achieved just from the effective action.
  Of course, there is no way to get the $ Z_2 $ factor just from the symmetry based effective action.
  Due to Eq.\ref{clessless}, then $ T_c \sim t a^2 $
  as claimed below Eq.\ref{vxvytc}. So the boost indeed increase the $ T_c $ on the SF side.
  However, despite its smallness, the second term  with the negative sign contains important physics
  which originates from the Doppler shift to the spectrum, so experimental detectable.

\subsubsection{  The emergent space-time from the $ z=2 $ theory }

  When $ \alpha > 0 $, the particle condenses first $ \Delta_p \to 0 $ at the QCP at $ t/U \sim 1/2- \alpha $,
  while the hole remains massive $ \Delta_h > 0 $, so can be integrated out.
  Using the expansion of the hole propagator:
\begin{align}
  \frac{1}{ \Delta_h + X} & = \frac{1}{\Delta_h} - \frac{X}{ \Delta^2_h} + \frac{X^2}{ \Delta^3_h} + \cdots,
     \nonumber      \\
  X &  =  -\partial_\tau + ic \partial_x -\frac{1}{2m} \nabla^2
\label{hprop}
\end{align}
  where we have reversed the sign of the first two terms in the $ h- $ propagator due to the integration by parts on the hole part.
  Then integrating out the $ h- $ order by order in $ u_h $ leads to:
\begin{align}
  {\cal L}[ p ] & =  p^{\dagger} ( \partial_\tau-ic \partial_x + \Delta_p-\frac{1}{2m} \nabla^2 ) p + u_p |p|^4
              -   \frac{ \lambda^2}{ \Delta_h } p^{\dagger} p    \nonumber      \\
              & +  \frac{ \lambda^2 }{ \Delta^2_h }   p^{\dagger} X p
              - \frac{ \lambda^2 }{ \Delta^3_h }   p^{\dagger} X^2 p  + \cdots
               \nonumber      \\
              & +  4 u_h ( \frac{ \lambda }{ \Delta_h } )^4  ( p^{\dagger} p )^2
              - 4 \frac{ u_h }{ \Delta_h } ( \frac{ \lambda }{ \Delta_h } )^4 | p|^2
                 p^{\dagger} X p
\label{pactionlong}
\end{align}
   where one can see the last term in the first line renormalize down the p- gap.
   The first term in the second line  renormalizes
   the particle propagator, the second leads to the second derivative term
   $ Z_2= \frac{\lambda^2}{\Delta^3_h}  $ in the imaginary  time presented in Eq.\ref{Z1Z2} in Sec.VI-C  ( see also Eq.\ref{VwZ2} below ).
   While the third line involves the h- interaction $ u_h $:
   the first term just renormalizes the p- interaction, the second is the very important dangerously irrelevant term
   which leads to the Doppler shift term in the SF to be explored in the following.

   Dropping the second and the third line leads to:
\begin{equation}
  {\cal L} [ p ]= p^{\dagger} ( \partial_\tau-ic \partial_x + \tilde{\Delta}_p-\frac{1}{2m} \nabla^2 ) p
                + u |p|^4  + \cdots
\label{paction00}
\end{equation}
    where $ \tilde{\Delta}_p =\Delta_p- \lambda^2 t^2/\Delta_h $ is the renormalized p- gap.
    It is nothing but the
    effective action Eq.\ref{boostz2}. Because the high energy hole is projected out, so
    the conservation of $ p^{\dagger} p- h^{\dagger} h $ reduces to  $  p^{\dagger} p $.
    Even so, the $ p $ operator is still different than the original $ b_i $,   despite both have the $ U(1) $ symmetry.
    Very similarly, one can derive the identical action when $ \alpha < 0 $ where the hole condenses first,
    it is just related to Eq.\ref{paction} by the C- transformation.

    One can also establish the connections between the phenomenological parameters in Eq.\ref{z2}
    and the microscopic parameter in Eq.\ref{boson} at the lattice scale as
\begin{align}
   v^2_{y,z=2} & = \frac{1}{2m} \sim  t a^2 =t_{b1} a^2 \sqrt{v^2_c + v^2 },
        \nonumber   \\
   \mu & =- ( \Delta_p- \lambda^2 t^2/\Delta_h )=- \tilde{\Delta}_p
\label{relationz2}
\end{align}
   where  $ v_{y,z=2} $ is independent of the interaction at the lattice scale and
   does not carry the dimension of a velocity anymore as in the $ z=1 $ case shown in Eq.\ref{relationz2}.
   $ \mu=0 $ controls the QPT boundary shift for $ z=2 $ which contributes to the Step-I shift in the chemical potential in Fig.\ref{phaseslattice}.
   As said below Eq.\ref{vxvytc}, $ v^2_{y,z=2} $ control the KT transition temperature in the SF side.
\begin{equation}
   k_B T_c \sim v^2_{y,z=2} \sim t a^2=t_{b1} a^2 \sqrt{v^2_c + v^2 }
\label{Tcz20}
\end{equation}
 As explained in Sec.V, $ v_{y,z=2} $ is also exactly marginal under the RG, so
 Eq.\ref{Tcz1} simplifies for the $ z=2 $ case as:
\begin{equation}
   \frac{T_c}{T_{c0}}|_{z=2} = \sqrt{ 1 + (\frac{v}{v_c})^2 }
\label{Tcz2}
\end{equation}
  where $ T_{c0} $ is the KT transition temperature when the sample is static.

   Setting $ \mu=0 $  leads to the equation determining the critical value of $ (t/U)_c $ as:
\begin{equation}
    ( U/2t -4 )^2 - ( U/t)^2 \alpha^2= \lambda^2
\label{curvez2}
\end{equation}
    which indicates one can tune the QFT from the SF to Mott  by either $ t/U $ or $ \alpha= \mu-1/2 $.
    Setting $ \alpha=0 $ gives back to the $ z=1 $ result Eq.\ref{curvez1}.
    When taking into account the effective $ t $ in Eq.\ref{vxvytc}, we find the Step-I shift of the QPT boundary to the Mott side \cite{sloppy}
    away from the tip  with $ z=2 $ in  Fig.\ref{phaseslattice}a. So the Step-I shift at both $ z=1 $ and $ z=2 $ case is given by
    the scale $ v_c $ which will be estimated   in the experimental section X-2.

   Remembering the reverse sign of the boost in the second and third line in Eq.\ref{pactionlong}
   than the original p- propagator in the first line
   and performing the transformation  $ p= \tilde{p} e^{i k_0 x } $ with the $ k_0 $ listed in Eq.\ref{z2k0}
   and Eq.\ref{z2k0lattice}, one obtains the $ z=2 $ effective action describing the Mott to SF transition driven by the  p-
   BEC in the presence of the boost in the $ \tilde{p} $ basis:
\begin{align}
  {\cal L}[ \tilde{p} ] & = \tilde{p}^{\dagger} ( \partial_\tau + \tilde{\tilde{\Delta}}_p -\frac{1}{2m} \nabla^2 ) \tilde{p}
                + u |\tilde{p}|^4 - i V |\tilde{p}|^2 \tilde{p}^{\dagger} \partial_x \tilde{p}
                     \nonumber   \\
                & + i w \tilde{p}^{\dagger} \partial^3_x \tilde{p}-i\tilde{Z}_2\partial_\tau\tilde{p}^*\partial_y\tilde{p} + \cdots
\label{paction}
\end{align}
    where $ \tilde{\tilde{\Delta}}_p =\tilde{\Delta}_p - \frac{1}{2} m^2 c^2 $
    ( or equivalently  $ \tilde{\tilde{\mu}}_p =- \tilde{\tilde{\Delta}}_p =
    \mu + \frac{1}{2} m^2 c^2  $ in Eq.\ref{z2inv} )  is the renormalized particle gap  and
\begin{equation}
     V= 8 c \frac{ u_h }{ \Delta_h } ( \frac{ \lambda }{ \Delta_h } )^4,~ w = c,~\tilde{Z}_2=-2c Z_2=  -2c \frac{\lambda^2}{\Delta^3_h}
\label{VwZ2}
\end{equation}
    where $ V $ is the dangerously irrelevant term which also breaks the GI explicitly.
    It is also the term leading to the Doppler shift in the SF.
    We also added  the third-order  $ \partial^3_x $ term from Eq.\ref{phenergy}, so $ w = c$.
    For the completeness, we also listed the $ Z_2 $ term in Eq.\ref{Z1Z2} ( or $ \tilde{Z}_2 $ term in Eq.\ref{tildeZ2} ).
    Then it leads to Eq.\ref{z2primeleading} and Eq.\ref{tildeZ2}.
    Because both $ V, w $ and $ \tilde{Z}_2 $ are proportional to $ c $,
    so they change sign at $ v=v_c $ as explained below Eq.\ref{clessless}.

    Plugging the microscopic values in Eq.\ref{vxvytc} into Eq.\ref{z2inv} leads to:
\begin{equation}
    k_0= -\frac{ \alpha_b v }{ 2 a \sqrt{ v^2_c + v^2 } }
\label{z2k0lattice}
\end{equation}
    Eq.\ref{z2left} in the Mott side $ \mu < 0 $  contributes to the Step-I shift in the chemical potential in Fig.\ref{phaseslattice} 
    and becomes:
\begin{align}
    - \mu & =\frac{ \alpha^2_b v^2 t^2_{b1} }{ t_{b1} \sqrt{ v^2_c + v^2  } }=  ( \frac{t_{b2}}{t_{b1}} )^2 t_{b1}f(v),
     \nonumber   \\
    f(v) & = ( \frac{v^2_c -v^2 }{ v^2_c + v^2 } )^2 \frac{ v^2  }{ \sqrt{ v^2_c + v^2 } }
\label{z2boundary}
\end{align}
    where $ f(v) \sim v^2, (v-v_c)^2,  v $ when $ v \ll v_c,  \sim v_c,  \gg v_c $ shown in Fig.\ref{phaseslattice}b.
    Remembering the relation between this $ \mu $ and that in the BH model Eq.1 listed below Eq.\ref{z2},
    this lead to the step-II boundary shift at  a given $ v $  away from the tip with $ z=2 $ in Fig.\ref{phaseslattice}a.
    This Step-II shift is determined by the scale  $ \frac{t_{b2}}{t_{b1}} $ which will be estimated
    in the experimental section X-2.
    This non-monotonic behaviour in the boundary shift is also experimental detectable in Sec.X-2.

    For the $ z=2 $ effective theory, combining Eq.\ref{paction00} with Eq.\ref{paction} leads to
    the relation between the original bosons on the lattice and the order parameters in the effective action:
\begin{equation}
    p =e^{- i ( k_{0L} + k_0 ) x } \tilde{p}
\label{z2order}
\end{equation}
   which can be detected in the scattering experiments outlined in Sec.X.

  In fact, the derivations of Eq.\ref{boostz1} and  Eq.\ref{boostz2} rely on just one important fact that the hopping term is P, T even, but
  the boost term is P, T odd, but still PT even. It is independent of many microscopic details such as how many terms
  we keep in the hopping terms in Eq.\ref{pathprime} and the boost terms in Eq.\ref{Hblattice}.
  it establishes the connections between the phenomenological parameters in Eq.\ref{boostz1} and  Eq.\ref{boostz2}
  and the microscopic parameters in Eq.\ref{boson} plus Eq.\ref{Hblattice}, also
  the relation between the original bosons on the lattice and the order parameters in the effective actions Eq.\ref{z1order} and Eq.\ref{z2order}.
  Furthermore, it also bring additional insights to the emergent space-time of $ z=1 $ and $ z=2 $ QCP.

In a short summary of the results achieved in Sec.VII, there are two step I and step II to reach the global phase diagram Fig.\ref{phaseslattice}:
 Step 1: These effects are due to the $ H_{b1} $ and easily captured by the direct exact treatments on the lattice.
 The introduction of the BEC momentum $ k_{0L} $ of the ground state in Eq.\ref{k0phase}. There is also
         an increase of the hopping strength from $ t_0 $ to $ t=t_{b1} \sqrt{  v^2_c + v^2  } $ in Eq.\ref{vxvytc}.
         which, in turn, leads to the shift of the Mott-SF QPT into the Mott side.
         It also modifies the effective strength of the $ H_{b2} $ term in Eq.\ref{alphasign}.
 Step 2. These effects are due to the $ H_{b2} $ and can only be resolved by constructing the effective actions in the low energy limit.
 For $ z=1 $, there is no more BEC momentum shift and no more boundary shift either. The residual effect of $ H_{b2} $ is just
 to introduce a Doppler shift in Eq.\ref{clessless} to the excitation spectrum in both the Mott
 Eq.\ref{Mottz1} and the SF phase Eq.\ref{GoldHiggs}.
 For $ z=2 $, there is one additional BEC momentum shift $ k_0 $ of the ground state in Eq.\ref{z2k0}, Eq.\ref{z2inv}
 and Eq.\ref{z2k0lattice},  one more boundary shift $ \tilde{\mu} $ in Eq.\ref{z2inv}
 and Eq.\ref{z2boundary}. The shape of the boundary is given by the non-monotonic
 $ f(v) $ in Eq.\ref{z2boundary} and shown in Fig.\ref{phaseslattice}b.
 Even so, there are still residual effects of $ H_{b2} $ which come from
 the dangerously irrelevant $ i V $ term in Eq.\ref{z2cubic} and Eq.\ref{paction}
 and also the $ Z_2 $ cross metric term in Eq.\ref{Z1Z2}. Their values at the lattice scale are listed in Eq.\ref{VwZ2}.
 The $ iV $ term leads to the Doppler shift in Eq.\ref{z2SFV} in the excitation spectrum in the BSF phase.
 The $ Z_2 $ term leads to the Doppler shift in both the Mott and the BSF term as presented in Sec.II-C.
 Combining the results at $ z=1 $ and $ z=2 $ leads to Fig.\ref{phaseslattice}a,b.
 The practical size at step-1 is determined by $ v_c= t_0/t_{b1} $,
 while that of step-2 is determined by $ \alpha_b \sim t_{b1}/t_{b2} $.
 As shown in Eq.\ref{Tcz1}, Eq.\ref{Tcz2} and Fig.\ref{phaseslattice}c,
 the KT transition temperature $ T_c $ increases when the sample is moving for both $ z=1 $ and $ z=2 $.

 We ended this section by proposing  two theorems which apply to any lattice systems:

{\bf Theorem 1 }:
For any order to disorder quantum phase transition at $ T=0 $
in a static sample with $ P $ or  $ T $    symmetry,  the QCP
always moves to the symmetric side in a moving sample.

{\bf Theorem 2 }:
For any  order to disorder classical phase
transition at $ T > 0 $ in a static smaple with $ P $ or  $ T $  symmetry,
the critical temperature $ T_c $ in the symmetry breaking side  must increase in a moving
sample.

How does it moves in Theorem 1 depends on specific lattice models.
How does it increase in Theorem 2 depends on specific lattice models. This is also due to the fact stressed in the introduction
that there is an absolute lab frame in any materials where the lattice is static, so the critical $ T_c $ reaches the
absolute minimum when the sample is static.  This is also the frame where most experiments are carried over.
It would interesting to have a mathematical proof of the two theorems independent of any microscopic lattice models.

\begin{figure}[tbhp]
\centering
\includegraphics[width=.9\linewidth]{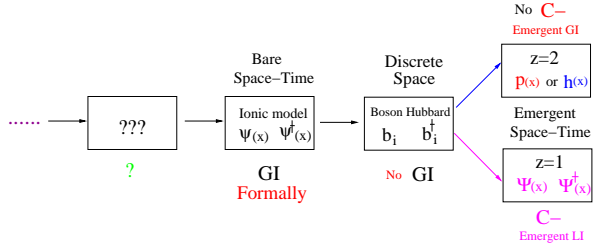}
\caption{ The hierarchy  of the energy scales and emergent space-time near a QPT.
The emergent space-time in the effective action could be very much different than the bare one in the ionic model.
The dots on the left means that it come from the low end of Fig.\ref{towerlevelm}.
The ionic model means Eq.\ref{Seq0Nlattice}. C- means the charge conjugation symmetry.
Performing the GT at the lowest level leads to Eq.\ref{boostz1} and Eq.\ref{boostz2} respectively.
But one must also perform the GT on the ionic and Boson Hubbard model to establish the connections between
the order parameters and the phenomenological parameters in the two effective actions and those bare ones at the lattice scale in the ionic model.
The question marks in the first box means we do not even know what is the Hamiltonian at this level and if
it has GI or not. The emergent LI in the $ z=2 $ effective theory really means the pseudo-LI as stressed in the introduction.
In general, the internal symmetry can only increase along the arrow, but the space-time symmetry may not.
Another example is the bulk FQH and its associated edge mode to be presented in appendix G and H.
For its possible implications on effective theories in high energy physics, see \ref{towerlevelm}. }
\label{towerlevel}
\end{figure}

\subsection{  On the hierarchy of energy scales  and  emergent space-time from a QCP  }

  The above analysis shows that the effective tight-binding second quantization  Boson-Hubbard  model Eq.\ref{boson} is not Galileo invariant anymore. This observation can also be reached intuitively:
 in Eq.\ref{Seq0Nlattice2prime}, the boson coordinate $ \vec{x} $ and the lattice site $ \vec{R} $ are two separate coordinates.
 in the tight-binding limit in Eq.\ref{boson}, the continuous boson coordinate $ \vec{x}_i $ is pinned to be the same as the discrete lattice site $ i $,
 so that $ \vec{x}=\vec{R} $. One may also call this procedure as the lattice compactification.
 The lattice just discretizes the space  into a square lattice which confines the bosons to hop on.
 However, only on such a discretized lattice, one can define the P-H ( C ) symmetry stressed in the introduction \cite{Cdefine}.
 The Eq.\ref{z1} is an even more effective action  than the boson Hubbard model Eq.\ref{boson}. It
 describes the Mott-SF transition  and
 is not GI either. Then when the sample is moving, the effective action should be Eq.\ref{boostz1}. Due to the pinning,
 the GT of the lattice site $ x_i \to x^{\prime}_i+ c t^{\prime} $  is also automatically encoded in Eq.\ref{boostz1}.
 Both the $ z=1 $ and $ z=2 $ SF-Mott transition happens at integer fillings which
 has the C- and no C-symmetry respectively. It is this C-symmetry which plays the crucial roles in Figs.\ref{phasesz1},
 \ref{phasesz2}, \ref{phaseslattice}a and also the emergent space-time encoded in Eq.\ref{boostz1} and Eq.\ref{boostz2},
 but it is not even defined in the bare action Eq.\ref{Seq0Nlattice2} which defines the bare space-time.

  One could start from the microscopic Hamiltonian Eq.\ref{Seq0Nlattice} which, of course, is also an approximation to real materials.
  It does not include many possible other terms such as the lattice vibrations ( or phonons ), SOC, etc.  The relevant fields are
  the electron or boson field operators $ b(\vec{x}), b^{\dagger}(\vec{x}) $. It has the GI.
  But it is even hard to see any emergent phenomena, especially if there is any QPT at this level (see Fig.\ref{towerlevel} ).
  It reduces to the boson Hubbard model Eq.\ref{boson} after many truncations in Eq.\ref{pathprime}. The relevant fields are
  the boson field creation or annihilation operators $ b_i, b^{\dagger}_i $ attached to a given lattice site $ i $.
  It does not have the GI anymore. But it has the QPT from the SF to the Mott phase shown in Fig.\ref{phaseslattice}.
  Then  near the QPT, one can construct the effective action Eq.\ref{z1} and Eq. \ref{z2} to describe the QPT
  at the tips and away from the tips respectively. The relevant fields are either the particle $ p $ or the hole $ h $ for the $ z=2 $ case, or the field $ \Psi $ in Eq.\ref{unitaryt} which is a linear combination of $ p $ and $ h^{\dagger} $.
  Both $ p $ and $ h $ are defined with respect to the Mott state Eq.\ref{Mottb},
  can be considered as the two emergent particles from, but  not simply related to the $ b_i, b^{\dagger}_i $.
  It does not have the GI either, but the C- symmetry can be readily
  and explicitly constructed in this $ p $ or $ h $ representation ( Fig.\ref{towerlevel} ).

  It is important to observe that  Eq.\ref{z1} and Eq. \ref{z2} are completely determined from the symmetry principle, independent of many microscopic details. For example, we can add many terms to the Boson Hubbard model Eq.\ref{boson}
  consistent with the symmetries, without changing the resulting effective actions Eq.\ref{z1} or and Eq. \ref{z2}.
  However, to establish the connections of the order/phenomenological parameters with the microscopic  parameters in a lattice, one must also
  perform the GT starting from the ionic model in the tower in Fig.\ref{towerlevel}.




\section{ Contrasts to the Doppler shift, Temperature shifts, Unruh effects, emergent $ AdS $ geometry  in  relativistic QFT  }

In relativistic QFT, different inertial frames are just related by Lorentz transformations.
relativistic Doppler shift is just a direct consequence of Lorentz transformations.
For relativistic QFT, this is just the end of the story, Doppler shift will not lead to no new phases and new QPT.
For non-relativistic quantum many body systems,  Doppler shift is just a direct consequence of Galileo transformation.
However, as shown in the previous sections, Doppler shift Eq.\ref{GoldHiggs},\ref{sfdis},\ref{bsfdis},\ref{Mottz2},\ref{BSFz2}
is far from being the end of the story. When the shift goes beyond the intrinsic velocity of the matter,
it may trigger QPT and  leads to new quantum phases.
This is another explicit demonstration of P. W. Anderson's view on quantum many body systems
" More is different  " which can be expanded to include emergent space-time as shown in Fig.\ref{moretwo}.



\begin{figure}[tbhp]
\centering
\includegraphics[width=.8 \linewidth]{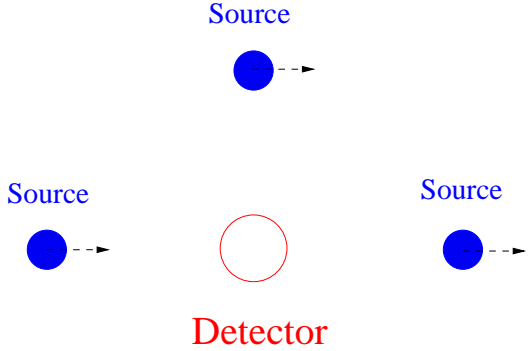}
\caption{ The Doppler shift  in  the special relativity. 
Due to the absence of the reservoir, One can exchange the role of the source with the detector.
This is in sharp contrast to the grand canonical ensemble in Fig.\ref{frames} and Fig.\ref{moretwo},  }
\label{source}
\end{figure}

\subsection{ Contrast to Doppler shift  in  the special relativity }

In relativistic quantum field theory, different inertial frames are related by Lorentz transformation, so they are completely equivalent.
Even so, changing to a different inertial frame,
the frequency changes as follows:
\begin{equation}
 \omega^{\prime}= \gamma ( \omega - \vec{v} \cdot \vec{k} ),~~~\gamma=1/\sqrt{1-\beta^2},~~~\beta=v/c_l
\label{qftboost}
\end{equation}
 which is the frequency in the boosted frame. For the corresponding momentum shift in the 4-vector $ ( \vec{k}, i \omega/c_l ) $,
 see Eq.\ref{qftboostk} in the appendix A.
 Because $ \omega^2= c^2_l \vec{k}^2  + m^2 c^4_l $ and $ v < c_l $, so
 $\omega^{\prime} $ always remains the same sign as $ \omega $.
 For the light $ m=0 $, then $ k_1 = \omega/c_l \cos \theta,  k_2 = \omega/c_l \sin \theta $ in the lab frame, then
\begin{equation}
  \omega^{\prime}= \omega \frac{ 1-\beta \cos \theta }{ \sqrt{1 - \beta^2 } }
\label{dslight}
\end{equation}
 When $ \theta= 0, \pi $, it simplifies to $ \omega^{\prime}= \omega \sqrt{ \frac{1- \beta}{1+\beta} } = \omega ( 1- \beta+ \cdots ) $
 where $ \beta=v/c_l $ takes positive ( negative )  when the source is moving away ( towards ) from the detector 
 in Fig.\ref{source}   ( which correspond to the moving sample in Fig.\ref{frames}a ).
 This is nothing but well known relativistic Doppler shift.
 Positive ( negative ) frequency stays invariant in different inertial frames, no QPT.

 When $ \theta=\pi/2 $, it reduces to the transverse  Doppler shift  in Fig.\ref{source}:
\begin{equation}
 \frac{ \omega^{\prime} }{\omega }=  \frac{ 1 }{ \sqrt{1 - \beta^2 } }
 =   ( 1 + \beta^2/2 + \cdots )
\label{tranlight}
\end{equation}
  which in facts, hold for any transverse motion $ \vec{v} \cdot \vec{k} =0 $ in Eq.\ref{qftboost}, not just for the light.
  This transverse Doppler shift for any massless or massive QFT can be contrasted to Eq.\ref{tt0T}.

 Eq.\ref{qftboost} can be contrasted with the non-relativistic Doppler shift term in  Eq.\ref{GoldHiggs},\ref{sfdis},\ref{bsfdis},\ref{Mottz2},\ref{BSFz2}
 ( see also \cite{dopplerhigh,thermalhigh} ).
 For a massive particle,
 when taking the non-relativistic limit $ k \ll m c_l $ ( or $ k^2/2m \ll m c^2_l $ ) limit, as expected,
 they become identical with $\omega^{\prime}=  \omega - \vec{v} \cdot \vec{k} $ and
\begin{align}
  \omega=mc^2_l + \vec{k}^2/2m -k^4/8 m^3c^2_l+ \cdots
\label{vc2}
\end{align}
  where the rest mass $ m c^2_l $ is still important ( Appendix A ) to keep $\omega^{\prime} $ always positive in taking the non-relativistic limit. However, the crucial difference between Eq.\ref{qftboost} and our case is that:
 In the former, despite the frequency of a mode depends on the choice of the inertial frame,
 the decomposition into positive and negative frequencies
 is invariant. While, in the latter, the positive frequency can turn into a negative one driven by the boost,
 therefore trigger the QPTs shown in Fig.\ref{phasesz1} and  Fig.\ref{phasesz2}.

\subsection{ Contrast to temperature shift in  the thermal relativistic QFT }

\begin{figure}[tbhp]
\centering
\includegraphics[width=.5 \linewidth]{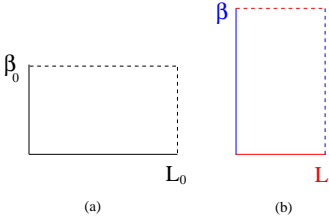}
\caption{ In the relativistic thermal QFT at a finite temperature $ \beta=1/T $,
the finite size is invariant under the LT: $ \beta L= \beta_0 L_0 $. This should close its long dispute on how
the $ T $ transforms under the LT ( see appendix L ).
Unfortunately, it is still unknown how to experimentally measure such a temperature effect in a moving frame. }
\label{arealaw}
\end{figure}

  The relations Eq.\ref{Tcz1} and Eq.\ref{Tcz2} can be contrasted  to the well known classical relativistic effect:
\begin{equation}
   \frac{\Delta t}{ \Delta t_{0}} = \frac{1}{ \sqrt{1 - (\frac{v}{c_l})^2 } } >1
\label{tt0}
\end{equation}
 where $ \Delta t_0 $ is the time elapse in the co-moving frame.
 If one use the relation between the imaginary time $ \tau= i t= 2 \pi/ k_B T $, then one reach:
\begin{equation}
   \frac{T}{T_{0}} =  \sqrt{1 - (\frac{v}{c_l})^2 }  < 1
\label{tt0T}
\end{equation}
  In the non-relativistic limit $ c_l \to \infty $, it reduces to $ T=T_0 $.
  When $ v/c_l \to 1^{-} $, $ T \to 0 $ approaches to zero temperature.
  It would be also important to extend this to the relativistic QFT at both finite temperature and finite chemical potential
  which may find some applications in the quark-gluon plasma in RHIC experiments ( see appendix L ).

  In fact, one can also derive Eq.\ref{tt0T} just using the simple fact that the measure $ dx^{\mu} dx_{\mu} $ is LI
  and the relativistic quantum field theory in Euclidean signature at a finite temperature:
\begin{equation}
 {\cal Z}= e^{ - {\cal S} },~~~{\cal S}[ \beta, L ]= \int^{\beta}_0 d \tau \int^{L} d^d x {\cal L }[\phi]
\label{actionfiniteT}
\end{equation}
  where the Lagrangian $  {\cal L } $ is assumed to be Lorentz invariant. Then the total finite size of the system  $ \beta L= \beta_0 L_0 $
  is LI Fig.\ref{arealaw}, then $ L=L_0 \sqrt{1- (\frac{v}{c_l})^2 } $ implies $ \beta/\beta_0 = 1/\sqrt{1- (\frac{v}{c_l})^2 } $
  which leads to Eq.\ref{tt0T}. Despite the facts that the LT  Eq.\ref{spin012} does not change any parameters of the Lagrangian $  {\cal L } $
  and  the integration regime in Fig.\ref{arealaw} has the same area, but the regime is different, so in general
\begin{equation}
 {\cal Z}[ \beta, L ] \neq  {\cal Z}[ \beta_0, L_0 ]
\label{SBL}
\end{equation}
  which tells the partition function  is not a  LT scalar in any two initial frames.

  One may also understand Eq.\ref{SBL} directly from the definition  of the partition function
  $ {\cal Z}= Tr e^{- \beta (H- \mu N) } $ where the Hamiltonian $ H $ and the total number of particle $ N $ are
  conserved quantities at any given initial frames, but not a LT scalar,
  so does $  {\cal Z} $. So the free energy $ f= -\ln Z/\beta V $   is not a  LT scalar either.
  Note the four momentum $ ( i E/c, \vec{p} ) $ and the four currents $  ( i c \rho , \vec{j} ) $.



 One may see possible interesting connections between
 Eq.\ref{Tcz1}, Eq.\ref{Tcz2} and  Eq.\ref{tt0T}. However, despite these superficial resemblances, the physics is quite different:
 the former is completely non-relativistic ( namely in sending $ c_l \to \infty $ limit ), $ v_c $ is the intrinsic velocity determined by
 the Wannier function of the lattice system, also sending $ \hbar \to 0 $ classical limit,
 but it is on the emergent classical phase transition of a macroscopic lattice system.
 It reflects P. W. Anderson's "more is different".
 The latter is completely relativistic, $ c_l $ is the light velocity in the vacuum, still sending $ \hbar \to 0 $ classical limit,
 Originally, it is on a single fast moving classical particle only, here, we also extended to non-relativistic QFT at a finite temperature.

  In the low velocity limit $ \frac{v}{c_l} \ll 1 $ of  Eq.\ref{tt0T}
\begin{equation}
   \frac{T-T_{0} }{T_{0}} =   - \frac{1}{2} (\frac{v}{c_l})^2  < 0
\label{tt0Tminus}
\end{equation}
  which can be contrasted to the difference between Eq.\ref{Tcz1} and Eq.\ref{Tcz2}:
\begin{align}
  \frac{T_c}{T_{c0}}|_{z=1} - \frac{T_c}{T_{c0}}|_{z=2} = -  \frac{1}{2}  ( \frac{\alpha^2_b t^2_{b1} }{\lambda t t_0 } ) v^2 <0
\label{Tcminus}
\end{align}
  As can be seen from Eq.\ref{tranlight},
  both down shifts can be interpreted as coming from the same source as the Doppler shifts in both cases.

  As explained in the experimental section X-3, the experimental detection of Eq.\ref{tt0Tminus} need to get to a moving body at a speed not too below
  the speed of the light $ c_l $. However, Eq.\ref{Tcminus} is a completely non-relativistic effect, so can be tested
  at a very low velocity $ v_c \sim 1 cm/s $.

\subsection{ Contrast to the Unruh effects in the general relativity }

 However, the in-equivalence may come from a non-inertial frame such as an uniformly accelerating frame.
 Even an observer at rest in the Minkowski space-time see a true vacuum $ |0\rangle_M $ with no particles.
 An uniformly accelerating observer would see a quite different vacuum $ |0\rangle_R $ with no particles in the  Rindler space-time
 $ _R\langle 0 | n_R(k)|0\rangle_R=0 $, due to the Einstein's equivalence principle.
 However, it does have the  excitations in $ |0\rangle_M $:
\begin{equation}
  _M\langle 0 | n_R(k)|0\rangle_M = \frac{1}{e ^{2\pi \omega/a} -1 }
\end{equation}
  where $ \omega= \sqrt{c^2_l \vec{k}^2 + m^2 c^4_l }  $ also listed below Eq.\ref{qftboost}.

 This is the well known Unruh effects \cite{unRev1,unRev2}:
 the uniformly accelerating observer ( with a constant 4-acceleration $ a $ ) will see a thermal bath of particles with the temperature:
\begin{equation}
  k_B T_U= \frac{  \hbar a }{ 2 \pi c_l}
\label{TU}
\end{equation}
 Namely a pure state at $ T=0 $ in Minkowski space-time becomes a mixed state at $ k_B T_U $.
  Unfortunately, the Unruh effect is so small that it is extremely difficult to detect.
 A proper acceleration of $ a= 2.47 \times 10^{20} m/s^2 $ corresponds to $ T_U \sim 1 K $.
 Or conversely, $ a = 1 m/s^2 $ corresponds to $ T_U \sim 4 \times 10^{-21} K $ which is beyond any current available cold atom experiment.

 The observer draws a worldline ( trajectory ) $ x^2=t^2+ (1/a)^2  $ in the  Minkowski space-time.
 The reference frame where an  uniformly accelerating observer is at rest is called Rindler space-time
 which is related to the Minkowski space-time by $ x= r cosh \eta, t= r sinh \eta $.
 It is confined to the wedge $ x \geq |t| $ separated by the " Rindler horizon " at $ x= \pm t, x >0  $ from the rest of the space-time.
 The entanglement entropy between A and B in the Minkowski space is equal to the thermal entropy of B in the Rindler space
 with $ T_U $ in Eq.\ref{TU}.
 The uniformly accelerating observer follows the trajectory with $ r=1/a, \eta= a \tau $ where $ \tau $ is the proper time.

 In relativistic QFT, the Unruh effect tells us that two different sets of observers such as
 inertial and Rindler will describe the same quantum state $ |0\rangle_M $ in very different ways.
 Here in a non-relativistic quantum field theory on a lattice, we showed that even two different inertial observers will also
 see the same quantum system in very different ways. It is constructive to compare
 Eq.\ref{TU} with Eq.\ref{Tcz1} and Eq.\ref{Tcz2}: the main difference is that (1) the former is for an accelerating system, so
 $ a $ is the acceleration, the latter is for an inertial frame, so $ a $ is just a lattice constant,
 $ v $ is the velocity of the moving frame relative to the lattice.
 (2) the former is for the change from $ T=0 $ in any inertial frame to $ T_U $ in a non-inertial frame,
 so no way to form a universal ratio,
 the latter is for the change in the critical temperature $ T_c $ of a classical phase transition in  a moving frame,
 so one can form a universal ratio $ T_c/T_{c0} $.
 (3) The former is a quantum effect which vanishes as $ \hbar \rightarrow 0 $.
 It is also a relativistic effect  which vanishes as
 $ c_l \rightarrow \infty $ ( namely when taking a non-relativistic limit by sending $ c_l \rightarrow \infty $ ).
 While the latter survives not only the $ c_l \rightarrow \infty $, but also the classical limit $ \hbar \to 0 $.
 (4) Especially, as estimated in experimental section  X-2, the KT transition temperature $ T_c $
 increases substantially when the sample is moving for both $ z=1 $ and $ z=2 $, so it may be more robust than the Unruh effect.
 This shows that non-relativistic systems
 show much richer effects than their relativistic counter-parts, which can be much more easily detected experimentally in condensed matter or cold atom set-ups to be presented in Sec.XI.

\begin{figure}[tbhp]
\centering
\includegraphics[width=.9\linewidth]{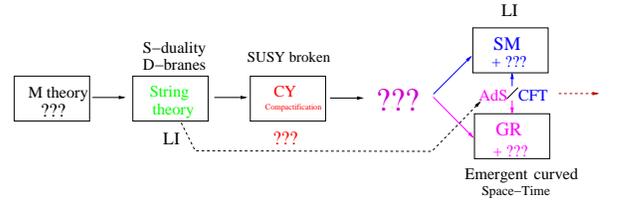}
\caption{ The hierarchy  of the energy scales and emergent space-time starting from the Planck scale $ \sim 10^{19} $ Gev
down to the Standard Model (SM) scale $ \sim 10^{2} $ Gev.
The question marks in the first box means we do not even know what is the global structure of the 11 dimensional M theory.
A string propagating in a  9+1  dimensional flat space-time has the LI (
A string propagating in a curved 10 dimension was not well studied yet ).
The question marks in the Calabi-Yau (CY) compactifications mean we are far from being know how to achieve such a dimensional reduction
from 9+1 to 3+1 dimension.
The big question marks mean there are huge barriers from such a dimensional reduction to
the Standard Model and the classical general relativity plus any possible quantum corrections denoted by the small question marks
which could face any experimental tests.
The arrow on the right means that it moves to the direction of Fig.\ref{towerlevel} on non-relativistic QFT.
The AdS/CFT comes from the low energy limit of open string on D-branes and closed strings respectively.
Because it circumvents the CY compactifications ( denoted by the dashed line ),  so it is far from
the real possible duality between SM and GR yet. The D-branes are similar to some topological defects in materials such as the
domains walls in a Ising ferromagnet, vortices in the SF, monopoles in a FM, dislocations or disclinations in topological crystalline insulators, etc.
It plays the remarkable roles of unifying 5 different string theory and 11d supergravity into
the M theory on the most left box, it also leads to the AdS/CFT.
One may see the one box to one box correspondence between Fig.\ref{towerlevel} and here.
For example,  the ionic lattice model box there looks like exact, but not effective and lacks any predictive power,
so it may be compared to the string theory box here which, so far, also lacks any predictive power on the real world.
The lattice projection box there may be compared to the CY compactification box here,
the $ z=1 $ and $ z=2 $ effective actions there may be compared to the SM and GR here, etc.
For an interesting unification of various versions of  the SM under the
framework of QPT, see a recent work \cite{ultra}.}
\label{towerlevelm}
\end{figure}

\subsection{ Emergent curved space-time from the boundary in low dimensional $ AdS_{d+1}/CFT_d $.  }

High Energy physics such as String (M) theory and quantum gravity
always starts from the (build in) high symmetries which look quite unusually large groups  such as $ SO(32) $, $ E_8 \times E_8 $, etc.
also some usual symmetries such as the Lorentz invariance, Supersymmetry, conformal invariance, etc.
Then try to understand how these symmetries are broken explicitly or spontaneously. Its low energy limit naturally leads to
the  $ AdS_5/CFT_4 $ duality with a $ SO(4,2) $ isometry group structure ( Appendix A )
where $ CFT_4 $ stands for the 4-dimensional  $ {\cal N}=4 $ SUSY $ SU(N) $ Yang-Mills
gauge theory in the large $ N $ limit.
The planck length $ l_p = ( \hbar G/c^3_l )^{1/2}\sim 1.6 \times 10^{-33} cm $ is the natural length in such a low energy limit.
Obviously  $ l_p \to 0 $, by taking $ \hbar \to 0 $ or $ G \to 0 $, especially the non-relativistic limit $ c_l \to \infty $.
This fact indicates the quantum gravity disappears not only in the $ \hbar \to 0 $ limit, but also
in the non-relativistic limit $ c_l \to \infty $. Of course, $ G \to 0 $ limit simply means the gravity disappears.
This maybe called Top-down approach in Fig.\ref{towerlevelm}.

On the other hands, condensed matter physics always starts from
the real materials which in most cases, living on a lattice and have much lower exact symmetries:
No Lorentz invariance, no Supersymmetry,  no conformal symmetries, etc.
Then try to understand how these symmetries and other bigger symmetries such as $ SO(8), SO(10), E_8 $ etc.  emerge.
This maybe the basic ingredient of  P. W. Anderson's great insight: More is different !
In fact, as argued in \cite{NPwitten}, even the curved space-time can emerge from some condensed matter systems on the boundary.
This is called Down-top approach in Fig.\ref{towerlevel} which is complementary to the above Top-down approach in Fig.\ref{towerlevelm}.
The renormalization group (RG) in the materials living on the boundary is closely related to the general relativity (GR) in the bulk.
The $ RG=GR $ means that the energy scale in the boundary is related to the radial coordinate  in the Poincare coordinate
in the bulk as
\begin{equation}
 RG=GR,~~~~E \sim 1/z
\label{E1z}
\end{equation}
So the cutoff of the boundary layer $ \epsilon $ corresponds to the UV cutoff the boundary CFT $ \Lambda \sim 1/\epsilon $,
while the $ z \to \infty $ bulk geometry corresponds to the IR limit of the CFT.

The Sachdev-Ye-Kitaev (SYK) model \cite{SY,kittalk,syk2} is a natural product of such
a down-up approach. The SYK is an infinite range random interacting 4-fermion model, so it is an effective zero-space dimensional model
which of course, is neither Lorentz nor Galileo invariant. It has an emergent parametrization invariance ( which can also be called 1d
conformal invariance ) which is both explicitly and spontaneously broken.
The IR limit of the SYK model is  described by the Schwarzian
in terms of the gapless re-parametrization mode which, in turn, corresponds to the $ c= \infty $ limit of the 2d Liouville CFT.
The identical Schwarzian also emerges in the IR geometry of the 2d Jackiw-Teitelboim (JT) gravity.
This connection between the SYK in the boundary and the 2D JT gravity  naturally leads to the $ NAdS_2/NCFT_1 $ duality
where $ NCFT_1 $ stands for the  emergent parametrization invariance with a leading irrelevant operator $ \partial_\tau $,
while $ NAdS_2 $ stands for the dilaton field added to the pure $ AdS_2 $ gravity.
Obviously, the speed of light $ c_l $ does not even appear in the microscopic SYK model at the first place.

Now we move to one dimension higher   $ AdS_3/CFT_2 $ where the $ CFT_2 $ stands for the 2d CFT with a central charge $ c $
which is related to the $ AdS_3 $ geometry with a radius $ R $  by $ c= 3R/2 G^{(3)}_N  $.
Some typical $ CFT_2 $ is the 1d Luttinger liquid  and 1d chiral edge state of FQH \cite{wen}.
This correspondence holds best in the central charge $ c \to \infty $ limit \cite{stanford}.
Here again in the $ CFT_2 $, the space-time is related by the intrinsic velocity $ v $ as  $ z= x+ivt, \bar{z}= x-ivt $,
the speed of light $ c_l $ is also irrelevant and plays no role in such a correspondence.
This is in sharp contrast to the CFT in the worldsheet
$ ( \sigma, \tau ) $ of the string theory in the second box of Fig.\ref{towerlevelm}
where the space-time is related by the light velocity $ c_l $.
Just like SUSY is not necessary, the Lorentz invariance is also not necessary in $ AdS_{d+1}/CFT_d $ with the isometry
group $ SO(d, 2) $ at low $ d $.
As stressed in the previous paragraph,  the bulk space-time structure can be a low energy product emergent from the boundary \cite{NPwitten}.
Indeed, as shown by the RT formula \cite{RT}, the entanglement entropy (EE) in the  boundary can be evaluated by the minimum surface
in the bulk with a suitable AdS metric \cite{lovelock}. This fact may also be interpreted as the bulk space-time
geometry emerge from the boundary.
A concrete generation of the 3d AdS bulk geometry in terms of  2d boundary Liouville action, application to $ NAdS_2/NCFT_1 $
in the context of SYK
and its extension to higher dimension can be found in \cite{optimal}.

\section{ Experimental detections in a moving sample }

\begin{figure}[tbhp]
\centering
\includegraphics[width=0.8 \linewidth]{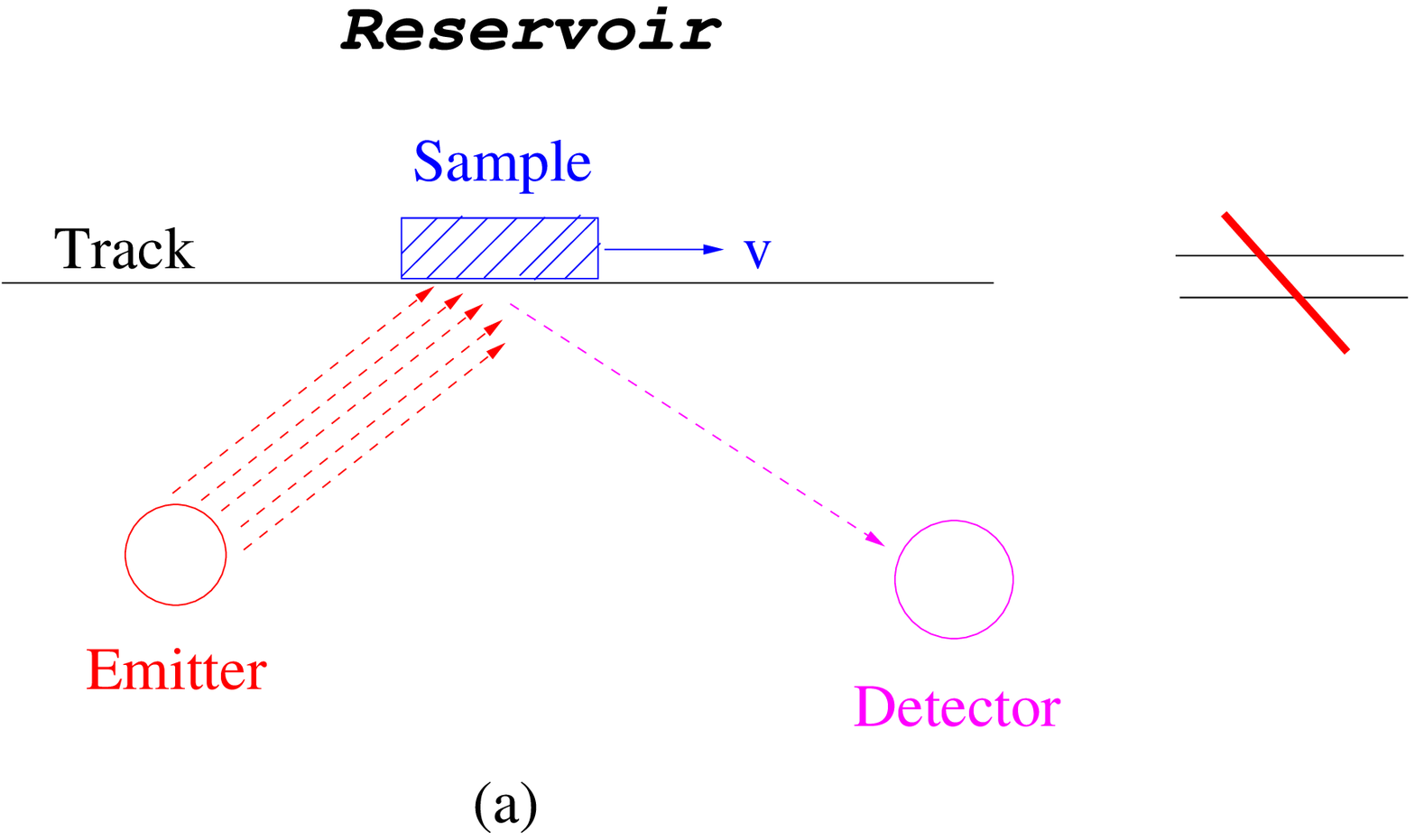}
\includegraphics[width=0.45 \linewidth]{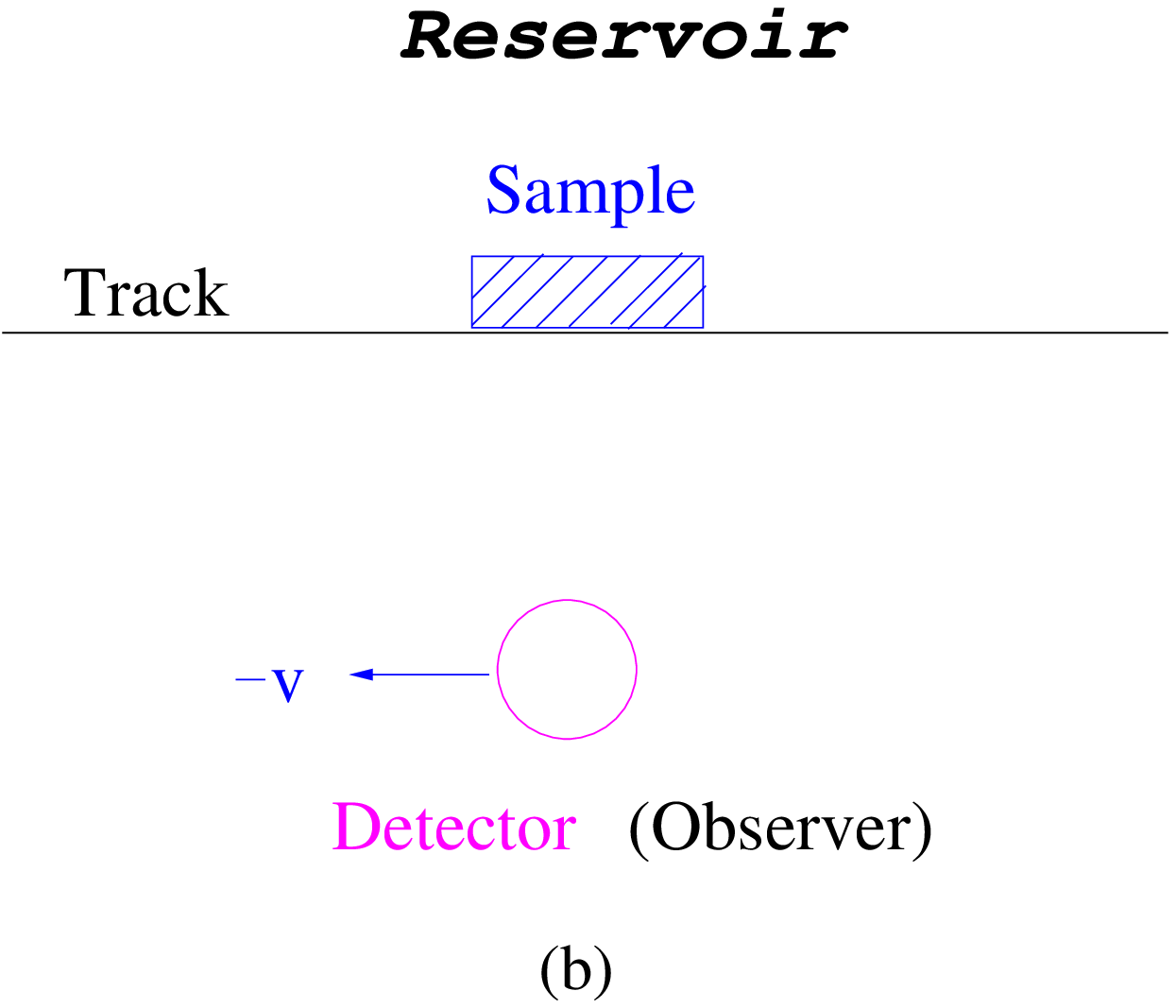}
\hspace{0.8cm}
\includegraphics[width=0.4 \linewidth]{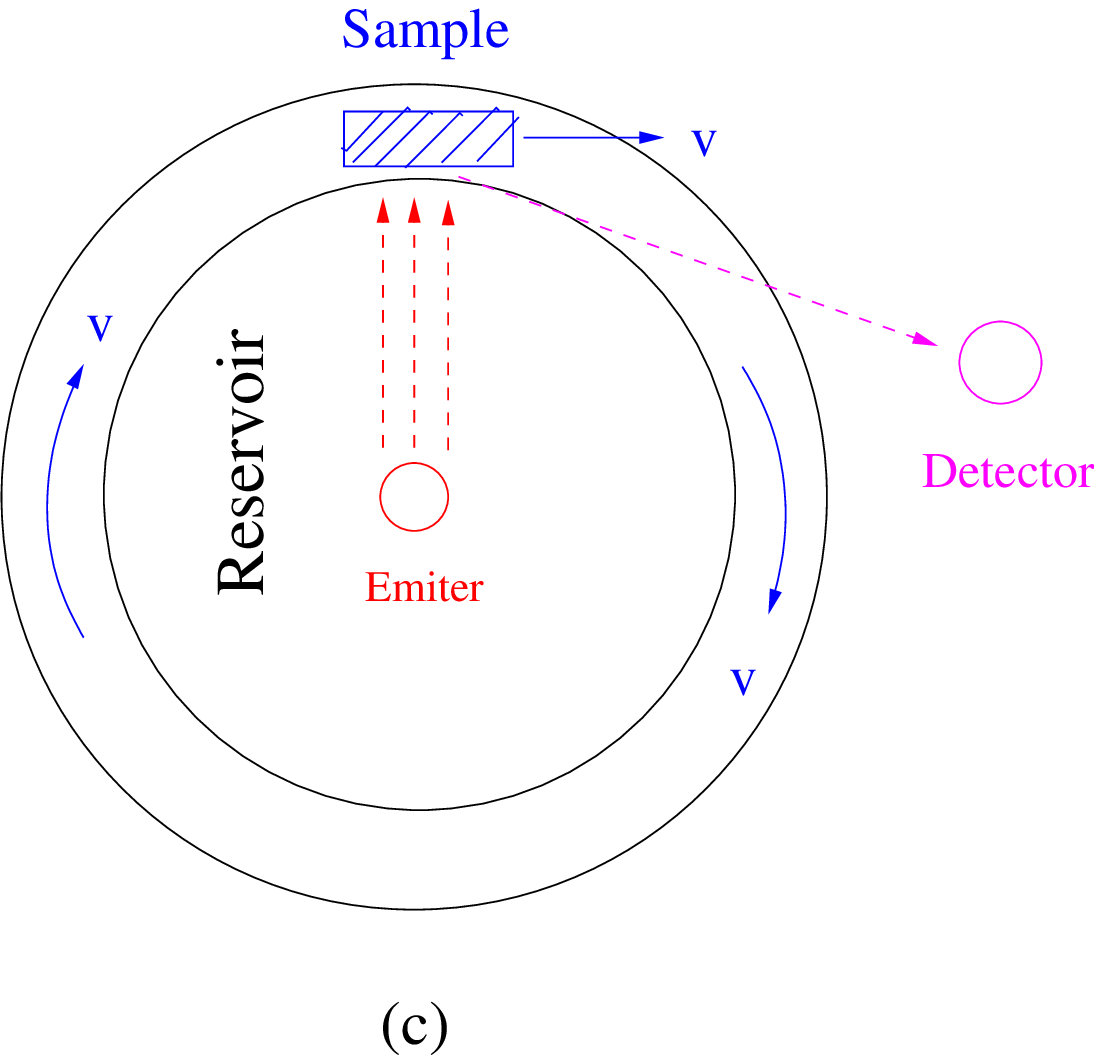}
\caption{ (a)  Light, atom or neutron scattering on a moving sample with a velocity $ \vec{v} $.
The multiple irradiating lines from the emitter declinate the irradiation regime where the sample enter and leave, then the scattered beams
can be detected by the receiver.
If the intrinsic velocity is too high, then one can perform the experiment
in a fast moving train or even a satellite and perform the scattering measurements on the ground. This set-up
may also a practical way to probe the emergent space-time structure near a QPT ( see Fig.\ref{towerlevel} ).
(b) The sample is static, but the detector(or Observer ) is moving with an opposite velocity $ - \vec{v} $.
As stressed in Fig.\ref{frames}, (a) and (b) are dramatically different.
The results achieved in this manuscript only applies to (a). (b) will only be briefly commented in the appendix L.
(c) is just the annulus version of (a) which could be more easily experimentally implemented than (a).
The small centripetal acceleration $ a=v^2/r $ should have no detectable effects in the annulus case.  }
\label{detector}
\end{figure}

\begin{figure}[tbhp]
\centering
\includegraphics[width=.8 \linewidth]{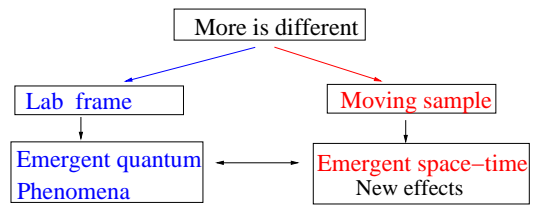}
\caption{ P. W. Anderson's " More is different "\cite{anderson} contains two parts:
the first part is the emergent quantum phenomena in the static ( lab ) frame.
The second part is the emergent space-time structure leading to new effects in a moving sample.
The second part is initiated in this work.
The lab frame is the absolute frame where the experimentalists are comfortable to do experiments on a static sample.
But the moving sample may be used to probe the flat space-time structure.
While the possible curved space-time may emerge in an extra-dimension such as the JT gravity emerges from the SYK model in a boundary.  }
\label{moretwo}
\end{figure}

 In a practical scattering detection experiment shown in Fig.\ref{detector},
 it is more convenient to set the emitter and the receiver static and set the sample moving with a constant velocity $ \vec{v}  $.
 Due to the small size of the sample, it is not easy to focus the beam on the sample when it is moving.
 To overcome this difficulty, one may just continuously shine the emitting beam, only when the sample move into its
 shadow, it will be scattered and collected by the detector. When it moves out of the shadow, there is no scattered beam anymore.

  For $ z=1 $ in a bosonic system, in principle, only when the moving velocity $ c $ reaches the intrinsic velocity of the matter,
  the full phase diagram Fig.\ref{phasesz1} can be explored.
  However, as shown in Sec.VIII, in practice, just moving the sample alone can never reach the $ z=(3/2,3) $ line.
  As shown in \cite{SSS}, one need to get inside the sample and boost the SF directly to explore the SF to the BSF transition
  with $ z=(3/2,3) $. See \cite{SSS} for the experimental realization of directly driving a SF beyond a critical velocity in Fig.\ref{frames}c.
  In the following, we focus on Fig.\ref{phaseslattice}.


{\sl 1. Limitations and applicability of detection techniques on a moving sample }

 When a sample is static, many experimental techniques can be applied. For example, various light,
 atom, X-ray and neutron scattering  can be used to detect all the excitation spectrum in all the phases in
 Fig.\ref{phasesz1}, \ref{phasesz2}.
 The conventional 2-terminal or 4- terminal  transport measurements can be applied to detect the currents in Sec.IV,
 The compressibility $ \kappa $ and the specific heat $ C_v $, the superfluid density $ \rho_s $ in Eq.\ref{rhos}
 in Sec.V can be separately measured by various techniques \cite{heat1,heat2,sfdensity}, also by {\sl In-Situ} measurements \cite{dosexp}.

 However, when setting the sample moving with a constant velocity $ v $ in Fig.\ref{detector}, some of the measurements
 such as transports may become hard to implement. Fortunately, all kinds of scattering experiments sketched in  Fig.\ref{detector} remain
 applicable. Because just by measuring the difference of the energy and momentum between the  scattering beam and the incidence beam,
 one can map out the whole excitation spectrum inside the moving sample.
 One simply perform all the measurements in the same reference frame $ S^{\prime} $, so for the light scattering,
 there is no need to consider the  Doppler shift of the photon in Eq.\ref{dslight}.

 In the cold atoms, the Bragg spectroscopies and photo-emissions are easily applicable when setting the trap holding the optical lattice in a constant motion. Then the single-particle Green functions, the density and density correlation functions investigated in Sec.V,
 the Goldstone and Higgs modes in Fig.\ref{phasesz1} and \ref{phasesz2} can be detected by all kinds of Bragg spectroscopies
 such as dynamic or elastic, energy or momentum resolved, longitudinal or
 transverse Bragg spectroscopies \cite{lightatom1,lightatom2,braggbog,braggangle,braggeng,braggsingle}.
 The compressibility $ \kappa $  automatically follow. Unfortunately, it seems difficult to measure the free energy,
 therefore the specific heat in the moving sample.
 The currents carried by the BSF Eq.\ref{z1current0} may also be hard to measure.

{\sl 2. Estimates of the two relevant scales in materials and cold atom systems }

 As summarized at the end of Sec.VIII-B and also said in the caption of Fig.\ref{phaseslattice},
 the Doppler shifts in the excitation spectrum in the BSF phase is given in Eq.\ref{GoldHiggs} and Eq.\ref{vxvytc}
 for $ z=1 $ and  in Eq.\ref{z2SFV} and Eq.\ref{VwZ2} for $ z=2 $ respectively.
 Here we re-group Eq.\ref{k0phase}, Eq.\ref{alphasign} and  Eq.\ref{vxvytc} as:
\begin{align}
    t & = t_{b1} \sqrt{  v^2_c + v^2  },~~ \tan k_{0L}=- \frac{v}{v_c},~~v_c= t_0/t_{b1}      \nonumber  \\
    c &= v \alpha_b t_{b1}a, ~~ \alpha_b=  ( \frac{v^2_c -v^2  }{ v^2_c + v^2  } ) \frac{t_{b2}}{t_{b1}} \ll 1,
\label{regroup}
\end{align}
  which shows that $ -\pi/2 <  k_{0L} < 0 $ reaches $  k_{0L}= -\pi/4 $ when $ v=v_c $.
  As stressed in Eq.\ref{Tcz1} for $ z=1 $ and Eq.\ref{Tcz2} for $ z=2 $,
  $ K_B T_c = t a^2 $ controls the superfluid density in the SF phase,
  therefore the KT transition temperature.
  The increase of the $ T_c $ when the sample is moving is determined by the scale $ v_c= t_0/t_{b1} $,
  while the Doppler shift in the Mott and SF phase when the sample is moving is controlled by the scale $  \alpha_b \sim \frac{t_{b2}}{t_{b1}} $,
  the sign is by $ v-v_c $.

  They depend on the 4 microscopic quantities on the lattice listed in Eq.\ref{tijtbij}:
  the NN hopping $ t_0 $, the NN and NNN Wannier functions overlap $ t_{b1}, t_{b2} $ and the boost $ v $.
  To proceed further, in the tight binding limit, we use the atomic wavefunctions on a lattice site to
  approximate the Wannier functions in Eq.\ref{tijtbij} ( for simplicity, we still use the same notations ):
\begin{align}
  t_0  & =\int d^2 x \phi( |\vec{x}| )( \Delta V(\vec{x}) )  \phi( |\vec{x} + a \hat{x} | ),
                         \nonumber  \\
  t_{bn} & = \hbar \int d^2 x \phi( |\vec{x}| ) \frac{ \partial}{\partial x } \phi( |\vec{x} + na \hat{x} | ),~~~n=1,2
\label{tttt}
\end{align}
  where $ \Delta V(\vec{x})= V_1(\vec{x})- v_a(\vec{x}) = \sum_{ \vec{R} \neq 0 } v ( \vec{x}- \vec{R} ) $
  is the lattice potential with the ion potential at the site $ \vec{R} =0 $ removed.
  They lead to the estimates of the critical velocity $ v_c= t_0/t_{b1} $ and the dimensionless ratio $ t_{b2}/t_{b1} $.
  The chemical potential is renormalized by ( see below Eq\ref{tijtbij} )
  $ -\mu \to -\mu + E_{s}( \vec{R}=0 ), E_{s}( \vec{R}=0 )=  E_{s}+ \int d^2 x \phi( |\vec{x}| )( \Delta V(\vec{x}) )  \phi( |\vec{x} | ) $
  where $ E_{s} $ is the S-wave atomic energy.

  As stressed in Sec.VIII, the relations Eq.\ref{z1order} and Eq.\ref{z2order} between the original boson on the lattice
  and the order parameters in the $ z=1 $ and $ z=2 $ effective actions
  can only be established from the microscopic calculations. They are reflected in various 2- and 4- point boson correlation functions, so
  can be detected in scattering experiments outlined in the last subsection.

  For 3d electronic materials in a lattice, one may estimate their numerical values by taking
  $ \phi( |\vec{x}| ) $ as the 3d Hydrogen atom S-wave electronic ground state  wavefunction
  and also taking the NN lattice spacing $ a=5 a_0 $ where $ a_0 $ is the Bohr radius.
  Then we find $ v_c= t_0/t_{b1} \sim 8.51 \times 10^{-4 } c_l \sim 255 km/s, t_{b2}/t_{b1} \sim 2.47 \times 10^{-2 } $.
  When taking the NN lattice spacing $ a=10 a_0 $ both change quickly to
  $ v_c= t_0/t_{b1} \sim 2.22 \times 10^{-3 } c_l \sim 666 km/s,  t_{b2}/t_{b1} \sim 1.73 \times 10^{-4 } $.
  For a material, $ v_c \sim 255 km/s  $  is even much faster than the escape velocity of a satellite
  $ v_{esc} = 11.2km/s $, so the effects may be too tiny to be detected in any electronic materials.
  Even so, as explained in the appendix A, it is still much larger than the relativistic corrections
  at the order of $ (v/c_l )^2 \sim 10^{-6}  $.

  Of course, the SF-Mott QPT here is for bosonic systems.
  So far, the cold atoms in an optical lattice is  the only experimental system which has
  realized the SF-Mott transition in the first place. For the optical lattice
  with a depth $ V_0 $ and a recoil energy $ E_r= \hbar^2 k^2_r/2m $,
  after defining $\alpha=V_0/E_r$, the 1d harmonic oscillator ground-state
\begin{align}
    \psi_0(x)
        =\Big[ \alpha^{1/2} \frac{\pi}{d^2}\Big]^{1/4}
        e^{-\frac{\pi^2}{2}\alpha^{1/2}(\frac{x}{d})^2}
\end{align}
where $k_r=\frac{2\pi}{\lambda}=\frac{\pi}{d} $ with $ d=\lambda/2$ as the optical lattice spacing.

 After setting the NN lattice spacing $ a=d $ in Eq.\ref{tttt}, one can
 evaluate $ t_0, t_{b1}, t_{b2} $ analytically:
 \begin{align}
    t_0 & =\frac{4}{\sqrt{\pi}}E_r\alpha^{3/4}e^{-2\sqrt{\alpha}}
                   \nonumber  \\
    t_{b,n}&=-n\pi\frac{\hbar k_r}{2}\sqrt{\alpha} e^{-\frac{n^2\pi^2}{4}\sqrt{\alpha}},~~~n=1,2
\end{align}
   where $ t_0 $ is taken from Ref.\cite{coldrev}. They lead to the two ratios
\begin{align}
    t_0/t_{b1}&=\frac{\hbar}{2m}\frac{4\alpha^{1/4}}{\pi^{3/2}}e^{(\frac{\pi^2}{4}-2)\sqrt{\alpha}}
                       \nonumber  \\
    t_{b2}/t_{b1}&=2e^{-\frac{3}{4}\pi^2\sqrt{\alpha}}
\end{align}
where $\pi^2/4=2.467$.

We first estimate $
    \frac{\hbar k_r}{m}
    =\frac{\hbar}{m}\frac{2\pi}{\lambda}
    =0.00761 \text{ m/s} $
where we use $\lambda=6\times 10^{-7}$ m (visible light 400 nm -- 800 nm),
$\hbar=1.05\times 10^{-34}$ m\textsuperscript{2}kg/s,
the mass for $^{87}$Rb $m=87\times 1.661\times 10^{-27}$ kg.

For $\alpha=1,2,3$, we obtain
\begin{align}
  v_c & =  t_0/t_{b1} =0.88, 1.26, 1.62,~~~\text{cm/s}
  \nonumber  \\
    t_{b2}/t_{b1}&=1.2\times 10^{-3}, 5.7\times 10^{-5}, 5.4\times 10^{-6},
\end{align}
   which are the two relevant scales to determine the sizes of various effects when the sample is moving.

{\sl 3. Detections of  the raise of the finite temperature KT transition }

   The effects when the sample is moving controlled by the scale $ v_c $ ( which, in turn,  is determined by $ H_{b1} $ ) are:
   the shift of the BEC momentum $ k_{0L} $, the shift of the phase boundary towards the Mott side in the Step-I,
   the increase of the KT transition temperature $ T_c $ in the SF side  are clearly within the current experimental reach.
   Note that this $ v_c $ somehow is quite close to the critical velocity $ v^{SF}_c \sim 1 cm/s $
   in a weakly interacting BEC . We consider this as an coincidence instead of any intrinsic connections\cite{twovc}.
   Especially, according to Eq.\ref{Tcz2}
\begin{align}
   T_c/T_{c0}=\sqrt{2}, 2, 3,.....;~~~~ v/v_c=1, 3, 8,......
\label{twiceTc}
\end{align}
so the critical temperature
   can be raised substantially from its value when the sample is static $ T_{c0} \sim mK $. As alerted in the last subsection,
   it seems difficult to measure the free energy, therefore the specific heat in the moving sample.
   For the KT transition in Fig.\ref{phaseslattice}c, the specific heat $ C_v $ has no singularity at $ T_c $ anyway, but there is a universal jump in
   the superfluid density  $ \Delta \rho_s /k_B T_c= 2/\pi $, it is not clear how to measure this universal jump in a moving sample.

   The subleading effects when the sample is moving  controlled by the scale $ \alpha_b \sim t_{b2}/t_{b1} $ ( note $ t_{b1} a \sim 1 $ )
   ( which, in turn,  is determined by $ H_{b2} $ ) are:
   the shift of the BEC momentum $ k_{0} $,
   the shift of the phase boundary towards the Mott side in the Step-II in Eq.\ref{z2boundary},
   the Doppler shift in the Mott and SF phase at both $ z=1 $ and $ z=2 $ are more challenging to measure.
   However, the fact that the Doppler shift reverse its sign at $ v=v_c $ still make the challenge not too hard to overcome in
   the current cold atom technology. Of course, as explained in the appendix A, it is still much much larger than the relativistic corrections
   at the order of $ (v/c_l )^2 \sim 10^{-18}  $.

   However, setting the whole sample moving could also cause heating, in order to observe the increase
   in Eq.\ref{twiceTc}, one need to control the heating to be much smaller than  $ T_{c0} \sim mK $.

{\sl 4. Time of flight (TOF) detection of the BEC momentum $ k_{0L}, k_0 $ near the SF-Mott transition }

 The cold atom condensation wavefunction can be directly imaged through Time of flight (TOF) images \cite{coldrev} which
 after a time $ t $ since opening the trap is given by:
 \begin{equation}
     n( \mathbf{x} )= ( M/\hbar t )^3 f(\mathbf{k}) G(\mathbf{k})
 \end{equation}
   where $ \mathbf{k}= M \mathbf{x}/\hbar t $,  $ f(\mathbf{k})= | w(\mathbf{k}) |^2 $ is the form factor due to the
   Wannier state of the lowest Bloch band of the optical lattice and
   $ G(\mathbf{k}) = \frac{1}{N_s} \sum_{i,j} e^{- \mathbf{k} \cdot ( \mathbf{r}_i- \mathbf{r}_j ) }
    \langle \Psi^{\dagger}_i \Psi_j \rangle $ is the equal time boson structure factor.
   At the mean field level,
   $ \langle \Psi^{\dagger}_i \Psi_j \rangle \sim \langle \Psi^{\dagger}_{0i} \Psi_{0j}  \rangle$
   where $ \Psi^{\dagger}_{0i} $ is the condensate wavefunction Eq.\ref{z1order}  for $ z=1 $ and Eq.Eq.\ref{z2order}  for $ z=2 $
   respectively.  So the TOF can detect the BEC condensation momentum of the SF state wavefunction directly.
   While the SF density determines the magnitude of the condensation peaks at the BEC momentum $ k_{0L} $ and $ k_{0L} + k_0 $
   for $ z=1 $ and $ z=2 $ respectively.

   Especially, the Mott phase in the regime I and II in Fig.\ref{phaseslattice}a when the sample is static turns into a SF phase when the sample is moving.
   So the TOF in these two regimes will not show any diffraction peaks at  $ k_{0L} $ and $ k_{0L} + k_0 $  when the sample is static, but does show
   the diffraction peaks at $ k_{0L} $ and $ k_{0L} + k_0 $ when the sample is moving whose magnitude is proportional to the small SF density. As estimated in the last subsection, the sizes of the regime I and II are controlled by the two scales
   $ v_c $ and $ t_{b2}/t_{b1} $  respectively. So the regime-I is clearly within the experimental reach, but the regime-II is
   much smaller and more challenging to detect.
   However, the effects of the confining harmonic traps could make the detections intriguing \cite{TOFrecent,lightatom2}.


{\sl 5. The re-arrangement of the wedding cake inside a harmonic trap }

  By comparing the Mott lob in a static sample in Fig.8 with that in Fig.9a in a moving sample,
  one can see that the SF regime expands at the expanse of the Mott  regime.
  Under a local density approximation $ \mu(r)=\mu- V(r) = \mu -\frac{1}{2} K r^2 $ where $ \mu $ is the chemical potential at the
  center of the harmonic trap, then especially the SF annulus around the edge of the trap will gets larger.
  In a static sample, the wedding cake structure can be precisely mapped out by an In-situ STM measurement \cite{dosexp}.
  In a moving sample, as explained in the first subsection, it seems only scattering experiments may be used to map it out.

{\sl 6. Experimentally detect the Doppler shifted Goldstone and Higgs mode by the scattering experiment }

 In reality, only the path I in Fig.\ref{phasesz1} is reachable. So we still focus on Fig.\ref{phaseslattice}.
  There was a report that the Higgs amplitude mode was detected in a three dimensional superfluid of
  strongly interacting bosons in an optical lattice by Bragg spectroscopy \cite{higgs3dboson}.
  The Higgs mode has also been detected \cite{Higgsexp}  near the 2d  $ z=1 $ SF-Mott QPT with $ ^{87} Rb $ by slightly  modulating the
  lattice depth within a linear response regime in the lab frame. They especially mapped out both the Mott gap and the Higgs map near the $ z=1 $ QCP.
  So the Doppler shifted Goldstone and Higgs mode along the $ z=1 $ line in Fig.\ref{phaseslattice},
  especially in the new regime (I), can be detected.
  But the Higgs mode should disappear away from the $ z=1 $ line, especially in the new regime (I)+(II).



\section{ Conclusions and perspectives  }


Despite there are enormous number of previous works on emergent phenomena, but very little work on the associated emergent space-time.
Here we  explore the outstandingly emergent space-time problem by systematic and concrete effective action
approach and the microscopic approach.
Practically,
we start from the basic single particle Schrodinger equation,
then to many body  Schrodinger equation in  both first and second quantization, in both canonical and grand canonical ensemble.
Then we apply it to many body systems in a periodic lattice which  break the GI explicitly.
We perform the GT in the 3 boxes in Fig.\ref{towerlevel} from the IR to the UV: Eq.\ref{boostz1}  and Eq.\ref{boostz2} in the effective actions,
Eq.\ref{BoostedBHform} in the boson-Hubbard model and Eq.\ref{Hblattice12}  in the ionic model which are consistent and complementary to each other.
During the process, the emergent space-time comes out which is quite different than the bare space-time.
As a byproduct, we also push
our systematic approach to many-body systems in an external magnetic systems such as FQH and its associated edge states.

{\sl 1. Intuitive understandings of the weak Mott phase turns into a weak SF phase in a moving sample } 

We start from the microscopic model of boson-Hubbard model of interacting bosons at integer fillings
in a square lattice which shows  the Mott to Superfluid (SF) transitions in Fig.\ref{phaseslattice}.
In the continuum limit, the low energy effective actions have either the emergent pseudo-Lorentz or the emergent Galileo invariance
with the dynamic exponent $ z=1 $ and $ z=2 $ respectively.
Then by the combination of constructing effective actions and performing microscopic  calculations in the lattice when the sample is moving,
we obtain the effective phase diagrams Fig.\ref{phasesz1} and \ref{phasesz2} and finally the global phase
diagram Fig.\ref{phaseslattice}a when the sample is moving. The phase boundary shifts by the two steps (I) and (II) in
Fig.\ref{phaseslattice}a  show that
  the Mott insulating state in the regimes (I)+(II)
  in  a static sample  may turn into a BSF state when the sample is moving, but not the other way around.
  So in the regime near the QPT boundary,  a weak Mott insulator can tune into a weak dissipation-less  SF in a moving sample.
  One may understand this exotic and counter-intuitive quantum phenomena from a physical point of view \cite{MottBSF}:
  along the $ z=2 $ vertical line  in Fig.\ref{phaseslattice0} when the sample is static, one increases the chemical potential  $ \mu $
  to drive the QPT from the Mott state to a SF state. Here, as shown in Eq.\ref{tildemuz2} for $ z=1 $ and  in Eq.\ref{z2inv} for $ z=2 $,
  one increases the kinetic energy of the fast moving train which  plays the similar role as increasing $ \mu $ to drive
  the QPT from the Mott state to the BSF state with $ z=2 $, so a moving sample makes the gap of a weak Mott state
  vanishing and drives it into a weak BSF state. It will not affect too much the strong Mott phase
  deep inside the QPT, but will turn a weak Mott phase into a weak BSF \cite{MottBSF}.
  More intuitively, one may understand this counter-intuitive phenomenon by just viewing that the shift of the QCP from the SF to Mott
  to the Mott side when the sample is moving. Indeed, the location of any QCP is never universal even in a given frame,
  it can not be invariant under a GT either. Of course, the finite critical temperature $ T_c $ is never universal either.
  So there is no reason to expect it remains the same in different inertial frame. Our specific microscopic calculations in Sec.VIII
  shows that it increases when the sample is moving. In fact, it substantially increases  as double, triple, etc in a moving sample when
  $ v $ gets past the intrinsic velocity $ v_c= t_0/t_{b1} $ which is
  solely determined by the Wannier functions and easily reachable in the cold atom systems.


 {\sl 2. The internal $ U(1) $ gauge field  in a moving sample  }


 The charge-vortex duality presented in Sec.VI is formulated in the continuum.
 The effects of the gauge fields under a GT are given in Eq.\ref{GTdual} in Sec.IV-B-3 and also the appendix F-H.
 It can also be formulated in a dual lattice where the vortices are hopping subject to a dual Abelian gauge field on the links.
 Then injecting vortex currents in the presence of the dual gauges field is interesting on its own right,
 it belongs to a new class of problems: boosting a lattice gauge theory \cite{lattice1,lattice2} with both matter on the lattice
 site and the dynamic gauge fields on the link.
 So it  will also bring important insights to quantum phases and transitions with matter and gauge fields in the dual lattice.
 One can also add an Abelian gauge field \cite{slowgold} to Eq.\ref{Seq0Nlattice}, namely, the combination of  Eq.\ref{Seq0Nlattice} and Eq.\ref{Seq0NQH}.  Then in going from the first quantization to the second quantization, then projecting to the tight binding limit,
 the matter fields go to the lattice sites, the Abelian gauge field go to the links.
 Because the single particle Hamiltonian in Eq.\ref{hphiE}
 becomes $ \hat{h}= -\frac{1}{ 2 m} [ -i \hbar \nabla -\frac{e}{c} \vec{A}( \vec{x} ) ]^2 + V_1( \vec{x} ) $ ( See Eq.\ref{Seq0NQH} ),
 the Wannier functions  in  Eq.\ref{hphiE} may also depend on the Abelian gauge fields.
 It is important to derive the form of the boost on a lattice in the presence of Abelian gauge field
 in the tight -binding limit.
 However, putting a gauge field in a lattice is not guaranteed. For example, it is still not known how to
 put topological QFT such as the Abelian Chern-Simon theory to be discussed in appendix G and H
 in a lattice. In this case, the continuum limit is the only way to proceed, but may run into many technical problems ( see appendix G ).

 {\sl 3. The internal Non-Abelian  gauge field ( SOC )  in a moving sample  }

 One can also add Non-Abelian gauge field leading to the spin-orbit coupling (SOC).
 Then one need to distinguish two different kinds of SOC \cite{QCD}. The first kind is that appearing  in a material \cite{SOC}
 which comes from the relativistic effects at the order of $ (v/c)^2 $. In 3d, it takes the form $ \vec{S} \cdot \vec{L} $ which respects both time-reversal
 and the parity. In 2d non-centrosymmetric materials, it may take the Rashba form $ \vec{k} \cdot \vec{S} $ which keeps the  time-reversal,
 but breaks the parity.
 Then as stressed in the appendix A,  one may use
 the low velocity limit of the LT upto the order $ (v/c)^2 $. Then it is beyond the GT and generalized GT where one may need to consider the
 the relativistic time contraction effects.
 Because the SOC is a major factor leading to various  topological phases \cite{kane,zhang,tenfold},
 one may need also to consider  the relativistic time contraction effects to study how to observe
 all the topological phases due to the SOC \cite{kane,zhang,tenfold,wenrev} when the sample is moving.
 The second kind is created artificially in charge neutral cold atom systems by laser beams, so not a relativistic effect,
 one can just apply the GT to such SOC systems \cite{weyl,soc1,soc2,soc3}.
 It may take the form $ \vec{k} \cdot \vec{S} $  which keeps the time-reversal, but breaks the parity in both 2d and 3d
 where $ \vec{S} $ is the pseudo-spin representing the two components of the hyper-fine states.
 As mentioned at the beginning of Sec.II, this kind of SOC plays a very similar role as the boost, so
 the results achieved in Sec.II can be directly applied to the SOC in a Zeeman field \cite{response}.
 Some preliminary results in the presence of both the artificial SOC and the boost are given in the appendix D-4.
 This kind of SOC may also lead to many topological phases \cite{kane,zhang,tenfold,wenrev}.
 
 {\sl 4. Entanglement entropy (EE) and quantum chaos in a moving sample  }

  It is known that the Entanglement entropy (EE) at $ T=0 $ dramatically increases near a QCP.
  Fig.\ref{phasesz1} and \ref{phasesz2}  at $ T=0 $ show that the EE
  at $ T=0 $ will also change when the sample is moving. Indeed, the EE may also be used to describe the emergent space-time
  structure \cite{NPwitten,RT}. The out of time correlation (OTOC) to describe the quantum information scramblings \cite{SY,kittalk,syk2}
  at a finite $ T $ also  dramatically increases. Despite new quantum phases may emerge when the sample is moving, the
  $ T=0 $ remains, so a pure state remains pure. In a sharp contrast, the Unruh effects  reviewed in Sec.IX-A lead to the construction of
  the thermo-field double (TFD) state which is the purification of the thermal state in the Rindler space-time.
 As shown in Sec.III,  the Mott to SF transition along path-I in  Fig.\ref{phasesz1} is the new universality class
 of boosted 3D XY with $ z=1 $, its finite temperature properties also depend on the boost $ c $ explicitly, especially
 the finite critical temperature $ T_c $ in the SF side increases when the sample is moving.
 Because any 3D CFT has only global conformal symmetry, so no local parametrization invariance in 1d SYK model or 2d CFT\cite{SY,kittalk,syk2} are expected, so one must perform specific calculations at a finite $ T $  to study
 how the Lyapunov exponent $ \lambda_{L} = f(c) T \leq 2 \pi T $ depends on $ c $ at the QCP, especially near the critical velocity $ v_y $
 ( or the M point ) in Fig.\ref{phasesz1}.
 The quantum Lifshitz transition from the  SF to BSF  along the path II in Fig.\ref{phasesz1}  provides a rare example of a Gaussian theory with highly non-trivial exact dynamic exponent  $ z=(3/2, 3 ) $,
 the $ [Z]=-2 $ term is still quadratic, so will not lead to any quantum chaos,
 but the $ [b]=-2 $ term in Eq.\ref{ab} is non-linear and will lead to
 quantum chaos, so it remains important to evaluate the OTOC due to this leading irrelevant operator.

{\sl 5. The applications to other systems in a moving sample  } 

  Obviously, the GT is only for the space-time, does not care about if the
  microscopic degree of freedoms are bosons, fermions or spins.
  The methods developed in this work can be applied to study  any other QPTs such as the magnetic phase transitions \cite{socmag1}, lattice vibrations ( phonons ) in a solid or
  topological phase transitions in interacting bosonic or non-interacting/interacting fermionic systems \cite{QAHinj}, FQH plateau-plateau transitions.
  We expect the two general theorems listed at the end of Sec.VIII-B still hold.
  How to boost a quantum spin model in a lattice is interesting to pursue.
  Adding artificially generated SOC to the quantum spin model will also lead to novel physics
  which will be explored in future publications.
  This work focus on charge neutral SF, but it still be directly applied to the superconductor-insulator transition (SIT) in a thin film where the Coulomb interaction plays an important role \cite{field2}. It leads to
  a time-component of the gauge field $ A_0 $ in Eq.\ref{z1} which breaks its emergent Lorentz invariance \cite{retardation}.
  From Eq.\ref{GTdual} ( or Eq.\ref{GTgauge} ), the time component remains the same $ \tilde{A}_0= A_0 $ and the spatial components remain vanishing
  $ \tilde{A}_x= A_x=0, \tilde{A}_y= A_y=0 $ holding up to the order of $ (v/c)^2 $. This result holding up to the order of $ (v/c)^2 $
  based on the time-component $ A_0 $ is consistent with
  the invariance of the Coulomb interaction under the GT presented in Sec.VII, VIII and also Appendix D-1 of the Coulomb potential in the hydrogen atom.
  In fact, this fact can be contrasted to that the time $ t $ is invariant under the GT  up to the order of $ (v/c)^2 $.

{\sl 6. The applications to topological phases and TPT in a moving sample  }

 This work also raises a general question: how topological ordered phases defined in a lattice system viewed when the sample is moving ?
 especially when two topological ordered phases A and B near a topological transition  between them.
 Wen's string net condensation picture in \cite{wen} in a lattice model produces several emergent
 particle such as spin-1/2 fermions, spin-1 photons ( or light ), Yang-Mills non-Abelian gauge fields ( or Gluons )...
 except spin-2 gravitons which is forbidden by the WW no-go theorem \cite{WW}. Then one left question is how these emergent particles differ from the corresponding  elementary particles
 in terms of transformation in different inertial frames ?
 For elementary particles, this is trivial, they just transform co-variantly under the LT ( namely LI ):
 they just suffer Doppler shifts as shown in Sec.IX-A.
 But in Wen's picture, due to the existence of the lattice at very beginning, as demonstrated in this work,
 the transformation laws of there emergent particles under different inertial frames could be highly non-trivial,
 certainly very different than
 those of the corresponding elementary particles.
 Now if one view two gapped topological phases A and B near a Topological phase transitions (TPT ) between the two,
 then both gaps are small, could be closed in an inertial moving frame simply due to the Doppler shift. Then in the
 moving frame, will the topological A phase viewed when the sample is static change to the B phase when the sample is moving or vice versa ?
 In the QPT from the SF to the Mott  discussed in the main text, the Mott gap near the QPT when the sample is static
 can be driven to vanish when the sample is moving, then turn into a SF phase when the sample is moving.
 So the QCP always moves to the gapped Mott side, engulf some weak Mott phase in favoring of the gapless SF.
 If both topological phases A and B are gapped, it is highly non-trivial to expect to which side it will move.



{\bf Acknowledgements}

In the time order during various stages of completing this work,
J. Ye also thank Bo-Zheng Wang, Xianhui Ge, M. Novotny, Li Li, Z. Q. Hu, Y. S. Wu, C. Gu, X. J. Liu, Biao Wu, Guowu Meng, Chong Wang, Jun Nie, Haitang Yang
and Jiangqiang You
for helpful discussions,  Prof. Xinsheng Lin and Prof. Wei Ku for sharp  criticisms and Prof. W. M Liu for a long time encouragement,  also
inspiring questions from Ph.D. students Jiahou Yang, Xiao Wang from T.D. Lee institute and  C. L Chen from West Lake university.
We also thank C. L. Chen for very helpful discussions when writing the Appendix F,G and H \cite{chen}.

We thank Prof. Congjun Wu and Prof. Gang Tian for the hospitality during our visit at
the Institute for Theoretical Sciences. We also thank Prof. Congjun Wu  for helpful discussions.



\appendix

{\bf Appendix }

 The important role of Lorentz transformation (LT) and invariance (LI) in relativistic quantum field theory is well appreciated and understood.
 At a low velocity, the LT just reduces to the Galileo transformation (GT).
 While many body systems in materials or cold atom systems break the LI spontaneously or explicitly, therefore are non-relativistic. Unfortunately, the possible effects of GT in these non-relativistic quantum many-body systems were poorly understood.
 In the main text, we explore its dramatic effects, especially near  quantum phase transitions (QPT) which are among the most fantastic phenomena in Nature. In fact, there exist also quite confusing and contradicting discussions in the previous literatures on Galileo invariance.
 In the several appendices, we discuss it in a systematic way and also make connections to the new results established in the main text.


\section{ A few remarks on Galileo transformation (GT), generalized GT, low velocity $ v/c $ expansion of
the Lorentz transformation (LT). }

In fact, we should call such the LI in Eq.\ref{z1} emergent "Lorentz" invariance as pseudo-Lorentz invariance.
Because it is the intrinsic velocity $ v $ of the $ z=1 $ system  which plays the role of the speed of light $ c_l $
in real Lorentz invariance in relativistic QFT. So when applying the real-Lorentz transformation to such a  pseudo-Lorentz system,
the system is not invariant. Instead the real-Lorentz transformation reduce to the Galileo transformation
in the $ c_l \rightarrow \infty $ limit.
It is the main goal of this work to study the dramatic effects of such a Galileo transformation.

In the relativistic QFT at a finite $ T $ and a finite chemical potential $ \mu $,
called thermal field theory which may describe the quark-Gluon plasma  phases  in RHIC or LHC,
the Lorentz invariance was explicitly broken by both $ T $ and $ \mu $.  Setting $ T=0 $ and $ \mu=0 $ in the theory recovers the
Lorentz invariance, the light velocity $ c_l $ remains the typical velocity scale, so one must use Lorentz transformation between different inertial frames. In principle, one must also use Lorentz transformation between different inertial frames in Eq.\ref{boson} too.
However, the typical velocity scale is $ v_x, v_y \ll c_l $, so the LT just reduces to the GT. In fact, the speed of light $ c_l $
does not even appear in Eq.\ref{boson}, so we can simply set $ c_l \to \infty $ where the LT becomes identical to the GT.
However, in the presence of an external magnetic field, as shown in the appendices F,G,H,
one may need to consider the low velocity limit of the LT upto the linear order
of $ v/c_l $. One may call this transformation as "Generalized" GT, because one may still drop the time contraction effects
at this linear order. However, if going to the order $ (v/c_l)^2 $ or beyond, then one must consider the time contraction effects,
then it can not be called GT anymore, because Galileo himself did not even know the time contraction effects at the first place.

 The Lorentz group  $ SO(3,1) $ in the Lorentizian signature has $ 3 + 3 $ generators:
 The three boosts $ K_i, i=1,2,3 $ and 3 rotations $ J_i, i=1,2,3 $ which satisfy the commutation
 relations
\begin{align}
  [J_i,J_j]&=i \epsilon_{ijk} J_k,~~~~  [K_i,K_j]=-i \epsilon_{ijk} J_k,  \nonumber  \\
  [J_i,K_j]&=i \epsilon_{ijk} K_k
\label{JK}
\end{align}
 so two  Lorentz boosts generate a rotation instead of closing on itself.
 The Poincare group includes 4 space-time translation $ R^4 $ which do not commute with the 6 Lorentz generators,
 so can be written as a semi-direct product $ R^4 \rtimes  SO(3,1) $.
 In the non-relativistic limit $ c_l \to \infty $ ( also called Wigner contraction or Wigner
 degeneration ), $ K_i \to C_i $, then the algebra reduces to
\begin{equation}
  [J_i,J_j]=i \epsilon_{ijk} J_k,~ [C_i,C_j]=0,~ [J_i,C_j]=i \epsilon_{ijk} C_k
\label{JC}
\end{equation}
 The Lorentz group reduces to the Galileo group $ R^3 \rtimes  SO(3) $. The Poincare group reduces to $ R^4 \rtimes ( R^3 \rtimes  SO(3) )  $.
 Now two Galileo boosts close on itself and form a subgroup which is a Abelian group  $ R^3 $ on its own,
 So it can be discussed independent of the rotation group $ SO(3) $.
 It is this subgroup $ R^3 $  which is the focus of this work.
 In fact, the Poincare group can also be taken as the Wigner contraction limit of the AdS group $ SO(3,2) $
 in the large radius $ R \to \infty $ or zero curvature ( flat space-time ) limit.
 The AdS isometry group in $ AdS_5/CFT_4 $ in Sec.X-C is $ SO(4,2) $.
 We thank Prof. Guowu Meng for discussions leading to this classification.
 However, as shown in the appendix F,G, H, in the presence of an external magnetic field,
 one only need to go to the linear order in $ v/c_l $  to study the GT, which can be called generalized GT.

  Lorentz invariance means the Lagrangian takes the same form in two inertial frames,
  only a scalar under $ SO(4) $ in Euclidean signature (  or $ SO(3,1) $ in the Lorentizian signature )
  remains the same in all the initial frames. For example, the proper time is such a scalar, the causality is also preserved.
  But many other physical quantities such as a 4-vector does depend on the reference frame.
  The relativistic Doppler shift Eq.\ref{qftboost} is just one example on the 4-vector $ (\vec{k}, i \omega/c_l ) $.
  The oldest example for the classical mechanics is that the moving stick shorter and moving clock runs slower
  on the 4-vector $ (\vec{x}, ic_l t) $. A scalar can always be constructed from a 4-vector.
  Unfortunately, the scalar always goes to $ - \infty $ as  $ c_l \to \infty $ limit in any such construction, so it becomes
  trivial and carries no useful information. Therefore, no scalar can be constructed in a GT.
  Of course, one can still have many invariants under the GT such as
  the position difference $ \vec{x}-\vec{R}, \vec{x}_1-\vec{x}_2 $, etc.
  Here, we explore possible new quantum phases through novel quantum phase transitions for a non-relativistic lattice system when the sample is moving.
  In a grand canonical ensemble, the chemical potential in Eq.\ref{tildemuz2} and Eq.\ref{z2inv} clearly depends on the inertial frame,
  so different quantum phases and phase transitions
  are observed in different inertial frames should be expected just from the general principle.
  For non-relativistic classical system, one may also discuss Sub-sonic and Supersonic case as mentioned in the appendix B.

Obviously, the resulting single-particle low energy theory Eq.\ref{vc2} is neither Lorentz nor Galileo invariant.
However, this is sorely due to the low energy approximation, because
we knew the UV theory is the Dirac equation which is Lorentz invariant.
Indeed,  the rest mass $ m c^2_l $ can be absorbed into the energy. But as explained below Eq.\ref{dslight},
one still need to keep it to ensure the system is always positive definite.  The three extra terms  are of the order
$ \sim (v/c_l)^2 $ where $ v=k/m $
and do not survive the $ c_l \to \infty $ limit anyway. It can be shown that
despite the non-relativistic Schrodinger equation plus the three extra
terms break the GI, but still keeps the low velocity expansion of the LT up to the order $ \sim (v/c_l)^2 $ where the time contraction effects become important.
In fact, the Dirac equation should be understood as QED which
can be used to understand the Lamb shift due to the quantum fluctuations in the photon propagator. 
The QED at $ \mu=0, T=0 $  is clearly Lorentz invariant
The many-body boson-Hubbard model Eq.\ref{boson} is a UV finite theory which explicitly breaks the Lorentz and Galileo invariance,
the light velocity $ c_l $ does not even appear in the theory.
Lorentz invariance is a must for relativistic QFT, but neither Lorentz nor Galileo invariance is necessary for
quantum many-body systems in materials or AMO systems ( Fig.1). This maybe the starting point of P.W. Anderson's `` More is different ''
which are extended to the emergent space-time here, see Sec.VII-VIII, Fig.\ref{towerlevel} and appendix F-H.

 It is worth to note that under a GT, the speed of light changes as $ c_l \to c_l +v= c_l( 1 + v/c_l ) $
 which cause an error to the linear order in the speed of light. The time contraction effects upto the order of
 $ O( (v/c_l)^2 ) $ is needed to correct this error on the speed of light.
 Fortunately, this linear error in $ c_l $ only causes the damage to all the other physical quantities in the
 FQH in the second order in $ (v/c_l)^2 $. For example,
\begin{align}
  vt^{\prime} & = \frac{vt- (v/c_l)^2 x }{ \sqrt{1- (v/c_l)^2 } }
 = ( vt- (v/c_l)^2 x )(1+ \frac{1}{2} (v/c_l)^2 + \cdots)
              \nonumber  \\
 & =vt + O( (v/c_l)^2 )
 \label{vtO2}
\end{align}
 which gets back to the GT $ t^{\prime}=t $  after dropping $ O( (v/c_l)^2 ) $ corrections.
 Namely, one may ignore the time contraction effects to the linear order of $ v/c_l $.
 Otherwise, it would not be called the GT anymore.
 By GT, one mean one ignores the  time contraction effects because Galileo himself did not know this effect.
 Obviously,
\begin{align}
  x^{\prime}  = \frac{x-vt }{ \sqrt{1- (v/c_l)^2 } }=x-vt  + O( (v/c_l)^2 )
\end{align}
  whose first correction also comes at the order of $ O( (v/c_l)^2 ) $.

 One may also look at the drift velocity in Eq.\ref{driftvelocityx}
\begin{align}
  E_y& = \frac{v}{c_l} {B_z} \to \frac{v}{c_l}( 1-\frac{v}{c_l} + \cdots )  {B_z}
                    \nonumber  \\
  &= \frac{v}{c_l} {B_z} - (v/c_l)^2B_z + \cdots
\end{align}
 So one still So one gets back to the GT  Eq.\ref{driftvelocityx} $ E_y= \frac{v}{c_l} {B_z} $ after dropping $ O( (v/c_l)^2 ) $ corrections.

 One may compare with the relativistic Doppler shift discussed in Sec.IX-A.
 In fact, in addition to the frequency shift Eq.\ref{qftboost}, there is also a momentum shift:
\begin{equation}
  k^{\prime}_1=\gamma ( k_1- \frac{v}{c^2_l} \omega )
\label{qftboostk}
\end{equation}
 If the speed of light $ c_l $ is the characteristic velocity such as in a relativistic QFT, then
 $ c_l k^{\prime}_1= \gamma ( c_l k_1- \frac{v}{c_l} \omega ) $. Then
 $ k^{\prime}_1=  k_1 + O(v/c_l) $  upto the linear order in $ v/c_l $.
 For a light ray, this may also be seen from the angle of the light ray making with the x-axis:
 $ \tan \theta^{\prime} = \frac{\sin \theta}{ \gamma ( \cos \theta - v/c_l ) } $.
 This is the also case with the external magnetic field $ B $.

 However, if the moving velocity $ v $ is the characteristic velocity such as in a non-relativistic QFT, then
 $ v k^{\prime}_1= \gamma ( v k_1- (\frac{v}{c_l})^2 \omega ) $.
 Then $ k^{\prime}_1=  k_1 + O((v/c_l)^2 ) $ upto the quadratic orders in $ (v/c_l)^2 $, similar to Eq.\ref{vtO2}.
 So in non-relativistic QFT, one can simply drop the Doppler shift in the momentum upto the quadratic orders in $ (v/c_l)^2 $.
 So the momentum shifts when the sample is moving in Fig.\ref{phaseslattice}a can not be interpreted as the Doppler shift in the momentum,
 they can only be understood as the relation between the original boson field and the order parameter in the effective action.
 This is the also case with the fermionic or bosonic Hubbard model with only the Coulomb interaction.

 For the boson Hubbard model, the non-relativistic effects are at the zeroth order $ (v/c)^0 $ which were  shown in Sec.X to
 be much much larger than the relativistic corrections.
 While for the FQH, the non-relativistic effects are at the 1st order $ v/c $ which were  shown in the appendix F,G and H to
 be still much larger than the relativistic corrections.

By generalized GT, we mean the space-time GT to the order of $ v/c $, but it is vanishing anyway at this order, so practically, it is
still at the order of $ (v/c)^0 $, so the GT algebra Eq.\ref{JC} remains.
However, the corresponding LT in the EM filed need also to the order of $ v/c $ which is non-vanishing.
This is because $ c_l $ is the intrinsic velocity of EM field in the vacuum, so  $ v/c_l $ is the lowest order.


{\sl 1. Analogy with $ 1/N, 1/S, 1/\hbar $ expansion }

One may think the present problem in terms of $ v/c $ expansion.
There are two things to do the expansion.
 (A) The first is the Hamiltonian of the system itself
    If no external magnetic field, then Hamiltonian is a $ (v_c/c)^2 $ expansion with the leading relativistic correction at
    the order of $ (v_c/c)^2 $ which includes the SOC in a material. Here $ v_c $ is the characteristic velocity of the system.
    If there is an external magnetic field as is the case in FQH system,
    then the leading relativistic correction becomes  the order of $ v_c/c $ as shown in Appendix F,G,H.
 (B) The second is the expansion of the LT. As shown in the appendix A, it is always a $ (v_m/c)^2 $  where $ v_m $ is
     the mechanical velocity of the moving sample.  Of course, the leading term is nothing but the GT. Then  one can simply drop
     the time contraction effects which appears first in the order of $  (v_m/c)^2 $.
 Now applying the LT to the system, to probe the system's properties, then $ v_c \sim v_m $, then the two expansions need to be consistent.
 In the SF-Mott transition, one can simply keep only the leading term by simply sending $ c \to \infty $. This is what is done in the main text.
 For the FQH, one must keep to the linear  order of $ v/c $. Fortunately, one can still use the GT, because the next order comes in the order of $ (v/c)^2 $.
 However, if the system contains an SOC, then one must consider the order of $ (v/c)^2 $ which contains the time contraction effects,
 so  is beyond the GT.
 It is constructive to compare this $ v/c $ expansion with the well known $ 1/N $ expansion, $ 1/S $ expansion, or  $ 1/c $  where $ c $ is the central charge of a CFT.
 The $ N \to \infty, S  \to \infty $ or  $ c  \to \infty $ limit or leading order is nothing but a classical limit.
 While the next order stands for the quantum fluctuations. Here the $ v/c_l \to 0 $ is the GT, the next order $ (v/c_l)^2 $  stands for the time contraction effects. In practice, $ N=3 $ or $ S=2 $ in most applications, while here $ v/c_l $ is indeed a very small quantity.

\section{ Boosted Hamiltonians for $ z=1 $ and $ z=2 $. }

Here we will first derive the effective Hamiltonians from the effective actions, then we will
derive them from the microscopic Boson Hubbard model. It is constructive to compare the two complementary approaches.

\subsection{  The Boosted effective Hamiltonians of $ z=1 $ and $ z=2 $  }

 In the main text, we only used the boosted Lagrangian approach, here we applied the corresponding Hamiltonian.
 Despite the two approaches are completely equivalent, the  Hamiltonian approach may be more intuitive in the lattice approach
 on the boson Hubbard model Eq.\ref{boson} mentioned in the conclusion section.
 Similar quantization method is useful to quantize fields in a curved space-time \cite{wittencurve}.

{\sl (a). The boosted Hamiltonian in the $ z=1 $ case }

 From Eq.\ref{boostz1}, one can find the conjugate momentum,
\begin{eqnarray}
    \Pi & = &
    \frac{ \partial {\cal L}_M} { \partial (\partial_{\tau} \psi )}= \partial_\tau\psi^*-ic\partial_y\psi^*    \nonumber  \\
    \Pi^{*} & = &
    \frac{ \partial {\cal L}_M } { \partial (\partial_{\tau} \psi^{*} )}= \partial_\tau\psi-ic\partial_y\psi
\label{Piz1}
\end{eqnarray}

  After imposing the equal-time commutation relations:
\begin{align}
   [ \psi( \vec{x}, t ), \psi( \vec{x}^{\prime}, t ) ] & =  0    \nonumber  \\
   [ \Pi( \vec{x}, t ), \Pi( \vec{x}^{\prime}, t ) ]  & =  0     \nonumber  \\
   [ \psi( \vec{x}, t ), \Pi( \vec{x}^{\prime}, t )] & =    i \hbar \delta( \vec{x}- \vec{x}^{\prime} )
\label{commz1}
\end{align}
  where there is also an identical  set for  $ \psi^{*} $ and $ \Pi^{*} $. The two sets commute with each other.

  One can find the corresponding Hamiltonian:
\begin{align}
  {\cal H}_{M,z=1} & =  - \Pi \partial_{\tau} \psi -  \Pi^{*}\partial_{\tau} \psi^{*} + {\cal L_M}    \nonumber  \\
              & =   - \partial_{\tau} \psi^{*} \partial_{\tau} \psi - c^2 \partial_{y} \psi^{*} \partial_{y} \psi
              +v_x^2|\partial_x\psi|^2+v_y^2|\partial_y\psi|^2
                 \nonumber  \\
              & + r |\psi|^2+ u |\psi|^4+ \cdots
\label{Hz10}
\end{align}
  Now it is important to express  $ \partial_{\tau} \psi^{*}=\Pi + ic\partial_y\psi^*,
  \partial_{\tau} \psi=\Pi^{*} + ic\partial_y\psi $ in terms of the conjugate momentum
  and substituting them into the Eq.\ref{Hz10} leads to:
\begin{align}
   {\cal H}_{M,z=1} & =-\Pi \Pi^{*} -ic \Pi \partial_y\psi-ic \Pi^{*}\partial_y\psi^* + v_x^2|\partial_x\psi|^2
        \nonumber  \\
   & +v_y^2|\partial_y\psi|^2 + r |\psi|^2+ u |\psi|^4+ \cdots
\label{hZ1}
\end{align}
  which plus the commutation relations Eq.\ref{commz1} gives the complete quantized boosted Hamiltonian for $ z=1 $.
  Putting $ c=0 $ recovers the  $ {\cal H}_{M} $ when the sample is static. It can be shown that it is Hermitian in the real time formalism \cite{wrong}.
  It is interesting to put Eq.\ref{hZ1} in a lattice using the Hamiltonian formulation ( namely keeping the imaginary time continuous
  and putting both conjugate variables in a lattice )
  reviewed in \cite{lattice1,lattice2}.
  Eq.\ref{hZ1} can also be derived from the lattice model in the appendix B-1 below  where
  the physical meanings of conjugate variables $ \psi( \vec{x}, t ), \Pi( \vec{x}^{\prime}, t) $ become more transparent
  in Eq.\ref{HPP} and Eq.\ref{phactionH}.

 {\sl (b). The boosted Hamiltonian in the $ z=2 $ case }

 From Eq.\ref{boostz2}, one can find the conjugate momentum,
\begin{equation}
    \Pi  =    \frac{ \partial {\cal L_M}} { \partial (\partial_{\tau} \psi )}= \psi^{\dagger}
\label{Piz2}
\end{equation}
  After imposing the equal commutation relations:
\begin{align}
   [ \psi( \vec{x}, t ), \psi( \vec{x}^{\prime}, t )] & = 0    \nonumber  \\
   [ \psi^{\dagger}( \vec{x}, t ), \psi^{\dagger} ( \vec{x}^{\prime}, t )] & = 0     \nonumber  \\
   [ \psi( \vec{x}, t ), \psi^{\dagger} ( \vec{x}^{\prime}, t )] & = i \hbar \delta( \vec{x}- \vec{x}^{\prime} )
\label{commz2}
\end{align}
  One can find the corresponding Hamiltonian:
\begin{align}
 & {\cal H}_{M,z=2}  =  - \Pi \partial_{\tau} \psi  + {\cal L_M}    \nonumber  \\
              & =  -ic  \psi^{\dagger} \partial_y\psi + v_x^2|\partial_x\psi|^2+v_y^2|\partial_y\psi|^2 - \mu |\psi|^2+ u |\psi|^4+ \cdots
              \nonumber  \\
&  =  \psi^{\dagger} [ -ic  \partial_y - v_x^2 \partial^2_x - v_y^2\partial^2_y - \mu ]\psi+ u (\psi^{\dagger}\psi)^2+ \cdots
\label{Hz20}
\end{align}
   which is automatically in terms of the conjugate momentum $ \psi^{\dagger} $.
   It plus the commutation relations Eq.\ref{commz2} gives the complete quantized boosted Hamiltonian for $ z=2 $.
    Putting $ c=0 $ recovers the  $ {\cal H}_{M} $ when the sample is static.

   Eq.\ref{Hz20} can be written as
\begin{equation}
   {\cal H}_{M,z=2}=  \psi^{\dagger} [   v_x^2 (-i \partial_x )^2 + v_y^2 (-i \partial_y + k_0 )^2- \tilde{\mu} ]\psi
   + u (\psi^{\dagger}\psi)^2+ \cdots
\label{Hz20shift}
\end{equation}
  where $ k_0=- \frac{c}{ 2 v^2_y },  \tilde{\mu}= \mu + v^2_y k^2_0 $ as listed in Eq.\ref{z2inv}.
  Then by introducing $ \tilde{\psi}= \psi e^{-i k_0 y } $ which also satisfies the commutation relation Eq.\ref{commz2}.
  Then Eq.\ref{Hz20shift} can be re-written in terms of  $ \tilde{\psi} $ as:
\begin{equation}
   {\cal H}_{M,z=2}=  \tilde{\psi}^{\dagger} [   v_x^2 (-i \partial_x )^2 + v_y^2 (-i \partial_y )^2- \tilde{\mu} ] \tilde{\psi}
   + u (\tilde{\psi}^{\dagger} \tilde{\psi} )^2+ \cdots
\label{Hz20shifttilde}
\end{equation}
    which takes exactly the same form as that when the sample is static.
    It is nothing but the Hamiltonian version of the path-integral in Eq.\ref{z2prime}.


   Following the procedures in Appendix D-2 and combining the $ z=1 $ and $ z=2 $ case, one can also derive the Hamiltonian corresponding
   Eq.\ref{Z1Z2}

\subsection{  The Boosted Hamiltonians of $ z=1 $ and $ z=2 $ from the lattice models.  }

 The $ z=2 $ Hamiltonian corresponding to the  $ z=2 $  Lagrangian Eq.\ref{paction} is nothing but Eq.\ref{Hz20}.
 Furthermore, it provides the physical meaning of the $ p $ operator.

 One can also write down immediately the $ z=1 $ Hamiltonian corresponding to the  $ z=1 $  Lagrangian Eq.\ref{paction} as:
\begin{align}
  {\cal H}[ \Psi, \Pi] &= \Psi^{\dagger} (-ic \partial_x) \Pi + \Pi^{\dagger} (-ic \partial_x) \Psi
                        \nonumber  \\
    & +   \Psi^{\dagger}(-\frac{1}{2m} \nabla^2 +  \Delta_0 - \lambda t ) \Psi
                         \nonumber  \\
    &+  \Pi^{\dagger}(-\frac{1}{2m} \nabla^2 + \Delta_0 + \lambda t ) \Pi + \cdots
\label{HPP}
\end{align}
  where $  ( \Psi^{\dagger}, \Pi) $ or $  ( \Pi^{\dagger}, \Psi) $ are conjugate variables.
  One can see that by $ \Pi \leftrightarrow \Pi^{\dagger} $ and setting
  $ \Pi \to \Pi/\sqrt{\Delta_0+ \lambda t }, \Psi \to \sqrt{\Delta_0+ \lambda t } \Psi $ to keep the commutation relations,
  it is nothing but Eq.\ref{hZ1}. After this re-scaling, one gives back to Eq.\ref{relationz1}.
  Furthermore, it provides the physical meaning of the  $  ( \Psi, \Pi)  $ operator as written in Eq.\ref{unitaryt}.

  Then it is instructive to revert back to the Hamiltonian corresponding to Eq.\ref{phaction}:
\begin{align}
  {\cal H}_{p/h}&= p^{\dagger} (-ic \partial_x + \Delta_0-\frac{1}{2m} \nabla^2 ) p
                   \nonumber  \\
                &+ h^{\dagger} (-ic \partial_x + \Delta_0-\frac{1}{2m} \nabla^2 ) h
                 \nonumber  \\
                & - \lambda t ( p^{\dagger} h^{\dagger} + p h )+ \cdots
\label{phactionH}
\end{align}
  where $ ( p , p^{\dagger} ) $ and  $ ( h , h^{\dagger} ) $  are just two pairs of conjugate variables.
  Eq.\ref{phactionH} maybe more practical than Eq.\ref{HPP} in performing a lattice calculation to study the boost effects.

\section{ The excitation spectrum of the SF phase away from integer fillings in a lattice under some injecting currents }

 In the main text, we focused on the Mott-SF transition at integer fillings.
 If away from an integer filling with only onsite interaction, then the system is always in a SF phase.
 There is no Mott phase, no Mott-SF transitions anymore. Because the particle or hole in Eq.\ref{Mottb} is defined with respect to
 the Mott phase only, so is the emergent complex order parameter $ \Psi(x) $ in Eq.\ref{unitaryt}.
 Away from the integer fillings, there is no need to define $ p $ or $ h $ anymore, one just directly deal with
 the original bosons $ b_i $ directly. One may understand the SF phase at the weak coupling limit $ U/t \ll 1 $ just by
 Bogoliubov theory. We expect the calculations may also apply
 at the joint point between $ n $ and $ n+1 $ lobe in Fig.\ref{phaseslattice} where the boson number per site could be any one
 between $ n $ and $ n+1 $, any small hopping also induces a SF.

   Then one can inject three different kinds of currents which stand for three different
   subgroups of the GT put on the square lattice in Eq.\ref{BoostedBHform}:
\begin{align}
    H_{b1}&=-it_{b1}\sum_{i} (b_{i}^\dagger b_{i+y}-h.c.),    \nonumber \\
    H_{b2} & =-it_{b2}\sum_{i} (b_{i}^\dagger b_{i+2y}-h.c.)  \nonumber \\
    H_{b3}&=-it_{b3}\sum_{i} (b_{i}^\dagger b_{i+x+y}-h.c.)
\label{threeb}
\end{align}
   which stands for the current along the $ y $ bonds with $ n=1 $, or with $ n=2 $,
   along the diagonal $ (1,1) $ bond,  $t_b$  stands for the strength of the injected currents. Note that all these currents
   ( or any of these linear combinations ) still keeps the translational symmetry of the lattice.
   Of course, one can add infinite number of terms which still keep the translational symmetry of the lattice.
   Eq.\ref{threeb} just gives the three simplest ones. From Eq.\ref{Hblattice12}, one can see the boost $ H_{bx}= v( H_{b1}+ H_{b2} ) $.
   But here we will study their effects separately.

\subsection{ The calculations in the original basis }

 In the absence of the injections, the BEC occurs at $k=K=0$.
However, the injection drives the BEC momentum to $ K $ which also means
that the single particle spectrum $\epsilon_k$ develops a single minima at $k=K$ with
$ \min_k \epsilon_k=\epsilon_{K} $. In the following, it is convenient to define $k=K+q$.
Applying the Bogoliubov theory and rewriting
\begin{equation}
   b_{K+q}=\sqrt{N_0}\delta_{q,0}+\psi_{K+q}
\label{awaylattice}
\end{equation}

   The Hamiltonian $ H= H_0 + H_b $ becomes
\begin{align}
   & H=E_0+\sqrt{N_0}[\epsilon_K-\mu +Un_0](\psi_K+\psi_K^\dagger)\nonumber\\
    &+\sum_q[(\epsilon_{K+q}-\mu) \psi_{K+q}^\dagger\psi_{K+q}
    +Un_0(2\psi_{K+q}^\dagger\psi_{K+q}
    \nonumber \\
    &+\frac{1}{2}\psi_{K+q}\psi_{K-q}
    +\frac{1}{2}\psi_{K+q}^\dagger\psi_{K-q}^\dagger)]+\cdots
\end{align}
  Setting the linear term vanishing leads to
\begin{align}
  \mu=n_0U+\epsilon_K
\end{align}
   which also fix the total number of bosons.

  The quadratic part can be diagonalized by a Bogoliubov transformation,
  which gives the collective Goldstone mode:
\begin{align}
    E_{q}= \epsilon_{-}(K;q)
    +\sqrt{\Big[ \epsilon_{+}(K;q)-\epsilon_K\Big]
    \Big[\epsilon_{+}(K;q)-\epsilon_K+2n_0U\Big]}
\label{Ek}
\end{align}
  where $ \epsilon_{\pm}(K;q)=\frac{\epsilon_{K+q}\pm \epsilon_{K-q}}{2} $.
  The $  \epsilon_{-}(K;q) $ term outside the square root mimics the Doppler shift term in the continuum.
  In the following, we apply the formalism to the three injections in Eq.\ref{threeb}.

 (a). Injecting $ H_{b1} $.

  When considering $H_{b1}$ in Eq.\ref{threeb}, the single particle spectrum takes the form
\begin{align}
    \epsilon_k
    & =-t_0e^{ik_x}+(-t_0-it_{b1} )e^{ik_y}+h.c.
  \nonumber\\
    & =-2t_0(\cos k_x+\cos k_y)+2t_{b1} \sin k_y
\label{by}
\end{align}
  which develops a single minimum at $k=K=(0,k_0),\quad $ with
\begin{align}
    \epsilon_K=-2(t_0+\sqrt{t_0^2+t_{b1}^2}),\quad
    k_0=-\arctan(t_{b1}/t_0)
\end{align}
Note that $t_0\sin k_0+t_{b1} \cos k_0=0$  can be used to simplify many expressions.

One can calculate
\begin{align}
    \epsilon_{+}(K;q)
        &  =-2t_0\cos q_x-2\sqrt{t_0^2+t_{b1}^2}\cos q_y     \nonumber \\
    \epsilon_{-}(K;q)
        &=2(t_{b1}\cos k_0+t_0\sin k_0)\sin q_y
        =0
\end{align}
   Thus Eq.\ref{Ek} can be written as
\begin{align}
    E_{q}=\sqrt{[\epsilon_{K+q}+2T][\epsilon_{K+q}+2T+2n_0U]}
\label{Hb1E}
\end{align}
  where $ T=t_0+\sqrt{t_0^2+t_{b1}^2} $,
  the lattice Doppler shift term just vanishes.
  Why is this so will be re-examined below in the subsection {\sl 2.} from the tilted basis Eq.\ref{k0phase}.

   In the long-wavelength limit, $q\ll 1$ ( BZ size ), the Goldstone mode becomes:
\begin{align}
    E_{q}=\sqrt{2n_0U(t_0 q_x^2+\sqrt{t_0^2+t_{b1}^2}q_y^2)}
\label{bosonBEC}
\end{align}
    which has no associated Doppler shift term.
    When comparing this equation with Eq.\ref{z2SFV}  and considering the difference between
    $ n_0 $ here is the original boson density, while $ \rho_0 $ in Eq.\ref{z2SFV} is the density for the $ p $ or $ h $,
    one may identity  $ v_x^2=v_y^2= \sqrt{t_0^2+t_{b1}^2} $ in the boosting along both x- and y- bond.
    This is the same as that listed in Eq.\ref{vxvytc}.

 (b) Injecting $H_{b2}$

The single particle spectrum takes the form
\begin{align}
    \epsilon_k
    &=-t_0e^{ik_x}-t_0e^{ik_y}-it_be^{i2k_y}+h.c.
    \nonumber\\
    &=-2t_0(\cos k_x+\cos k_y)+2t_b\sin 2k_y
\label{2by}
\end{align}
 which develops a single minimum at $k=K=(0,k_0),\quad $ with (assuming $t_0>0$)
\begin{align}
    \epsilon_K&=-2t_0-\frac{3t_0+\sqrt{t_0^2+32t_b^2}}{2}
    \sqrt{\frac{1}{2}+\frac{t_0}{t_0+\sqrt{t_0^2+32t_b^2}}},
    \nonumber \\
    k_0&=\arcsin\Big(\frac{t_0-\sqrt{t_0^2+32t_b^2}}{8t_b}\Big)
\end{align}
Note that $t_0\sin k_0+2t_b\cos 2k_0=0$ which can be used to simplify many expressions
and $|k_0|\leq \frac{\pi}{4}$ holds for any ratio of $t_b/t_0$.

One can calculate
\begin{align}
    \epsilon_{+}(K;q)
        &=-2t_0(\cos q_x+\cos k_0\cos q_y)+2t_b\sin 2k_0\cos2q_y
        \nonumber \\
    \epsilon_{-}(K;q)
        & =t_0\sin k_0(2\sin q_y-\sin2q_y)
\end{align}

In the long-wavelength limit, $q\ll 1$, Eq.\ref{Ek} can be simplified  as
\begin{align}
    E_{q}&=\sqrt{2n_0U[t_0 q_x^2+(t_0\cos k_0-4t_b\sin 2k_0)q_y^2+O(q^4)]}
    \nonumber \\
    & +t_0\sin k_0 q_y^3+O(q^5)
\end{align}
  where the Doppler shift term does not vanish, but goes as $ q_y^3 $ instead of $ q_y $
  in the continuum. Again, there is no chance to drive a QPT in this special case.

 (c)  Injecting $H_{b3}$.

The single particle spectrum takes form
\begin{align}
    \epsilon_k
   & =-t_0e^{ik_x}-t_0e^{ik_y}-it_be^{i(k_x+k_y)}+h.c.
   \nonumber \\
   & =-2t_0(\cos k_x+\cos k_y)+2t_b\sin (k_x+k_y)
\label{bxbydia}
\end{align}
 which develops a single minimum at $k=K=(k_0,k_0),\quad $ with (assuming $t_0>0$)
\begin{align}
    \epsilon_K&=-\frac{1}{2\sqrt{2}t_b}
    (3t_0+\sqrt{t_0^2+8t_b^2})\sqrt{4t_b^2-t_0^2+t_0\sqrt{t_0^2+8t_b^2}},\quad
    \nonumber \\
    k_0&=\arcsin\Big(\frac{t_0-\sqrt{t_0^2+8t_b^2}}{4t_b}\Big)
\end{align}
 where $ k_0 $ satisfies:
\begin{equation}
t_0\sin k_0+t_b\cos 2k_0=0
\label{2bycon}
\end{equation}
 which can be used to simplify many expressions
and $|k_0|\leq \frac{\pi}{4}$ holds for any ratio of $t_b/t_0$.

  One can calculate
\begin{align}
     \epsilon_{+}(K;q)
        &=-2t_0\cos k_0(\cos q_x+\cos q_y)+2t_b\sin 2k_0\cos(q_x+q_y)
        \nonumber \\
     \epsilon_{-}(K;q)
        &  =2t_0\sin k_0[\sin q_x+\sin q_y-\sin(q_x+q_y)]
\end{align}

In the long wave-length limit $q\ll 1$ ( BZ size ), Eq.\ref{Ek} can be simplified as
\begin{align}
    E_{q}&=\sqrt{2n_0U[t_0\cos k_0(q_x^2+q_y^2)-t_b\sin 2k_0(q_x+q_y)^2+O(q^4)]}
    \nonumber \\
    &-\frac{1}{3}t_0\sin k_0[q_x^3+q_y^3-(q_x+q_y)^3]+O(q^5)
\end{align}
   By using the constraint Eq.\ref{2bycon}, one can show the quantity inside the square root is positive, while
   that outside vanishes as $ t_b $ as $ t_b \to 0 $.

  One can see the Doppler shift term does not vanish, but goes as $ q^3 $ instead of linear   in the continuum.
  The first term remains linear. So the Doppler shift term is always sub-leading to the first term in the long wavelength limit.
  There is no chance to drive a QPT in this special case either.

\subsection{ The calculations in the new basis with $ H_{b1} $. }

 It is also tempting to study the BEC of the bosons in the new basis Eq.\ref{k0phase}
 where the BEC condensation momentum gets back to $ 0 $.
 Rewriting $\tilde{b}_k=\sqrt{N_0}\delta_{k,0}+\tilde{\psi}_k$  and applying the Bogoliubov theory leads to:
\begin{align}
  &  H=E_0+\sqrt{N_0}[-2(t_0+\sqrt{t_0^2+t_{b1}^2})-\mu +Un_0](\tilde{\psi}_0+\tilde{\psi}_0^\dagger)
  \nonumber \\
    &+\sum_k[(\tilde{\epsilon}_k-\mu) \tilde{\psi}_k^\dagger\tilde{\psi}_k
    +Un_0(2\tilde{\psi}_k^\dagger\tilde{\psi}_k
    +\frac{1}{2}\tilde{\psi}_{k}\tilde{\psi}_{-k}
    +\frac{1}{2}\tilde{\psi}_{k}^\dagger\tilde{\psi}_{-k}^\dagger)]+\cdots
\end{align}
 Setting the linear term vanishing leads to $\mu=n_0U-2(t_0+\sqrt{t_0^2+t_{b1}^2})$.
 The quadratic Hamiltonian can be diagonalized by a Bogoliubov transformation:
\begin{align}
    \tilde{E}_k=\sqrt{[\tilde{\epsilon}_k+2T]
    [\tilde{\epsilon}_k+2T+2n_0U]}
\end{align}
   where $ T=t_0+\sqrt{t_0^2+t_{b1}^2} $ was defined below Eq.\ref{Hb1E}.
   It has no Doppler shift term. This basis explained the exact vanishing of the Doppler shift term in the $ H_{b1} $ case,
   but not the other two cases with $ H_{b2} $ and $ H_{b3} $.

 Note that the above Bogoliubov calculation of the SF phase automatically break the C- symmetry at $ z=1 $, so no Higgs mode can be found.
 This is expected, because the above calculations only apply to away from integer fillings  where there is no C- symmetry anyway.
 But as mentioned below Eq.\ref{bosonBEC},  it may apply to the $ z=2 $ case where either $ p $ or $ h $ condenses,
 after considering the difference between the density of bosons $ n_0 $ and that $ \rho_0 $ of either $ p $ or $ h $,
 the interaction between the bosons $ U $ and that of $ u $  between $ p $ or $ h $.
 This is the main  difference between $ b_i $ and the order parameter $ \psi $ in Eq.\ref{boostz1} and Eq.\ref{boostz2}.

\section{ The Galileo invariance, enlargement of a Fermi surface  and its explicit breaking in artificially generated 2d/3d SOC case }

 It was known that the Electro-Magnetic (EM) fields satisfy Maxwell equations which is intrinsically Lorentz invariant.
 The Klein-Gordan (KG) equations for spin-0 bosons $ \phi $ and Dirac equations  for spin-1/2 fermions $ \psi $ are also Lorentz invariant.
 So the KG equation  and Dirac equation with a rest-mass $ m $ couple to external EM fields
 in the minimal coupling scheme are both gauge invariant and  Lorentz invariant.
 The intrinsic velocity is the speed of light $ c_l $.
 Under the Lorentz transformation
\begin{align}
  x^{\prime} = \Lambda x,~~\Lambda=e^{ i \omega_{\alpha \beta} L_{\alpha \beta} }
\end{align}
 where $  L_{\alpha \beta}= - L_{\beta \alpha} $ stands for  the three rotation $ J_i,i=1,2,3 $ and the
  three boosts $ Ks_i,i=1,2,3 $ ( Eq.\ref{JC} ), then KG spin-0 bosons
 and Dirac spin-1/2 fermions or spin-1 gauge field $ A_\mu $ give its scalar, spinor and vector representation of the Lorentz group respectively:
\begin{align}
  \phi^{\prime}( x^{\prime} ) & =  \phi(x),   \nonumber   \\
  \psi^{\prime}( x^{\prime} ) & = \Lambda_{1/2}  \psi(x),~~\Lambda_{1/2}= e^{ i \omega_{\mu \nu} S_{\mu \nu} }   \nonumber   \\
  A^{\prime}_{\mu}( x^{\prime} ) & = \Lambda A_{\mu}(x),
\label{spin012}
\end{align}
 where $  S_{\mu \nu}=\frac{i}{4}[ \gamma_{\mu}, \gamma_{\nu} ] ,  \Lambda^{-1}_{1/2} \gamma_{\mu} \Lambda_{1/2}= \Lambda_{1/2} \gamma_{\mu} $
 and $\gamma_{\mu} $ are the 4 Dirac $ \gamma $ matrix.

 When contrasted to Eq.\ref{z2inv} for the non-relativistic case,  one can see the chemical potential
 is kept to be zero $ \mu=0 $ in Eq.\ref{spin012}. This is because any $ \mu \neq 0 $ breaks Lorentz invariance.
 So the  chemical potential shift in Eq.\ref{z2inv} is a unique feature for the Galileo boost.
 Indeed, it is this shift which is responsible for the step (II) shift in the QPT  boundary
 from the Mott to the BSF  in Fig.\ref{phasesz2}.

 Taking the non-relativistic limit $ v/c_l \ll 1 $ where $ v=k/m $ limit,
 the KG  or the Dirac equation in an EM field just reduces to the S-equation with a mass $ m $ in the
 EM field ( or the spin-1 gauge field $ A_\mu $ in Eq.\ref{spin012} ) which is GT ( see Eq.\ref{Seq0NQH} ).
 For the Dirac equation, there is an additional Zeeman field term $ -\mu_B \vec{\sigma} \cdot \vec{B} $
 where $ \mu_B = \frac{ e \hbar }{2 m c } $ is the Bohr magneton and the $ \vec{\sigma} $ is the Pauli matrix.
 This Zeeman term is obviously GI. When pushing to higher orders,
 the Dirac equation in a scalar potential $ A_0 $ reduces to S-equation
 with a mass $ m $ in the scalar field ( See Eq.\ref{Seq0Nlattice} )   plus
 several new terms such as the, $ p^4 $ term, the spin-orbit coupling (SOC) term
 and Darwin term which break the GI at the order of $ (v/c)^2 $. The gauge invariance is always kept in such a limit ( Appendix A ).
 Indeed, as shown below, the SOC breaks the Galileo invariance.

 In this appendix, we first discuss the GT in the Schrodinger equation in the first quantization which clarify a lot of confusions
 in the existing literatures or even textbooks, then
 show that  the Fermi Surface of non-interacting fermions ( in the second quantization ) breaks Lorentz invariance, but owns Galileo invariance.
 Then we show that in the second quantization language, any Luttinger liquid in 1d or the  spin-orbit coupling \cite{SOC,QCD}
 break the Galileo invariance.

{\sl 1. Galileo transformation on the single particle Schrodinger equation: first quantization }

 It was known that the non-relativistic Schrodinger equation ( 1st quantization form )
\begin{equation}
   i \hbar \frac{ \partial \psi }{ \partial t } = [ - \frac{ \hbar^2}{ 2 m} \frac{ \partial^2}{ \partial x^2} + V(x, t) ] \psi
\label{Seq0}
\end{equation}
  where the single-body potential is usually time independent when the sample is static $ V(x,t)=V(x) $.

  Eq.\ref{Seq0} transforms under the Galileo transformation $ x^{\prime}= x-vt, t^{\prime}=t $ as:
\begin{equation}
   i \hbar \frac{ \partial \psi^{\prime} }{ \partial t^{\prime} } =
   [ - \frac{ \hbar^2}{ 2 m} \frac{ \partial^2}{ \partial x^{\prime 2}} + V^{\prime}( x^{\prime},t^{\prime} ) ] \psi^{\prime}
\label{Seq1}
\end{equation}
   where the wavefunction and the potential when the sample is static are related to those when the sample is moving by:
\begin{align}
   \psi ( x,t) & =e^{ i ( k_0 x^{\prime}- E_0 t^{\prime}/\hbar )  }
   \psi^{\prime} ( x^{\prime},t^{\prime} )   \nonumber  \\
    V^{\prime}( x^{\prime},t^{\prime} )&= V( x^{\prime}+ v t^{\prime}, t^{\prime} )
\label{psi0}
\end{align}
 where  $ \psi ( x,t) =\psi ( x^{\prime} - v t^{\prime},t^{\prime} ) $ and
 $ k_0= -\frac{ m v}{\hbar}, E_0= \frac{ \hbar^2 k^2_0 }{ 2 m }= \frac{1}{2} m v^2 $.
 Of course, if $ V=0 $, it is Galileo invariant (GI).
 For the hydrogen problem,
 the central potential is the Coulomb interaction
 $  V( \vec{x} -\vec{R} )= - \frac{Ze^2}{| \vec{x}-\vec{R} | } $,
 because the atomic center is also moving under the GT, then $  V( \vec{x} -\vec{R} )
 = V( \vec{x}^{\prime} -\vec{R}^{\prime} ) $.
 So it remains GI.  Its many-body generalization in a periodic lattice potential  was shown in Eq.\ref{Seq0Nlattice} and Eq.\ref{psi0Nion}.

 In general, due to  $ V^{\prime}( x^{\prime},t^{\prime} ) \neq V( x^{\prime}, t^{\prime} ) $
 Schrodinger equation is not  Galileo invariant, because it is time-independent when the sample is static,
 but becomes time-dependent when the sample is moving, so can not be GI.
 For example, if one takes the harmonic potential $ V_h(x)= \frac{1}{2} m \omega^2 x^2 $ and the center of the potential
 is fixed at $ x=0 $, so not moving under the GT. Then the potential becomes time
 dependent  $ V^{\prime}( x^{\prime},t^{\prime} )= V( x^{\prime}+ v t^{\prime}, t^{\prime} )
 =\frac{1}{2} m \omega^2  ( x^{\prime}+ v t^{\prime} )^2 $ when the sample is moving, therefore  breaks the GI.
 For a typical optical lattice potential  $ V_o(x) = \cos^2 \pi x/a= \cos^2 \pi ( x^{\prime}+ v t^{\prime} )/a $, then
 it becomes time-dependent which is  periodic not only in the space with $ x^{\prime} \to x^{\prime} + a $, but also
 in the time $ t^{\prime} \to t^{\prime} + na/v $, or the joint space-time symmetry.



{\sl 2. A moving FS in second quantization: Galileo invariance and enlargement of the FS in a moving sample }

 Using the prescription presented in the main text, we immediately find the FS observed when the sample is moving with the velocity
 $ v\hat{y} $ in the second quantization form:
\begin{equation}
  H_M =\psi^{\dagger}(\vec{k} ) [ \frac{ \hbar^2 k^2 }{2m} - \mu -  c k_y] \psi(\vec{k} ), ~~~\mu= \frac{ \hbar^2 k^2_{0F} }{2m}
\label{FS}
\end{equation}
 where $ k_{0F} $ is the Fermi momentum when the sample is static. In a grand canonical ensemble, one need also introduce the chemical potential
 $ \mu $  which  also plays the role of the energy $ E $ in the canonical ensemble.

 Eq.\ref{FS} can be written as:
\begin{equation}
  H_M =\psi^{\dagger}(\vec{k} ) [ \frac{ \hbar^2 }{2m}[( \vec{k}-\vec{k}_0 )^2 - ( k^2_{0F} + k^2_0 ) ] \psi(\vec{k} )
\label{FSk0}
\end{equation}
 where $ \vec{k}_0= mc/\hbar \hat{y} $ is the FS center shift due to the boost.
 In the fermion case, one can perform the similar set of transformation as listed in Eq.\ref{z2inv}, one can show
 Eq.\ref{FSk0} is also Galileo invariant, but no TPT. Of course, the GI is always emergent, because
 there always exist higher order derivative terms such as $ k^4, k^6,..... $ which break the GI explicitly.

 For 1d case, one can identify the two Fermi points when the sample is moving:
\begin{eqnarray}
  k^{+}_F= k_0 + \sqrt{ k^2_0 + k^2_{0F} } > 0  \nonumber   \\
  k^{-}_F= k_0 - \sqrt{ k^2_0 + k^2_{0F} } < 0
\label{1dFS}
\end{eqnarray}
  which shows the Fermi momentum is shifted from $ \pm k_{0F} $. Because, no change of topology of FS, so no topological phase transition (TPT)
  tuned by the boost. However, the FS gets enlarged $
   k^{+}_F- k^{-}_F = 2 \sqrt{ k^2_0 + k^2_{0F} } > 2 k_{0F} $.
  This should not be too surprising: because there is a relative motion between the sample and the reservoir
  in a grand canonical ensemble, then the extra fermions just comes from the reservoir.
  This interesting phenomenon can be easily extended to 2d or 3d.


 In fact, Eq.\ref{FSk0} can also be viewed as the kinetic term in the interacting boson case in Eq.\ref{Hz20}.
 In the boson case, one focus on the BEC momentum, so the shift can also be transformed away
 by introducing the new field  $ \tilde{\psi} $ and then the new effective chemical potential $ \tilde{\mu} $ in Eq.\ref{z2inv}.
 So the boson case in Eq.\ref{Hz20}  also has  the Galileo invariance.
 However, in contrast to the fermion case, the effective chemical potential $ \tilde{\mu} $ can tune
 the QPT from the Mott to SF with $ z=2 $ as shown in Sec.VI.
 Both the boson and fermion case  are also due to the Galileo invariance of the non-relativistic Schrodinger equation
 at $ V=0 $ presented in the first quantization language in the last subsection.

{\sl 3. 2d/3d artificial Spin-orbit coupling and split FS with opposite helicities in second quantization:
breaking the Galileo invariance by an effective Zeeman field. }

  As stressed in the conclusion section, here we only focus on the artificially generated SOC here \cite{SOC}, so one can simply apply the GT.
  One can add a 3d Weyl or 2d Rashba like SOC \cite{weyl} term  $  \lambda \vec{k} \cdot \vec{S} $ to Eq.\ref{FS}:
\begin{equation}
  H_{M,SOC} =\psi^{\dagger}(\vec{k} )[ \frac{ \hbar^2 k^2 }{2m} - \mu  -  c k_y + \lambda \vec{k} \cdot \vec{S}] \psi(\vec{k} ), ~~~\mu= \frac{ \hbar^2 k^2_{0F} }{2m}
\label{FSsoc}
\end{equation}
  In the helicity basis\cite{weyl,soc1,soc2,soc3}, the SOC alone \cite{SOC} without the boost can also be written as a linear
  $ k $ term with $ \pm $ sign ( see Eq.\ref{kFSOC} ).
  So it is interesting to look at the combined effects of SOC and the boost.

  Eq.\ref{FSsoc} can be written as
\begin{equation}
  H_{M,SOC} =\psi^{\dagger}(\vec{k} )[ \frac{ \hbar^2 }{2m}[( \vec{k}-\vec{k}_0 )^2 - ( k^2_{0F} + k^2_0 ) ]+ \lambda \vec{k} \cdot \vec{S} ] \psi(\vec{k} )
\label{FSk0soc}
\end{equation}
 where $ \vec{k}_0= mc/\hbar^2 \hat{x} $ is the FS center shift due to the boost.

 After performing the similar set of transformation as in Eq.\ref{z2inv}, Eq.\ref{FSk0soc} becomes:
\begin{equation}
  H_{M,SOC} = \psi^{\dagger}(\vec{q} )[ \frac{ \hbar^2 }{2m} [ q^2  - ( k^2_{0F} + k^2_0 ) ] + \lambda \vec{q} \cdot \vec{S} +\lambda \vec{k}_0 \cdot \vec{S}] \psi(\vec{q} )
\label{FSk0socq}
\end{equation}
  where the extra term $ \lambda \vec{k}_0 \cdot \vec{S} $ breaks the Galileo invariance.
  It plays the role of a Zeeman field which is due to the combination of both the SOC $ \lambda $ and the boost $ \vec{k}_0 $.
  So we conclude the SOC breaks the Galileo invariance.

It is easy to find its two eigen-energies at 2d:
\begin{equation}
   E(q_x,q_y) = \frac{ \hbar^2 }{2m} [ q^2  - ( k^2_{0F} + k^2_0 ) \pm 2 k_R \sqrt{k^2_0 + 2 k_0 q_x +q ^2 } ]
\label{FSEq}
\end{equation}
    where $ k_R= m \lambda $  corresponds the recoil momentum \cite{soc1}. Setting $ k_0 =0 $ recovers
    the SOC when the sample is static  $ \frac{ \hbar^2 }{2m}[ q^2 \pm 2 k_R q- k^2_{0F} ] $ which leads to
    one large ( and one small) FS with negative ( positive ) helicity and the Fermi momentum
\begin{equation}
     k_{F\pm}= \sqrt{ k^2_{0F} + k^2_R } \pm  k_R
\label{kFSOC}
\end{equation}
  which is due to the SOC, indeed can be contrasted with Eq.\ref{1dFS} which is due to the boost.
  Of course, any boost breaks the joint spin-momentum rotation symmetry, so the helicity is not a good quantum number anymore.

  As shown in the last subsection, there is no TPT tuned by the boost in the absence of SOC.
  However, as $ k_0 $ increases in Eq.\ref{FSk0socq}, there is a competition between the enlargement of the FS
  and  the presence of SOC  which plays the role of the Zeeman field $ \lambda \vec{k}_0 \cdot \vec{S} $.
  It would be interesting to see if such a competition will lead to any TPT.
  For the dramatic effects of Weyl type SOC  in interacting fermionic systems in a continuum, see  \cite{weyl,soc1,soc2}.
  Similarly, one can show that the conventional
  $ \lambda  \vec{S} \cdot  \vec{L} $ in materials \cite{SOC} also breaks the Galileo invariance.
  But as alerted in the conclusion and the appendix A, one need to consider the relativistic effects at the order of $ (v/c)^2 $.
  For the conventional SOC, see the review \cite{SLrev1}.


\section{ Emergent Galileo invariance of many-body systems in an external magnetic field  }

 The GT in Eq.\ref{Seq0Nlattice} discussed in Sec.VII in the main text only contains the lattice potential, no external gauge field.
 Here we discuss the role of the external gauge field which shows quite different behaviours than just the lattice scalar potential in Eq.\ref{Seq0Nlattice}.
 Following the strategies used in the last section, we will first present the GI in the first quantization in a canonical ensemble, then
 analyze its form in the second quantization in either a canonical or a grant canonical ensemble.

\subsection{ Galileo transformation on many body Schrodinger equation: the first quantization  }

 In a quantum Hall system subject to a uniform magnetic field, assuming electron spins are all polarized due to the large Zeeman splitting,
 so one only need to consider the orbital effects of the magnetic field for these spinless fermions,  the lattice potential $ V_1(x) $
 can be ignored in such an effective continuum system, then  Eq.\ref{Seq0Nlattice} changes to:
\begin{align}
   & i \hbar \frac{ \partial \Psi(x_1,x_2,\cdots x_N,t) }{ \partial t } =
   [ \sum_{i} \frac{1}{ 2 m} [ -i \hbar \frac{\partial }{ \partial x_i }- \frac{e}{c_l} \vec{A}( x_i) ]^2
      \nonumber   \\
   & + \sum_i e A_0( x_i) + \sum_{i<j} V(x_i, x_j ) ] \Psi(x_1,x_2,\cdots x_N,t)
\label{Seq0NQH}
\end{align}
 where $ \vec{B}= \nabla \times \vec{A} = B \hat{z} $. Because of the possible involvement of the time component $ A_0 $ of
 the gauge field, one may also need to put it into the Hamiltonian in any case. As shown below, even $ A_0 =0 $ at beginning,
 it may be generated by a GT. The crucial difference than Eq.\ref{Seq0Nlattice} is the appearance of the speed of light $ c_l $
 in the minimal coupling to $ \vec{A} $ which need to be considered carefully when performing the GT.
 So Eq.\ref{Seq0Nlattice} and Eq.\ref{Seq0NQH} are two different class of quantum MBS in Fig.1.
 Setting $ A_0=0,~~~\vec{A}=0 $ recovers the Helium 4 case which is charge neutral and  clearly Galileo invariance.

 Similarly, $  V( x^{\prime}_1- x^{\prime}_2 )= V( x_1- x_2 )= e^2/|x_1-x_2| $ is
 translational and Galileo invariant \cite{retardation}.
 Note that Eq.\ref{Seq0NQH} also applies to the charge neutral  interacting bosons in a rotating trap\cite{rotation}
 where the artificial gauge field $ \vec{A}= \frac{1}{2} \vec{\omega} \times \vec{r} $ comes from the rotation close to the trapping frequency
 $ \omega= \omega_t $. In this artificial gauge field case, then the speed of light $ c_l $ does not appear either.

 The gauge potential $ \vec{A}(x) $ usually breaks the translational invariance, however,
 the magnetic field does not.
 For example, in the Landau gauge $ A_x=0, A_y=Bx $ or $ A_x=-By, A_y=0 $, the gauge potential breaks translational invariance
 along a given direction $ x $ or $ y $. In the symmetric gauge $ \vec{A}= \frac{1}{2} \vec{B} \times \vec{r} $, namely
 $ A_x=-\frac{1}{2} B y, A_y=\frac{1}{2}Bx $,
 it breaks translational invariance completely, but keeps rotational symmetry around an
 arbitrarily chosen origin chosen at $ z=0 $.

 However, in a moving frame,
 the space-time, the many-body wavefunction and the gauge potential in the  (symmetric gauge) in the lab frame are related to
 those in the moving frame
\begin{align}
  x^{\prime} & = x+vt,~~~ t^{\prime}=t     \nonumber   \\
  \partial_t & = \partial_{t^{\prime}} + v \partial_{x^{\prime}},~~~\partial_x=\partial_{x^{\prime}}
\label{GTQH}
\end{align}
   The many-body wavefunction changes as the same as in Eq.\ref{psi0Nion}
\begin{align}
   \Psi ( x_i,t)  =  e^{ i ( k_0 \sum_{i} x^{\prime}_i- N E_0 t^{\prime}/\hbar )  } \Psi^{\prime} ( x^{\prime}_i,t^{\prime} )
\label{GTwave}
\end{align}
    where $ k_0= -\frac{ m v}{\hbar}, E_0= \frac{ \hbar^2 k^2_0 }{ 2 m }= \frac{1}{2} m v^2 $.
    While the gauge field changes as
\begin{align}
    A_0 & = 0, A_{x^{\prime}}( y^{\prime} ) =  - \frac{1}{2} B y^{\prime},
    A_{y^{\prime}}( x^{\prime}, t^{\prime} )= \frac{1}{2} B x^{\prime}-\frac{1}{2} B v  t^{\prime}   \nonumber  \\
    A^{\prime}_0 & = A_0 + \frac{v}{c_l} A_{x^{\prime}}= - \frac{v}{2c} B y^{\prime}, A^{\prime}_{x^{\prime}} = A_{x^{\prime}},
    A^{\prime}_{y^{\prime}}=A_{y^{\prime}}
\label{GTgauge}
\end{align}
  where $ [A_0]= [A_x] $, $ c_l $ is the intrinsic velocity of the EM field.
  In a contrast, the periodic lattice ( Coulomb ) potential listed  below Eq.\ref{psi0Nion} is GI.

   Then one can check the changes in electric and magnetic field ( be careful of $ A^0=-A_0 $ in the Lorentz signature )
\begin{align}
    \vec{E}^{\prime}& = -\frac{1}{c_l} \frac{ \partial \vec{A} }{\partial t } - \nabla A_0
    = \frac{ v}{c_l} B \hat{y}^{\prime} \ll B,
                            \nonumber  \\
    \vec{B}^{\prime} & = \nabla \times \vec{A} = B \hat{z}^{\prime}
\label{GTnonEB}
\end{align}
   which turns out to be the LT of $ E $ and $ B $ to the order of $ v/c_l $, but dropping  $ (v/c_l)^2 $ and higher orders.
   When taking the non-relativistic limit $ v/c_l \ll 1 $, naively, one may simply drop this electric field generated by the
   GT. However, the above manipulations demonstrate that one must keep this small electric field to the order of $ v/c_l $
   to keep the Galileo invariance \cite{Lforce} ( Appendix A )
   This should also be expected because the speed of light $ c_l $ does
   appear in the minimal coupling in the many-body Schrodinger equation Eq.\ref{Seq0NQH}. Of course, it also appears in
   the magnetic length $ l_0= \sqrt{ \frac{\hbar c_l}{eB} } $.
   This is dramatically different than the charge neutral systems
   such as the Boson Hubbard model and the ionic lattice model Eq.\ref{Seq0Nlattice} discussed in the main text
   where there is simply no external magnetic
    field, so one can safely  take $ c_l \rightarrow \infty $ limit.


 In the moving frame, one can perform the $ U(1) $ gauge transformation to get rid of the time component $ A_0 $ in Eq.\ref{GTgauge},
\begin{align}
   \Psi \to \Psi e^{ -i \frac{1}{2} Bv t^{\prime} y^{\prime} }
\label{GTgaugeT}
\end{align}
  where $ \chi= -\frac{1}{2} Bv t^{\prime} y^{\prime} $.
  Then the gauge field in Eq.\ref{GTgauge} becomes:
\begin{align}
    A^{\prime}_0 & \to  A^{\prime}_0 - \frac{1}{c_l} \partial_{0} \chi=0 ,
      \nonumber  \\
    A^{\prime}_{x^{\prime}} & \to  A^{\prime}_{x^{\prime}} + \partial_{x^{\prime}} \chi = - \frac{1}{2} B y^{\prime},
         \nonumber  \\
    A^{\prime}_{y^{\prime}} & \to A^{\prime}_{y^{\prime}} + \partial_{y^{\prime}} \chi =\frac{1}{2} B x^{\prime} -B v  t^{\prime}
\label{GTgauge2}
\end{align}
   One can check that $ \vec{E}^{\prime} $ and $ \vec{B}^{\prime} $ stay the same.
   In real time, one need to consider $ A^0=-A_0 $ to get the sign right.

  Similarly, one can perform the $ U(1) $ gauge transformation to recover the spatial component of the gauge field,
\begin{align}
   \Psi \to \Psi e^{ i \frac{1}{2} Bv t^{\prime} y^{\prime} }
\label{GTgaugeS}
\end{align}
  where $ \chi=  \frac{1}{2} Bv t^{\prime} y^{\prime} $.
  Then the gauge field in Eq.\ref{GTgauge} becomes:
\begin{align}
    A^{\prime}_0 & \to  A^{\prime}_0 - \frac{1}{c_l} \partial_{0} \chi=-\frac{v}{c} B y^{\prime} ,
                     \nonumber  \\
    A^{\prime}_{x^{\prime}} & \to  A^{\prime}_{x^{\prime}} + \partial_{x^{\prime}} \chi = - \frac{1}{2} B y^{\prime},
                     \nonumber  \\
    A^{\prime}_{y^{\prime}} & \to A^{\prime}_{y^{\prime}} + \partial_{y^{\prime}} \chi =\frac{1}{2} B x^{\prime}
\label{GTgauge3}
\end{align}
   So in the combination of the GT Eq.\ref{GTQH},\ref{GTwave},\ref{GTgauge} and the gauge  transformation Eq.\ref{GTgaugeS}, one can recover
   the spatial component $ \vec{A} $, but still generates a time component
   $ A^{\prime}_0 =\frac{v}{c} B y^{\prime} $ which leads to the electric field  $ \vec{E}^{\prime}= \frac{ v}{c_l} B  \hat{y}^{\prime} \ll B $.
   One can also see the differences between the GT and gauge transformation, the former changes $ E $ or $ B $,
   but the latter  must keep both the same.

   One can repeat the calculations for the two Landau gauges $ A_x=0, A_y=Bx $ or $ A_x=-By, A_y=0 $.


  The Laughlin ground state wavefunction in the symmetric gauge  indeed breaks the
  translational invariance,  but keeps rotational symmetry. In the lab frame, it takes the form:
\begin{equation}
   \Psi(z_1, z_2, \cdots, z_N,t)=\Pi_{i < j } ( z_i-z_j)^{m} e^{ - \sum_i \frac{ |z_i|^2}{4 l^2_0} }e^{- i E_G t/\hbar}
\label{lau}
\end{equation}
 where $ E_{G} $ is the ground state energy,
 $ m $ is odd ( even) for fermion ( boson ) respectively, $ l_0= \sqrt{ \frac{\hbar c_l}{eB} } $ is the magnetic length.
 In the moving frame, the Laughlin wavefunctions change as Eq.\ref{GTQH} in its $ N $ coordinate and
 Eq.\ref{GTwave} under the GT. The Jastraw factor  keeps the Galileo invariance, but the Gaussian factor does not.
 It was known that the Jastraw factor may act as the conformal blocks of the chiral edge theory,
 so it maybe interesting to explore how does it relate
 to the edge reconstruction to be discussed in the appendix H-3 and Fig.\ref{wenedge}.
 Similar GT maybe applied to Jain's CF wavefunctions also \cite{jain}.
 The bulk-edge correspondence can be stated as that the ground state wavefunction in the bulk can be expressed as the conformal
 blocks of the CFT in the edge. More generally, the bulk ground state wavefunction at a given time slice ( namely 2d spacial dimension )
 can be viewed as the  partition function of the CFT.

\subsection{ Galileo transformation on many-body QFT: the second quantization  }

 Eq.\ref{Seq0NQH} can be written in the second quantization language:
\begin{align}
  {\cal H} & = \int d^2 x \psi^{\dagger}( \vec{x} )
   [ \frac{1}{ 2 m} [ -i \hbar \frac{\partial }{ \partial x }- \frac{e}{c_l} \vec{A}( x) ]^2  - e A_0 (x ) - \mu ] \psi( \vec{x} )
               \nonumber  \\
   & +  \int d^2 x_1 d^2 x_2  \psi^{\dagger}( \vec{x}_1 )\psi( \vec{x}_1 )
   V_2(x_1-x_2 ) \psi^{\dagger}( \vec{x}_2 )\psi( \vec{x}_2 )
\label{Seq0NQH2}
\end{align}
  where the single-body gauge potential $ \vec{A}( \vec{x} ), A_0( \vec{x} ) $
  and the two-body interaction $ V_2(x_1-x_2 ) $ are automatically
  incorporated into the kinetic term and the interaction term respectively.
  By adding the chemical potential $ \mu $, we also change the canonical ensemble with a fixed number of particles
  $ N $ in the first quantization to
  the grand canonical ensemble in the second quantization.

  Following the step leading from Eq.\ref{z1} to Eq.\ref{boostz1}, one can obtain the effective action in the moving frame
  with the velocity $ v \hat{x} $ ( for notational simplicity, we drop the $ \prime $ in the moving frame ):
\begin{align}
  &{\cal H} = \int d^2 x \psi^{\dagger}( \vec{x} )
   [ \frac{1}{ 2 m} [ -i \hbar \frac{\partial }{ \partial x }- \frac{e}{c_l} \vec{A}( x) ]^2
    \nonumber  \\
  &  -iv  \partial_x -  e A_0(x) - \mu  ] \psi( \vec{x} )
    \nonumber  \\
  & +  \int d^2 x_1 d^2 x_2  \psi^{\dagger}( \vec{x}_1 )\psi( \vec{x}_1 )
   V_2(x_1-x_2 ) \psi^{\dagger}( \vec{x}_2 )\psi( \vec{x}_2 )
\label{Seq0NQH2prime}
\end{align}

 It is easy to show that Eq.\ref{Seq0NQH2prime} takes the same form as Eq.\ref{Seq0NQH2}
 after formally making the same GT as in Eq.\ref{z2inv}
\begin{equation}
   \tilde{\psi} = \psi e^{-ik_0x},~~~~   \tilde{\mu}= \mu + \frac{m v^2}{2}
\label{z2invboth}
\end{equation}
 in a grand canonical ensemble. While the gauge field ( in the symmetric gauge ) changes as Eq.\ref{GTgauge}.
 The wavefunction exponential factor in Eq.\ref{GTwave} in the first quantization is equivalent to the shift of the momentum and
 the chemical potential as listed in Eq.\ref{z2invboth} in the second quantization. Indeed,
 $ - \mu \int d^2 x \psi^{\dagger}( \vec{x} ) \psi( \vec{x} ) = - \mu N $ where $ N $ is the number of particles
 in Eq.\ref{GTwave} in the canonical ensemble.

  For a canonical ensemble with a fixed number of particles $ N $ which is more convenient for the FQH ( See Eq.\ref{CFFQH} )
  to be discussed in the following  appendix:
\begin{equation}
   \tilde{\psi} =  \psi e^{-i( k_0 x - E_0 t/\hbar) }
\label{z2invboth2}
\end{equation}
 where $ k_0= -\frac{ m v}{\hbar}, E_0= \frac{ \hbar^2 k^2_0 }{ 2 m }= \frac{1}{2} m v^2 $

  For a small number of $ N $, this is the end of story.
 Unfortunately, it is still difficult to see what are the effects of such a GT in the present second quantization scheme
 in the thermodynamic limit $ N \to \infty $. This should not be too surprising, because as advocated by P. W. Anderson,
 `` More is different '', there are many emergent quantum or topological phenomena such as quantum magnetism,
 superconductivity, superfluidity, fractional Quantum Hall effects etc which can nowhere be seen in such a formal model Eq.\ref{Seq0NQH}.
 This kind of formal model looks complete, exact to some level, but not effective to see any emergent phenomena.
 So only when proceeding further and deeper to get an effective model to see the signature of emergent quantum or topological phenomenon,
 one may start to see the real effects of such a GT on such emergent phenomenon.
 In the present case, the emergent phenomenon is the fractional Quantum Hall effects which can be best seen by
 introducing the Chern-Simon gauge fields to be discussed in the following appendix.

\section{ Emergent  Galileo invariance in the Chern-Simon effective action of FQH in the bulk:
comments on HLR  theory with $ z=2 $ and Son's Dirac theory  with $ z=1 $ at $ \nu=1/2 $ filling. }

  The new quantum phases and novel quantum phase transitions discovered in the main text still fall into Ginsburg-Landau symmetry breaking picture.
  It was known that topological transitions without accompanying symmetry breaking are beyond  Ginsburg-Landau scheme.
  The Fractional Quantum Hall systems (FQH) \cite{CB,hlr,blqh} mentioned in the introduction have  Galileo invariance. It dictates that the mass appearing in the Landau level spacing  ( or cyclotron frequency $ \omega_c= eB/mc_l $  )
  is simply the bare mass $ m $ which is not renormalized by any inter-particle interaction
  $ V(x_i-x_j) $ in Eq.\ref{Seq0NQH} \cite{hlr,CB}.
  This result may be considered as the counter-part in a strongly interacting bosonic system $ \rho_s(T=0) = \rho $,
  namely, it is 100 \% superfluid at $ T=0 $ despite many atoms are kicked out of the condensate
  at zero momentum to high momentum states due to the atom-atom interaction $ V(x_i-x_j) $.

  As alerted in the last appendix, it is impossible to see how the FQH emerge from the microscopic Hamiltonian
  Eq.\ref{Seq0NQH} in its first quantization  or Eq.\ref{Seq0NQH2} in its second quantization.
  One may need to construct effective actions to study this emergent quantum phenomena.
  Both bosonic \cite{CB,blqh,BLQHwave} and fermion CS effective action \cite{fradkin,hlr} have been constructed to study the FQHE
  by performing a singular gauge transformation to attach even or odd number of fluxs to electrons.
  It is important to check if the GI has been kept in such a singular gauge transformation
  which despite was claimed to be exact formally,
  due to its singularity, could be very dangerous in practice.
  One may perform a singular gauge transformation by attaching  $ \tilde{\phi}=2 $ flux quanta to the electron operator $ \psi $:
\begin{equation}
   \psi_{CF} ( \vec{x} ) = \psi  ( \vec{x} )
   e^{ i \tilde{\phi} \int d^2 \vec{x}^{\prime}  arg( \vec{x}- \vec{x}^{\prime} ) \rho(\vec{x}^{\prime} ) }
\label{sing}
\end{equation}
 where  $ \rho(\vec{x} )= \psi^{\dagger}( \vec{x} ) \psi  ( \vec{x} )=  \psi^{\dagger}_{CF}( \vec{x} ) \psi_{CF} ( \vec{x} )
 $ is the electron or composite fermion (CF) density.

  This flux attachment to the electron operator $ \psi $ can be formally achieved by introducing the Chern-Simon (CS)
  gauge field $ a_{\mu} $ coupled to the CF. The Lagrangian corresponding
  to Eq.\ref{Seq0NQH2prime} in the canonical ensemble  becomes \cite{CB,hlr,onehalf0,blqh}
\begin{align}
 & {\cal L}_{CF}[ \psi ]  =  \psi^{\dagger}_{CF} ( -i \partial_t - \frac{e}{\hbar} A_0- a_0 ) \psi_{CF}
        \nonumber  \\
  & -\frac{1}{2m} | ( \partial_i -i \frac{e}{\hbar c_l} A_i -i a_i ) \psi_{CF} |^2
    +  \frac{i}{ 4 \pi \tilde{\phi} }  \epsilon_{\mu \nu \lambda}a_{\mu} \partial_\nu a_{\lambda}   \nonumber  \\
   & +  \int d^2 x_1 d^2 x_2  [\psi^{\dagger}( \vec{x}_1 )\psi( \vec{x}_1 )- \bar{n}]
   V_2(x_1-x_2 ) [\psi^{\dagger}( \vec{x}_2 )\psi( \vec{x}_2 )- \bar{n}]
\label{CFFQH}
\end{align}
 where $ ( A_0, A_i ) $ is the external magnetic field and  $ ( a_0, a_i ) $ are the Chern-Simon gauge field,
 $ \bar{n} $ is the average density of the electrons.
 One may also note that the former is a real EM field, so the speed of light $ c_l $ must be kept,
 while the latter has nothing to do with the light speed $ c_l $.

 In the kinetic term, it may be convenient to absorbing the external gauge field into the CS field by defining
\begin{equation}
   a_{\mu} + A_{\mu}=( a_0 + \frac{e}{\hbar} A_0, a_i + \frac{e}{\hbar c_l} A_i )
\label{comb}
\end{equation}
 which is a suitable combination of the dynamic CS field $ a_\mu $ having no intrinsic velocity with the external gauge field
 $ A_{\mu} $ having the speed of light as the intrinsic velocity. This right combination make the two different type of gauge fields transform
 the same under the GT.

 Considering the functional integral measure
 $ \int {\cal D } \psi_{CF} {\cal D } \psi^{\dagger}_{CF}= \int {\cal D } \psi {\cal D } \psi^{\dagger} $,
 then it is straightforward to show Eq.\ref{CFFQH} is GI under the space-time Eq.\ref{GTedge},
 the CF operator Eq.\ref{z2invboth2}, the external gauge field Eq.\ref{GTgauge} and the CS gauge field Eq.\ref{GTCS}.
 In fact, all the three terms in Eq.\ref{CFFQH} are separately GI under the GT.


  At the half filling case $ \nu=1/2 $, the average value of the CS field $ a_{\mu} $ and the
  external gauge field $ A_{\mu} $ just cancels each other $  \langle a_{\mu} + A_{\mu} \rangle =0  $ in Eq.\ref{comb}.
  Eq.\ref{CFFQH} becomes the starting point of the HLR theory \cite{hlr}.
  Unfortunately, Eq.\ref{CFFQH} does not respect the particle-hole (PH)  symmetry at $ \nu=1/2 $
  expected in the $ m \to 0 $ limit \cite{hlr,MS,onehalfexp,onehalf0,onehalf1,onehalf2,onehalf3,MSrev}.
  When an effective action does not get the right symmetry inherited from the microscopic Hamiltonian,
  it will problems in many physical quantities \cite{onehalf0,MS,MSrev}.
  Some ad hoc ways could be developed to fix some of these problems \cite{onehalf2,onehalf3}, but
  to fix all these problems at a fundamental level must come up with a theory which respects the fundamental PH symmetry.
  The root of such fundamental problems may just come from the singular gauge transformation Eq.\ref{sing}.
  It is non-perturbative and formally exact, but mix up all the energy scales of the original Hamiltonian Eq.\ref{Seq0NQH2prime}.
  This maybe problematic because the low energy sectors in the transformed Hamiltonian Eq.\ref{CFFQH} may
  actually correspond to some high energy levels in the original Hamiltonian Eq.\ref{Seq0NQH2prime}, so screw up the PH symmetry at the LLL.
  However, the CS action can be used to re-derive all the global topological properties of the wavefunctions, anyon excitations and the fractional statistics from  the second quantization, this is because these topological properties are robust against
  these local energy level re-distributions ( This is also the underlying mechanism of topological quantum computing which is robust against
  local perturbations ).  Unfortunately, it may not be used to calculate correctly all the local properties such as the energy gaps, correlation functions etc.  This is because these local properties are usually sensitive to these local energy level re-distributions, so they
  may only be accurately calculated in the first quantization using the wavefunctions \cite{jain,MSfield}.
  As mentioned in the conclusion, these difficulties maybe directly related to the fact that in contrasts to the Maxwell theory, it is still unknown how to regularize a CS theory on a lattice.
  In this regard, the status of FQH maybe similar to string theory which is  a first quantization theory, while
  the string field theory which is a second quantization theory is just too difficult to formulate so far.

  This fact motivated Son to develop an effective 2-component Dirac theory \cite{onehalf1,onehalf2}
  with a finite chemical potential  also coupled to the external gauge field $ A_{\mu} $
  and the CS field $ a_{\mu} $, but with no CS term for $ a_{\mu} $. This Dirac fermion theory has the exact PH symmetry
  in the zero Dirac mass limit, therefore can be used to address several difficult problems \cite{onehalf0}
  suffered in the original HLR theory \cite{hlr,onehalf3}.
  However, it breaks the GI of the microscopic Hamiltonian of interaction electrons in an external magnetic field Eq.\ref{Seq0NQH2}.
  As shown here, the HLR theory does have the GI inherited from the  microscopic Hamiltonian.
  Therefore, in terms of the PH symmetry, the Son's Dirac fermion theory ( can be called $ z=1 $ ) has a clear advantage over the HLR theory.
  However, in terms of the GI,  the HLR theory  ( can be called $ z=2 $ ) seems has an edge over the  Son's Dirac fermion theory.
  It has been quite difficult to distinguish the two competing theory in any experiments \cite{onehalfexp,onehalf3}.
  On the face value, the lack of GI in Son's Dirac theory may be its deficiency.
  In reality, it may not. As concluded in the caption of Fig.\ref{towerlevel}, the PH symmetry is a must for any sensible low energy theory
  in the LLL, but the space-time symmetry like the GI may not.

  In fact, Dirac fermion interacting with Coulomb interaction coupled the CS gauge theory has been developed to investigate
  the QH to QH transition \cite{field1,field2,Ginvye} theory. It is well known a lattice system is a common one leading to Dirac or Weyl fermions.
  Son's Dirac fermion may emerge from a low energy effective theory under the LLL projection.
  If so the LLL projection may give a completely new mechanism leading to Dirac fermion.
  In addition to the Kohn theorem on the inter-Landau level spacing dictated by the GI,
  the GI does not have any other implications in the LLL physics which has the exact PH symmetry  at $ \nu=1/2 $.
  So an effective theory at the LLL may not even need to care about the GI anymore.
  Unfortunately, despite some numerical evidences to support the nature of Dirac fermion with the $ \pi $ Berry phase, such a microscopic derivation leading to a possible Dirac fermion with a finite potential is still lacking.
  In the following appendix H, we show that the bulk gapped  CS theory has the GI, the low energy excitations along the edge
  derived from it inherits the GI from the bulk, in contrast to a recent claim made in \cite{QHedge1}.

\section{ Observing the FQH chiral edge state with $ z=1 $ in a moving Hall bar. }

As alerted at the beginning of Sec.II, the main difference between the space-time transformation such as the GT
and an internal transformation such as the gauge transformation is that in the former, the parameters in the Hamiltonian
may need also make the corresponding changes as shown in the context of VII and VIII in the context of boson Hubbard model,
while in the latter, the parameters do not change. This is the main difference between the two dramatically different transformations.
This subtly is  used in this section to show the FQH Chiral edge has emergent  GI.
Following the strategies used in the main text, we will explore its various experimental detectable
consequences from both effective actions and  the microscopic calculations.


\subsection{ Effective chiral edge description }

 In the following, we take the real time formalism, so the bulk CS term describing the
 Abelian FQH state at $ \nu=1/m $ is
\begin{equation}
 {\cal L}_{CS}  = \frac{m}{ 4 \pi  }  \epsilon^{\mu \nu \lambda}a_{\mu} \partial_\nu a_{\lambda}
\label{CSonlyreal}
\end{equation}
  where $ \nu=1/m $ is the filling factor.

  It is important to observe that the bulk CS term has no intrinsic velocity relating the space $ x $ and the time $ t $.
  Dimensional analysis $ [ a_0 \partial_{\alpha} ]= [ a_{\alpha} \partial_{0} ] $ shows that
  $ [ a_0  ]= [ c a_{\alpha} ] $ where $ c $ is any velocity.

  It can be shown that the bulk CS term is invariant under the GT with any boost velocity $ c $  Eq.\ref{GTedge} and Eq.\ref{GTCS}
  which are the real time version of Eq.\ref{GTy} and Eq.\ref{GTdual} adapted to the boost along x-direction:
\begin{align}
   t^{\prime} & =t,~~~~x^{\prime} = x+ c t,     \nonumber   \\
  \partial_t & = \partial_{t^{\prime}} + c \partial_{x^{\prime}},~~~\partial_x=\partial_{x^{\prime}}
\label{GTedge}
\end{align}
   The CS field  transforms as :
\begin{equation}
   a^{\prime}_0 = a_0 + c a_x,~~~~   a^{\prime}_x= a_x,~~~~a^{\prime}_y= a_y
\label{GTCS}
\end{equation}
  That is expected, as shown in the last appendix G, the bulk FQH systems is GI.

  When deriving the chiral edge effective action, we take a similar approach to
  study a moving SF as done in \cite{SSS}.
  Assuming the chiral edge state moves along the x-edge  with a drift velocity  $ v $
  ( See Eq.\ref{driftvelocityx} from the  microscopic evaluations ),
  then we take the co-moving $ S^{\prime} $ frame moving together with the edge mode, so that $ c=v $
  which is the intrinsic velocity of the edge mode.
  In this co-moving frame, we also  choose the temporal gauge
\begin{equation}
   a^{\prime}_0 = 0
\label{CS0}
\end{equation}
  which vanishes in the co- moving frame only.

  As shown in \cite{wen,wenedge}, this gauge imposes the constraint $ f^{\prime}_{ij}= 0 $ which can be solved by
  $  a^{\prime}_i= \partial^{\prime}_i \phi $. Substituting the solution into Eq.\ref{CSonlyreal} leads to the chiral
  edge action in the co-moving frame:
\begin{equation}
  S_{Co-}= \frac{m}{ 4 \pi} \int dx^{\prime} dt^{\prime}  \partial^{\prime}_t \phi \partial^{\prime}_x \phi
\label{cedgeprime}
\end{equation}

  Now one need to get back to the lab frame by performing the GT Eq.\ref{GTedge} and Eq.\ref{GTCS}.
  Then Eq.\ref{GTCS} becomes
\begin{equation}
   a^{\prime}_0 = a_0 + v a_x=0,~~~~   a^{\prime}_x= a_x,~~~~a^{\prime}_y= a_y
\label{GTCSv}
\end{equation}

  Then Eq.\ref{cedgeprime} transfers back to that in the lab frame;
\begin{equation}
  S_L= \frac{m}{ 4 \pi} \int dx dt ( \partial_t - v \partial_x ) \phi \partial_x \phi
\label{cedgelab}
\end{equation}
  which is the chiral edge state of the Abelian QH state at the filling $ \nu=1/m $  along the edge along x-direction.
  Then one only need to change the sign of $ v $ of the other edge.

 Eq.\ref{cedgelab} was first derived by Wen  \cite{wen,wenedge}. What is new here is we derive it by GT which relates
 Eq.\ref{cedgeprime} in the co-moving frame to Eq.\ref{cedgelab} in the lab frame.
 Our derivation establishes a deep connection between the GT and the chiral edge state and also brings additional insights
 on the connections between the bulk theory Eq.\ref{CSonlyreal} which is GI to the chiral edge
 action Eq.\ref{cedgelab} which is not GI.

\begin{figure}[tbhp]
\centering
\includegraphics[width=.9 \linewidth]{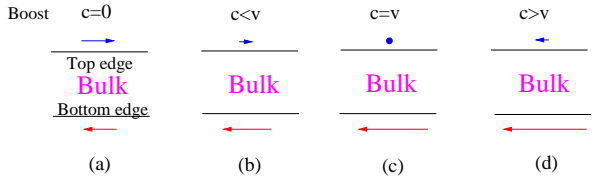}
\caption{ How are the Abelian chiral edge modes viewed in different moving frames.
 $ c $ is the longitudinal boost, $ v $ is the intrinsic drift velocity.  (a) the Lab frame  (b)
 in a frame moving slower than the drift velocity (c) co-moving frame with the edge mode. The dot means zero velocity in the top edge.
 Then one need to go beyond the linear order to see $ \omega \sim k^2_x $ with the edge dynamic exponent $ z=2 $.
 These higher order terms do break the emergent GI of the chiral edge action.
 (d)
 in a frame moving faster than the drift velocity. The top edge reverses its direction,
 then the two edge modes at the top and the bottom move along the same direction.
 The bulk stays the same FQH state. The bulk- edge correspondence is enriched in the moving frame. }
\label{wenedge}
\end{figure}

 It can be pushed further to get some new results.
 First,  one can perform the GT Eq.\ref{GTedge} with any boost velocity $ c $ along the x-direction
 on the effective action Eq.\ref{cedgelab} again to reach the action in $ S^{\prime} $ frame
 ( still drop the $ \prime $ for simplicity ):
\begin{equation}
  S_{B, l}= \frac{m}{ 4 \pi} \int dx dt ( \partial_t +c \partial_x - v \partial_x ) \phi \partial_x \phi
\label{cedgeboost}
\end{equation}
   Setting $ c=v $ recovers that in the co-moving frame Eq.\ref{cedgeprime}.
   When $ c < v $, the edge moves along the same direction as in the lab frame.
   When $ c =v $, it becomes zero in the co-moving frame.
   Then one need to go beyond the linear order to see $ \omega \sim k^2_x $ with the edge dynamic exponent $ z=2 $.
   When $ c > v $, it reverses
   the direction.  So despite the edge action Eq.\ref{cedgeboost} is GI,
   the behaviour of the chiral edge modes does depend on which frame it is observed.
   The edge velocity in the other side will be $ -(v+c ) $ whose magnitude increase monotonically ( Fig.\ref{wenedge} ).
   In fact, the longitudinal boost can be simply interpreted as a longitudinal Doppler shift term.

 Second, one can also study the edge modes when boosting along the normal direction to the x-edge \cite{chen}.
 Then one may need to study from the bulk Chern-Simon action
 Eq.\ref{CSonlyreal} and  re-derive the effective action under such a transverse boost along y-direction.
 Extending Eq.\ref{GTedge} to a transverse boost leads to:
\begin{align}
  t^{\prime} & =t,~~~~x^{\prime}  = x+ v t, y^{\prime} = y+ c t     \nonumber   \\
  \partial_t & = \partial_{t^{\prime}} + v \partial_{x^{\prime}} + c \partial_{y^{\prime}},
  ~~~\partial_x=\partial_{x^{\prime}}, \partial_y=\partial_{y^{\prime}}
\label{GTedgetran}
\end{align}
   The CS field  transforms as :
\begin{equation}
   a^{\prime}_0 = a_0 + v a_x + c a_y,~~~~   a^{\prime}_x= a_x,~~~~a^{\prime}_y= a_y
\label{GTCStran}
\end{equation}
   Eq.\ref{CS0} and Eq.\ref{cedgeprime} remain the same. While Eq.\ref{GTCSv} changes to
\begin{equation}
   a^{\prime}_0 = a_0 + v a_x+ c a_y =0,~~~~   a^{\prime}_x= a_x,~~~~a^{\prime}_y= a_y
\label{GTCSvtran}
\end{equation}
  Eq.\ref{cedgeboost} changes to
\begin{equation}
  S_{B,t}= \frac{m}{ 4 \pi} \int dx dt ( \partial_t - v \partial_x + c  \partial_y ) \phi \partial_x \phi|_{y=ct}
\label{cedgelabtran}
\end{equation}
  which can be transformed back to Eq.\ref{cedgelab} after making the substitution Eq.\ref{drifttran}.
  Unfortunately, the physical interpretation of this effective action under the transverse boot is not clear from
  itself. However, it becomes clear from the microscopic calculations as to be shown in the following.
  In fact, the transverse boost can be simply interpreted as a transverse Doppler shift term.

\subsection{ The microscopic calculations. }

  In fact, the chiral edge state moves along the x-edge  with the drift velocity given by:
\begin{equation}
   v_x=c_l \frac{E_y}{B_z}
\label{driftvelocityx}
\end{equation}
  where one notes the three perpendicular directions: the x-edge, the electric field $ E_y $  normal to the edge and the
  external magnetic field perpendicular to the sample.

  Now we will try to understand the physical meanings of the velocity $ v-c $ in the effective action  Eq.\ref{cedgeboost}.
  According to Eq.\ref{GTnonEB}, $ E^{\prime}_y=E_y - \frac{c}{c_l} B_z $, so
  the drift velocity in Eq.\ref{driftvelocityx} changes under the GT as
\begin{equation}
   v_x=c_l \frac{E_y}{B_z} \to v^{\prime}_x= v_x-c
\label{driftvelocityprime}
\end{equation}
   So $ v-c $ is nothing but the edge velocity in the $ S^{\prime} $ frame, so one can rewrite Eq.\ref{cedgeboost} as:
\begin{equation}
  S_{b,l}= \frac{m}{ 4 \pi} \int dx dt ( \partial_t - v^{\prime} \partial_x ) \phi \partial_x \phi
\label{cedgeboostprime}
\end{equation}
   which takes the identical form as Eq.\ref{cedgelab} in the lab frame when one replace the edge velocity $ v $ in the lab frame
   by that $ v^{\prime} $ in the $ S^{\prime} $ frame.
   In the co-moving frame $ v^{\prime}=0 $, then  one need to go beyond the linear order to see $ \omega \sim k^2_x $ with the edge dynamic exponent $ z=2 $. In fact,
   these higher order terms always exist, they do break the emergent GI of the chiral edge action.
   Eq.\ref{driftvelocityprime} can be compared to the shift of the chemical potential in Eq.\ref{z2inv} for the
   $ z=2 $ SF-Mott transition.

 One can also study microscopically the edge states when boosting along the normal direction to the x-edge \cite{chen}.
 According to Eq.\ref{GTnonEB}, $ E^{\prime}_x=E_x + \frac{c}{c_l} B_z, E^{\prime}_y=E_y $, so
 the drift velocity in Eq.\ref{driftvelocityx} changes under the GT as
\begin{equation}
  \vec{v}^{\prime}= c_l \frac{ (E_x \hat{x} + E_y \hat{y} ) \times \hat{z} }{ B }
  =c_l \frac{E_y}{B_z} \hat{x} - c \hat{y} = \vec{v}- \vec{c}
\label{drifttran}
\end{equation}
 where $ \vec{c}=  c \hat{y} $ stands for the transverse boost. It shows it is nothing but the GT of the edge velocity in the
 $ S^{\prime} $ frame.

  Now the physical interpretation of the effective action Eq.\ref{cedgelabtran} under the transverse boot
  is quite straightforward from the microscopic perspective: the drift velocity along the x-edge remains the same as Eq.\ref{driftvelocityx}.
  Its y-component is nothing but the boost velocity. So the two perpendicular motions just decouple from each other.

 It is constructive to compare the Abelian Chiral edge states with that of a Quantum Anomalous Hall ( QAH )
 under a transverse boost \cite{QAHinj}. The QAH can be considered
 as an integer Quantum Hall, but due to the spin-orbital coupling which breaks the GI explicitly.
 Under a transverse boost,  the chiral edge state of the Chern insulator transforms non-trivially due to the SOC:
 For example, its edge velocity becomes $ \sqrt{v^2 -c^2 } $ which indicates that the two
 perpendicular motions: the spinor edge wave propagation along the x-edge and the boost along the y- axis are coupled to each
 other through the SOC. Indeed, its 2-component spinor $ \chi_{\pm} $ also depends on the transverse boost  sensitively.
 There is a chiral edge state when $ c < v $, but not anymore when $ c > v $.
 This is because the SOC in the QAH violates the GI explicitly.
 While the spin is completely polarized in the FQH, it has a GI.

 It is also interesting to extend the chiral edge theory of the Abelian FQH to non-Abelian such as the Read-Moore state or
 Read-Rezayi state which have multi-channels\cite{chen} with different edge drift velocities.
 Gravitational anomaly \cite{tenfold} has been studied in the bulk, but not in the edge yet.
 Here, we only study the flat bulk and flat boundary, so may not need to worry about the Gravitational anomaly yet.
 In the curved space with a generic matric, it could be a different class of problems \cite{chen}.

\subsection{ Implications in the bulk through the  bulk-edge correspondence  }

 There are recent works called holography mapping or dual between topological order at dimension $ d $ and
 the symmetry at one dimensional lower at $ d-1 $, or reversely, the symmetry is a shadow of topological order of one
 dimension higher \cite{wengapless,holgraphicorder}. It could be also called topological order$_{d+1}$/symmetry$_d$ duality
 in analogy to $AdS_{d+1}/CFT_d $ duality.
 It aims to establish the connection between the classification  of symmetry or symmetry breaking
 at dimension $ d $ and the classification of topological orders at one higher dimension $ d+ 1$.
 This holographic point of view maybe taken  just as an extension of bulk-edge correspondence
 in FQH or TQPT: For example,
 the $ 1+1 $d Wen's chiral edge theory ( a $ c=1 $ CFT with Virasora algebra symmetry ) is the edge theory
 of the Abelian Chern-Simon theory which describes the topological orders of Abelian FQH states.
 Similarly, the $ 1+1 $d chiral WZW CFT  ( with $ SU(2)_k $ KM symmetry )  is the edge theory of the Non-Abelian Chern-Simon theory
 which describes the topological orders of Non-Abelian FQH such as Moore-Read state or Read-Razayi state, etc \cite{wen}.
 So in the context of FQH, it is just the CFT/TQFT correspondence.

 It is well known that the edge determines the bulk, but not the other way around. The same bulk may correspond to different
 boundaries. Namely, the bulk is the center of the boundary. So this holographic point view from projecting topological orders to symmetries at one dimensional lower,
 or the symmetry can be viewed as the image of topological orders at one higher dimension is surjective instead of injective.
 This fact may makes its application subtle.
 As shown in Fig. \ref{wenedge}, the edge undergos edge reconstruction transition in different inertial frames.
 How do the bulk topological orders change in different inertial frames ?
 It is possible that the edge reconstruction in a moving frame may just have  changed the edge without affecting the bulk qualitatively.
 Namely, as long as the bulk gap remains open, the bulk stay in the same topological phase.
 The bulk is always more robust than its corresponding edge.
 However, if the bulk gap also vanish, the system may get to some gapless topological ordered phases.
 It remains interesting to find how the bulk gap changes in a moving frame.

\subsection{ Discussions }
 A recent work \cite{QHedge1} pointed out a striking result:
 the fact that Wen's chiral edge theory is not invariant under the GT along the straight edge is an intrinsic deficiency of theory
 which need to be fixed in a refined theory.
 Our results show that there is probably no deficiencies in the chiral edge theory \cite{bert1982,wen,stone}, in contrast to the claim made in \cite{QHedge1}. The Wen's Chiral edge theory is \ref{cedgeboost} is emergent GI already.
 Only at the edge re-construction QCP in Fig.\ref{wenedge}c, one need to consider the higher order terms in the edge mode dispersion
 which break the  emergent GI explicitly.

 Another important example is that it was known that the Coulomb interaction is responsible for the fact that
 QPT from one FQH phase to another FQH phase has the dynamic exponent $ z=1 $ which seems
 in-compatible with the Galileo invariance. The $ z=1 $ may imply that
 both the quasi-particle and quasi-hole excitation gaps vanish at the QCP, but may have different gaps away from it.
 Indeed, the bare mass $ m $ can be sent to $ 0 $ in the chiral edge states and near the $ z=1 $ bulk QCP.
 It hints new emergent space-time structure near the TPT in the bulk and also in the chiral edge theory
 which may be interpreted as a TPT in the real space.

 We studied two class of problems. The first class is the  SF-Mott transition in a lattice
 where the lattice potential plays an important role.
 One must pay special attention on how the underlying lattice transforms under the GT.
 Due to the $ U(1) $ symmetry breaking in the SF side, the number is not conserved, one must use the second quantization approach to construct
 the effective actions to study the symmetry breaking QPT from the Mott to SF phase.
 As shown in Sec.VII-VIII, the first quantization wavefunction approach is not convenient to explore
 the effects of the exponential factor  $ e^{ i ( k_0 \sum_{i} x^{\prime}_i- N E_0 t^{\prime}/\hbar )  } $
 near such a QPT in different inertial frames.
 As shown in the main text, it becomes quite involved to explore how this QPT changes in a moving frame from the effective actions in the second
 quantization.

 The second class is the FQH in an external magnetic field where the external magnetic field also plays an important role.
 One must pay special attention on how the external magnetic field transforms under the GT.
 FQH has no symmetry breaking, the particle number $ N $ is conserved.
 The first quantization in terms of the wavefunction at a given number of particles $ N $ works best
 for the bulk gapped  topological phases.
 However, one may still need to construct effective actions  to describe the gapless edge excitations.
 In this case, the exponential factor  $ e^{ i ( k_0 \sum_{i} x^{\prime}_i- N E_0 t^{\prime}/\hbar )  } $  in Eq.\ref{GTwave} has a clear physical meaning:
 it just stands for the global center of mass (COM) motion of the gapped edge excitations,
 topological ordered phase: $ k_0/m $ is  nothing but the COM motion, $ E_0 $ is the shift of the ground state energy density.
 This is indeed the case by our specific calculations from Wen's chiral edge theory: its edge velocity along a straight edge just transforms as
 a velocity under the GT.
 In a sharp contrast, despite it is quite powerful to study the global bulk topological bulk properties,
 the second quantization in terms of topological quantum field theory is not successful yet
 to explore any local bulk properties. However, when one study the $ z=1 $ bulk TQP from one FQH to any FQH by tuning the
 filling or the external magnetic field, the wavefunction approach is the first quantization may not be useful anymore,
 one is forced to use effective action approach in the second quantization \cite{field1,field2}.
 Then the exponential factor    could play quite non-trivial important roles near such a TPT.

 The third class of problem which we only mention in the conclusion section is the combination of the two classes:
 in the presence of both lattice and the external magnetic field, namely, the Abelian flux problem \cite{slowgold}.
 It remains outstanding to explore the emergent space-time in this class.


\section{ The evolution of the Hamiltonian From High energy physics to Condensed matter physics }

Standard model: $ SU(3) \times SU(2) \times U(1) $, strong interaction and electro-weak interaction, gluons/quarks, $ W^{\pm}, Z_0 $ and photons.
Lorentz invariance at $ (T=0, \mu=0 ) $ and CPT symmetry.
Einstein's special relativity dictates that Different inertial frames are related by the Lorentz transformation (LT).
The standard model must be Lorentz invariance ( Covariance ) at  $ (T=0, \mu=0 ) $ and CPT symmetry.
There is No quantum phase transition at $ T=0 $. Different inertial frames are related by LT.
The ground state or vacuum stay the same in different inertial frames
The speed of light $ c_l $ is the only relevant velocity scale.
In particle/nuclear physics experiments such as Fermi lab, J-lab or RHIC, what one can tune is:
increase the scattering energy, tune the scattering angle, what is the temperature ?
One may just take it to be zero.

Of course, any $ T > 0 $ breaks the LI. Any $ \mu > 0 $ also breaks the LI.
In the RHIC experiments at  $ (T>0, \mu>0 ) $ there is a
quark Confinement and de-confinement ( first-order )  transition at a finite $ T_{dec} \sim 144 $ Mev $ \sim 10^{12} $ K
to the Quark-gluon  plasma phase in the QCD.
When the quarks start to detach from the nucleons, one must introduce a finite chemical potential $ \mu > 0 $.

In the Electro-weak interaction, there is a symmetry breaking $ SU(2) \times U(1) \to U(1)_{EM} $
which leads to the Higgs particle, the massive $ W^{\pm}, Z_0 $ and the massless photons $ A_{\mu} $.
So it also has a finite temperature phase transition $ T_{EW} \sim 10^{16} $ K above which the symmetry breaking is restored.


How does $ T $ depend on different inertial frames was found in Eq.\ref{tt0T}.
How do $ T_{dec}, T_{EW} $ depend on different inertial frames should follow the same equation.
Unfortunately, It is extremely difficult to go to a different inertial frame at a speed close to $ c_l $
to measure any such effects.

How do the particle physics evolve into condensed matter ?
As the energy gets lower and lower, the ions start to detach from electrons, then one must introduce a finite chemical potential $ \mu > 0 $.
It breaks the LI. Then the ions may start to form a lattice to break translational invariance:
Then What one can tune in a solid  is much more than in a particle accelerator:
Temperature T, magnetic field B, electric field E, Pressure ( strain ) P, electron density, n, by gate voltage.....
Then there are quantum phase transitions at $ T=0 $ tuned by these parameters, of course, also corresponding
finite temperature phase transitions. This is the essence of P. W. Anderson: More is different !

There are two kinds of condensed matter systems:

{\sl 1. Continuous systems where the underlying lattice is not important.}

 Superfluid Helium 4: one  tune the pressure P to drive a 1st order transition from  SF to a solid transition at $ T=0 $ and $ p=p_c $
 Translational symmetry breaking, the relevant velocity is the sound velocity $ v_s \ll c_l $.
 LT reduce to the Galileo transformation (GT) in the $ c_l \to \infty $ limit.

 Fractional quantum Hall systems:  tune B or n  to change the filling, there are TPT from one quantum Hall Plateau
 to another, TPT without symmetry breaking, the relevant velocity is the Fermi velocity $ v_F \ll c_l $.

 As said in Appendix F,G,H, the LT reduce to a generalized Galileo transformation upto the linear order in $ v_F/c_l \ll 1 $

{\sl 2. Lattice systems which are  much more involved }

 A lot of QPT in lattice systems at $ T=0 $: SF-Mott transition, AFM to VB transition,......
 Intrinsic velocity is determined by the Wannier functions in a  lattice structure $ v_L \ll  c_l $.

 In the $ c_l \to \infty $ limit, the LT reduce to GT.
 How does the QPT at $ T=0 $ or the thermal phase transition at $ T > 0 $ change under the GT or any low velocity limit of LT ?
 A missing component of P. W. Anderson: More is different !
 This question was addressed in the present work in the context of the SF-Mott transition.
 In contrast to the particle scattering experiments,  it is not too difficult to go to a different inertial frame  or in a moving sample at a low speed
 $ c \ll c_l $  to measure any such effects. So the moving velocity $ c $ become a new tuning parameter to drive new QPT.

\section{ Combining statistical mechanics with  Quantum mechanics,  special relativity, quantum field theory and general relativity }

 In standard undergraduate courses, we learned Quantum mechanics, statistical mechanics, special relativity, then
 then the combination of Quantum mechanics and  statistical mechanics  leads to Quantum  statistical mechanics which resolved
 many paradoxes suffered in classical statistical mechanics such as the third law of thermodynamics, the specific heat of a solid $ C_v \sim T^3 $ which leads to
 the support for a quantum theory. Then the combination of Quantum mechanics and the special relativity leads to the Dirac equation whose low velocity limits
 leads to the electron spin, SOC, etc. But so far, no combination of  statistical mechanics with the special relativity  yet.

 This difficulty can be circumvented by promoting Dirac equation to relativistic quantum field theory.
 There are two ways to do the quantization: Canonical quantization and path integral, the latter seems more powerful in the
 relativistic quantum field theory. Then when getting to a finite temperature, one can start to write a partition function of the QFT
 by going to imaginary time, this way directly and naturally incorporate the special relativity, QFT and statistical mechanics.
 It also establishes the deep connections between the imaginary time and the inverse temperature.
\begin{equation}
  \tau=it= \hbar/ k_B T
\end{equation}
  which is, in fact, compatible with the Heisenberg uncertainty relation $ \Delta E \cdot \Delta t \sim \hbar $ with $ \Delta E \sim k_B T $
  in an equilibrium system with a temperature $ T $.
 This connection seems can only be established  by the path integral instead of canonical quantization.
 This fact shows that the path integral approach is beyond the canonical quantization when combing quantum field theory with
 the statistical mechanics. It has been used to establish the LT law  of the temperature in Eq.\ref{tt0T}.

 In classical general relativity or the Unruh radiation discussed in Sec.IX-C, the temperature can be identified in the metric part containing the time which is nothing but that of a black hole solution of the Einstein equation.
 It leads to the black-hole Thermodynamics.  The Ads/CFT automatically leads to the identification of the $ T $ in the boundary CFT with that of the bulk black hole in the AdS geometry.
 The  Hawking-page transition in the bulk corresponds to the confinement to de-confinement transition in the quark-gluon plasma phase in the boundary QCD.
 The SYK models show that the quantum information scrambling rate with the Lyapunov exponent $ \lambda_L= 2 \pi/\beta $ in the SYK model in the boundary
 saturates that of the bulk black hole in the JT gravity. So Ads/CFT, with the relativistic CFT on the boundary in general, 
 the SYK which is non-relativistic in particular, establish the one to one connection between the
 quantum statistical mechanics in a temperature $ T $ on the boundary with that in the bulk in the AdS geometry.

\section{  Relation and difference between a closed system with a reservoir and an open system with a bath }

It is important to distinguish the Reservoir from the environment ( or called bath sometimes ).

In the former which is called a closed system, the system is an equilibrium and unitary system which exchange energies and 
particle numbers with the reservoir ( Fig.\ref{frames},\ref{detector} ).
One only need to introduce a chemical potential $ \mu $  to take care of the effects of the reservoir.
Its density matrix evolution is just the Heisenberg equation of motion.
The systems's energy is always conserved. This belongs to the grand-canonical ensemble in the text-book statistical mechanics.

Taking the simplest example, the free fermi gas at a temperature $ T $. How do the non-interacting fermions reach a temperature $ T $,
if they do not interact with each other ? This is due to the Reservoir which gives the temperature $ T $ in either canonical or grand canonical ensemble.
All the electrons interact with
the reservoir, but not with each other, to acquire the same temperature $ T $.
Will the reservoir mediate  the mutual interactions between or among fermions ? No, the R is so large than it will not do it.
Just like two persons standing on the earth will not interact with each other due to both kicking the ground.

While in the latter which is called an open system, it is an non-equilibrium and non-unitary system.
Its density matrix evolution has a Linblard term
There are all kinds of interactions between the system and the bath degree of freedoms:
for example, in a cavity system, the cavity photons may leak into the photon bath outside the cavity.
or qubits couple to their environments.
This process leads to the de-coherence which may be described by a Master equation.

Taking a simple example:  how a moving condensate  such as an exciton superfluid in semi-conductor bilayer emit photons 
\cite{ehbl1,ehbl2,ehbl3}.
Then the photons act as the environment.  Obviously, when the condensate is moving, the pattern of emitted photons also change.
Then one can just detect the emitted photons instead of performing the scattering experiments in Fig.\ref{detector}.
Another example is the qubit-bath interaction discussed in \cite{mark}.

  Due to the existence of the reservoir, one must distinguish the case A with the case B.  Einstein's special theory of relativity
  is about two point particles, but a many-body system is a trinity with the reservoir playing an important role.
  So our theory only applies to the case A, but not the case B.
  Unfortunately,  the experimentalists may do case B easily, much more difficult to do the case A.

  It may be instructive to do a more fundamental re-derivation on the case A where there is a relative motion between the sample and the
  reservoir in both canonical and grand-canonic ensemble. What we did here is just perform  a Lorentz or Galileo transformation on the system
  without touching the reservoir, then this relative motion indeed shows many interesting effects, especially near a quantum phase transition in the system.
  We think this maybe the most direct way to look at the effects of this relative motion between the system and the reservoir.
  The existence of a lattice in the system makes a huge deal here.
  Indeed, it would be nice to have an independent way to do it in a future publication.



\section{  Brief comments on the case B: an inertial moving frame }

  In this work, we addressed the case A in Fig.\ref{frames} and Fig.\ref{detector} where there is a relative motion between the system and the reservoir.
  Then one only need to perform the LT on the degree of freedoms in the system.
  In the case B in Fig.\ref{frames} and Fig.\ref{detector}, there is no relative motion between the sample and the reservoir, only the observer is moving,
  so one must perform LT on the degree of freedoms in both the system and the reservoir.
  One knows how to perform a LT or GT to the system, but not the combined system of the system plus the reservoir yet.
  So case (B) remains unknown, but interesting to explore.

  However if the system has a Galileo invariance such as superfluid Helium 4 in the normal phase or the FQH,
  then we may assume that there is no particle transfer between the system and  the reservoir in the case (B).
  The chemical potential change in Eq.\ref{z2inv}
  need to be offset by the corresponding change in the reservoir, so Eq.\ref{z2inv} may be replaced by ( for a moving sample, see 
  the comment in \cite{evenso} )
\begin{equation}
   \tilde{\psi} = \psi e^{-ik_0y},~~~~   \tilde{\mu}= \mu + \frac{m c^2}{2} = \mu+ \frac{c^2}{4 v^2_y} \rightarrow \mu
\label{z2invsame}
\end{equation}

 If we apply this rule to Eq.\ref{Seq0Nlattice2prime}, then it seems except the simple plane-wave factor in the boson or fermion field
 which leads to the shift of the BEC momentum or the shift of the FS center, there is no other effects in a formally GI many-body lattice system,
 namely no wedding cake rearrangements in Fig.\ref{phaseslattice} or no the FS enlargement in the Appendix E-2.
 As alerted in the last appendix,  it may be instructive to do a more fundamental derivation on the case B also.
 The experiment in this case may be more easily achieved than the case A by just moving the detector.

  There are previous result  that Eq.\ref{tt0T} should be replaced by
\begin{align}
  \frac{T}{T_{0}} =  1/\sqrt{1 - (\frac{v}{c_l})^2 }  > 1
\label{tt0T1}
\end{align}
This can not be right. When $ v/c_l \to 1^{-} $, $ T \to \infty $ approaches to infinity. This can not be right, because
the QFT breaks down at such a high temperature.

  There are also previous result that Eq.\ref{tt0T} should be
\begin{align}
  \frac{T}{T_{0}} =  1
\label{tt0T2}
\end{align}
This can not be right either, because it is GT result holding only in the $ c_l \to \infty $ limit.

In fact, Eq.\ref{tt0T} was first achieved by the late giants such as Planck, Einstein, Pauli  and Laue \cite{Planck,Einstein,Pauli,Laue}
from completely different approach: namely by incorporating the special relativity into the Thermodynamics.
Unfortunately, at late stages of his life, Einstein became confused on what is the correct way to do such an incorporation, so not sure about the final answer.
Based on various different phenomenological assumptions on the LT transformation properties of different thermodynamic quantities,
all the three results Eq.\ref{tt0T}, \ref{tt0T1}, \ref{tt0T2} can be reached\cite{thermo1}. These confusions or disputes can only be resolved from the fundamental
( or microscopic ) statistical mechanics. The thermodynamic is just an effective ( or macroscopic )  description of the fundamental
( or microscopic ) statistical mechanics.
The quantum statistical mechanics naturally leads to all the zeroth to the third law of thermodynamics.
Especially, the third law is completely due to the quantum nature of matter at $ T=0 $.

As explained in appendix J, the confusions  suffered by Einstein in combining the special relativity with the Thermodynamics can be circumvented
by putting QFT at a finite $ ( T, \mu) $. of course, these modern developments happened after the late greats,
of Planck, Einstein, Pauli  and Laue. This modern approach naturally lead to the correct answer Eq.\ref{tt0T}.
The fundamental quantity is the partition function $ {\cal Z} $ in Eq.\ref{SBL} in Sec.IX-B.
Indeed, the GKP-W relation in the Ads/CFT is also through the equality of the partition function on both sides.
The Lorentz invariance of the $ {\cal N}=4  $ SUSY Yang-Millis $ SU(N) $ gauge field on the boundary just corresponds to the diffeomorphism
on the $ AdS^5 \times S^5 $ in the bulk.
The duality between the  SYK and the JT gravity is also through the equality of the partition function on both sides in the low energy limit,
both of which are described by the Schwarzian.

It is important to point out that at $ \mu=0 $, both case A and B are the same in the relativistic QFT case.
Because the PH symmetry dictates that there are nether particles nor holes in the vacuum state despite it is full of quantum fluctuations.
However, any $ \mu >0 $ represents the existence of a reservoir which exchange particles with the system, then A and B become different.
Eq.\ref{tt0T} only holds for the case A.
For a relativistic thermal QFT at a finite $ ( T, \mu) $ such as the quark-gluon phase in RHIC, one may also assume
$ ( T, \mu) $ do not change under the LT in case B. If so, Eq. \ref{tt0T2} may hold in case B.
But it is not known if this assumption is valid. We leave this for a future work.
Note that Eq.\ref{tt0T}, \ref{tt0T1}, \ref{tt0T2} become the same when taking $ c_l=\infty $ with the difference appearing only in $ (v/c)^2 $.

It is tempting to think A and B are same for the particle-hole symmetric $ z=1 $ in Fig.\ref{phaseslattice0}.
Indeed, fixed $ n $ and fixed $ \mu $ are in general different in the original BH model Eq.1 and  Fig.\ref{phaseslattice0}:
at fixed $ \mu $, then $ n $ changes, while at fixed $ n $, then $  \mu $ changes.
Only at the PH symmetry point fixing $ \mu $ and fixing $ n $ become the same.
So there is a tantalizing possibility that case A and B are the same at this enlarged PH symmetry line.
However, this is in-correct due to that the " vacuum " is the Mott state with $ \langle n \rangle =1 $ instead of a real vacuum.
The effective action to describe the P-H excitations above the Mott phase is a "pseudo-Lorentz invariant " instead of
"Lorentz invariant ", so our results still only apply to case A.
Intuitively, if it had also been alos applied to case B, then one would have reached a paradox: 
an insulating Mott phase in the lab frame turns into a dissipationless  SF in a moving frame.


\begin{thebibliography}{99}
\bibitem{light}  In this work, because $ c $ has been used for the boost, so we use $ c_l $ to stand for the speed of light.
\bibitem{addmore} Higher order time derivative terms respecting the exact PT and C symmetry, but breaking the emergent LI such as
$ [(\partial_\tau\psi^*-ic\partial_y\psi^*)(\partial_\tau\psi-ic\partial_y\psi)]^2 $ can be added without changing
 the final results qualitatively.
\bibitem{addmore0} One may also add terms like  $ c|\partial_x^2\psi|^2 +
 d ( \partial_x^2\psi^{\dagger} )(\partial_y^2\psi) +  e  |\partial_x \psi|^4 $ respecting the exact PT and C symmetry, but breaking the emergent LI.
 Because the QCP happens along the $ \hat{y} $ direction, theses terms will not play any important roles, so can be safely dropped
 in the following.
\bibitem{Mottsideaway}  Similar approach is applied in \cite{dual1,dual2} to study how the
 Mott insulating phase  such as the charge density wave (CDW) or valence bond solid (VBS)
 at the half-filling $ 1/2 $ response when the filling $ f $ moves
 slightly away from the half-filling $ \delta f = f-1/2 $  in the dual vortex approach.
 One need to find the effective action inside the CDW or VBS phase which is a vortex condensed phase and breaks the translational symmetry,
 then apply a dual magnetic field $ \delta f = f-1/2 $ to such an effective action. Then CDW supersolid or VB supersolid naturally results
 from such an approach. In fact, this may be the stratagy  of this work: one may start from
 the  effective action which explictly or spotaneously breaks the symmetries of the original action, then
 augumented by the microscopic calculations on the correspodning lattice models.
 One may also extract the emergent space-time from such a combination.
\bibitem{putative}
If $c_1c_2 \neq 0$, then the putative stripe state breaks both $ U(1) $ and the lattice translational symmetry, so
possess both off-diagonal and diagonal symmetry, could be named a  stripe super-solid state.
Despite it is excluded here in the boosted $ z=1 $ SF-Mott QPT, it does happen in a driving SF in  the density channel induced by the roton dropping
discussed in \cite{SSS}.
\bibitem{morederivative} Originated from the terms listed in \cite{addmore0},  there are always higher derivative terms
such as  $ v_{xx} (\partial_x^2\phi)^2 + v_{yy} (\partial_y^2\phi)^2 +  v_{xy} \partial_x^2 \phi \partial_y^2\phi $, but
including these extra terms will not change the results derived in the following, so we will not write them out explicitly.
\bibitem{groundexcitation} Here, we only use the mean field ground state for the SF or BSF. In fact, as shown in
\cite{BLQHwave} in the context of exciton SF in BLQH and \cite{SOCwave} in the context of a Y-y magnetic state in an fermionic SOC system,
the mean field ground state will be modified by the quantum fluctuations in the ground state which is determined by
the excitation spectrum even at $ T=0 $.
\bibitem{spd}  For fermionic Hubbard model in materails, it maybe p- or d- even f- bands. Then one need to make the corresponding
changes to Eq.\ref{pathprime}, Eq.\ref{Hblattice} and Eq.\ref{Hblattice12}.
\bibitem{CF}  This strategy maybe contrasted to the composite fermion approach \cite{hlr,onehalf0}: the even number of flux attachments transfer
the fermions to composite fermions is a non-perturbation treatments, then the residual Chern-Simon gauge field fluctuations will be treated
perturbatively.
\bibitem{twovc} This intrinsic velocity $ v_c=t_0/t_{b1} $ is completely different from the $ v_{y,z=1} $ listed in
Eq.\ref{relationz1}. The former is solely detrmined by the Wannier functions, while the latter also depends on the bare boost
valocity $ v $ through Eq.\ref{vxvytc}. This dramatic qualtitative difference has also been stressed in  the introduction.
\bibitem{currenttrick} One should not confuse this PH current or P ( or H ) current with 
 the moving velocity of the atoms loaded on the lattice in teh moving sample.
 For example, well inside the Mott state, every lattice site captures one atom, so the atom moves together with the lattice.
 So the atom in the Mott phase also moves with the same velocity $ \vec{v} $.
 However, due to the lack of PH excitations in the $ z=1 $ case and the
 particle or hole excitations in the $ z=2 $ case, the Mott phase carries no net ( or mean field ) PH current or P ( or H ) current respectively.
 See Sec.VIII for the differences between the P/H excitations and the original bosons.
\bibitem{MottBSF}  The Mott state in the moving sample turns into the SF phase should not be too counter-intuitive.
  As shown explicitly in
  Eq.\ref{tildemuz2}  for $ z=1 $ where the $ C $ symmetry was spontaneously broken by the BSF phase and  Eq.\ref{z2inv}
  for $ z=2 $ where the $ C $ symmetry was explicitly broken by the Hamiltonian,
  the moving just increases the effective chemical potential $ \tilde{\mu} $
  and leads the Mott to the BSF transition with $ z=2 $.
  This may be contrasted to the original  Mott to the SF transition  tuned by $ \mu $ in Fig.\ref{phaseslattice0}.
  See also the conclusion  Sec.XI-1.
\bibitem{CFT12} The 2d classical KT transition is described by the $ c=1 $ CFT at the crossing point of the circle
at $ r_c=\sqrt{2} $ and $ Z_2 $ orbifold  $ r_o=1/\sqrt{2} $. Here by classical, we mean the 2d dimension
are treated as space in a Euclidean space. If one treat it as a quantum system in $ 1+ 1 $ dimension with one space and one time,
then it could behave differently under the GT and will be studied in a separate publication.
\bibitem{SOC} The Weyl SOC in cold atom systems is the coupling between a momentum shift induced by external lasers
and the pseudo-spin which is two hyperfine structures, so it is just a quantum simulated SOC which is not a relativistic effect.
However, just like in the atoms explained in the  Appendix A,
the Rashba SOC in non-centro symmetric semi-conductor systems is a relativistic effect modified by the crystal field.
If so, one may need to consider the time contraction effects upto the $ (v/c_l)^2 $ order of the LT.
\bibitem{impuritypotential} Even for the  impurities randomly distributed over $ \vec{R}_i $, the potential due to the impurities
 $ V_1( \vec{x}   )= \sum_{\vec{R}_i} v( \vec{x} -\vec{R}_i ) $  completely breaks the translational invariance,
 but is still Galileo invariant. The moving impurity case is outlined in \cite{SSS}.
 This is not suprising, becasue the translation and GT belong to two different sectors of the Galileo group $ R^3 \rtimes  SO(3) $.
 See also  Appendix A.
\bibitem{retardation} In principle, there could also be long-range regarded interaction between electrons mediated by the
 spacial component of the dynamic gauge field $ \vec{A} $ in the Coulomb gauge.  These retarded interactions do violate the Galileo invariance,
 but they are much smaller than the instantaneous Coulomb interaction mediated by the dynamic  time component $ A_0 $,
 so can be neglected up to the order of $ ( v_F/c_l)^2 $ where $ v_F $ is the intrinsic Fermi velocity of the electrons. 
 So the Galileo invariant Eq.\ref{Seq0Nlattice}
 remains a very good approximation  up to the order of $ ( v_F/c_l)^2 $.
\bibitem{vbsdetect} The same transformation from the field operator to the tight-binding limit was
used in \cite{lightatom1,lightatom2}
to derive the coupling of photon scattering to the valence bond order parameter. This fact provides the mechanism for the possible light or atom scattering detection of VBS or VBS supersolid orders.
\bibitem{sloppy}  In the following, we take the first term $ H_{b1} $ as the boost along both $ x- $ and $ y- $ bond,
but only take $ H_{b2} $ as the boost along  $ x- $ bond only, for the notional convenience.
One can just repeat the calculation when taking $ H_{b2} $ as the boost along  $ y- $ bond only
\bibitem{currentx} For example, taking the kinetic term $ K=
 -t_1 \sum_{ \langle i j \rangle }  ( b^{\dagger}_i b_j + h.c. ) - t_2 \sum_{ \langle \langle i j \rangle \rangle }  ( b^{\dagger}_i b_j + h.c. ) $
 where $ t_1 $ and $ t_2 $ are the NN and NNN hopping, then the corresponding current along the x-direction  is
 $ J_x \propto i \sum_{  i  } ( t_1  b^{\dagger}_i b_{i+ x }  +  t_2   b^{\dagger}_i  b_{i+ 2x } ) + h.c. $.
 This can be extended to any range of hopping.
\bibitem{Cdefine} For the fermionic Hubbard model with spin-1/2, there is a C- symmetry at half-filling $ n=1 $.
  The PH transformation is simply $ f_{i \sigma} \to f^{\dagger}_{i \sigma} $.
  For the bosonic Hubbard model, there is also a C- symmetry at integer-filling $ n=1 $.
  Unfortunately, it is still not known what is the PH transformation in terms of the boson operator $ b_i $.
  This fact may also be related to new space-time emerge from such a $ z=1 $ QCP, see Sec.VIII and Fig.\ref{towerlevel} for
  the detailed discussions on this important point.
\bibitem{lovelock}
  Here, these higher derivative terms contribute to the Noether currents.
  One could also incorporate higher derivative terms into the Einstein gravity called Lovelock gravity.
  The RT formula may take a different form than that  for the Einstein gravity \cite{RT}.
\bibitem{Lforce} In fact, the classical Hall effects  $ R_H= -\frac{1}{n e c_l} $ come from
the balance between the electric force $ \vec{F} = q \vec{E} $ and the Lorentz force  $ \vec{F} = q \frac{\vec{v}}{c_l}  \times \vec{B} $.
So one must keep $ E \sim v/c_l B $ to get the Hall effect in the first place. See also Eq.\ref{driftvelocityx}.
\bibitem{QCD} For an Abelian gauge field theory like QED, dropping its own dynamics leads to the static external Abelian gauge field.
For a non-Abelian gauge field theory like QCD, dropping its own dynamics leads to the static external non-Abelian gauge field
for multi-component bosons or fermions which leads to the SOC. For their lattice version, see \cite{lattice1,lattice2}.
\bibitem{evenso} In the moving sample, Even so, we did not know what is the effects of this chemical potential shift
until we perform the  microscopic calculations in Sec.VIII.
However, when reaching the effective action Eq.\ref{z2}, then  the effects of this chemical potential shift in Eq.\ref{z2inv} 
is very clear, it  leads to the QPT shown in Fig.\ref{phasesz2}.
\end{thebibliography}
\end{document}